\documentclass{ws-ijmpa}

\usepackage{multirow}
\usepackage{hyperref}

\makeatletter

\def\addcontentsline#1#2#3{%
  \addtocontents{#1}{\protect\contentsline{#2}{#3}{\thepage}{}}}
\long\def\addtocontents#1#2{%
  \protected@write\@auxout
    {\let\label\@gobble \let\index\@gobble \let\glossary\@gobble}%
    {\string\@writefile{#1}{#2}}}

\newcommand\@pnumwidth{1.55em}
\newcommand\@tocrmarg{2.55em}
\newcommand\@dotsep{4.5}
\newcommand\tableofcontents{%
    \section*{\contentsname
        \@mkboth{%
           \MakeUppercase\contentsname}{\MakeUppercase\contentsname}}%
    \@starttoc{toc}%
    }
\newcommand*\l@section[2]{%
  \ifnum \c@tocdepth >\z@
    \addpenalty\@secpenalty
    \addvspace{1.0em \@plus\p@}%
    \setlength\@tempdima{1.5em}%
    \begingroup
      \parindent \z@ \rightskip \@pnumwidth
      \parfillskip -\@pnumwidth
      \leavevmode \bfseries
      \advance\leftskip\@tempdima
      \hskip -\leftskip
      #1\nobreak\hfil \nobreak\hb@xt@\@pnumwidth{\hss #2}\par
    \endgroup
  \fi}
\newcommand*\l@subsection{\@dottedtocline{2}{1.5em}{2.3em}}
\newcommand*\l@subsubsection{\@dottedtocline{3}{3.8em}{3.2em}}
\newcommand*\l@paragraph{\@dottedtocline{4}{7.0em}{4.1em}}
\newcommand*\l@subparagraph{\@dottedtocline{5}{10em}{5em}}
\newcommand\listoffigures{%
    \section*{\listfigurename}%
      \@mkboth{\MakeUppercase\listfigurename}%
              {\MakeUppercase\listfigurename}%
    \@starttoc{lof}%
    }
\newcommand*\l@figure{\@dottedtocline{1}{1.5em}{2.3em}}
\newcommand\listoftables{%
    \section*{\listtablename}%
      \@mkboth{%
          \MakeUppercase\listtablename}%
         {\MakeUppercase\listtablename}%
    \@starttoc{lot}%
    }
\let\l@table\l@figure
\newcommand\contentsname{Contents}
\newcommand\listfigurename{List of Figures}
\newcommand\listtablename{List of Tables}

\makeatother

\newcommand{\GeV}{\ensuremath{\rm\ GeV}} 
\newcommand{\TeV}{\ensuremath{\rm\ TeV}} 
\newcommand{\pbinv}{\ensuremath{\rm\ pb^{-1}}}
\newcommand{\fbinv}{\ensuremath{\rm\ fb^{-1}}}
\newcommand{\Lint}{\ensuremath{\rm\ L_{int}}}
 
\newcommand{\stat}{\ensuremath{\rm\ (stat.) \, }}
\newcommand{\syst}{\ensuremath{\rm\ (syst.) \,  }}
\newcommand{\statsyst}{\ensuremath{\rm\ (stat.+syst.) \, }}
\newcommand{\lumi}{\ensuremath{\rm\ (lum.) \,  }}
\newcommand{\exper}{\ensuremath{\rm\ (exp.) \, }}
\newcommand{\ther}{\ensuremath{\rm\ (th.)  \, }}
    
\newcommand{\ttbar}{\ensuremath{{t\bar{t}}}}
\newcommand{\MET}{\ensuremath{\not\mathrel{E}}_{T}}
\newcommand{\cts}{\ensuremath{ \cos\theta^* }}
\newcommand{\mtpole}{\ensuremath{ {  m_t^{(pole)}   }  }}
\newcommand{\mtmsbar}{\ensuremath{ {  m_t^{(\overline{MS})}   }  }}

\newcommand{\RE}{\ensuremath{ {\rm Re}  }}
\newcommand{\BR}{\ensuremath{ {\rm BR}  }}

\begin{document}

\markboth{F.-P. Schilling}
{Top Quark Physics at the LHC: A Review of the First Two Years}

%%%%%%%%%%%%%%%%%%%%% Publisher's Area please ignore %%%%%%%%%%%%%%%
%
\catchline{}{}{}{}{}
%
%%%%%%%%%%%%%%%%%%%%%%%%%%%%%%%%%%%%%%%%%%%%%%%%%%%%%%%%%%%%%%%%%%%%

\title{TOP QUARK PHYSICS AT THE LHC: \\
A REVIEW OF THE FIRST TWO YEARS}

\author{FRANK-PETER SCHILLING}

\address{
Institute of Experimental Nuclear Physics (EKP), \\
Karlsruhe Institute of Technology (KIT), D-76128 Karlsruhe,  Germany \\
\textit{frank-peter.schilling@cern.ch}}

\maketitle

\begin{history}
\received{1 June 2012}
%\revised{Day Month Year}
\end{history}

\begin{abstract}
This review summarizes the highlights in the area of top quark physics obtained with the two general purpose detectors ATLAS and CMS during the first two years of operation of the Large Hadron Collider LHC. It covers the 2010 and 2011 data taking periods, where the LHC provided $pp$ collisions at a center-of-mass energy of $\sqrt{s}=7 \rm\ TeV$. Measurements are presented of  the total and differential top quark pair production cross section in many different channels,  the top quark mass and various other properties of the top quark and its interactions, for instance the charge asymmetry. Measurements of single top quark production and various searches for new physics involving top quarks are also discussed. The already very precise experimental data are in good agreement with the standard model.
\end{abstract}

\ccode{PACS numbers: 12.38.Qk, 13.85.-t, 14.65.Ha}

\setcounter{tocdepth}{3}
\tableofcontents

%%%%%%%%%%%%%%%%%%%%%%%%%%%%%%%%%%%%%%%%%%%%%%%%%%%%%%%%%%%%%%%%%%%%%%%%%%%%%%%

\section{Introduction}
\label{sec:intro}

Since its discovery by the CDF and D0 experiments at the $p\bar{p}$ collider Tevatron in 1995~\cite{Abe:1995hr,Abachi:1995iq}, the top quark has remained in the focus of particle physics research, for many good reasons. With a mass $m_t=173.2 \pm 0.9 \GeV$~\footnote{Throughout this review, natural units are used, such that $\hbar=c=1$.}~\cite{Lancaster:2011wr}, it is by far the heaviest of all known quarks. This has posed many questions whether the top quark may play a special role in the standard model (SM), in particular in the electroweak symmetry breaking. 

The top quark is around 40 times heaver than the $b$-quark and has a mass which is comparable to the one of a Rhenium atom (atomic number $Z=75$).
As it is also much heavier than the $W$-boson, it can decay into two-body final states, $t \rightarrow Wq$, of which the mode $t\rightarrow W b$ has almost 100 percent branching fraction. The latter observation implies that, in the SM with three generations of fermions, the Cabibbo-Kobayashi-Maskawa (CKM) matrix element $|V_{tb}|$ is close to unity. Due to the very short lifetime of the top quark, it decays before it can hadronize. This offers the unique opportunity to study the properties of a bare quark, including effects due to its spin which are transferred into respective angular correlations among its decay products.

The top quark is the charge $Q= +2/3e$, weak isospin $T_3=+1/2$ partner of the $b$-quark in the third generation weak isospin quark doublet. Its existence was postulated already many years before its experimental evidence, in particular once the $b$-quark was discovered in 1977. In the following years, indirect evidence  that the top quark must exist was obtained from limits on flavor-changing neutral-current (FCNC) decays of the $b$-quark as well as from the absence of tree-level mixing in the $B^0_d-\bar{B}_d^0$ system, which indicated that the $b$-quark must be the member of an isospin doublet. In addition, its weak isospin $T_3 = -1/2$ was determined from measurements at LEP and SLC, leading to the conclusion that the postulated partner of the $b$-quark should have $T_3 = +1/2$.

At hadron colliders, top quarks are produced either in pairs, dominantly through the strong interaction, or singly through the weak interaction. Thus, top quark production  and decay allow important tests of the features of two important forces of the SM.

The large value of $m_t$ also implies a large coupling to the Higgs boson. The top quark Yukawa coupling $y_t = m_t / v$, where $v\sim 246 \GeV$ is the vacuum expectation value, is close to unity. Because of this observation, it has often been speculated that the top quark may play a special role in the electroweak symmetry breaking, either in the context of the Higgs model, or invoking alternative mechanisms through which elementary particles acquire mass.

The top quark appears in higher order loop diagrams of the electroweak theory, which implies that $m_t$ is a crucial parameter in this theory. Precise measurements of $m_t$ provide, together with other parameters of the electroweak theory, in particular the mass of the $W$-boson $m_W$, indirect constraints on the mass of the Higgs boson. The $W$-boson mass is presently known from LEP and Tevatron data as $m_W=80.385 \pm 0.015 \GeV$~\cite{TevatronElectroweakWorkingGroup:2012gb}.

Besides its potential role in electroweak symmetry breaking, the top quark plays an important role in many scenarios for new physics beyond the SM. This constitutes one of the main motivations for the top quark physics program at the Large Hadron Collider (LHC). Several models predict the existence of new particles which decay predominantly into top quark pairs. Therefore, it is attractive to search for resonances in the top quark pair invariant mass distribution. New particles may also be produced in top quark decays, for instance a charged Higgs boson $t \rightarrow H^+b$, provided that $m_{H^+}< m_t-m_b$. In addition, precise measurements of the properties of the top quark and its interactions may reveal effects from new physics. This concerns in particular the study of differential distributions, such as the asymmetry in the rapidity distributions of top quark and anti-quark, but also the search for FCNC in top quark  decays and for the production of same-sign top quark pairs. 

Experimental signatures involving top quark production often constitute an important background to various new physics scenarios, such as super-symmetry (SUSY). In particular, searches at the LHC for super-symmetric partners of quarks and gluons, namely squarks and gluinos, have not yet shown any hint at their existence. This gives rise to speculations that the first two generations of squarks may be very heavy, while the top and bottom squarks could still be comparatively light (see, e.g., Ref.~\refcite{Papucci:2011wy} for a recent study). The decays of stop quarks generally yield signatures containing top or bottom quarks and missing transverse energy due to the lightest SUSY particle, the LSP, which may be hard to distinguish from SM top quark decays. The increased interest in SUSY scenarios with a light third squark generation thus constitutes a further important motivation to study top quark production in detail at the LHC.

Since its discovery at the Tevatron, and using data collected up to the collider's shut-down in 2011 ($\sqrt{s}$ up to $1.96 \TeV$), the properties of the top quark and its interactions is being studied in detail~\cite{Wimpenny:1996dz,Campagnari1997,Bhat1998,Tollefson:1999wt,Chakraborty:2003iw,Wagner2005,Quadt:2007jk,Kehoe2008,Pleier:2008ig,Incandela:2009pf,Wicke:2010cg,Deliot:2010ey,Lannon:2012fp}. These studies are now being continued at the LHC, which is in operation since the end of 2009.  The LHC collides protons with protons, rather than with anti-protons as was the case at Tevatron, and does that at a higher center-of-mass energy ($\sqrt{s}=7 \TeV$ until the end of 2011). The large collision energy, as well as the large instantaneous luminosity of the LHC, result in top quarks being produced in very large quantities: In 2011, around 800\,000 top quark pairs were produced per LHC experiment. This allows to study many aspects of the top quark very precisely, and to search for new physics involving the top quark in a comprehensive way.

This review summarizes the experimental results on top quark physics which have been obtained at the LHC during the first two years of its operation (2010 and 2011), based on data samples corresponding to an integrated luminosity of around $\Lint \sim 5 \fbinv$ for each of the two general purpose experiments ATLAS and CMS.

The article is structured as follows. In section~\ref{sec:theory}, the foundations and theoretical status of top quark physics at the LHC are briefly discussed. In section~\ref{sec:exp}, a short summary about the LHC accelerator and the ATLAS and CMS detectors is given, supplemented by a more detailed discussion of various relevant experimental aspects, in particular the reconstruction of the various \textit{physics objects} such as leptons, jets and missing transverse energy, and the Monte-Carlo simulation. The following section~\ref{sec:xsection} presents the measurements of the total and differential cross sections of top quark pair production. Direct and indirect measurements of the top quark mass, as well as of the top quark - anti-quark mass difference are discussed in section~\ref{sec:mass}, while measurements of other properties of top quarks and their interactions are reported in section~\ref{sec:prop}. Section~\ref{sec:singletop} summarizes measurements of single top quark production. Section~\ref{sec:topbsm} highlights the status of searches for new physics in the top quark sector, including searches for a fourth generation of quarks. Finally, an outlook on future perspectives is given in section~\ref{sec:summary}.

%%%%%%%%%%%%%%%%%%%%%%%%%%%%%%%%%%%%%%%%%%%%%%%%%%%%%%%%%%%%%%%%%%%%%%%%%%%%%%%

\section{Theory Overview}
\label{sec:theory}

In the following, the foundations and current theory status of top quark physics at the LHC will be briefly summarized. While the information given here can only be brief\footnote{As this review is focusing on experimental results, no attempt is made to fully reflect all developments in the areas of the theory and phenomenology of top quarks, and references given do not aim at being complete.}, more detailed overviews can be found in Refs.~\refcite{Bernreuther:2008ju,Moch:2008qy,Kidonakis:2011ca,Weinzierl:2012tc}.

\subsection{Top quark pair production}
\label{sec:xstheo}

\begin{figure}[t]
\centering
\includegraphics[width=0.99\linewidth]{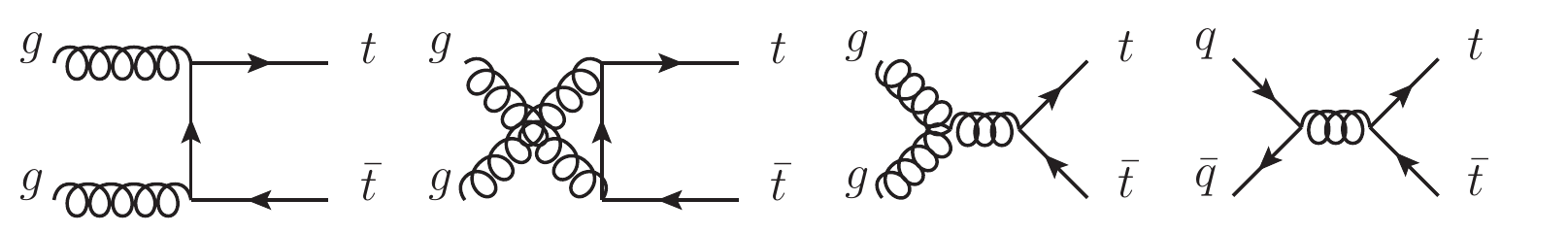}
\caption{Feynman diagrams for $\ttbar$ production at leading order QCD.}
\label{fig:feynttbar}
\end{figure}

In the SM, the dominant production mechanism for top quark pair production is mediated by the strong interaction. Thus, since the top quark mass $m_t$ is much larger than $\Lambda_{QCD}$, $\ttbar$ production at LHC can be successfully described in terms of quantum chromodynamics (QCD), the theory of the strong interaction. In the QCD-improved parton model, the inclusive production cross section of the process $pp \rightarrow \ttbar $, which depends on $m_t$ and the center-of-mass energy squared of the collider $s=4E_{\rm beam}^2$, can be expressed using the factorization theorem as a convolution of parton distribution functions (PDF) and a partonic cross section $\hat{\sigma}$ (at leading twist, i.e., up to terms suppressed by powers of $s$):
\begin{equation}
\sigma_{pp\rightarrow \ttbar}(s,m_t) = \sum\limits_{i,j=q,\bar{q},g} \int d x_i d x_j f_{i}(x_i,\mu_f^2) f_{j}(x_j,\mu_f^2) \cdot \hat{\sigma}_{ij \rightarrow \ttbar}(\hat{s},m_t,\mu_f,\mu_r,\alpha_s) \ .
\label{eq:ttsigma}
\end{equation}
The sum runs over all quarks and gluons contributing, $x_{i}$ are the parton momentum fractions with respect to the proton momenta, $f_{i}(x_{i},\mu_f^2)$ are the proton PDF, $\mu_{f(r)}$ are the factorization and renormalization scales, $\alpha_s$ is the strong coupling and $\hat{s}\sim x_i x_j s$ is the partonic center-of-mass energy.  At leading order (LO) QCD, i.e., $\mathcal{O}(\alpha_s^2)$, the processes $gg\rightarrow\ttbar$ and $q\bar{q}\rightarrow\ttbar$ contribute (Fig.~\ref{fig:feynttbar}), while at next-to-leading order (NLO) there are also partonic sub-processes with $gq$ ($g\bar{q}$) in the initial state.
The dependence on $\mu_r$ of the partonic cross section, computed in truncated perturbation theory, arises in particular from the definition of the renormalized coupling $\alpha_s$, which is usually done in the $\overline{MS}$-scheme. 
The top mass $m_t$ in Eq.~\ref{eq:ttsigma} may also depend on $\mu_r$, depending on the choice of renormalization scheme (see section~\ref{sec:masstheo}).
The dependence of the partonic cross section and the PDF on $\mu_f$ arises from absorbing uncanceled collinear initial state singularities into the PDF.
The renormalization and factorization scales are typically set to a hard scale of the process, and one often identifies $\mu = \mu_r = \mu_f$. In the case of the total cross section, one usually sets $\mu=m_t$. However, in the case of differential cross sections, other choices are more appropriate since additional hard scales may be given, for example by the transverse momentum of a jet $p_{T,jet}$, or by the top-quark pair invariant mass $M_\ttbar$. The variation of the cross section when the scale is changed within a certain range (often $\mu/2 - 2\mu$) is commonly used as an estimate of the uncertainty due to missing higher orders (so-called scale uncertainty), even though the range of variation chosen is in principle arbitrary.

\begin{figure}[t]
\centering
\includegraphics[width=0.44\textwidth]{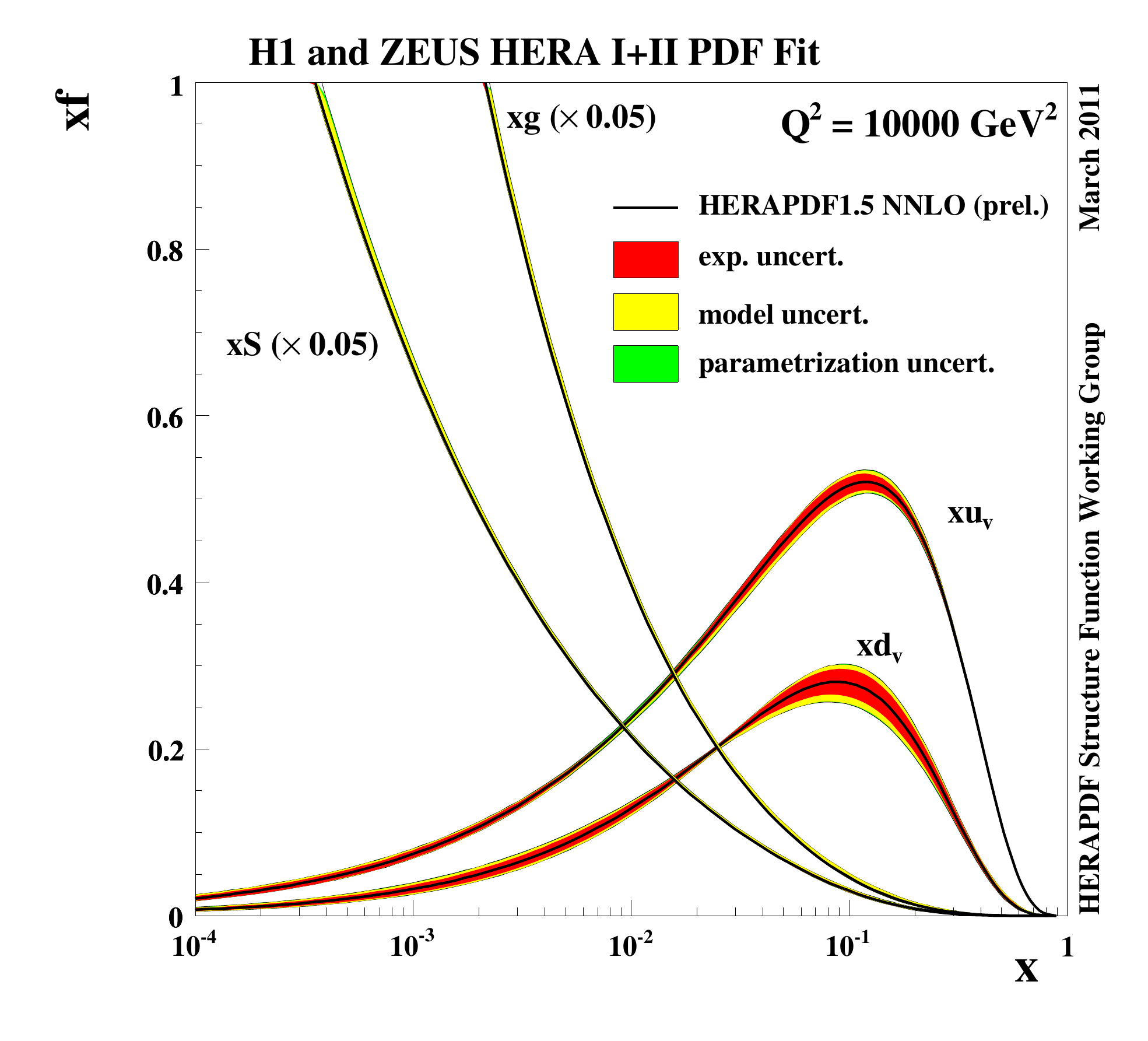}
\includegraphics[width=0.55\textwidth]{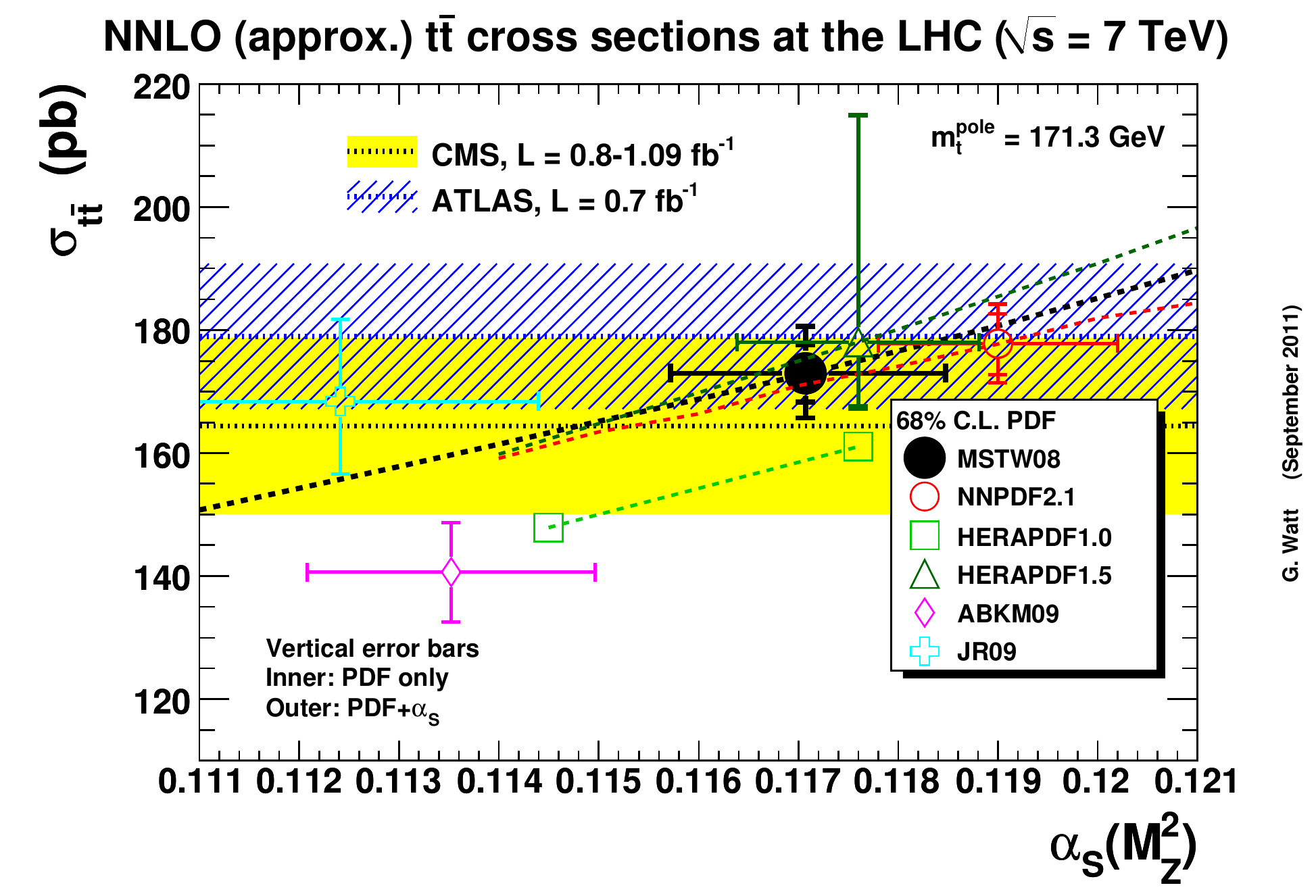}
\caption{Left: The HERAPDF1.5NNLO PDF~\protect\cite{herapdf15}, evaluated at a scale $\mu^2 = 10\,000 \GeV^2$. The behavior for $\mu^2 = m_t^2 \sim 30\,000 \GeV^2$ is qualitatively similar.
Right: Approximate NNLO $\ttbar$ total cross section as function of $\alpha_s(M_Z^2)$ for $m_t=171.3 \GeV$~\protect\cite{Watt:2012fj}
, evaluated for various choices of PDF sets using the HATHOR~\protect\cite{Aliev:2010zk} program, and compared with measurements from ATLAS~\protect\cite{ATLAS-CONF-2011-121} and CMS~\protect\cite{CMS-PAS-TOP-11-003}. }
\label{fig:pdf}
\end{figure}

The universal (i.e., process independent) proton PDF $f_i(x_i,\mu_f^2)$ are determined by several groups (see,  e.g., Refs.~\refcite{Martin:2009iq,:2009wt,Lai:2010vv,Ball:2011uy,Alekhin:2012ig}) from global fits to experimental data on deep-inelastic scattering (especially from the high precision HERA $ep$ data), but also on jet and heavy quark production at hadron colliders. At the LHC with $\sqrt{s}= 7 \ (14) \TeV$, around $80 \ (90)\%$ of the total cross section is due to the $gg$ induced contribution, while the remainder is mostly due to the $q\bar{q}$ initial state. This is due to the large gluon density in the proton at small $x$ (Fig.~\ref{fig:pdf}) and the fact that the typical value of $x=2 m_t / \sqrt{s}$ (due to the minimal energy needed of $\hat{s}> 4 m_t^2$ and setting $x_1=x_2$ ) is 0.05 (0.025) at  $\sqrt{s}= 7 \ (14) \TeV$. At the Tevatron $p\bar{p}$ collider, the situation was reversed with the $q\bar{q}$ contribution dominating and the PDF being probed at much larger $x$ values (around $x=0.2$). At both colliders, the $gq$ ($g\bar{q}$) contributions contribute only at the percent level, since they are suppressed by an additional factor $\alpha_s$.

The NLO QCD $\mathcal{O}(\alpha_s^3)$ corrections to the total $\ttbar$ cross section are known since more than 20 years~\cite{Nason:1987xz,Nason:1989zy,Beenakker:1988bq}.  The mixed QCD-weak corrections of $\mathcal{O}(\alpha_s^2\alpha)$ were computed in Refs.~\refcite{Beenakker:1993yr,Bernreuther:2005is,Bernreuther:2006vg,Bernreuther:2008md,Kuhn:2005it,Kuhn:2006vh}
and the mixed QCD-QED corrections were determined  in Ref.~\refcite{Hollik:2007sw}.
There are also calculations of $\ttbar$ production at NLO QCD which include the top quark decays and the correlations between production and decay, such as the information on the top quark spin, which have been performed in the narrow-width approximation for top quarks produced on-shell~\cite{Bernreuther:2001rq,Bernreuther:2004jv,Melnikov:2009dn,Bernreuther:2010ny,Campbell:2012uf}, as well as for the more general case of $WWb\bar{b}$ production including off-shell contributions and non-factorisable corrections~\cite{Denner:2010jp,Bevilacqua:2010qb}.
While the impact of the finite top quark width is expected to be small for the total cross section, it may be more important in the case of differential distributions~\cite{Maestre:2012vp}.
The NLO QCD differential cross sections for the production of $\ttbar$ in association with one ($\ttbar$+jet)~\cite{Dittmaier:2007wz,Dittmaier:2008uj,Melnikov:2010iu,Melnikov:2011qx} and two ($\ttbar$+2 jets)~\cite{Bevilacqua:2010ve,Bevilacqua:2011aa} extra jets are available. The sub-process of the latter consisting of $\ttbar + b \bar{b}$ production was calculated at NLO QCD as well, see, e.g., Ref.~\refcite{Bredenstein:2010rs} and references therein.
At the moment, work is ongoing towards the full next-to-next-to-leading order (NNLO) QCD calculation of the inclusive $\ttbar$ cross section, which is not yet available for all production channels. 
Very recently, the first complete NNLO calculation of the $q\bar{q} \rightarrow \ttbar$ sub-process has become available~\cite{Baernreuther:2012ws}.

The problem of logarithmic enhancements near threshold regions of the fixed order cross section due to the emission of soft gluons can be handled with techniques called \textit{soft gluon resummation}, first discussed in Ref.~\refcite{Laenen:1991af} and subsequently refined with increasing precision and using various techniques. In addition, resummation methods are also used to deal with Coulomb singularities.
Today, these techniques have been refined up to the next-to-next-to-leading logarithmic (NNLL) accuracy~\cite{Langenfeld:2009wd,Czakon:2009zw,Cacciari:2011hy,Kidonakis:2010dk,Ahrens:2010zv,Beneke:2011mq}. 
The existing ingredients of the fixed order calculation can be combined with corrections using soft gluon resummation to obtain an approximate result for the $\ttbar$ total cross section at NNLO QCD, which has been done by various groups~\cite{Kidonakis:2010dk,Aliev:2010zk,Cacciari:2011hy,Czakon:2011xx,Beneke:2011mq,Ahrens:2010zv,Ahrens:2011px,Moch:2012mk}. 
For a review and comparison of the various approaches, see Ref.~\refcite{Kidonakis:2011ca}. 

\begin{table}[t]
\tbl{Approximate NNLO QCD calculations of the $\ttbar$ total cross section in $pp$ collisions at $\sqrt{s}=7 \TeV$. The first uncertainty corresponds to the scale uncertainty, while the second uncertainty is the PDF uncertainty. For reference, also the NLO value is given.}
{
\begin{tabular}{lll}
\toprule
Calculation & $\sigma_\ttbar \ [{\rm pb}]$ & Comment \\
\colrule
NLO QCD~\protect\cite{Ahrens:2011px} & $160 \ ^{+20}_{-21} \ ^{+8}_{-9}$ & --  \\
\colrule
Kidonakis~\protect\cite{Kidonakis:2010dk} &  $163 \ ^{+7}_{-5} \ ^{+9}_{-9}$ & 1PI resummation \\
Aliev et al.~\protect\cite{Aliev:2010zk} & $164 \ ^{+5}_{-9} \ ^{+9}_{-9}$ & threshold resummation \\
Ahrens et al.~\protect\cite{Ahrens:2010zv} & $155 \ ^{+8}_{-9} \ ^{+8}_{-9}$ & SCET (1PI and PIM) resummation  \\ %of differential \\ & & cross section in momentum space \\
Beneke et al.~\protect\cite{Beneke:2011mq} & $163 \ ^{+7}_{-8} \ ^{+15}_{-14}$ & threshold resummation in momentum space \\
Cacciari et al.~\protect\cite{Cacciari:2011hy} & $159 \ ^{+12}_{-14} \ ^{+4}_{-4}$ & threshold resummation in Mellin space \\
Moch et al.~\protect\cite{Moch:2012mk} &  $175 \ ^{+10}_{-13} \ ^{+5}_{-5}$ & threshold resummation including high energy limit \\
\botrule
\end{tabular}
\label{tab:xs} 
}
\end{table}

Table~\ref{tab:xs} shows a compilation of the most recent results for the total cross section for $\ttbar$ production at LHC at $\sqrt{s}=7 \TeV$. The calculations differ in the way the resummation was performed, for instance the choice of the soft limit considered. Examples are threshold resummation ($\sqrt{1-4m_t^2/\hat{s}} \rightarrow 0$), as well as  the pair-invariant-mass (PIM, $1-M_\ttbar^2/\hat{s} \rightarrow 0$) and single-particle inclusive (1PI, $\hat{s}+\hat{t}_1+\hat{u}_1 \rightarrow 0$)\footnote{The variables $\hat{t}(\hat{u})_1 = (p_{1(2)}-p_3)^2-m_t^2$ are modified partonic Mandelstam variables of the process $q/g_i(p_1)+q/g_j(p_2) \rightarrow t(p_3)+\bar{t}(p_4)$.} kinematic schemes. While resummation is often performed in Mellin space, Ref.~\refcite{Ahrens:2010zv} uses a soft collinear effective theory (SCET) approach. Also, Coulomb singularities may be resummed or not.
The very recent calculation from Ref.~\refcite{Moch:2012mk} includes new insights from the high energy limit $s\gg m_t^2$.
The calculations were evaluated using the MSTW2008NNLO~\cite{Martin:2009iq} PDF set (except the NLO result which used the NLO PDF of the same set). The PDF uncertainty corresponds to the 90\% confidence level (CL) envelope of the used PDF set except for Refs.~\refcite{Cacciari:2011hy,Moch:2012mk} which used the 68\% envelope. Ref.~\refcite{Beneke:2011mq} also includes an $\alpha_s$ uncertainty.
The renormalization and factorization scales were set to $\mu_r=\mu_f=m_t$ with $m_t=173 \GeV$, except for Ref.~\refcite{Ahrens:2010zv} which used $m_t=173.1 \GeV$ and Refs.~\refcite{Cacciari:2011hy,Beneke:2011mq} which used $m_t=173.3 \GeV$. The scale uncertainty was evaluated by changing both $\mu_r$ and $\mu_f$ within factors of two up and down. The uncertainty of the result of Ref.~\refcite{Ahrens:2010zv} includes an additional contribution evaluated from the difference between the two kinematics schemes (PIM and 1PI). The results of the various approximate NNLO calculations are in good agreement with each other. In general, the scale uncertainty is significantly reduced with respect to the NLO result. However, the results from Refs.~\refcite{Cacciari:2011hy} and~\refcite{Moch:2012mk} quote a significantly larger scale uncertainty compared with the other approximate NNLO results. 
Finally, the authors of  Ref.~\refcite{Brodsky:2012sz} propose a scheme in which $\mu_r$ would be fixed, however it is not clear how such an approach relates to actual properties of $\mu_r$-dependent terms in the higher order perturbative expansion.

Fig.~\ref{fig:pdf} (right)~\cite{Watt:2012fj} shows the approximate NNLO QCD $\ttbar$ cross section as a function of $\alpha_s(M_Z^2)$ ($M_Z$ is the mass of the $Z$-boson) at the LHC for $\sqrt{s}=7 \TeV$, evaluated using the HATHOR program~\cite{Aliev:2010zk}, setting $m_t=171.3 \GeV$ and using various PDF sets. The vertical error bars of the theory predictions correspond  to the combined PDF and $\alpha_s$ uncertainty. It is evident that the cross section depends strongly on both $\alpha_s$ and the PDF, in particular on the gluon density. The predictions are compared with ATLAS~\cite{ATLAS-CONF-2011-121} and CMS~\cite{CMS-PAS-TOP-11-003} measurements which will be discussed in more detail in section~\ref{sec:xsection}. In general, reasonable agreement is observed, even though the prediction based on the ABKM09~\cite{Alekhin:2009ni} PDF is slightly disfavored. However, note that $\sigma_{\ttbar}$ also depends strongly on the used value for $m_t$. This dependence can be used to indirectly constrain $m_t$ using the measured cross section~\cite{Langenfeld:2009wd,Beneke:2011mq}, which will be further discussed in section~\ref{sec:massfromxsec}.

The measurements  of the top quark pair production cross section $\sigma_\ttbar$ at the LHC will be discussed in section~\ref{sec:xsection}.
In $\ttbar$ production, the top quarks are unpolarized, but their spins are correlated. See section~\ref{sec:spincorr} for more details.
At NLO QCD, an asymmetry occurs in the rapidity difference distribution of top quarks and anti-quarks in $\ttbar$ production at hadron colliders. This will be further discussed in section~\ref{sec:ac}.

%%%%%%%%%%%%%%%%%%%%%%%%%%%%%%%%%%%%%%%%%%%%%%%%%%%%%%%%%%%%%%%%%%%%%%%%%%%%%%%

\subsection{Production of single top quarks}
\label{sec:stoptheo}

\begin{figure}[t]
\centering
\includegraphics[width=0.99\linewidth]{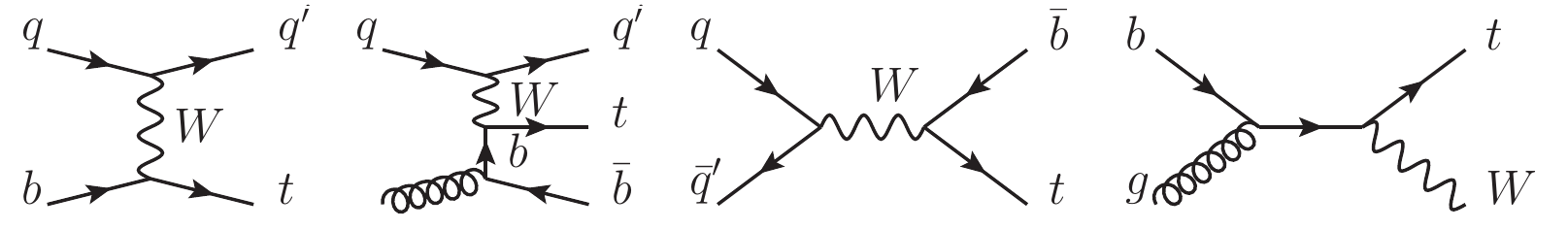}
\caption{Example Feynman diagrams for single top quark production at LO QCD. From left to right: $t$-channel production as flavor excitation and as $W$-gluon fusion; $s$-channel production; $tW$-channel production.}
\label{fig:feynsingletop}
\end{figure}

Single top quarks can be produced through the electroweak interaction and the $Wtb$ vertex (almost exclusively, since $|V_{tb}|\gg|V_{td}|,|V_{ts}|$). Three different production modes exist (Fig.~\ref{fig:feynsingletop}):
\begin{itemize}
\item In the $t$-channel mode, a space-like $W$-boson scatters off a $b$-quark, which is either considered through the $b$-quark PDF in the proton (flavor excitation, massless scheme) or produced via gluon splitting $g\rightarrow b\bar{b}$ ($W$-gluon fusion, massive scheme);
\item In the $s$-channel mode, a time-like $W$-boson is produced from two quarks belonging to an isospin doublet, e.g., $u\bar{d}$, and subsequently decays into $t\bar{b}$;
\item In the $tW$-channel mode, which is also called associated production, the top quark is produced in association with a close-to real $W$-boson.
\end{itemize}

Single top quark production is interesting for various reasons. Its proof of existence provides a relevant test of the standard model. It is important to measure all three production modes, since they are sensitive to the $Wtb$ vertex in different ways. Non-standard couplings would indicate the presence of contributions from new physics. Also, single top quark production allows to directly measure the CKM matrix element $|V_{tb}|$ (assuming $R=1$, see Eq.~\ref{eq:r} in section~\ref{sec:topdecay}), without making an assumption on the number of generations, and to verify the unitarity of the CKM matrix. Deviations from the SM expectation could indicate a possible fourth generation. The flavor excitation production allows constraints on the $b$-quark PDF, though this requires significant statistics. Standard model single top quark production constitutes a background in several new physics scenarios, for instance production of a new $W'$ or a charged Higgs $H^+$ boson ($tW$- or $s$-channel signature). New physics involving FCNC would lead to single top production via $ug\rightarrow t$ ($t$-channel signature).

\begin{table}[t]
\tbl{Approximate NNLO QCD calculations of the total cross sections for single top quark and anti-quark production in $pp$ collisions at $\sqrt{s}=7 \TeV$. The first uncertainty corresponds to the scale uncertainty, while the second one (where given)  is the PDF uncertainty.
}
{
\begin{tabular}{lll}
\toprule
Production mode (author) & $\sigma_t \ [{\rm pb}]$ & $\sigma_{\bar{t}} \ [{\rm pb}]$ \\
\colrule
$t$-channel (Kidonakis~\protect\cite{Kidonakis:2011wy}) & $41.7 ^{+1.6}_{-0.2} \pm 0.8$ & $22.5 \pm 0.5 ^{+0.7}_{-0.9}$ \\
$s$-channel (Kidonakis~\protect\cite{Kidonakis:2010tc}) & $3.17 \pm 0.06 ^{+0.13}_{-0.10}$ & $1.42 \pm 0.01 ^{+0.06}_{-0.07}$ \\
$s$-channel (Zhu et al.~\protect\cite{Zhu:2010mr}) & $2.81 ^{+0.16}_{-0.10}$ & $1.60 ^{+0.08}_{-0.05}$ \\

$tW$-channel (Kidonakis~\protect\cite{Kidonakis:2010ux}) & $7.8 \pm 0.2 ^{+0.5}_{-0.6}$ & $7.8 \pm 0.2 ^{+0.5}_{-0.6}$ \\
\botrule
\end{tabular}
\label{tab:singletop} 
}
\end{table}

The cross section for single top quark production in hadron collisions was calculated at NLO QCD ten years ago~\cite{Harris:2002md,Zhu:2002uj}. The most recent calculations also incorporate NNLL resummation~\cite{Kidonakis:2011wy,Kidonakis:2010tc,Zhu:2010mr,Kidonakis:2010ux}.  The numerical results are summarized in Table~\ref{tab:singletop}. The calculations from Refs.~\refcite{Kidonakis:2011wy,Kidonakis:2010tc,Kidonakis:2010ux} used $m_t = 173 \GeV$, and the first uncertainty corresponds to the scale uncertainty, while the second uncertainty is the PDF uncertainty using the MSTW2008NNLO PDF set at 90\% CL. The $s$-channel calculation from Ref.~\refcite{Zhu:2010mr} used $m_t=173.2 \GeV$ and quotes only the scale uncertainty.

The $t$-channel production mode is dominant at the LHC, followed by the $tW$-channel associated production. Note that, except for the $tW$-channel, the cross sections for top quark production are larger than that for top anti-quark production, due to the proton PDF.
At $\sqrt{s}=7 \TeV$, the ratio of the $t$-channel single top quark and anti-quark production cross section to the $\ttbar$ cross section, $\sigma_{t+\bar{t}}(t-ch.) / \sigma_{\ttbar}$ is around $40\%$.

Some higher order diagrams of $t$- and $s$-channel production have the same initial and final states. However, there is no interference at NLO QCD between the two production modes since the $t\bar{b}$ pair produced in the $t$-channel forms a color octet, while in the $s$-channel it forms a color singlet. On the other hand, there is interference at higher orders between $tW$-channel associated production and top quark pair production. This leads to the problem of unambiguously defining the two, which will be discussed further in section~\ref{sec:mcsingletop}.

Another feature of electroweak single top quark production is that the top quark is produced left-handed and in its rest frame, it is 100\% polarized along the direction of the light quark. Since top quarks decay before they can hadronize, the polarization information is transferred to their decay products. In particular, the distribution of the polar angle of the lepton from the $t\rightarrow Wb \rightarrow l \nu_l b$ decay and the spin axis, approximated by the direction of the light quark jet in the top quark rest frame, is expected to be proportional to $(1+\cts)$~\cite{Stelzer:1998ni}.

The current status of the measurements of single top quark production at LHC will be discussed in section~\ref{sec:singletop}.
See section~\ref{sec:fcnc} for results on FCNC anomalous single top quark production, and section~\ref{sec:bsmother} for $W'$ and charged Higgs boson searches.

%%%%%%%%%%%%%%%%%%%%%%%%%%%%%%%%%%%%%%%%%%%%%%%%%%%%%%%%%%%%%%%%%%%%%%%%%%%%%%%

\subsection{Top quark decays}
\label{sec:topdecay}

The top quark decays almost exclusively as $t\rightarrow Wb$. Since $|V_{tb}|\gg|V_{td}|,|V_{ts}|$, the decays $t\rightarrow W(d,s)$ are strongly suppressed and will be further discussed only at the end of this section.
Neglecting the decays $t\rightarrow W(d,s)$, the total width of the top quark in the SM at NLO QCD is~\cite{Jezabek:1988iv}
\begin{equation}
\Gamma_t = \frac{G_F m_t^3}{8 \pi \sqrt{2}} |V_{tb}|^2 
\left ( 1-\frac{m_W^2}{m_t^2}\right)^2  
\left ( 1+2\frac{m_W^2}{m_t^2}\right)
\left[1-\frac{2\alpha_s}{3\pi} \left(\frac{2\pi^2}{3}- \frac{5}{2}\right) \right ] \ ,
\end{equation}
where $G_F$ is the Fermi constant. Using $m_t=172.5 \GeV$ yields $\Gamma_t = 1.33 \GeV$. The large width of the top quark corresponds to a very short lifetime $\tau_t = 1 / \Gamma_t \sim 5 \cdot 10^{-25} \rm\ s$. A D0 measurement~\cite{Abazov:2012vd}, using the $t$-channel single top cross section and the branching fraction $\BR(t\rightarrow Wb)$ measurements, yielded $\Gamma_t = 2.00^{+0.47}_{-0.43}$ and $\tau_t = 3.29 ^{+0.90}_{-0.63} \cdot 10^{-25} \rm\ s$, in good agreement with the SM.

The lifetime of the top quark is one order of magnitude smaller than the typical formation time of hadrons $\tau \sim 1 {\rm\ fm} /c \sim 3 \cdot 10^{-24} \rm\ s $, which means that top quarks decay before they can hadronize. It also means that no {\em toponium} $\ttbar$ bound state can exist.
As a consequence, the spin information of the top quark is transferred to its decay products. The polarization of the $W$-boson from the top quark decay can be either longitudinal or transverse (neglecting the mass of the $b$-quark and at LO QCD with negative helicity only, due to angular momentum conservation), according to the V-A structure of the $Wtb$ vertex. This will be further discussed in the context of the measurements of the $W$-boson polarization in top decays and the top pair spin correlation, see sections~\ref{sec:spincorr} and~\ref{sec:whelicity}.

The decays of top quark pairs can be classified according to the decay of the $W$-bosons: 
\begin{itemize}

\item Di-lepton channel: both $W$-bosons decay into lepton (electron, muon, tau) and neutrino, $\ttbar \rightarrow W^+b W^-\bar{b} \rightarrow \bar{l} \nu_l b l' \bar{\nu}_{l'} \bar{b}$. The branching fraction is $\BR({\rm di-lepton};e,\mu,\tau)=10.3\% \ (\sim 9/81)$. Considering only decays to electrons or muons, it is $\BR({\rm di-lepton};e,\mu)=6.45\% \ (\sim 4/81)$.

\item Lepton+jets channel: one $W$-boson decays into lepton and neutrino, the other one into a quark - anti-quark pair, $\ttbar \rightarrow W^+b W^-\bar{b} \rightarrow q \bar{q}' b l \bar{\nu}_{l} \bar{b} + \bar{l} \nu_l b q \bar{q}' \bar{b}$. The branching fraction is $\BR({\rm lepton+jets},e,\mu,\tau)=43.5\% \ (\sim 36/81)$, approximately evenly split for $W$-boson decays into electron, muon and tau.

\item Hadronic channel: both $W$-bosons decay into a quark - anti-quark pair, $\ttbar \rightarrow W^+b W^-\bar{b} \rightarrow q \bar{q}' b q'' \bar{q}''' \bar{b}$. The branching fraction is $\BR({\rm hadronic})=46.2\% \ (\sim 36/81)$.

\end{itemize}

The decays of the top quark into a $W$-boson and a quark of another isospin doublet $t\rightarrow W s$ ($\BR\sim 0.2\%$), $t\rightarrow W d$ ($\BR\sim 0.005\%$)~\cite{Nakamura:2010zzi} are strongly suppressed in the SM. Unitarity of the CKM matrix implies that the denominator of the quantity $R$, defined as
\begin{equation}
R=\frac{\BR(t\rightarrow W b) }{ \BR(t\rightarrow W q) } = 
\frac{|V_{tb}|^2}{ |V_{tb}|^2+|V_{ts}|^2+|V_{td}|^2  } \ .
\label{eq:r}
\end{equation}
is unity. This means that, assuming unitarity of the CKM matrix, a measurement of $R$ provides a constraint on $|V_{tb}|$. A deviation of $R$ from one may hint at a fourth generation. Measurements of $R$ were performed by means of counting the $b$-jet multiplicity in a sample enriched in top pair events at Tevatron~\cite{Acosta:2005hr,Abazov:2011zk}. The most recent result from D0~\cite{Abazov:2011zk} yielded $R=0.90\pm 0.04$, somewhat lower than the SM value. Assuming unitarity of the CKM matrix, this corresponds to $|V_{tb}|=0.90-0.99$. A measurement of $R$ performed at LHC will be discussed in section~\ref{sec:r}.

In the SM, the FCNC decays $t\rightarrow Z q$ and $t\rightarrow \gamma q$ have negligible branching ratios, and a deviation from this would be a sign of new physics. At the Tevatron, the most precise upper limit on BR($t\rightarrow Z q$) has been set as 3.2\%~\cite{Abazov:2011qf}. Searches for FCNC in top quark decays and production at LHC will be discussed in section~\ref{sec:fcnc}.

%%%%%%%%%%%%%%%%%%%%%%%%%%%%%%%%%%%%%%%%%%%%%%%%%%%%%%%%%%%%%%%%%%%%%%%%%%%%%%%

\subsection{Mass and charge of the top quark}
\label{sec:masstheo}

The mass of the top quark, which is a fundamental parameter of the SM, is currently known as $m_t=173.2 \pm 0.9 \GeV$ from direct measurements at the Tevatron~\cite{Lancaster:2011wr}, which corresponds to a precision of 0.5\%. Thus, the top quark is not only the heaviest of the known quarks, but also the one for which its mass has been measured with the highest precision.

Indirect constraints on $m_t$ can be obtained from precision measurements of the parameters of the electroweak theory. The mass of the $W$-boson can be expressed as a function of the electro-magnetic coupling $\alpha(M_Z^2)$, the Fermi constant $G_F$ and the electroweak mixing angle $\theta_W$, where $\sin^2\theta_W = 1- m_W^2 / m_Z^2$, as $m_W^2 = \frac{\pi\alpha(M_Z^2) / \sqrt{2} G_F}{\sin^2\theta_W \cdot (1-\delta r)}$. The term $\delta r$ contains contributions from higher order electroweak loop diagrams involving the top quark which depend quadratically on $m_t$. The most recent indirect constraint on $m_t$ based on electroweak precision measurements is $m_t = 179.7 ^{+11.7}_{-8.7} \GeV$~\cite{:2010vi}, in good agreement with the direct measurements. Since the $\delta r$ term also contains a contribution from loop diagrams involving the Higgs boson depending logarithmically on its mass  $m_H$, it is possible to obtain indirect constraints on $m_H$ from global electroweak fits including direct measurements of $m_t$. The most recent determinations fits indicate that the Higgs boson should be light~\cite{:2010vi,Baak:2011ze}, consistent with the direct searches at Tevatron~\cite{CDFandD0:2011aa} and LHC~\cite{Chatrchyan:2012tx,:2012si}.

Renormalization relates the bare mass which appears in the Lagrange density with a renormalized mass, depending on the choice of renormalization scheme. Popular choices for the top quark mass are the on-shell scheme, where the pole mass $\mtpole$ corresponds to the location of pole of the propagator, and the $\overline{MS}$-scheme, where the $\overline{MS}$-mass $\mtmsbar(\mu_r)$ is scale dependent. Perturbation theory allows to convert between the two. Non-perturbative effects introduce an ambiguity in the definition of the pole mass of $\mathcal{O}(\Lambda_{QCD})$ (see, e.g., Ref.~\refcite{Skands:2007zg}). However, the direct measurements of $m_t$ usually identify the measured value with $\mtpole$.
The current status of direct and indirect top quark mass measurements, including the mass difference between top quarks and anti-quarks, will be discussed in section~\ref{sec:mass}.

The charge of the top quark, which is $+2/3e$ in the SM, can in principle be inferred from its decay products, which is however diluted in the case of quarks hadronizing to jets. It also requires matching the decay products to the top (anti-) quark. Measurements at the Tevatron have excluded the hypothesis that the top quark has an exotic charge of $-4/3e$ at the 95\% CL~\cite{Abazov:2006vd,Aaltonen:2010js}. Another possibility to get a handle on the top quark charge is the measurement of the cross section of the production of top pairs in association with a photon, which is sensitive to the top quark charge~\cite{Baur:2001si}.
See section~\ref{sec:charge} for constraints from LHC data on the charge of the top quark, and section~\ref{sec:ttgamma} for first results on the production of $\ttbar+\gamma$.

\section{Experimental Methods}
\label{sec:exp}

In this section, a brief overview of the LHC accelerator and the two general-purpose experiments ATLAS and CMS is given, including brief discussions on how the luminosity is measured and how events are selected at the trigger level. This is followed by a description of the reconstruction of the main physics objects relevant for top quark physics, as well as of the Monte-Carlo simulation of signal and background processes.

\subsection{Accelerator}

The Large Hadron Collider LHC is a proton-proton (alternatively lead ions are used) collider which has a circumference of 27~km and is located around 100~m underground at CERN in Geneva, Switzerland.
After providing initial $pp$ collisions at low center-of-mass energies in 2009 and early 2010,
the LHC started operating at an energy of 3.5~TeV per beam on March 30, 2010. The total integrated luminosity accumulated in 2010 at $\sqrt{s}= 7 \TeV$, with up to 368 bunches per beam, corresponds to $47 \pbinv$. In 2011, the maximum specific luminosity was increased from the 2010 value of $2 \cdot 10^{32} \rm\ cm^{-2}s^{-1}$ to $3.6 \cdot 10^{33} \rm\ cm^{-2}s^{-1}$ by raising the number of bunches per beam to 1380 and optimizing other beam parameters, resulting in around $5.7 \fbinv$ of integrated luminosity delivered to the experiments ATLAS and CMS. The large luminosity resulted in an average number of additional interactions per bunch crossing (so-called pile-up) of around 6 (11) for the first (second) part of the data taking, with a maximum of more than 20. This posed a significant challenge for the experiments with respect to triggering and reconstructing interesting physics processes.
Since April 2012, the LHC operates at the increased energy of 4 TeV per beam ($\sqrt{s}= 8 \TeV$).

\subsection{Detectors}

At the moment, top quark physics at the LHC is mostly studied using the two large multi-purpose detectors ATLAS~\cite{atlasdet} and CMS~\cite{cmsdet}. 
Both ATLAS and CMS are optimized in order to ensure that the complete final state of a collision event can be reconstructed. In particular, all physics objects relevant for top quark physics, such as leptons, jets (in particular also those originating from $b$-quarks) and missing transverse energy, the latter usually originating from undetected neutrinos, can be measured with high efficiency and very good energy and spatial resolution.

Both experiments use a right-handed coordinate system, with the origin at the nominal interaction point, the $x$-axis pointing to the center of the LHC, the $y$-axis pointing upwards (perpendicular to the LHC plane), and the $z$-axis along the counterclockwise-beam direction. The polar angle, $\theta$, is measured from the positive $z$-axis and the azimuthal angle, $\phi$, is measured in the $x$-$y$ plane. 
The pseudo-rapidity is defined as $\eta=- \ln \tan(\theta/2)$.
In the case of massive objects, the rapidity $y=\frac{1}{2}\ln[(E+p_z)/(E-p_z)]$ is also often used.
The transverse momentum $p_T$, the transverse energy $E_T$, and the missing transverse energy $\MET$ are defined in the x-y plane, unless stated otherwise. 
The angular distance $\Delta R$ in the pseudo-rapidity - azimuthal space is defined as $\Delta R=\sqrt{\Delta\eta^2+\Delta\phi^2}$.

\subsubsection{ATLAS detector}

The ATLAS detector~\cite{atlasdet} is 25~m high, 44~m long and has a total weight of 7\,000 tonnes.
The Inner Detector comprises a silicon (Si) pixel and micro-strip detector (SCT), as well as a straw-tube tracking detector (TRT). It is embedded within a 2~T thin superconducting solenoid magnet of 2.5~m diameter, and it provides tracking and vertexing capabilities within $|\eta|<2.5$, as well as electron identification within $|\eta|<2.0$.
The pixel detector consists of three barrel layers complemented by three end-cap disks on each side. The SCT consists of four double layers in the barrel region and nine end-cap disks per side. The TRT comprises many layers of gaseous straw tube elements interleaved with transition radiation material. It yields on average 36 hits per charged particle trajectory (or \textit{track}) and provides electron identification capability.
The Inner Detector provides a transverse impact parameter resolution of $\sim 35 \ (\sim 10) \rm\ \mu m$ for pions with $p_T=5 \ (100) \rm\ GeV$,
and a transverse momentum resolution of about 4\% for 100~GeV muons.

The high granularity liquid-argon (LAr) electromagnetic sampling calorimeter is located outside the solenoid magnet. It provides excellent energy and position resolution within $|\eta|<3.2$ using a barrel and two end-cap calorimeters. The thickness of a barrel (end-cap) module is in the range $22-30 \ (24-38)$ radiation lengths $X_0$.
In the region $|\eta|<1.8$, the calorimeter is complemented by pre-samplers, an instrumented argon layer which provides a measurement of the energy lost in front of the electromagnetic calorimeters.
An iron-scintillator tile calorimeter with radial depth of $7.4$ nuclear interaction lengths $\lambda$ provides hadronic energy measurements in the range $|\eta|<1.7$. 
The forward region is instrumented with the hadronic end-cap ($1.5<|\eta|<3.2$) and forward (FCAL, $3.1<|\eta|<4.9$) LAr calorimeters.
The energy resolution for photons of $p_T=100 \GeV$ is better than $1.5\%$,
while jet energies are measured with a resolution $\Delta E/E \sim 65\% / \sqrt{E\,[{\rm GeV}]} \oplus 2.5\% \oplus 5/E  \,\%$.

The calorimeter system is surrounded by a muon spectrometer, which comprises separate trigger and high-precision tracking chambers, measuring the momenta of muons in a magnetic field generated by superconducting
air-core toroids. The precision chamber system
covers the region $|\eta|<2.7$ with three layers of monitored
drift tubes, complemented by cathode strip chambers in the
forward region. The muon
trigger system covers the range $|\eta|<2.4$ with resistive plate
chambers in the barrel, and thin gap chambers in the end-cap
regions.
The muon transverse momentum resolution varies between 3 and 12\%, for $p_{\rm T}$ values between 10 and 1\,000 GeV.

A three-level trigger system is used to select the events of interest for subsequent  analysis. The level-1 trigger is implemented
in hardware and uses a subset of the detector information
to reduce the rate to a design value of at most
75~kHz, followed by two software based trigger levels, level-2 and the event filter, which reduce the event rate to about 200~Hz.

\subsubsection{CMS detector}

The CMS apparatus~\cite{cmsdet} has an overall length of 22~m, a diameter of 15~m, and weighs 14\,000 tonnes. The central feature of CMS is a superconducting solenoid, of 6~m diameter, providing a field of 3.8~T. Within the field volume are the tracking detector and both electromagnetic and hadronic calorimeter. Muons are measured in gas-ionization detectors embedded in the steel return yoke. In addition to the barrel and end-cap detectors, CMS has extensive forward calorimetry. 

The inner tracking detector measures charged particles in the range $|\eta| < 2.5$. It consists of 1440 Si pixel and 15\,148 Si strip modules. In order to deal with high charged particle multiplicities, the strip detector employs 10 layers in the barrel region. In addition, three layers of pixel detectors are placed close to the interaction region to improve the measurement of the impact parameter of charged-particle tracks, as well as the position of secondary vertices.
The tracking detector provides an impact parameter resolution of $\sim$\,15~$\mu$m and a $p_{\rm T}$ resolution of about 1.5\,\% for 100~GeV particles.

The electromagnetic calorimeter (ECAL) consists of nearly 76\,000 lead tungstate ($\rm PbWO_4$) crystals, which provide coverage in $\vert \eta \vert< 1.479 $ in the barrel (EB) and $1.479 <\vert \eta \vert < 3.0$ in two end-caps  (EE). The crystals are $25.8 \ (24.7) \, X_0$ thick in the barrel (end-caps).
A pre-shower detector consisting of two planes of silicon sensors interleaved with $3 X_0$ of lead is located in front of the EE and used for $\pi^0$ rejection.
The ECAL has an energy resolution of better than 0.5\,\% for unconverted photons with $E_T>100 \rm\ GeV$. 

The ECAL is surrounded by a brass/scintillator sampling hadron calorimeter (HCAL) with coverage up to $|\eta|<3.0$. 
In the $(\eta,\phi)$ plane, and for $\vert \eta \vert< 1.48$, the HCAL cells map on to $5 \times 5$ ECAL crystals to form calorimeter towers projecting radially outwards from the nominal interaction point. 
Within each tower, the energy deposits in ECAL and HCAL cells are summed to define the calorimeter tower energies.
The combined ECAL+HCAL jet energy resolution is $\Delta E/E \sim 100\,\% / \sqrt{E\,[GeV]} \oplus 5\,\%$.
This central calorimetry is complemented by a tail-catcher (HO) in the barrel region, ensuring that hadronic showers are sampled with nearly $11 \lambda$.
Coverage up to $|\eta|=5.0$ is provided by an iron/quartz-fiber forward calorimeter, and even higher forward coverage is obtained with additional dedicated calorimeters and with the TOTEM tracking detectors.

The return field of the magnet is large enough
to saturate 1.5 m of iron, allowing muons to be measured in four stations, to ensure robustness and geometric coverage. Each station consists of several layers of aluminum drift tubes in the barrel and cathode strip chambers in the end-caps, complemented by resistive plate chambers. Muons are measured in the range $|\eta|< 2.4$.
Matching the muons to the tracks measured in the silicon tracking detector results in a $p_T$ resolution between $1\%$ and $5\%$, for $p_{\rm T}$ values up to 1~TeV.

The first level  of the CMS trigger system, composed of custom hardware processors, uses information from the calorimeters and muon detectors to select the most interesting events. The High Level Trigger processor farm further decreases the event rate from around 100~kHz to around 300~Hz, before data storage.

\subsection{Luminosity measurement}

The instantaneous luminosity is measured in both experiments by counting the rate of collisions for a given number of bunch crossings using various techniques and detectors, which are either dedicated devices situated close to the beam and sensitive to minimum bias collisions, or alternatively elements of the main detector with acceptance for particles produced at small angles (typically forward calorimetry, or alternatively track and vertex counting using the tracking detectors). These counting methods provide information about relative changes in the luminosity. They need to take the pile-up properly into account in order to provide meaningful results.
The absolute luminosity calibration is obtained from measured accelerator parameters such as bunch intensities and beam profiles, where the latter are measured using dedicated \textit{van der Meer}~\cite{vanderMeer:1968zz} or beam-separation scans which are performed on a regular basis.

In the case of ATLAS, the systematic uncertainty on the measured luminosity currently corresponds to $3.4 \, (3.7) \, \%$ for the 2010 (2011) data taking period~\cite{Aad:2011dr,ATLAS-CONF-2011-011,ATLAS-CONF-2011-116}.  For CMS, the corresponding uncertainties are $4.0 \, (4.5) \, \%$ for the two years, respectively~\cite{CMS-DP-2011-002,CMS-PAS-EWK-10-004,CMS-PAS-EWK-11-001}.
Most recently, the CMS uncertainty for 2011 data was reduced to $2.2\%$~\cite{cmspas-smp-12-008} using a pixel cluster counting method, and benefiting from an improved knowledge of the LHC beam parameters. It also yielded an upward shift of the overall normalization of $6.6\%$ (from $\Lint = 4.7$ to $5.0 \fbinv$). These new numbers may or may not have been used in the results presented here, depending on the time at which they were made available.

%%%%%%%%%%%%%%%%%%%%%%%%%%%%%%%%%%%%%%%%%%%%%%%%%%%%%%%%%%%%%%%%%%%%%%%%%%%%%%%

\subsection{On-line selection of top quark events}

To select samples of top quark events at the trigger level, requirements which are matched to the chosen decay mode were used. 
In the $\ttbar$ lepton+jets channel, events were typically selected using inclusive lepton ($e$ or $\mu$) triggers with a $p_T$ threshold below the  requirement in the off-line analysis. To control the rate, lepton identification (ID) or isolation were used already at the trigger level. These requirements had to be successively tightened during 2011 data taking to cope with increasing luminosity and pile-up.
In the case of CMS, triggers in which one or more jets are required in addition to the lepton candidate were also used, in particular for the high luminosity phase of 2011 data taking.
Di-lepton ($ee$, $\mu\mu$, $e\mu$) final states were triggered requiring two lepton candidates. Due to the lower rates compared with the single lepton triggers, less stringent requirements on $p_T$, ID or isolation were needed. In order to recover efficiency losses from the double lepton requirement, single lepton triggers were also often used. Di-lepton final states where one lepton is a hadronically decaying tau and the other one is either an electron or a muon ($e\tau$, $\mu\tau$), used a single electron or muon trigger.
The hadronic and tau+jets channels used multi-jet triggers, typically requiring at least four or five jets and, in the case of CMS, supplemented with a requirement that at least one of them had to be identified as $b$-quark jet.
Single top quark events generally used similar triggers than the ones used in the $\ttbar$ di-lepton and lepton+jets channel. CMS also employed a dedicated trigger with a $b$-quark jet requirement to select $t$-channel single top events.

In addition to the signal triggers discussed above, so-called control triggers were often used. They employed looser selection requirements (for example lower $p_T$ thresholds or no lepton isolation or ID requirements). Their purpose was to collect samples which could be used to measure trigger or off-line selection efficiencies, or to estimate backgrounds.

Lepton trigger, reconstruction, ID and isolation efficiencies were typically measured from the data themselves, using the \textit{tag-and-probe} technique. Here, a pure sample of $Z$-bosons decaying into a pair of charged leptons was selected, but with tight requirements on only one of the leptons (the \textit{tag}). The efficiencies could then be determined on the second lepton, the \textit{probe}.

%%%%%%%%%%%%%%%%%%%%%%%%%%%%%%%%%%%%%%%%%%%%%%%%%%%%%%%%%%%%%%%%%%%%%%%%%%%%%%%

\subsection{Reconstruction of physics objects}

In the following, details are given on the reconstruction of physics objects relevant for top quark analyses, in particular charged leptons ($e$, $\mu$, $\tau$), hadronic jets including $b$-quark jet identification, and missing transverse energy $\MET$.
More information can also be found in Refs.\refcite{Bayatian:2006zz,Ball:2007zza} for CMS and Refs.~\refcite{atlastdr,Collaboration:2010knc} for ATLAS, respectively.

In CMS, a global event reconstruction called Particle Flow (PF) is used to reconstruct and identify each single particle with an optimized combination of all sub-detector information. Details on this approach can be found in Ref.~\refcite{CMS-PAS-PFT-10-002}.

Dedicated techniques can be  applied to cope with the presence of a large number of pile-up interactions overlaid with the primary event.  Charged hadrons identified as not originating from the primary vertex can be removed from the reconstruction. After this correction, only the neutral component of pileup remains aside from the true jet constituents, which can be removed by, e.g., applying a residual area-based correction described in Refs.~\refcite{Cacciari:2008gn,Cacciari:2007fd}. 

\subsubsection{Electrons}
\label{sec:electrons}

Electron candidates are reconstructed in both ATLAS~\cite{Aad:2011mk} and CMS~\cite{CMS-PAS-EGM-10-004,:2011nx} from energy deposits (clusters) in the electromagnetic calorimeter, which are associated with charged particle tracks reconstructed in the tracking detectors.
The reconstruction algorithms take into account the possibility of significant energy loss through bremsstrahlung as electrons traverse detector material.

Several ID criteria are applied to the electron candidates, based on, e.g., the shower shape in the calorimeter, the track-cluster spatial matching or the value of $E/p$, the ratio of the calorimeter energy and the track momentum.
In the case of ATLAS, the TRT is used due to its capability to separate electrons from charged hadrons.
Converted photons which are mis-identified as electrons are removed by dedicated requirements on the amount of lost pixel hits and using dedicated algorithms to find a partner conversion track of opposite charge.
The ID criteria are optimized in simulation for inclusive $W \rightarrow e \nu_e$ events, maximizing the rejection of non-prompt electrons or fakes, while maintaining high efficiency for electrons from the decay of $W/Z$-bosons.
While a simple approach in which just a few ID variables employed already results in a good performance, it can be improved further by optimizing separately various categories of electron candidates, defined by their location in the detector, $E_T$ range or the amount of bremsstrahlung emitted.
Sets of requirements on the values of the electron ID variables (also called working points) are defined, such that typical efficiencies are in the range $70-90\%$, depending on the background rejection power.
For the energy range relevant for the measurements presented here, the electron energy scale is known to better than 0.5\% (0.3\%) in ATLAS (CMS), while the energy resolution is 2\% (3\%) or better.
The probability to misidentify the electron charge is below 1\%.

Isolation requirements are applied on electron candidates to select leptons from the decay of $W/Z$-bosons, rather than leptons produced within jets. Energy sums are formed within a cone of typical size $\Delta R = 0.3$ around the lepton direction, excluding the lepton itself. These sums are based on the transverse momenta of charged particle tracks and/or on the energy of calorimeter objects.  Often one defines a requirement based on a relative isolation variable, where these energy sums are divided by the lepton (transverse) momentum. This allows for more energy in the isolation cone for high $p_T$ leptons.

\subsubsection{Muons}

Muons are reconstructed in ATLAS~\cite{Aad:2011dm} and CMS~\cite{CMS-PAS-MUO-10-002} combining the information from the muon chambers and the inner tracking detectors. 
First, tracks are reconstructed individually in the inner tracking detector as well as in the muon system. They are subsequently merged to form muon candidates, starting either from the muon (outside-in matching) or inner tracking detector (inside-out matching) track, and a combined track fit is performed. In most cases relevant here, the muons considered are found by both approaches. 
Muon tracks are required to be of good quality and to be consistent with the reconstructed primary vertex. In particular, the track associated with the muon candidate is required to have a minimum number of hits in the inner tracking detectors, and to have a high-quality global fit including a minimum number of hits in the muon detector. This selection greatly reduces the contribution from decays-in-flight, at the price of a small efficiency loss.
Isolation requirements are placed to select muons from $W/Z$-boson decays in a similar way as for electrons (see section~\ref{sec:electrons}).

The muon $p_T$ resolution is 3-4\% (1-2\%) in the case of ATLAS (CMS) in the kinematic range relevant here, dominated by the inner tracking detector resolution. 
The reconstruction efficiency is typically above 95\% for both experiments, while both the muon fake-rate as well as the charge mis-identification probability are at very small levels.

\subsubsection{Taus}
\label{sec:tauid}

Tau leptons have a mean lifetime of $2.9 \cdot 10^{-13}$ s (corresponding to a path length of $87 \ \mu\rm m$), such that most of them decay before leaving the beam pipe. They decay either leptonically to an electron or muon and two neutrinos ($\BR \sim 35\%$) or into hadrons ($\BR \sim 65\%$). The hadronic final states consist mostly of one (\textit{1-prong}, $\BR \sim 50\%$) or three (\textit{3-prong}, $\BR \sim 15\%$) charged pions, a neutrino and potentially additional neutral pions. Since the cross section for the production of hadronic jets is much larger than the one for tau leptons, the challenge lies in rejecting the jets faking tau candidates. Because of this, hadronic jets (discussed in section~\ref{sec:jets}) form the starting point of the tau reconstruction.

In the ATLAS tau reconstruction algorithm~\cite{ATLAS-CONF-2011-152,Aad:2011fu}, tracks passing certain quality criteria are associated with a calorimeter jet using a narrow cone criterion of $\Delta R=0.2$, where the leading track has to fulfill $p_T>4 \GeV$. Candidates overlapping with an electron or muon are removed, and electrons misidentified as taus are removed using a Boosted Decision Tree (BDT). The jet forming the tau candidate also must not be identified as originating from a $b$-quark. Another BDT is used to identify actual hadronic tau decays and reject fakes. The reconstruction efficiency is between 40 and 85\%, depending on the working point.

Also the CMS PF tau reconstruction algorithm~\cite{tau-11-001} considers the different hadronic decay modes of the tau individually.
Using a PF jet as a seed, first the $\pi^0$ components of the tau are reconstructed, which are then combined with charged hadrons to reconstruct the tau decay mode, and to calculate the tau four-momentum and isolation quantities.
A special feature of the algorithm is that it takes into account the broadening in $\phi$ of calorimeter signatures due to early showering photons.
The tau reconstruction efficiency of this algorithm is estimated to be approximately 37\% for the analyses discussed here.

In both experiments, the tau charge is taken as the sum of the charges of the charged hadron tracks (prongs) in the signal cone, and isolation criteria are applied in a similar way as for electrons and muons.

\subsubsection{Jets}
\label{sec:jets}

Hadronic jets are clustered with the infrared and collinear safe anti-kT algorithm~\cite{Cacciari:2008gp} as implemented using the FASTJET~\cite{Cacciari:2005hq,Cacciari:2011ma} package, using a size parameter R=0.5 (0.4) in CMS (ATLAS). The jet momentum is determined as the vectorial sum of all particle momenta in the jet.
ATLAS uses \textit{topological clusters} that group together neighboring calorimeter cells with energy deposits above certain thresholds. Additional correction factors are applied to correct the jet energy to the hadronic scale.
CMS is using reconstructed PF objects as input to the jet clustering, and the reconstructed momenta are found in the simulation to be within $5-10\%$ of the true momentum over the whole $p_T$ spectrum and detector acceptance. 

Jet energy corrections are derived from the simulation, and are confirmed with in situ measurements, for example using the energy balance in di-jet or photon+jet events. 
In the case of CMS, the jet four momenta are corrected for non-linearities in $\eta$ and $p_T$ with simulated data, with a residual $\eta$-dependent correction added to correct for the difference in simulated and true responses.
The jet energy scale (JES) uncertainty is around 2.6\% for jets with $p_T=30 \rm\ GeV$~\cite{Chatrchyan:2011ds} and generally smaller than 2\% for $p_T>45 \rm\ GeV$.
For ATLAS, the JES uncertainty is less than 2.5\% for central jets with $|\eta|<0.8$ and $p_T=60-800 \GeV$~\cite{Aad:2011he}, while it increases
at lower or higher $p_T$ (up to $4.5\%$) or in the forward region.
The JES depends on the flavor of the jet, due to differences in fragmentation between light and heavy quarks, and gluons. This is included as systematic uncertainty in the JES uncertainty. For jets originating from $b$-quarks, the response is observed to be between the one for light quarks and gluons in the case of CMS, and therefore the flavor uncertainty also applies to the case of b-jets~\cite{Chatrchyan:2011ds}. In the case of ATLAS, an additional uncertainty is assigned based on Monte-Carlo studies and validation with data~\cite{Aad:2011he}.
The JES can also be determined using $\ttbar$ events exploiting the known $W$-boson mass and the correlation with the top quark mass, see Refs.~\refcite{cmsnote-2006-025,cmspas-top-07-004,atlastdr} for previous simulation based studies. Simultaneous determinations of $m_t$ and JES will also be discussed in section~\ref{sec:mass}.

The jet energy resolution amounts typically to $13\% \ (15\%)$ at 50 GeV and $8\% \ (11\%)$ at 100 GeV in the case of CMS (ATLAS)~\cite{Chatrchyan:2011ds,ATLAS-CONF-2010-054}. Using PF jets, CMS was able to significantly improve the resolution compared with using calorimeter jets.

\subsubsection{Missing transverse energy}

Besides mis-measurements, a genuine source of missing transverse energy $\MET$ in an event is due to neutrinos being produced in the interaction. In general, $\MET$ is calculated as the negative of the vector sum of the $p_T$ of all final-state particles.
Cleaning algorithms are applied in order to remove anomalous signals in the calorimeters (e.g., due to detector noise), as well as beam-halo muons produced upstream from the detector.
$\MET$ can be mis-measured for a variety of reasons,  including the nonlinearity of the calorimeter response, neutrinos from semi-leptonic decays, minimum energy thresholds, as well as inefficient detector regions.

In ATLAS, $\MET$ is reconstructed~\cite{Aad:2011re} from the energy depositions in the calorimeter associated to the objects used in the analysis. The same reconstruction and identification algorithms as for the analysis objects are used to identify electrons and jets. The corresponding topological clusters in the calorimeters are then included in the calculation of $\MET$ at the energy scale of the associated object. Muon momenta are corrected for additional energy deposition in the calorimeter. Remaining energy depositions not associated to any object are included at the electromagnetic energy scale.

In CMS, $\MET$ is calculated~\cite{Chatrchyan:2011tn} using reconstructed PF objects as input, which results in a much improved resolution compared with using only calorimeter information. Despite the worse resolution of the CMS hadronic calorimeter, the $\MET$ resolution using PF is comparable to the one of ATLAS.
To remove the bias in the
$\MET$ scale, the jet energy scale corrections of jets above a certain $p_T$ threshold are used, followed by a correction for jets below this threshold as well as for objects not clustered into jets.

\subsubsection{Identification of jets from b-quarks}
\label{sec:btag}

Top quark events are rich in jets originating from $b$-quarks (also called $b$-jets). Identifying these $b$-jets is an important means to select signal and reduce  background not being due to top quark production. In addition, they are useful for reducing the combinatorial background during the full reconstruction of top quarks from their decay products.
Several methods exist to identify b-jets. Most of these methods rely on the fact that $B$-hadrons have a significant lifetime of $1.5 \rm\ ps$ (corresponding to a path length of $450 \rm\ \mu m$), resulting in a displacement of the $B$-hadron's decay vertex from the primary vertex of the event. Tracks emerging from this decay vertex provide an \textit{impact parameter} which is defined as the minimum distance of the linearized track from the primary vertex. Furthermore, a secondary vertex can be reconstructed from tracks belonging to the $b$-hadron decay.

ATLAS and CMS use several algorithms based on impact parameters~\cite{CMS-PAS-BTV-09-001,ATLAS-CONF-2010-091}. The \textit{track counting} algorithm sorts the impact parameters of tracks associated with a jet and uses the impact parameter of the N-th track (usually N=2 or 3) as discriminating variable.
The \textit{jet probability} algorithm uses the impact parameters of all tracks associated with a jet in order to calculate a probability that a jet is a $b$-jet.
Secondary vertex algorithms~\cite{CMS-PAS-BTV-09-001,CMS-PAS-BTV-11-002,ATLAS-CONF-2010-042} try to reconstruct a secondary vertex in the jet. In the simplest case, the discriminating variable is given by the decay length (distance between primary and secondary vertex) or its significance. In the case of CMS, this algorithm has been shown to be more robust with respect to tracking detector misalignment~\cite{CMS-PAS-BTV-07-003}.
More sophisticated, but also better performing, algorithms are also now available~\cite{ATLAS-CONF-2011-102,cmspas-btv-11-004}, in which for example information from impact parameters and secondary vertices is combined using multivariate methods, or additional discriminating variables are used. 
Finally, it is possible to identify $b$-jets by the presence of a high $p_T$ lepton originating from a semi-leptonic decay of a $B$-hadron (\textit{soft lepton tagging}). This method suffers from reduced efficiency due to the semi-leptonic branching fraction.

The performance of an algorithm is usually expressed as a $b$-jet tagging efficiency at a given false positive (or mis-tag) probability. Working points can be defined in terms of a certain value of the discriminating variable.
In CMS, these correspond to light flavor mis-tag rates of 10\% (loose), 1\% (medium) and 0.1\% (tight) in simulated QCD jet events for jets with $p_T=80 \GeV$.
In ATLAS, they instead are defined in terms of a given $b$-tagging efficiency of 50-85\% in simulated $\ttbar$ events. 
The performance is estimated from data using various methods~\cite{CMS-PAS-BTV-11-001,ATLAS-CONF-2011-089,ATLAS-CONF-2011-143}. One method uses the fact that the transverse momentum of the muon from a semi-leptonic $B$-hadron decay relative to the jet axis, $p_{T,rel}$, is larger for muons in $b$-jets than for muons in light flavor or charm quark jets. The $p_{T,rel}$ distribution is fitted with templates for the different jet flavors. Another method employs three weakly correlated tagging methods in events with a muon jet. The mis-tag efficiency is measured using a negative discriminator value for a given algorithm. As negative tagged jets are rich in light flavors, this can be used to measure the mis-tagging rate in data.
Typical values for the efficiencies (mis-tag rates) of the various algorithms are in the range $50-70\%$ ($0.5-5\%$).  The detailed numbers depend for example on the $p_T$ and  $\eta$ of the jet and the considered event sample.
The systematic uncertainty in the $b$-tagging efficiency data-to-simulation scale factors is typically below $5\%$~\cite{cmspas-btv-11-004,ATLAS-CONF-2012-043} for jets in the $p_T$ range most relevant here, increasing towards higher $p_T$, while for the mistag rate the systematic uncertainty is $9-16\%$ ($10-40\%$) for CMS (ATLAS)~\cite{cmspas-btv-11-004,ATLAS-CONF-2011-089}.

The $b$-tagging efficiency can also be measured using $\ttbar$ events. Feasibility studies based on simulation can be found in Refs.~\refcite{cmsnote-2006-013,atlastdr}.
One method, often referred to as \textit{tag counting}, relates the measured $b$-tagged jet multiplicity to the $b$-tagging efficiency. The situation is complicated because of limited detector acceptance, extra $b$-jets from gluon splitting, jets from light or charm quarks which are wrongly $b$-tagged and because of non-$\ttbar$ background.
ATLAS has applied the tag counting method in both the $\ttbar$ di-lepton and lepton+jets channels~\cite{ATLAS-CONF-2011-089}. The measured $b$-tagging efficiencies agree within the statistical and systematic uncertainties, which are of the order of $5-10\%$ each, with the values obtained from simulation.
Very similar methods have also been applied by CMS~\cite{CMS-PAS-BTV-11-003}. The scale factors are within $10\%$ of unity and have uncertainties of $3-5\%$.
Another method, the kinematic selection method, employs a high-purity sample of $\ttbar$ events in the lepton+jets channel and measures the fraction of $b$-tagged jets in this sample, which is then related to the $b$-tagging efficiency. It  has been applied by ATLAS~\cite{ATLAS-CONF-2011-089} and yields results consistent with the tag counting method.
CMS has also measured the $b$-tagging efficiency using two further methods~\cite{CMS-PAS-BTV-11-003}, one of which is based on a profile likelihood fit to the two-dimensional distribution of jet multiplicity versus the $b$-tagged jet multiplicity in the di-lepton channel, and another one is using a sample enriched in $b$-jets by means of a kinematic selection. Both methods give results consistent with the tag counting method.
Within the present level of uncertainties, the $b$-tagging efficiencies measured using $\ttbar$ events are consistent with the ones measured in samples of QCD jet events.

\subsection{Monte-Carlo simulation}

Monte-Carlo (MC) generators are used for the simulation of top quark production in pairs or singly, as well as of the most important background processes. In order to facilitate comparison with experimental data, the simulated samples are processed through detailed detector simulations based on the GEANT4~\cite{Agostinelli:2002hh} framework and are subjected to the same reconstruction algorithms and analysis chain as the real data.
The simulated MC samples are used for several purposes. The signal simulation is used to determine the selection efficiency. Simulations are also used to model differential distributions for signal and background. The theory cross sections of some of the backgrounds are used for normalization purposes only in the case of small and well understood backgrounds. Large or not well modeled backgrounds are instead estimated from data.

General purpose MC generators~\cite{Buckley:2011ms} such as PYTHIA6~\cite{Sjostrand:2006za}, PYTHIA8~\cite{Sjostrand:2007gs}, HERWIG~\cite{Corcella:2000bw,Corcella:2002jc} and HERWIG++~\cite{Gieseke:2011na} interface LO matrix elements (ME) of the hard scattering process with parton showers (PS) in the leading-logarithmic approximation to simulate additional initial-state and final-state radiation. They also include hadronization, secondary decays and simulate the underlying event. 

While parton showers are designed to model parton radiation at low $p_T$ and/or small angle, they are not well suited to model the radiation of additional partons at large angle and/or high $p_T$. Programs such as MADGRAPH~\cite{Alwall:2007st,Alwall:2011uj}, ALPGEN~\cite{Mangano:2002ea} and SHERPA~\cite{Gleisberg:2003xi,Gleisberg:2008ta} combine $2\rightarrow N$ LO tree-level matrix elements for the production of, e.g., $W/Z/\ttbar$+jets with parton showers (ME+PS approach). Care has to be taken to avoid double counting, usually by matching matrix elements and parton showers in a way that the matrix element is used for high $p_T$, large angle emissions, while the parton shower is used for low $p_T$, small angle radiation. Prominent examples are the MLM~\cite{Mangano:2006rw} (ALPGEN, HERWIG) and CKKW~\cite{Schalicke:2005nv} (SHERPA) matching algorithms. The ME+PS approach is well suited to model differential distributions for multi-parton final states, especially $W/Z/\ttbar$+jets, but the total cross section does only correspond to the LO estimate. ALPGEN and MADGRAPH can be interfaced to PYTHIA or HERWIG to perform the parton showering and hadronization, while this is already included in SHERPA.

A complementary approach is provided by the idea to combine the correct description of the total rate at NLO with the exact tree level LO prescription of the first emission of one extra parton (NLO+PS). This is done by subtracting from the exact NLO matrix element the $\mathcal{O}(\alpha_s)$ emission probabilities generated by the shower. Commonly used implementations of this idea are MC@NLO~\cite{Frixione:2003ei} and POWHEG~\cite{Nason:2004rx}, which are in general in good agreement with each other, despite the different implementation.  Both programs can simulate $\ttbar$ and single top quark production, including $\ttbar$ spin correlation. While MC@NLO interfaces with HERWIG for the parton shower, POWHEG can use both HERWIG and PYTHIA and has the advantage that no events with negative weights appear, as is the case for MC@NLO. More recently, the POWHEG-BOX~\cite{Alioli:2010xd} provides a framework for implementing NLO calculations in shower MCs according to the POWHEG method. Implementations of $\ttbar$+jets production exist~\cite{Alioli:2011as,Kardos:2011qa}.
The most recent developments go in the direction to fully automatize the NLO+PS approach (aMC@NLO~\cite{Frederix:2011zi}), as well as to combine the ME+PS and NLO+PS approaches~\cite{Hamilton:2010wh,Hoche:2010kg}.

While not being a full MC generator including parton shower and hadronization, the MCFM~\cite{Campbell:2012uf} program provides parton level predictions of top-quark processes at NLO QCD.

\subsubsection{Top quark pair production}

In CMS, the top quark pair signal is modeled using MADGRAPH v4, where the events containing $\ttbar$ are generated accompanied by up to three extra partons in the matrix-element calculation. 
The renormalization and factorization scales are set to $\mu^2_r = \mu^2_f = m^2_t + \sum p^2_T$, where $\sum p^2_T$ is the sum of the squared transverse momenta of all accompanying hard jets in the event. The CTEQ6~\cite{Pumplin:2002vw} PDF are used, and $m_t=172.5 \GeV$.
The parton configurations are matched with PYTHIA v6.4 for parton showering and hadronization using the MLM prescription.
Tau decays are handled with TAUOLA~\cite{Davidson:2010rw}, which correctly considers the tau polarization in the tau decay.
In ATLAS, the top quark pair signal is simulated employing the NLO+PS generator MC@NLO v3.41, using $m_t=172.5 \GeV$ and the NLO CTEQ6.6 PDF set.
The $\ttbar$ samples are typically normalized according to one of the existing approximate NNLO cross sections, see Table~\ref{tab:xs}.

\subsubsection{Single top quark production}
\label{sec:mcsingletop}

In CMS, single top quark events are simulated using MADGRAPH. To give a fair approximation of the full next-to-leading order properties of the signal in the $t$-channel, the dominant NLO contribution ($2\rightarrow3$ diagram
$qg\rightarrow q't\bar{b}$ and its charge conjugate) are combined with the LO
diagram ($2\rightarrow2$, $qb\rightarrow q't$) by a matching procedure based
on Ref.~\refcite{Boos:2006af}.
Both ATLAS and CMS also simulate single top events using MC@NLO or POWHEG.
Both generators use the massless scheme for the $t$-channel production mode, such that the spectator $b$-jet is modeled only with LO accuracy. ATLAS also uses ACERMC~\cite{Kersevan:2004yg} for the simulation of single top quark processes. The normalization is done using the NLO or approximate NNLO cross sections (see Table~\ref{tab:singletop}).

Single top quark production in the $tW$-channel faces the conceptual problem of its definition within perturbative QCD, as it mixes at NLO with $\ttbar$ production~\cite{Belyaev:1998dn,White:2009yt}. To overcome this, two schemes have been proposed to define the $tW$ signal. In the \textit{diagram removal} (DR) scheme~\cite{Frixione:2008yi}, all signal diagrams which are doubly resonant are removed, while in the \textit{diagram subtraction} (DS) scheme~\cite{Frixione:2008yi,Tait:1999cf}, a gauge invariant term is subtracted which locally cancels the contribution of top quark pair production diagrams. Both schemes are implemented in POWHEG, used by CMS. The DR scheme is also implemented in MC@NLO, as used by ATLAS.

\subsubsection{Vector boson production}

Background samples simulating $W/Z$-boson production in association with up to four extra jets (V+jets) are simulated in CMS using MADGRAPH interfaced to PYTHIA (using MLM matching), using the scales $\mu^2_r = \mu^2_f = m^2_{W/Z} + \sum p^2_T$.
In ATLAS, the production of $W$- and $Z$-bosons in association with up to five extra partons is simulated using ALPGEN v2.13 interfaced to HERWIG, also employing the MLM matching procedure.
While the V+jets background is typically estimated from data, for comparison purposes the overall sample is often normalized to the known inclusive NNLO $W/Z$-boson production cross section, maintaining the relative fractions of parton multiplicities as predicted by the ME+PS generator.

Quark jets produced in V+jets events can originate either from light ($u,d,s$) or heavy ($c,b$) quarks. The latter contribution is particular important for analyses using $b$-tagging. Usually, one defines the following categories of events: $V+q\bar{q}+X$ (V+light), $V+b(\bar{b})+X$ and $V+c(\bar{c})+X$ (V+heavy). Heavy quarks can be produced either directly from the matrix element, or from gluon splitting $g\rightarrow b\bar{b} \ (c\bar{c})$, avoiding double counting.

Di-boson production ($WW$, $WZ$, $ZZ$) is simulated using PYTHIA in CMS, and using HERWIG in ATLAS.

\subsubsection{Estimation of modeling uncertainties}

The most common procedures for the estimation of uncertainties related to the choices made in the modeling of signal and background are discussed in the following.
For $\ttbar$ and single top quark signal, one compares different MC generators to assess differences between the NLO+PS generators, as well as to compare ME+PS with NLO+PS event generation. PYTHIA and HERWIG are compared to assess variations in the parton showering and hadronization description.
The impact on the choice of scales used is studied by varying the renormalization and factorization scales by factors 0.5 and 2.0 with respect to their default values, both for signal as well as for important backgrounds such as V+jets. Additionally, the effect of increasing or decreasing the amount of initial-state (ISR) and final-state (FSR) radiation is evaluated using dedicated samples, using either PYTHIA (CMS) or ACERMC (ATLAS). However, the effect of this variation is expected to be partially correlated with the effect due to changing the scale. Work is ongoing in order to define a fully coherent procedure for estimating the uncertainty on the amount of generated QCD radiation.

For samples in which matrix elements are matched with parton showers, the dependence of the measurement on the choice of matching scale is studied by varying this scale by some amount with respect to the default value.
The dependence on the choice of PDF is evaluated using either error PDF sets, provided with, e.g., the CTEQ6 PDF as provided in the LHAPDF~\cite{Whalley:2005nh} package (CMS), or alternatively by following the PDF4LHC prescription~\cite{Botje:2011sn} (ATLAS).
Finally, signal samples with varied $m_t$, underlying event tune or choice of color reconnection (color rearrangement between decay products of the top quark pair, and with the beam remnants) model are used to assess the impact of these variations on the measurement.

\section{Top Quark Pair Production Cross Section}
\label{sec:xsection}

The $\ttbar$ cross section was measured in $p\bar{p}$ collisions at the Tevatron ($\sqrt{s} = 1.96 \TeV$), most precisely as $\sigma_\ttbar = 7.56 ^{+0.63}_{-0.56} \statsyst \rm\ pb$ (D0~\cite{:2011cq}, precision 8\%) and $\sigma_\ttbar = 7.50 \pm 0.48 \statsyst \rm\ pb$ (CDF~\cite{cdf-conf-9913}, precision 6.4\%), in good agreement with the SM prediction.
The different energy regime and production mechanism make measurements of the $\ttbar$ production cross section at LHC an important test of perturbative QCD. 

Top quark pair production can be experimentally classified according to the decays of the $W$-bosons from the decay of the two top quarks (see section~\ref{sec:topdecay}). In the di-lepton channel, the experimental signature consists of two high $p_T$ leptons, large missing transverse energy $\MET$ and at least two $b$-jets. The branching fraction is comparatively small, but the backgrounds, mostly $Z$+jets, are also fairly small, especially when applying $b$-tagging. This makes the di-lepton topology an ideal place to obtain a very clean sample of $\ttbar$ events. On the other hand, the hadronic channel, where the experimental signature is at least six jets, two of them $b$-jets, suffers from a huge background of QCD multi-jet events. This makes measurements of $\ttbar$ production in this channel difficult, despite the large branching fraction. Finally, the lepton+jets channel offers both a large branching fraction as well as moderate backgrounds (mostly $W$+jets), such that it is often referred to as the \textit{golden channel}. Its signature is one high $p_T$ lepton, $\MET$ and at least four jets. In both the lepton+jets and di-lepton channel, one typically considers only decays into electrons or muons (usually including those from leptonic tau decays), while final states with hadronically decaying taus are experimentally much more challenging and are often studied separately.

In the following, the measurements of the inclusive $\ttbar$ cross section in $pp$ collisions at $\sqrt{s} = 7 \TeV$ are summarized. After discussing measurements in the di-lepton and lepton+jets channels using electrons and muons, which provide the most precise results, measurements in the tau di-lepton, tau+jets and hadronic channels are presented. Combinations made using several individual channels as input are also shown.
In addition to total cross section measurements, the large amount of top quarks produced at LHC makes also differential cross section measurements possible, for which first results are reported. First, however, the observation of top quarks at the LHC is recalled.

\subsection{Observation of top quark pair production at the LHC}

\begin{figure}[t]
\centering
\begin{minipage}{0.49\linewidth}
\includegraphics[width=0.99\linewidth]{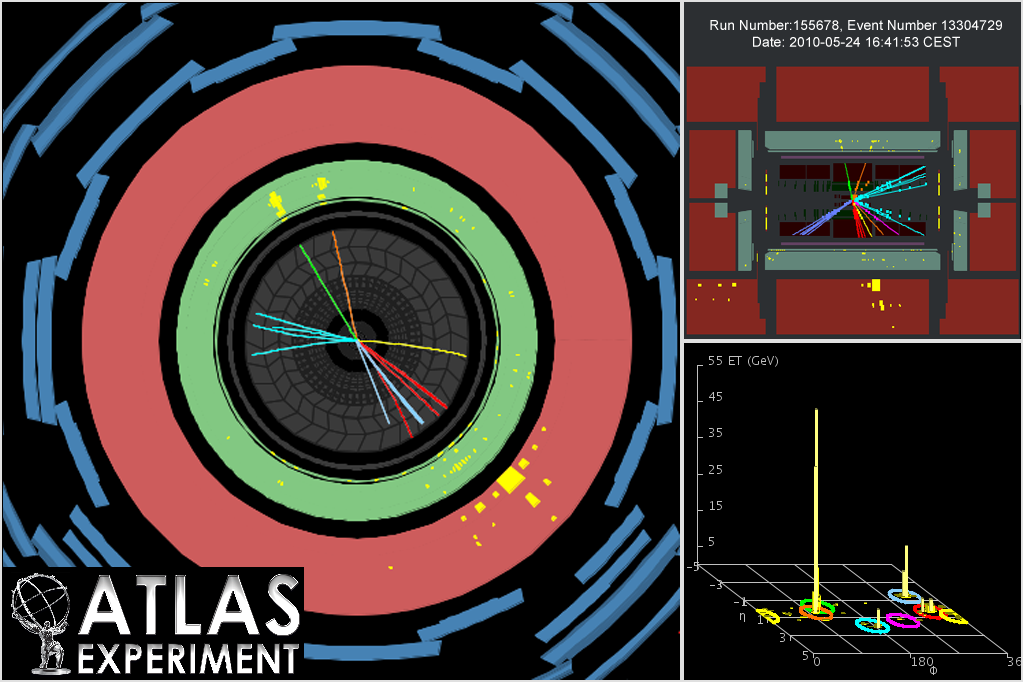}
\end{minipage}
\begin{minipage}{0.49\linewidth}
\centering
\includegraphics[width=0.99\linewidth]{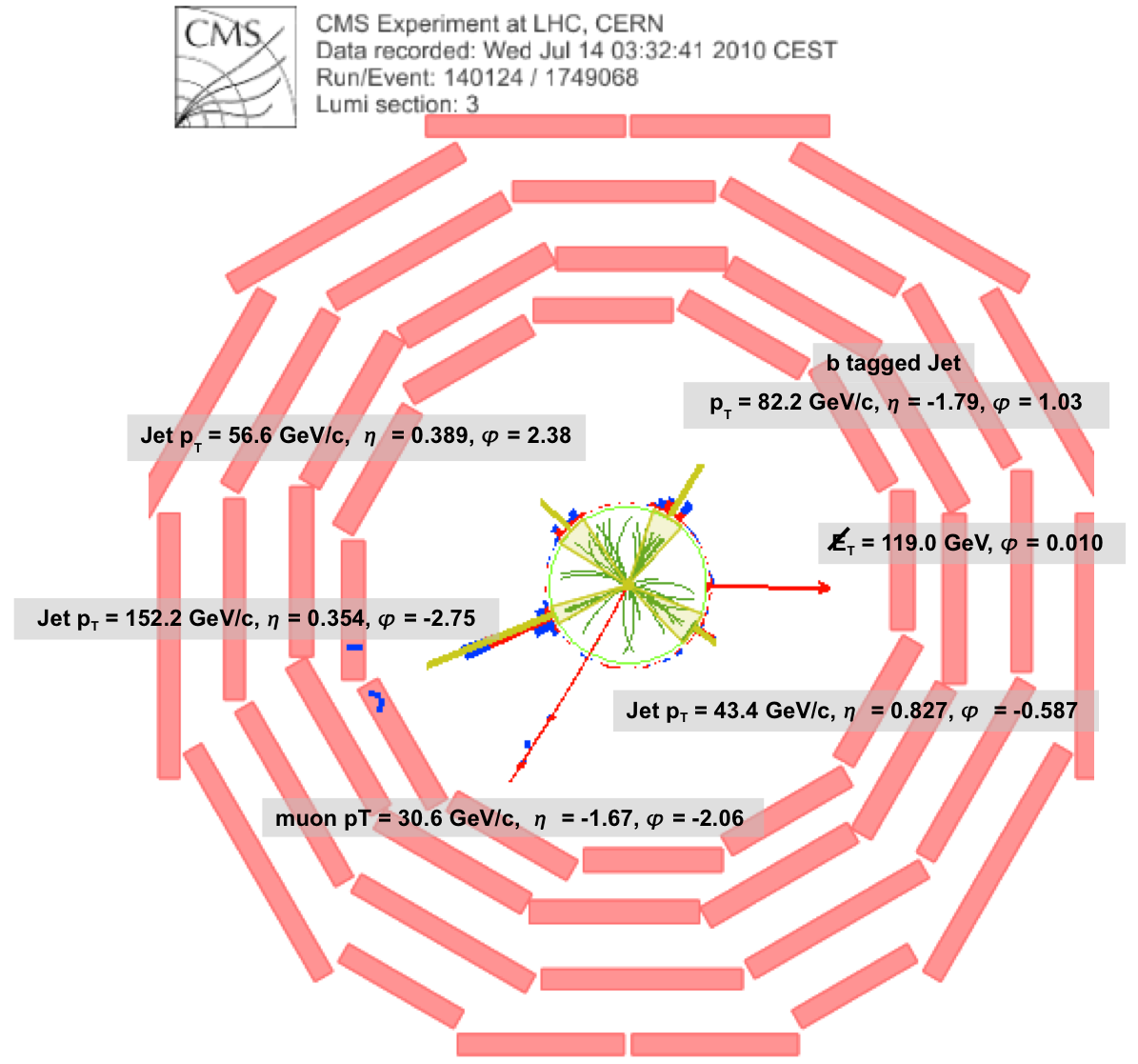}
\end{minipage}
\caption{Example event displays of the first top quark pair production candidates observed at the LHC. Left: A di-lepton candidate observed in ATLAS~\protect\cite{ATLAS-CONF-2010-063}. Right: A lepton+jets candidate observed in CMS~\protect\cite{CMS-PAS-TOP-10-004}.}
\label{fig:events}
\end{figure}

The first top quark pair candidate events at the LHC were reported already in the summer of 2010~\cite{ATLAS-CONF-2010-063,CMS-PAS-TOP-10-004,ATLAS-CONF-2010-087}, after just a few months of data taking at $\sqrt{s}= 7 \TeV$, and corresponding to a few hundred $\rm nb^{-1}$ of integrated luminosity. Two examples are presented in Fig.~\ref{fig:events}. 
The left plot shows a di-lepton candidate observed in ATLAS. It contains two high-$p_T$ electrons with invariant mass inconsistent with being due to $Z$-boson decay, large $\MET$ and three jets of which one is $b$-tagged.
The right plot shows a lepton+jets candidate observed in CMS. It contains one isolated muon, four high-$p_T$ jets of which one is $b$-tagged, and large $\MET$. 

The re-discovery of the top quark at LHC, culminating in the first cross section measurements as discussed in the next sections, represented a major milestone, due to the complexity of the experimental signature which involves essentially all physics objects. It also showed that, already at an early stage, the detectors were very well understood and calibrated.

\subsection{Di-lepton channel with electrons and muons}
\label{sec:xsdilemu}

In this section, measurements of the top quark pair production cross section using the di-lepton final state with electrons and muons are discussed. The corresponding event topology consists of a pair of high $p_T$ leptons ($ee$, $\mu\mu$, $e\mu$), large $\MET$ and at least two $b$-jets.
Feasibility studies for this channel based on simulation and for center-of-mass energies of 10 and 14 TeV can be found in Refs.~\refcite{cmsnote-2006-077,cmspas-top-08-001,cmspas-top-08-002,cmspas-top-09-002} for CMS and in Refs.~\refcite{atlastdr,atl-phys-pub-2009-086} for ATLAS, respectively.

\subsubsection{CMS measurements}
\label{sec:cmsdil}

CMS performed measurements of $\sigma_\ttbar$ in the di-lepton channel using luminosities of $\Lint =3.1 \pbinv$~\cite{Khachatryan:2010ez}, $36 \pbinv$~\cite{Chatrchyan:2011nb} and $1.14 \fbinv$~\cite{CMS-PAS-TOP-11-005}. Only the most recent result will be discussed in more detail, while the preceding ones are only briefly mentioned.

The first measurement of the top quark pair production cross section at the LHC was performed by CMS using a dataset of just $\Lint =3.1 \pbinv$ of 2010 data~\cite{Khachatryan:2010ez}. Events containing a pair of isolated, oppositely charged leptons ($l=e,\mu$) were selected with $p_{T}>20 \GeV$ and $|M(ll)-m_{Z}|>15 \GeV$ (the latter only for $ee$, $\mu\mu$ events) to reduce $Z/\gamma^* \rightarrow l l$ Drell-Yan background. At least two reconstructed jets with $p_T>30 \GeV$ were required, as well as $\MET>30 \ (20) \GeV$ for $ee,\mu\mu$ ($e\mu$) events. The selection yielded 11 events, of which $7.7\pm1.5$ were expected to be due to signal and $2.1\pm 1.0$ due to background. The main backgrounds were estimated using data-driven techniques as explained below. The cross section was measured as
\begin{equation}
\sigma_{t\bar{t}} = 194 \pm 72 \stat \pm 24 \syst \pm 21 \lumi \rm\ pb \ .
\end{equation}
This first measurement was clearly statistically limited, but within the uncertainties in good agreement with the perturbative QCD calculations discussed in section~\ref{sec:xstheo}.

An updated measurement using the full 2010 dataset ($\Lint=36\pbinv$) was presented in Ref.~\refcite{Chatrchyan:2011nb}. The main difference with respect to the previous measurement was the explicit use of $b$-tagging in the event selection.  The cross section was determined as
\begin{equation}
\sigma_{t\bar{t}} = 168 \pm 18 \stat \pm 14 \syst \pm 7 \lumi \rm\ pb \ .
\end{equation}
The ratio of the $\ttbar$ and $Z/\gamma^*$ production cross sections, expected to be less sensitive to certain systematic uncertainties than the measurement of $\sigma_\ttbar$ itself, for example due to luminosity and lepton efficiencies, was also measured as
$\sigma_\ttbar \ / \ \sigma_{Z/\gamma^* \rightarrow ee / \mu\mu} = 0.175 \pm 0.018 \stat. \pm 0.015 \syst$.
The uncertainty of $14\%$ on the ratio could only be marginally improved with respect to the one on the absolute $\sigma_\ttbar$ measurement, since most of the dominating systematic uncertainties actually did not cancel. Future, more precise measurements of such ratios may be used to constrain parameters of the theory, for instance the PDF.

A measurement of the top quark pair cross section in the di-lepton final state based on $\Lint = 1.14\fbinv$ of data collected in 2011 was presented in Ref.~\refcite{CMS-PAS-TOP-11-005}. Events containing a pair of isolated leptons of opposite charge with $p_T>20 \GeV$  and $|M(ll)-m_{Z}|>15 \GeV$, at least two jets with $p_T>30 \GeV$ and $\MET>30 \GeV$ (for $ee$, $\mu\mu$ only) were required. 
At least one of the jets was required to be $b$-tagged using the track counting algorithm and a working point corresponding to about 80\% efficiency and $10\%$ false positive rate. The event selection yielded about 3000 events.
The backgrounds from Drell-Yan $Z/\gamma^* \rightarrow ll \ (l=e,\mu)$ production, as well as from events containing non-$W/Z$ leptons (mostly $W$+jets and QCD multi-jet events) were estimated using data-driven methods.

Events rejected by the $Z$-boson mass veto were used to estimate residual contributions from Drell-Yan events in the selected sample. This background is difficult to model in the relevant region of phase space with requirements on additional jets and significant $\MET$.
The number of events surviving the $Z$-boson veto (``off peak") was assumed to be equal to the estimated number of Drell-Yan events near the $Z$-boson peak after subtraction of the non-Drell-Yan contribution estimated from $e\mu$ events, scaled by the ratio of off-peak to near-peak events in simulation. 
The systematic uncertainty of this method was estimated to be 50\%, dominated by detector calibration effects and changes of the fraction of vetoed Drell-Yan events with increasingly stringent event selection requirements.

The contributions to the selected sample from isolated lepton candidates from non-$W/Z$-boson decays were also derived from data, using a \textit{matrix method}. A set of equations relates the number of events in a ``Signal", ``$W$+jets" and ``QCD multi-jet" category (containing two, one and zero prompt isolated leptons respectively) to the number of events each passing a loose, medium and tight selection (with two loosely, at least one tightly and two tightly isolated leptons respectively), given the efficiencies for each event category to pass from the loose to the medium or tight selection. The latter were obtained from data, allowing the system of equations to be solved and the number of $W$+jets and QCD multi-jet background events to be determined. The systematic uncertainty on the data-driven efficiencies was taken conservatively as 50\% of the difference between data and MC.

\begin{figure}[t]
\centering
\begin{minipage}{0.48\linewidth}
\includegraphics[width=0.99\linewidth]{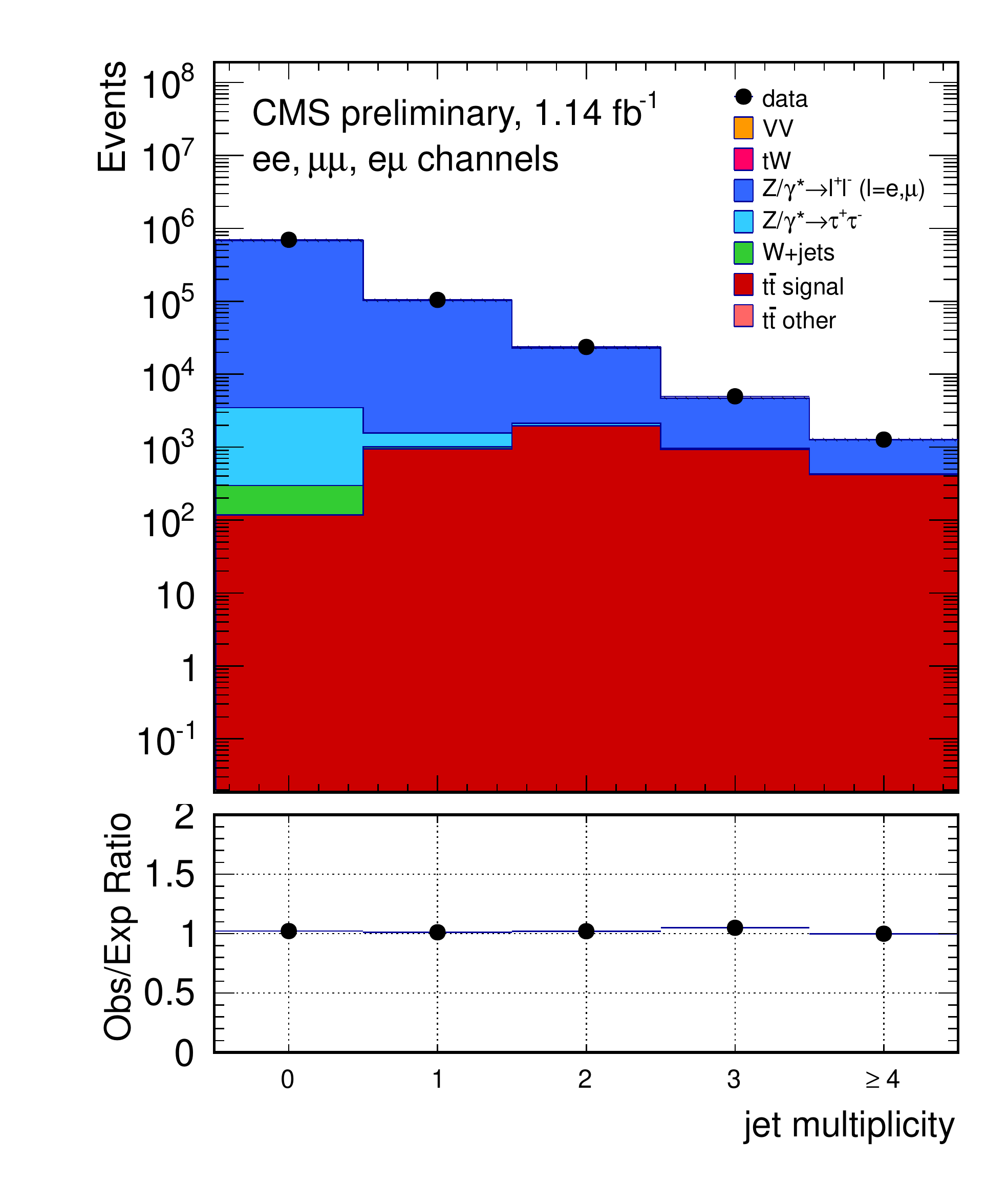}
\end{minipage}
\begin{minipage}{0.50\linewidth}
\vspace{1.5mm}
\includegraphics[width=0.99\linewidth]{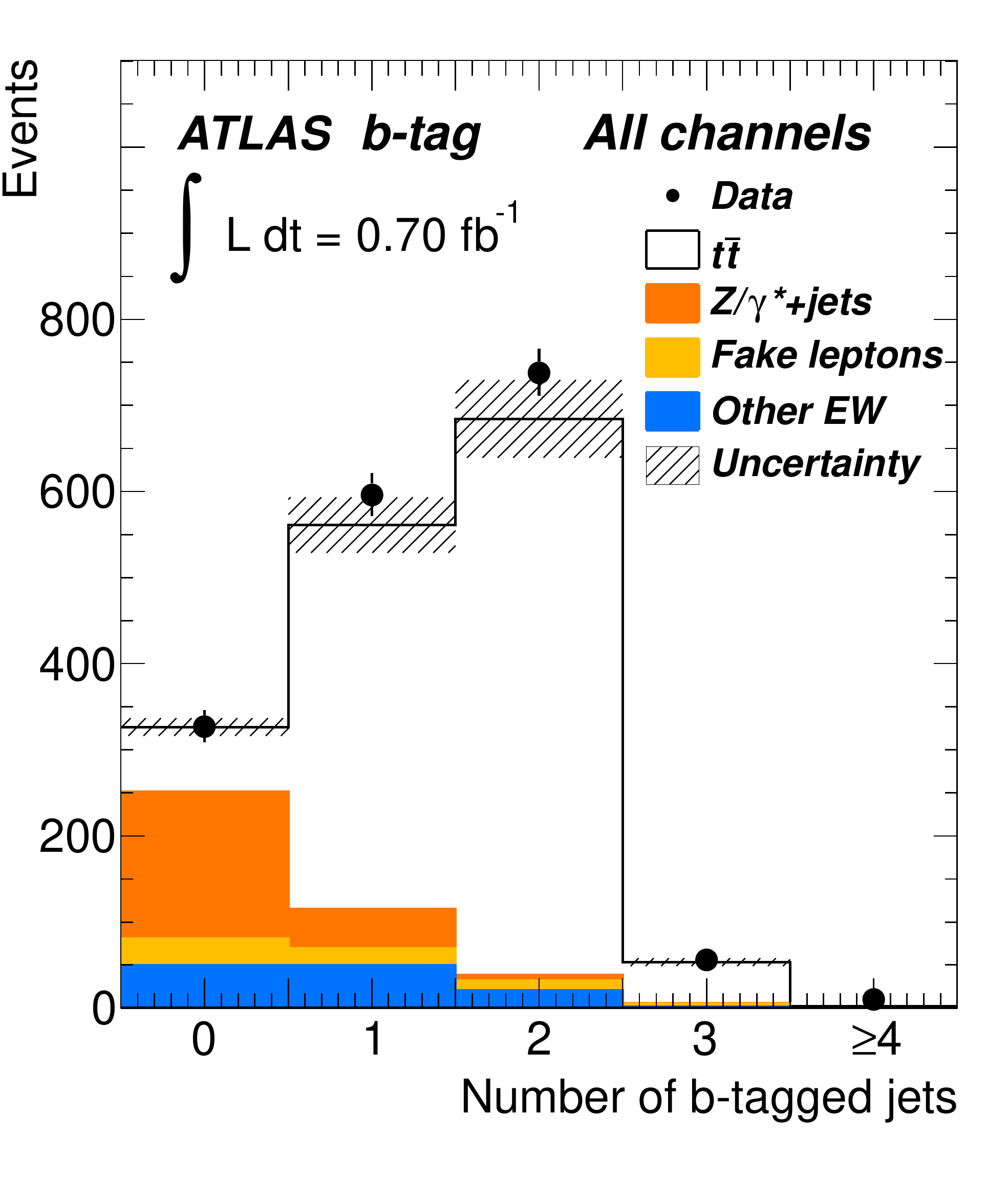}
\end{minipage}
\caption{Left: Reconstructed jet multiplicity without $\MET$ and $b$-tagging requirements in the CMS measurement of $\sigma_\ttbar$ in the di-lepton channel~\protect\cite{CMS-PAS-TOP-11-005}.
Right: $b$-tagged jet multiplicity in the corresponding ATLAS measurement~\protect\cite{:2012kg}.
Here and in following similar figures, the various signal and background contributions were normalized either to the known (N)NLO theory cross sections, or to the result of a data-driven background estimation method.
}
\label{fig:dil2011}
\end{figure}

Other backgrounds, such as di-boson or single top quark production and Drell-Yan $Z/\gamma^* \rightarrow \tau \tau$ events were estimated from simulation.
The jet multiplicity before the $\MET$ and $b$-tagging requirements is shown in Fig.~\ref{fig:dil2011} (left). The cross section was measured individually in the $ee$, $\mu\mu$ and $e\mu$ channels, and the combined result was
\begin{equation}
\sigma_{t\bar{t}} = 169.9 \pm 3.9 \stat \pm 16.3 \syst \pm 7.6 \lumi \rm\ pb \ .
\end{equation}
The precision of this result was already limited by the systematic uncertainty, which was somewhat increased compared to the previous result, mostly due to the larger amount of pile-up. Besides that, the other most important contributions to the systematic uncertainty were originating from the $b$-tagging and lepton efficiencies, the lepton selection model, jet and $\MET$ energy scale uncertainties, as well as the signal modeling.
The main task of future measurements of this kind will be to improve upon the knowledge of the systematic uncertainties.

%%%%%%%%%%%%%%%%%%%%%%%%%%%%%%%%%%%%%%%%%%%%%%%%%%%%

\subsubsection{ATLAS measurements}
\label{sec:atlasdil}

ATLAS performed measurements of $\sigma_\ttbar$ in the di-lepton channel using data samples corresponding to $\Lint = 2.9 \pbinv$~\cite{Aad:2010ey}, $35 \pbinv$~\cite{Aad:2011yb} and $0.70 \fbinv$~\cite{:2012kg}. In the following, the first two measurements will be mentioned only briefly, while the most recent result will be discussed in more detail.

The first ATLAS measurement of $\sigma_\ttbar$ in the di-lepton final state was made using $\Lint = 2.9 \pbinv$ of 2010 data~\cite{Aad:2010ey}.  After the full selection, demanding the presence of two isolated, opposite charge leptons, at least two jets as well as of significant $\MET$ or $H_T$, the latter being defined as the scalar sum of the transverse energies of the two leptons and all selected jets, nine candidate events were observed, of which approximately seven events were expected for signal. The cross section was measured as 
\begin{equation}
\sigma_{t\bar{t}} = 151 ^{+78}_{-62} \stat ^{+37}_{-24} \syst \rm\ pb \ ,
\end{equation}
dominated by the statistical uncertainty. 

An updated measurement using the full 2010 dataset of $\Lint = 35 \pbinv$ was presented in Ref.~\refcite{Aad:2011yb}. The main result did not use $b$-tagging, while for a secondary result at least one of the jets was required to be identified as a $b$-jet using the jet probability algorithm.
A novel feature of this analysis was the use of leptons reconstructed only in the tracking detector (so-called \textit{track leptons}), which were used to increase the acceptance with respect to the standard lepton selection. 
From the 154 (98) events which passed all selection requirements without (with) $b$-tagging, the cross section was extracted using a profile likelihood fit as
\begin{equation}
\sigma_{t\bar{t}} = 177 \pm 20 \stat \pm 14 \syst \pm 7 \lumi \rm\ pb \ .
\end{equation}
The cross check result using $b$-tagging was found to give a consistent result.

The most recent measurement in this channel, based on $\Lint = 0.70 \fbinv$ of 2011 data, was presented in Ref.~\refcite{:2012kg}.  Events were selected requiring two opposite charge, well identified leptons with $p_T> 25 \ (20) \GeV$ for electrons (muons), and at least two jets with $p_T>25 \GeV$. In addition, pairs of a single high quality electron or muon together with a track lepton were also considered.  To suppress vector meson production, $M(ll)>15 \GeV$ was required in the $ee$, $\mu\mu$ and track lepton channels. In the $ee$, $\mu\mu$ channels, Drell-Yan and QCD multi-jet background was suppressed requiring  $\MET > 60 \GeV$, $|M(ll)-m_{Z}|>10 \GeV$, while in the $e\mu$ channel,  a $H_T>130 \GeV$ requirement was imposed. In the track lepton channels, $\MET > 45 \GeV$, $H_T>150 \GeV$ and $|M(ll)-m_{Z}|>10 \GeV$ were required. A selection in which at least one jet had to be $b$-tagged using a working point with around $80\%$ efficiency was also made, relaxing the $\MET$ requirement to $\MET > 40 \GeV$.
The $b$-jet multiplicity for the channels not using track leptons is shown in Fig.~\ref{fig:dil2011} (right).

The backgrounds containing leptons not originating from the decay of a $W/Z$-boson as well as from Drell-Yan production were estimated from data, using methods very similar to the ones used by CMS as discussed in section~\ref{sec:cmsdil}. Other backgrounds (di-boson, single top and Drell-Yan $Z/\gamma^* \rightarrow \tau \tau$ production) were estimated using simulation.
The selection yielded 1920 (1400) events in the analysis without (with) the use of $b$-tagging. The $\ttbar$ cross section was measured using a profile likelihood technique in the individual lepton channels, and a combination was performed using the non-overlapping measurements without using $b$-tagging as well as the measurements using $b$-tagging, but with $\MET<60 \GeV$. The combined result was
\begin{equation}
\sigma_{t\bar{t}} = 176 \pm 5 \stat ^{+14}_{-11} \syst \pm 8 \lumi \rm\ pb \ .
\end{equation}
The dominating sources of systematic uncertainty are due to the lepton, jet energy and $\MET$ measurements, as well as the signal modeling.
As was the case for CMS, better understanding of the systematic uncertainties is required in order to improve upon the precision of such measurements in the future.

%%%%%%%%%%%%%%%%%%%%%%%%%%%%%%%%%%%%%%%%%%%%%%%%%%%%%%%%%%%%%%%%%%%%%%%%%%%%%%%

\subsection{Lepton+jets channel with electron or muon}

In this section, the top pair cross section measurements in the lepton+jets channel, where the lepton is either an electron or a muon, are discussed.
Previous studies using simulation and at higher center-of-mass energy can be found in Refs.~\refcite{cmsnote-2006-024,cmsnote-2006-064,cmspas-top-08-005,cmspas-top-09-003,cmspas-top-09-004,cmspas-top-09-010} for CMS, and in Refs.~\refcite{atlastdr,atl-phys-pub-2009-087} for ATLAS.

\subsubsection{ATLAS measurements}
\label{sec:atlasljets}

The first measurement of $\sigma_\ttbar$ at LHC in the lepton+jets channel was performed by ATLAS~\cite{Aad:2010ey}, using $\Lint = 2.9 \pbinv$ of 2010 data.
The event selection required exactly one isolated lepton with $p_T>20 \GeV$ and at least four jets with $p_T>25 \GeV$, one of which was required to be $b$-tagged using a secondary vertex algorithm. QCD multi-jet background was reduced by requesting $\MET>20 \GeV$ and $\MET + M_T>60 \GeV$, where the transverse mass $M_T$ squared of lepton $l$ is defined as $M_T^2=2 p_T^l \MET [ 1 - \cos ( \phi_{\MET} - \phi_l)] $. 37 events were selected.

The most important backgrounds due to $W$+jets and QCD multi-jet events were estimated using data-driven methods, while the  other backgrounds such as single top quark or $Z$+jets production were obtained from simulation.
The background from QCD multi-jet production was estimated either using a matrix method similar to the one described in section~\ref{sec:cmsdil} ($\mu$+jets), or from a template fit to the $\MET$ distribution ($e$+jets).
The template shapes were taken from simulation, except the one for QCD multi-jets, which was obtained from control regions in data enriched in background by loosening or inverting some of the electron identification criteria.
The background from $W$+jets production was estimated by convoluting the background estimated in the corresponding untagged sample with a $b$-tagging fraction estimated in the 2-jet sample and corrected for different flavor composition and tagging probabilities. The background in the untagged sample with $\geq 4$ jets was estimated from the event yields in lower jet multiplicities using the assumption that Berends-Giele scaling~\cite{Berends:1990ax} holds, which means that the ratio of $W$ + (N+1) jets to $W$ + N jets is expected to be approximately constant as a function of $n$.
The cross section was determined in a counting experiment as
\begin{equation}
\sigma_{t\bar{t}} = 142 \pm 34 \stat ^{+50}_{-31} \syst \ pb \ ,
\end{equation}
where the systematic uncertainty was dominated by contributions originating from jet energy scale uncertainty, data driven background estimation and signal modeling. The combination of this measurement with the one in the di-lepton channel based on the same dataset~\cite{Aad:2010ey} discussed in section~\ref{sec:atlasdil} yielded
\begin{equation}
\sigma_{t\bar{t}} = 145 \pm 31 \stat ^{+42}_{-27} \syst \ pb \ .
\end{equation}

A measurement of $\sigma_\ttbar$ using $\Lint = 35 \pbinv $ of 2010 data  without the use of $b$-tagging was presented in Ref.~\refcite{Aad:2012qf}.
The event selection and QCD multi-jet background determination  were similar to the measurement described above. A likelihood discriminant was constructed from several variables which  discriminate between signal and background and also profit from reduced sensitivity to systematic uncertainties. These were the lepton $\eta$, the aplanarity $\mathcal{A}$
\footnote{The aplanarity $\mathcal{A}$ is defined as 3/2 times the smallest eigenvalue of the momentum tensor $M_{ij}= \sum_{k=1}^{N} p_{ik} p_{jk} / \sum_{k=1}^{N} p_k^2$ ($p_{ik}$ is the $i$-th momentum component of the $k$-th lepton or jet and $p_k$ is the modulus of its momentum).}
and the lepton charge. 
The cross section was obtained by a likelihood fit of the discriminant distributions in the $N_{\rm jets}=3$ and $N_{\rm jets}\geq 4$ samples. The result was
\begin{equation}
\sigma_{t\bar{t}} = 173 \pm 17 \stat ^{+18}_{-16} \syst \pm 6 \lumi \ pb \ ,
\end{equation}
where the systematic uncertainty was dominated by the jet energy scale uncertainty, QCD multi-jet and $W$+jets background estimation as well as signal modeling.

A measurement with the use of $b$-tagging using the same dataset was also performed~\cite{Aad:2012qf}. In contrast to the measurement without $b$-tagging, the lepton charge was not used in the likelihood discriminant. Instead, the two variables $H_{T,3p}$\footnote{
$H_{T,3p}$ is defined as the sum of the $p_T$ of the third and fourth leading jets, normalized to the sum of the absolute values of the $p_z$ of the four leading jets, the lepton and the neutrino (the latter obtained using the W-mass constraint): $H_{T,3p} = \sum_{i=3}^{4} | p_{T,i} | / \sum_{j=1}^{N} | p_{z,j} | $.}
and the average of $w_{JP}=-log_{10} P_j$ for the two jets with lowest $P_j$ in the event (where $P_j$ is the probability for a jet to be a light-quark jet)
were used.
The cross section extraction was performed simultaneously in the 
$N_{\rm jets}=3,4$ and $\geq 5$ samples.
In the fit, the backgrounds  were allowed to vary within their corresponding uncertainties by means of Gaussian constraints and systematic uncertainties were accounted for in normalization and shape by adding nuisance terms to the fit. 
The resulting cross section was
\begin{equation}
\sigma_{t\bar{t}} = 187 \pm 11 \stat ^{+18}_{-17} \syst \pm 6 \lumi \ pb \ ,
\end{equation}
where the systematic uncertainty received dominant contributions due to signal modeling, $b$-tagging efficiency determination as well as the normalization of the heavy flavor contribution of the $W$+jets background.

The most recent ATLAS measurement in this channel, which used $\Lint = 0.70 \fbinv$ of 2011 data~\cite{ATLAS-CONF-2011-121}, did not use $b$-tagging, since the associated systematic uncertainties due to the $b$-tagging efficiency measurement and the heavy flavor fractions in the $W$+jets background turned out to be a limiting factor. 
The event selection was very similar to the one using 2010 data without $b$-tagging, yielding around 41\,000 events. The QCD multi-jet background was estimated using a matrix method, while the $W$+jets background was estimated using a method which exploits the charge asymmetry in $W$-boson production.

\begin{figure}[t]
\centering
\includegraphics[width=0.7\linewidth]{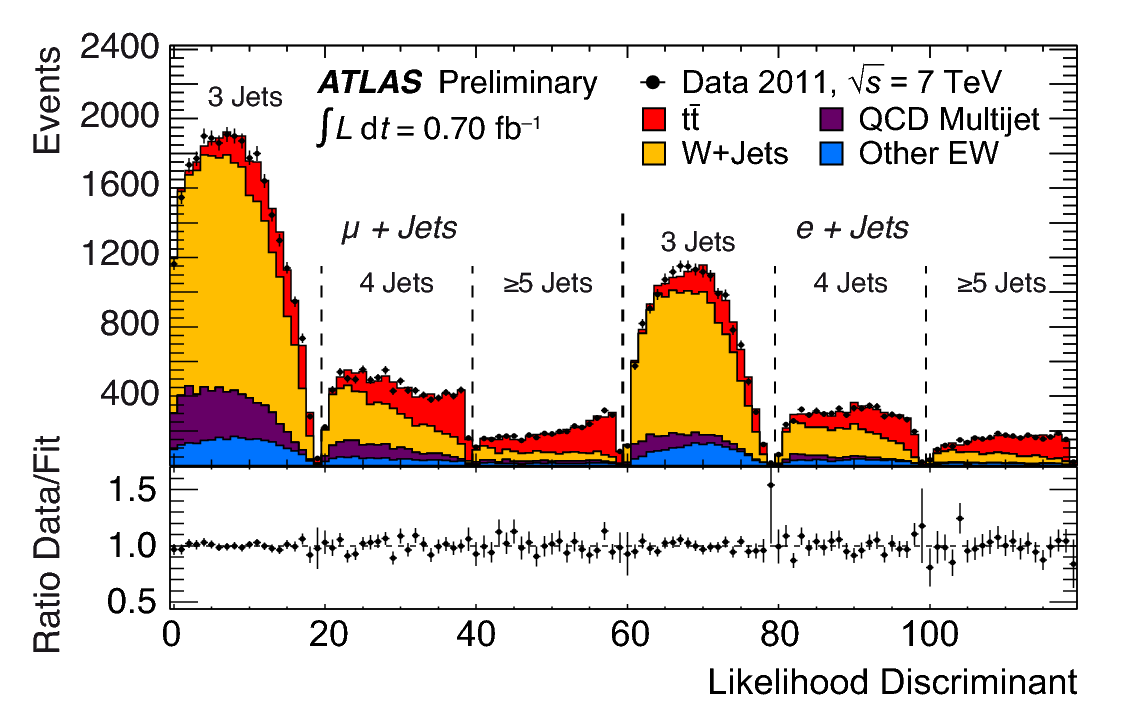}
\caption{Result of the combined fit to the likelihood discriminant, defined in the 3,4 and $\geq5$ jet bins in both the $e$+jets and $\mu$+jets channels, used to measure the $\ttbar$ cross section in the lepton+jets channel based on $\Lint = 0.70 \fbinv$ by ATLAS~\protect\cite{ATLAS-CONF-2011-121}.
\label{fig:atlasljets}}
\end{figure}

A likelihood discriminant was constructed using four variables: the $p_T$ of the leading jet, the $\eta$ of the lepton, and the variables $\mathcal{A}$ and  $H_{T,3p}$, as defined above.
Six discriminant distributions corresponding to lepton flavor ($e$ or $\mu$) and jet multiplicity (3, 4 and $\geq 5$ jets) were subjected to a simultaneous maximum likelihood fit using signal and background templates, where the background templates were a priori normalized to the expected data-driven ($W$+jets and QCD multi-jets) or theory (others) estimates, but allowed to float  using Gaussian constraints. Several sources of systematic uncertainty, such as lepton efficiencies, jet energy scale uncertainty and ISR/FSR modeling, were incorporated via nuisance parameters. The result was
\begin{equation}
\sigma_{t\bar{t}} = 179.0 \pm 3.9 \stat \pm 9.0 \syst \pm 6.6 \lumi \ pb \ .
\end{equation}
The systematic uncertainty was dominated by contributions arising from the choice of signal MC generator, the jet energy scale calibration as well as the modeling of ISR/FSR. The jet energy scale calibration uncertainty was constrained by the fit to $20\%$ to $70\%$ of its original value, and the ISR/FSR modeling uncertainty could be reduced to $20\%$ of its original value.
The result of the fit is shown in Fig.~\ref{fig:atlasljets}. With a precision of 6.6\%, the result represents the most precise measurement of $\sigma_\ttbar$ at the LHC to date.

%%%%%%%%%%%%%%%%%%%%%%%%%%%%

\subsubsection{CMS measurements}
\label{sec:cmsljets}

The first CMS measurement of $\sigma_\ttbar$ in the lepton+jets channel did not use $b$-tagging and was based on a dataset corresponding to $\Lint = 36 \pbinv $ of 2010 data~\cite{Chatrchyan:2011ew}. Events were required to have at least one isolated lepton with $p_T> 30 \ (20) \GeV$ for electrons (muons), respectively, and three or more jets with $p_T>30 \GeV$. Around 3100 events were selected. The cross section was extracted by means of a simultaneous binned likelihood fit of the $\MET$ and $M3$ distributions in the  $N_{\rm jets}=3$ and $N_{\rm jets}\geq 4$ sample, respectively, where $M3$ is defined as the invariant mass of the three jets which maximize the vectorially summed $p_T$ and is also an estimator of the hadronic top quark mass. Templates were obtained from simulation with the exception of the one for QCD multi-jet background, which was derived from data using inverted lepton selection requirements. In the fit, the single-top contribution as well as the ratio $W$+jets/$Z$+jets were constrained to within 30\% of their SM expectations. The cross section was measured as
\begin{equation}
\sigma_{t\bar{t}} = 173 \pm 14 \stat ^{+36}_{-29} \syst \pm 7 \lumi \rm\ pb \ .
\end{equation}
The dominant sources of systematic error were due to the jet energy scale as well as the signal modeling, most importantly those related to the choices of factorization scale and matching threshold.

A measurement using the same dataset but with the use of $b$-tagging was presented in Ref.~\refcite{Chatrchyan:2011yy}. Events were selected requiring exactly one lepton and $\MET>20 \GeV$. Jets with $p_T>25 \GeV$ were considered in the analysis, at least one of which was required to be $b$-tagged using a secondary vertex algorithm which had an efficiency of 55\% and mis-tagging rate of 1.5\% in simulated QCD multi-jet events.
The cross section was determined by means of a simultaneous maximum likelihood fit to the number of jets, the number of $b$-tagged jets and the invariant mass of the tracks associated with the secondary vertex. The latter allowed to discriminate light and heavy quark contributions, and the simultaneous fit in the 2D plane spanned by the jet and $b$-tag multiplicities made it possible to simultaneously constrain the $\ttbar$ signal as well as the $W$+light flavor and $W+$heavy flavor jets background contributions.  Again, the template shapes were obtained from simulation, except for the QCD multi-jet background.
The effects of the systematic uncertainties due to $b$-tagging efficiency, light jet mis-tagging rate, jet energy scale calibration and $W$+jets factorization scale, which were expected to dominate the total systematic uncertainty of the measurement, were included in the likelihood fit via nuisance parameters. Other systematic uncertainties were considered outside the fit.  The simultaneous maximum likelihood fit in both electron and muon channels, each covering nine jet-tag multiplicities, resulted in a cross section measurement of 
\begin{equation}
\sigma_{t\bar{t}} = 150 \pm 9 \stat \pm 17 \syst \pm 6 \lumi \rm\ pb \ .
\end{equation}
The systematics due to jet energy scale calibration, $b$-tagging efficiency and $W$+jets factorization scale were still found to be dominating, but were reduced with respect to their prior values during the fit. The contributions from $W$+heavy flavors were determined as somewhat larger than predicted, with the scale factors for $W+b$-jets ($W+c$-jets) derived as $1.9^{+0.6}_{-0.5}$ ($1.4\pm 0.2$).
The combination of this measurement with the measurement in the di-lepton channel using the same dataset and discussed in section~\ref{sec:cmsdil} yielded
\begin{equation}
\sigma_{t\bar{t}} = 154 \pm 17 \statsyst \pm 6 \lumi \rm\ pb \ .
\end{equation}
Several cross check analyses, for example one based on a soft muon requirement to identify $b$-jets, were performed and yielded consistent results.

\begin{figure}[t]
\centering
\begin{minipage}{0.45\linewidth}
\includegraphics[width=0.99\linewidth]{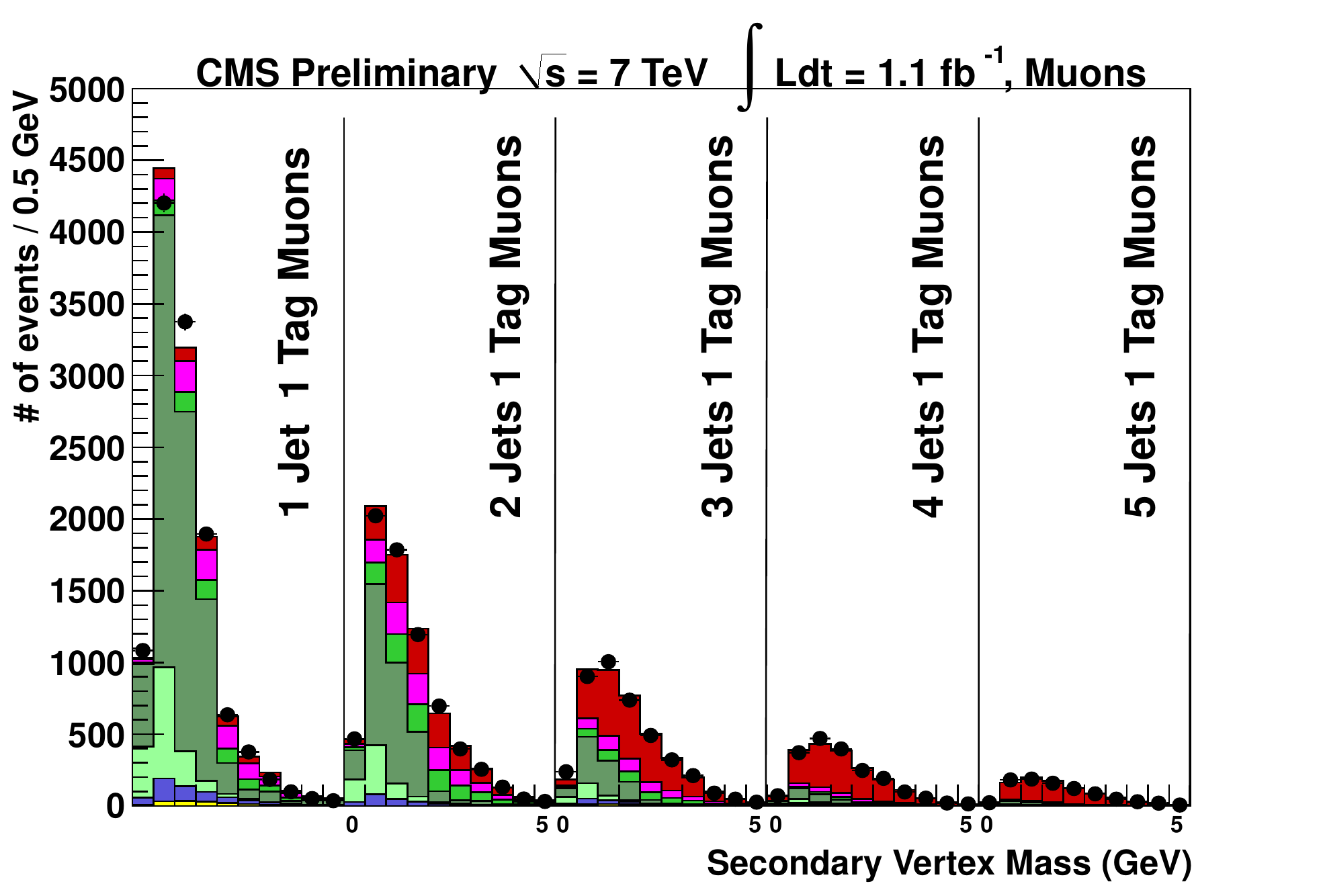}
\includegraphics[width=0.99\linewidth]{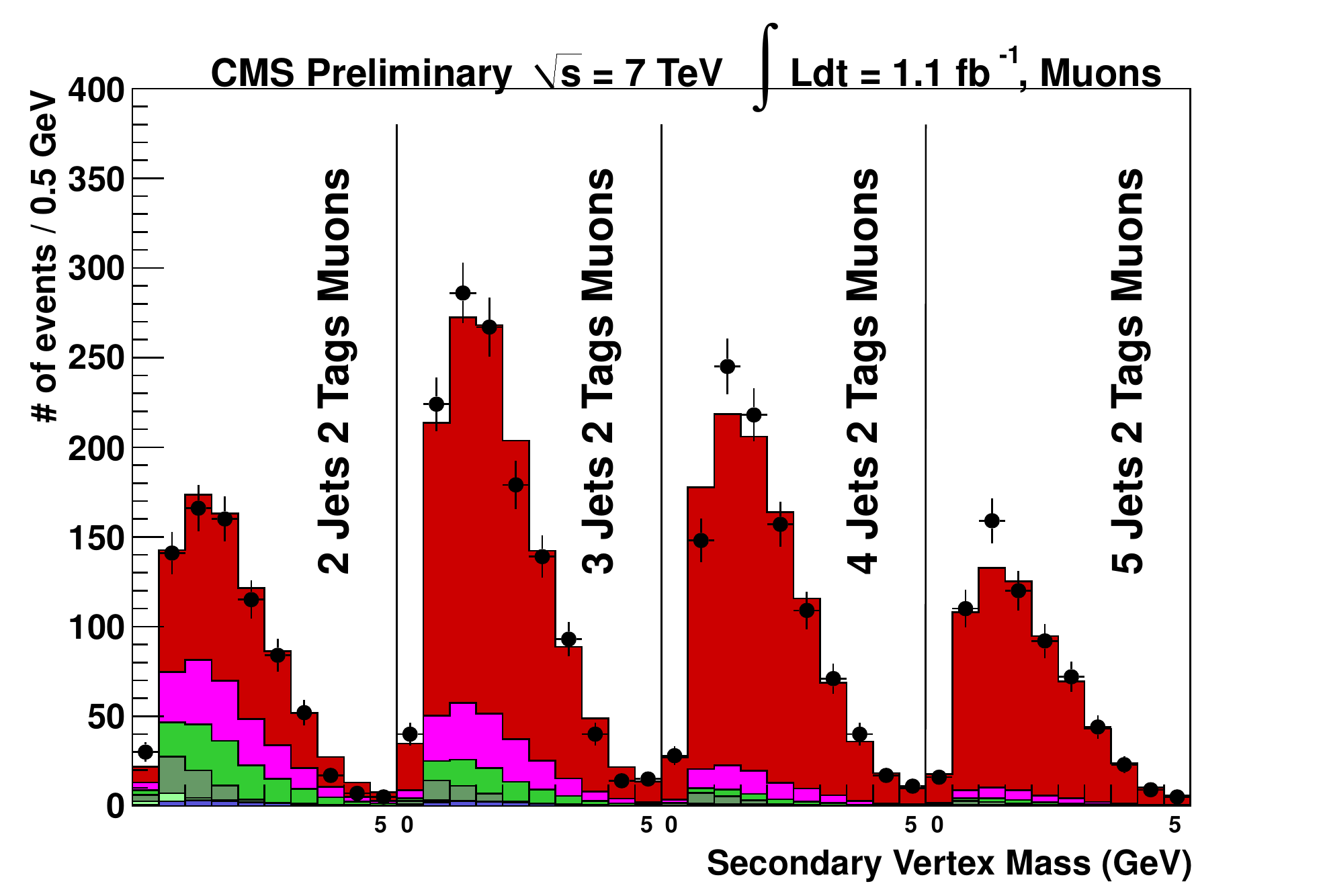}
\end{minipage}
\begin{minipage}{0.45\linewidth}
\includegraphics[width=0.99\linewidth]{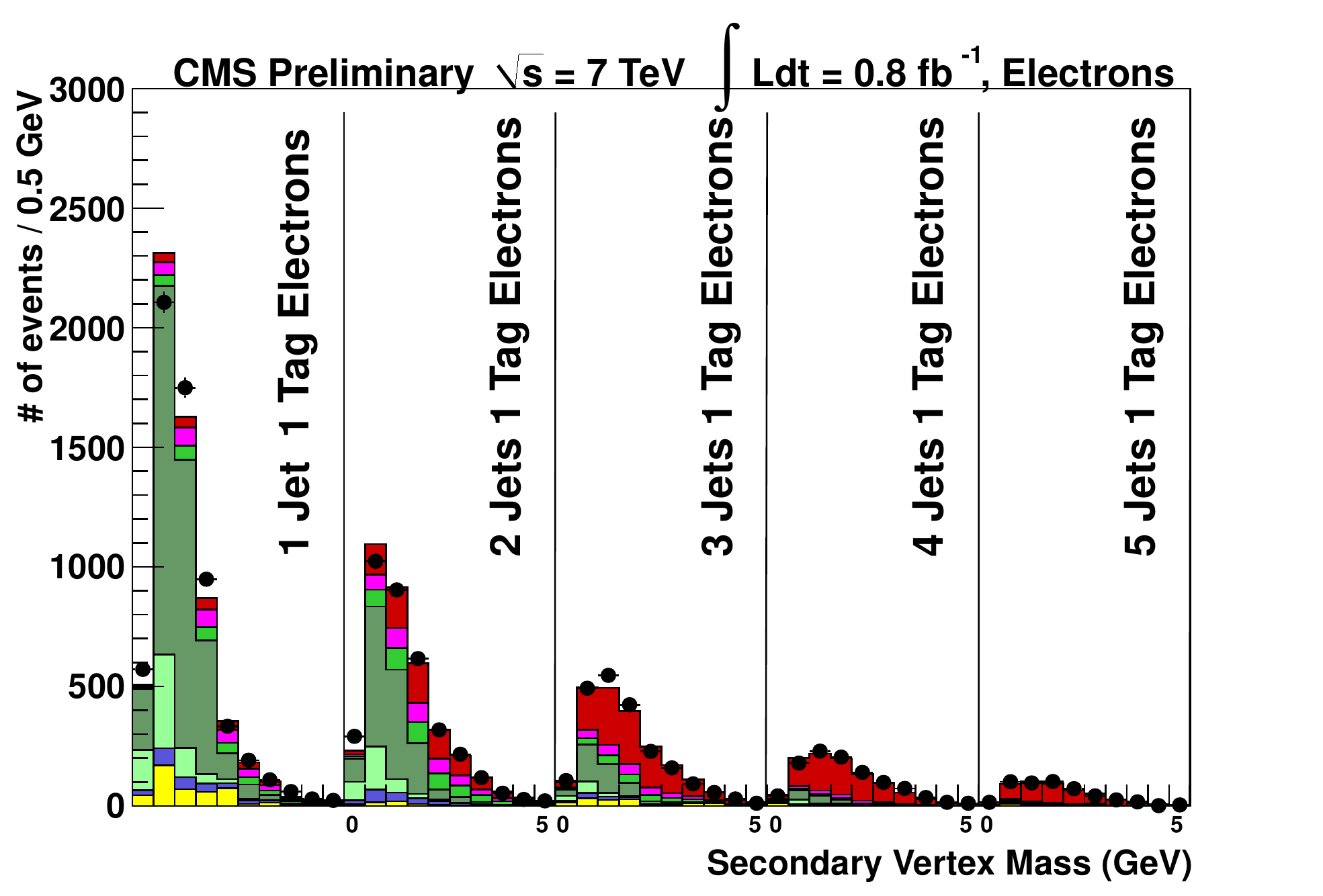}
\includegraphics[width=0.99\linewidth]{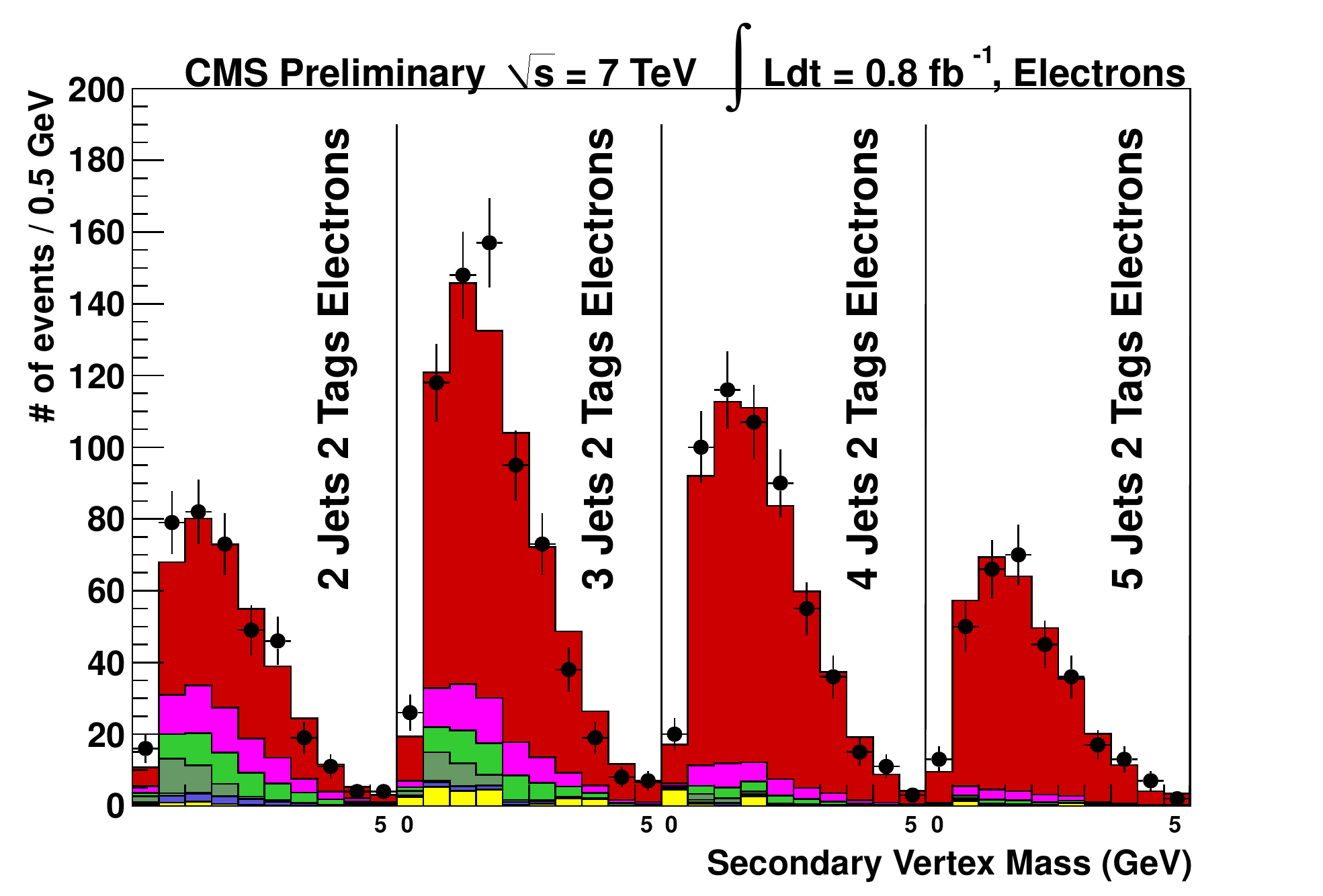}
\end{minipage}
\begin{minipage}{0.08\linewidth}
\includegraphics[width=0.99\linewidth]{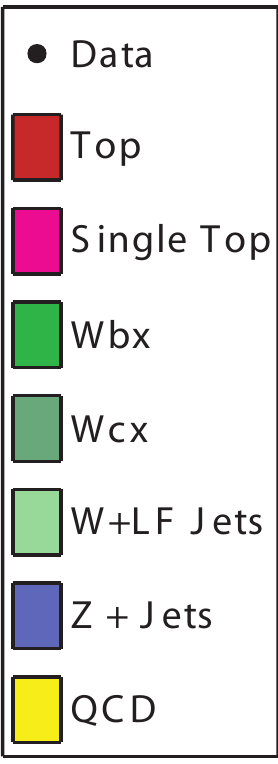}
\end{minipage}
\caption{Result of the combined fit to the secondary vertex mass in nine bins of jet and $b$-tag multiplicity for both the $\mu$+jets (left) and $e$+jets (right) channels, as used in the CMS $\ttbar$ cross section measurement based on up to $\Lint=1.1 \fbinv$~\protect\cite{CMS-PAS-TOP-11-003}.}
\label{fig:cmsljets}
\end{figure}

An update of the measurement discussed above, corresponding to a dataset of $\Lint = 0.8 \ (1.1) \fbinv$ in the $e$+jets ($\mu$+jets) channel was presented in Ref.~\refcite{CMS-PAS-TOP-11-003}. 
The lepton $p_T$ thresholds were increased to $p_T>45 \ (35) \GeV$ for electrons (muons) due to tighter trigger requirements during the 2011 data-taking.
The jet $p_T$ threshold was increased to $p_T>30 \GeV$, and $\MET > 30 \ (20) \GeV$ was required in the electron (muon) channel.
The fit result is shown in Fig.~\ref{fig:cmsljets} and yielded a cross section of
\begin{equation}
\sigma_{t\bar{t}} = 164.4 \pm 2.8 \stat \pm 11.9 \syst \pm 7.4 \lumi \rm\ pb \ .
\end{equation}
The $b$-tagging scale factor was determined as $97\pm1\%$ with respect to its prior value, which agrees well with Ref.~\refcite{CMS-PAS-BTV-11-001}.
The contributions from $W$+heavy flavors were determined as somewhat larger than predicted, with the scale factors for $W+b$-jets ($W+c$-jets) derived as $1.2\pm0.3$ ($1.7\pm 0.1$).
The precision of this measurement is 8.7\%, making this result the most precise measurement of $\sigma_\ttbar$ by CMS to date.

Both CMS and ATLAS performed their most sensitive $\ttbar$ cross section measurements in the lepton+jets channel using in-situ determination of backgrounds as well as simultaneous constraints on important sources of systematic uncertainty by including their effects in the measurement by means of nuisance parameters. These techniques have brought clear advancements in terms of precision in the case of systematically limited measurements.

%%%%%%%%%%%%%%%%%%%%%%%%%%%%%%%%%%%%%%%%%%%%%%%%%%%%%%%%%%%%%%%%%%%%%%%%%%%%%%%

\subsection{Di-lepton channel with a tau lepton}
\label{sec:taudilxs}

In the following, $\sigma_\ttbar$ measurements in the di-lepton final state, where one lepton is a tau and the second one is either an electron or a muon, are discussed. The branching fraction of this final state is approximately 5\% (4/81)  of all $\ttbar$ decays, which corresponds to the one for the di-lepton channel with $ee$, $\mu\mu$ or $e\mu$. Besides providing additional acceptance, the tau di-lepton channel is also interesting because the existence of a charged Higgs boson $H^\pm$  with $m_H<m_t-m_b$ would give rise to anomalous tau lepton production in top decays via $t \rightarrow H^\pm + b$, with the same final state signature (see section~\ref{sec:bsmother}). In addition, this channel provides information on lepton coupling universality.
Here and in the following section, only hadronic tau decays will be considered, since tau decays into electron or muon are usually included as signal in the corresponding $e/\mu$ di-lepton and lepton+jets channel analyses.
A previous CMS feasibility study at $\sqrt{s}=14 \TeV$ can be found at Ref.~\refcite{cmspas-top-08-004}.  Both ATLAS and CMS performed measurements in this final state using 2011 data.

\begin{figure}[t]
\centering
\begin{minipage}{0.49\linewidth}
\centering
%\vspace{1mm}
\includegraphics[width=0.99\linewidth]{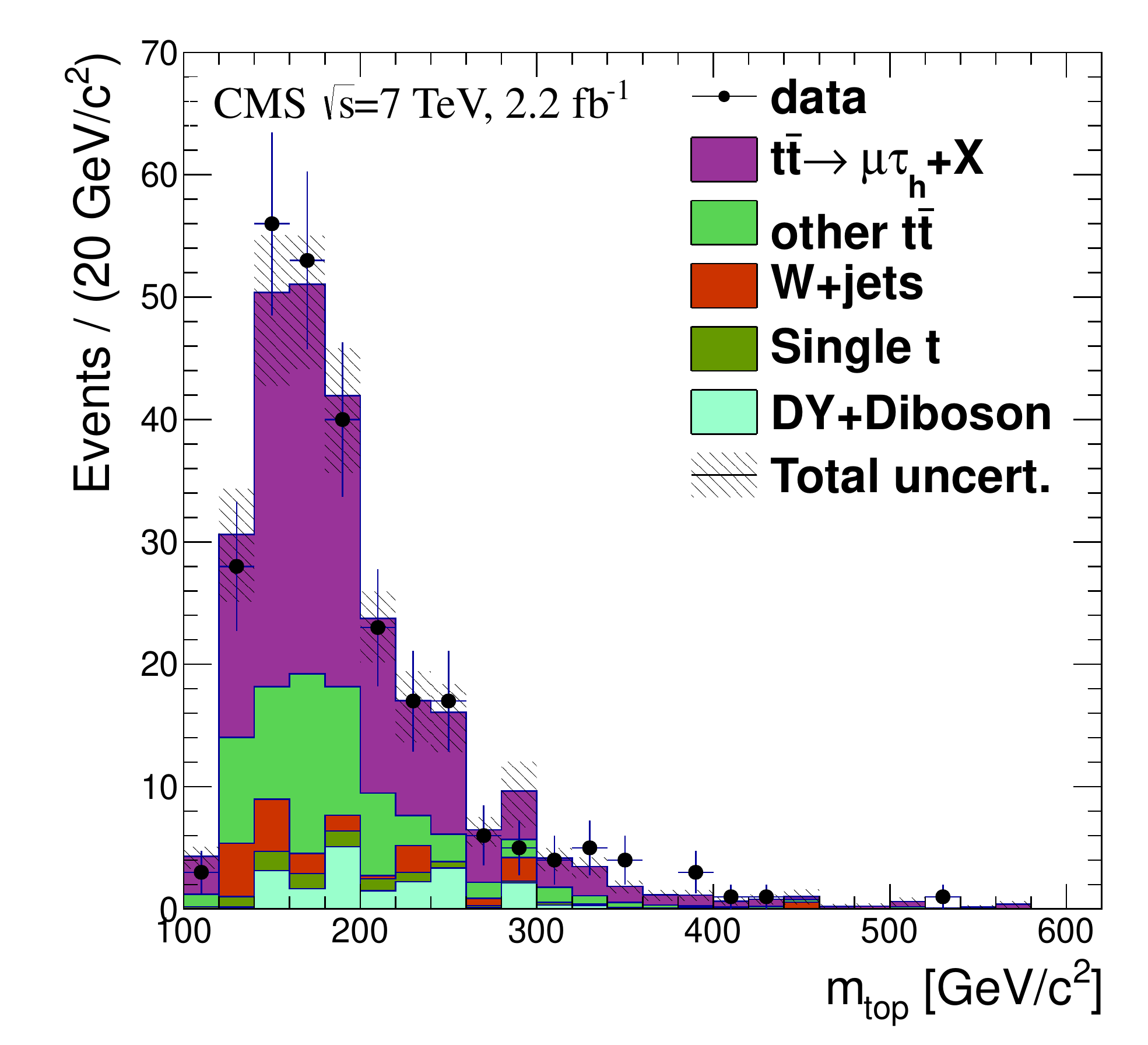}
\end{minipage}
\begin{minipage}{0.49\linewidth}
\centering
\vspace{1mm}
\includegraphics[width=0.99\linewidth]{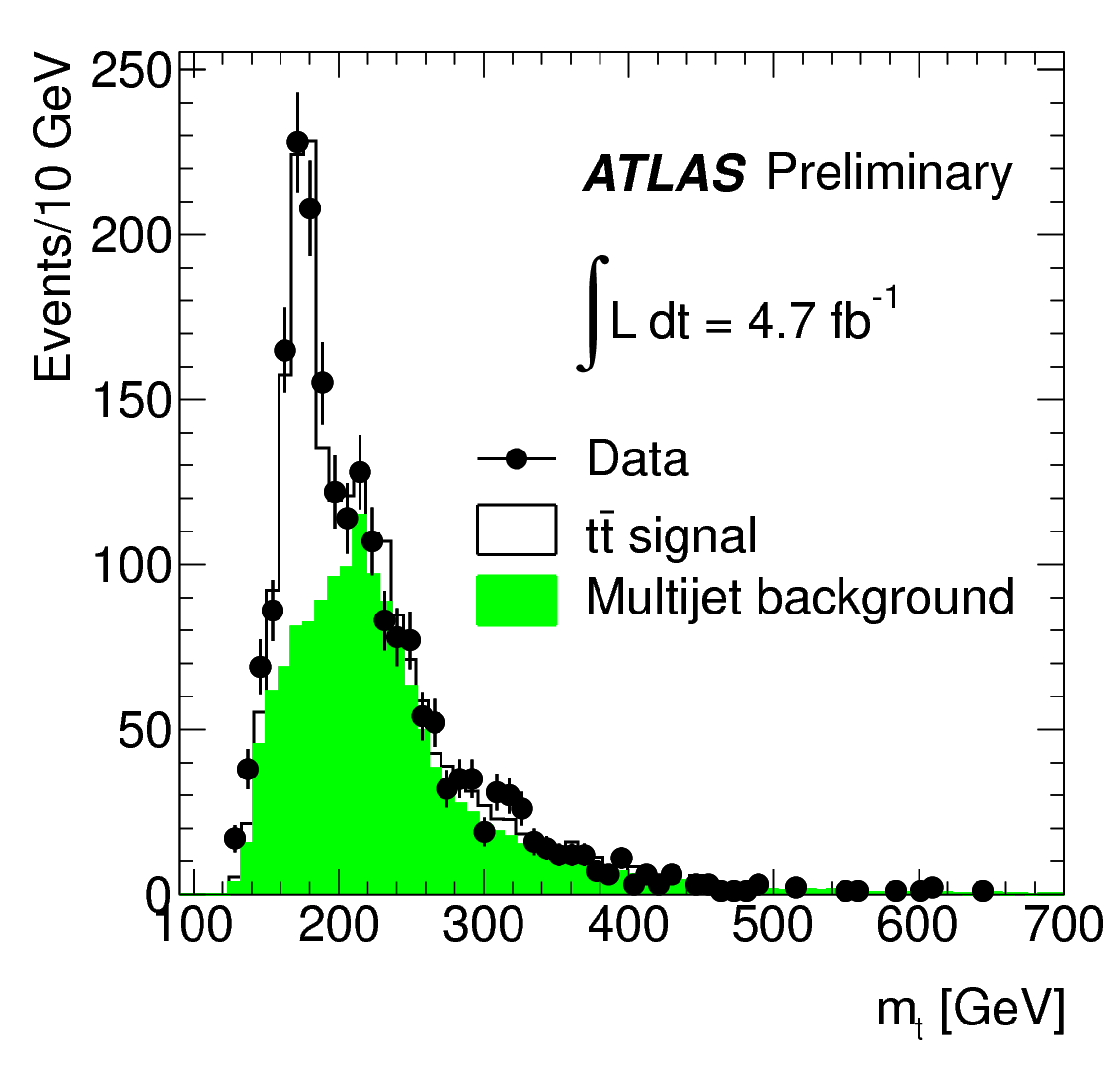}
\end{minipage}
\caption{Reconstructed $m_t$ distributions in the  $\ttbar$ cross section measurements in the tau di-lepton channel from CMS~\protect\cite{Collaboration:2012vs} (left),
and in the hadronic channel from ATLAS~\protect\cite{ATLAS-CONF-2012-031} (right).}
\label{fig:taudil}
\end{figure}

The CMS measurement~\cite{Collaboration:2012vs} considered both $e\tau$ and $\mu\tau$ final states and was based on data samples corresponding to $\Lint=2.0 \ (2.2) \fbinv$, respectively.
Events were selected containing a hadronic tau lepton candidate with $p_T>20 \GeV$, an isolated muon (electron) of opposite charge with $p_T>30 \ (35) \GeV$, at least two jets with $p_T>30 \ (35)\GeV$, one of which was required to be $b$-tagged using the track-counting algorithm, and $\MET>40 \ (45) \GeV$. Fig.~\ref{fig:taudil} (left) shows the reconstructed top quark mass after the event selection using the KINb method, which will be discussed in section~\ref{sec:cmsmassdil}.
The main contribution to the background, which originates from events with one muon or electron, large $\MET$ and several jets, one of which is faking the hadronic tau, was estimated from data. It is mostly due to $W$+jets or semi-leptonic $\ttbar$ production. The tau fake probability, estimated in and averaged between samples enriched in QCD multi-jet (mostly gluon jets) and $W$+1 jet (mostly quark jet) events and parameterized in $p_T$, $\eta$ and jet radius, was applied to muon + $\MET$ + 3 jet events to obtain this background contribution. Other backgrounds which contain real taus, such as $Z/\gamma^*\rightarrow \tau \tau$, single top quark, di-boson and other $\ttbar$ production (the latter involving different decay modes), were estimated from simulation.
The cross section was measured in a counting experiment as
\begin{equation}
\sigma_{t\bar{t}} = 143 \pm 14 \stat \pm 22 \syst \pm 3 \lumi \rm\ pb \ .
\end{equation}
The dominant systematic uncertainties arose from the tau fake background estimation, tau reconstruction and $b$-tagging efficiency measurements, as well as from the jet energy scale uncertainty.

The ATLAS measurement was based on $\Lint=2.05\fbinv$~\cite{arxiv:1205.2067}.
Events were selected which contained one isolated muon or electron with $p_T>20 \ (25) \GeV$ ($\mu,e$), $\MET>30 \GeV$, at least two jets with $p_T>25\GeV$ one of which was $b$-tagged using a secondary vertex algorithm, $H_T>200 \GeV$ (to suppress $W$+jets background), and one loose tau candidate with $p_T>20 \GeV$.
The tau fake background was estimated by fitting the sum of templates to the distribution in data of the discriminant of the BDT optimized for tau identification (see section~\ref{sec:tauid}). The signal template was obtained from simulation, while the background template was modeled in a data driven way. The fits were performed separately to the 1-prong and 3-prong samples whose BDTs were independently optimized.
The combined result for the cross section was
\begin{equation}
\sigma_{t\bar{t}} = 186 \pm 13 \stat \pm 20 \syst \pm 7 \lumi \rm\ pb \ .
\end{equation}
The dominant contributions to the systematic uncertainty that affect the signal acceptance in simulation originated from the tau identification and $b$-tagging efficiency, as well as from the signal modeling.

%%%%%%%%%%%%%%%%%%%%%%%%%%%%%%%%%%%%%%%%%%%%%%%%%%%%%%%%%%%%%%%%%%%%%%%%%%%%%%%

\subsection{Tau+jets channel}

The $\ttbar$  final state where one of the $W$-bosons decays into a hadronically decaying tau lepton and a neutrino and the other one decays into jets, usually referred to as the tau+jets channel, has a branching fraction of $\sim 10\%$. However, the cross section measurement in this channel is difficult due to the large background from QCD multi-jet production and other processes which fake the experimental signature of a hadronic tau.

The first cross section measurement in this channel at the LHC was carried out by ATLAS using $\Lint = 1.67 \fbinv$ of data~\cite{ATLAS-CONF-2012-032}. Events with five reconstructed jets were selected, of which two were required to be $b$-tagged and one was identified as a hadronically decaying tau with $p_T>40 \GeV$. 
Further requirements on the $\MET$ significance as well as the absence of additional high $p_T$ leptons were applied to improve the signal to background ratio. The tau signal was extracted using a template fit to the observed multiplicity distribution of well measured tracks within a cone around the axis of the jet considered a tau candidate. 
The signal template was taken from simulation, while the templates for the dominating backgrounds ($\ttbar$ events with a jet mis-identified as tau and QCD multi-jet events) were obtained from data. The amount of $W+b\bar{b}$ and single top production background was taken from simulation. The cross section was measured as
\begin{equation}
\sigma_{t\bar{t}} = 200 \pm 19 \stat \pm 42 \syst \pm 7 \lumi \rm\ pb \ .
\end{equation}
Besides the uncertainties in the shapes of the fit templates, important systematic uncertainties were due to the jet and $\MET$ measurements, $b$-tagging efficiency, as well as signal modeling.

CMS performed a $\ttbar$ cross section measurement in the tau+jets channel with $\Lint = 3.9\fbinv$ of data~\cite{cmspas-top-11-004}, using events with one hadronic tau candidate and at least four jets, one of which was required to be $b$-tagged. The shape of the dominating QCD multi-jet background was taken from data by using a sample without $b$-tagging requirement, weighted by the per-jet $b$-tag mis-identification probability. A neural network (NN) based on seven kinematic variables was used to discriminate the $\ttbar$ tau+jets signal from the backgrounds. The cross section was measured by means of a template fit to the NN output distribution. The result was
\begin{equation}
\sigma_{t\bar{t}} = 156 \pm 12 \stat \pm 33 \syst \pm 3 \lumi \rm\ pb \ ,
\end{equation}
where important contributions to the systematic uncertainty were due to the jet energy scale uncertainty, tau trigger and ID efficiency and energy measurement, as well as the QCD multi-jet background modeling.

%%%%%%%%%%%%%%%%%%%%%%%%%%%%%%%%%%%%%%%%%%%%%%%%%%%%%%%%%%%%%%%%%%%%%%%%%%%%%%%

\subsection{Hadronic channel}
\label{sec:xshadronic}

Despite the large branching fraction, the $\sigma_\ttbar$ measurement in the hadronic channel poses the challenge of a huge multi-jet QCD background, which requires well performing $b$-tagging algorithms for its reduction, and data-driven techniques in order to estimate the remaining contribution. In the following, the existing measurements at LHC using this channel are discussed. For a previous study based on simulation, see Ref.~\refcite{cmsnote-2006-077}.

CMS performed a measurement~\cite{CMS-PAS-TOP-11-007}  based on $\Lint = 1.1 \fbinv$ of 2011 data. Events were selected on-line by requiring the presence on five, or, at higher luminosities, six jets. Off-line, at least six jets were required with $p_T>40\GeV$, of which five (four) were required to fulfill $p_T>50 \ (60) \GeV$, respectively. In order to reduce the overwhelming QCD multi-jet background, at least two jets were required to be $b$-tagged using a secondary vertex algorithm with a working point corresponding to the small mis-tagging probability of only $\sim0.1\%$, at the price of a reduced $b$-tagging efficiency of $\sim38\%$.
A kinematic fit was employed to identify the association of jets to top quark decay products, using $m_W$ and $m_t-m_{\bar{t}}$ constraints, and choosing the jet permutation resulting in the smallest $\chi^2$ fit value.
The cross section was obtained from an un-binned maximum likelihood fit to the $m_t$ distribution obtained from the kinematic fit, 
based on a sample of 1620 events. 
While for the signal shape simulation was used, the QCD multi-jet background shape was obtained from data by selecting a sample with six jets but without identified $b$-jets, in which the signal contribution was negligible. This template was weighted with a parameterized $b$-tagging probability applied to each jet assumed to originate from a $b$-quark in the kinematic fit. The cross section was estimated from the signal fraction obtained in the fit ($\sim 25\%$) as
\begin{equation}
\sigma_{t\bar{t}} = 136 \pm 20 \stat \pm 40 \syst \pm 8 \lumi \rm\ pb \ .
\end{equation}
The systematic uncertainty was dominated by the contributions from the $b$-tagging efficiency, the jet energy scale and the background shape.

Following initial analyses using smaller integrated luminosities~\cite{ATLAS-CONF-2011-066,ATLAS-CONF-2011-140},
ATLAS presented a cross section measurement using the full 2011 data sample of $\Lint=4.7 \fbinv$ in Ref.~\refcite{ATLAS-CONF-2012-031}. Events were selected at the trigger level by requiring at least five jets with $p_T>30 \GeV$, while off-line the requirement was $p_T>55 \GeV$, supplemented by an additional jet with $p_T>30\GeV$. At least two jets were required to be identified as $b$-jets using an algorithm which provided 60\% efficiency and a light jet rejection factor of $\sim 500$. The events were also required to not contain isolated high $p_T$ leptons, and to have small $\MET$ significance
to reduce the contribution from electroweak processes. In addition, the $\Delta R$ between the $b$-jets was required to be above 1.2 to remove contributions from $g\rightarrow b \bar{b}$ gluon splitting. The selection yielded around 16\,400 events. The top quark mass was reconstructed using a kinematic likelihood fit method which took $b$-tagging information into account.
The shape of the QCD multi-jet background was modeled using data events without the $b$-tagging requirement, applying MC-based corrections for the effect of $b$-tagging on the $m_t$ shape.
The cross section was measured using an un-binned maximum likelihood fit to the reconstructed  $m_t$ distribution, yielding a signal fraction of $31.4\%$ (Fig.\ref{fig:taudil}, right). The result was
\begin{equation}
\sigma_{t\bar{t}} = 168 \pm 12 \stat ^{+60}_{-57} \syst \pm 7 \lumi \rm\ pb \ .
\end{equation}
The largest sources of systematic uncertainty were due to the jet energy scale, trigger and $b$-tagging efficiency measurements, as well as signal modeling.

%%%%%%%%%%%%%%%%%%%%%%%%%%%%%%%%%%%%%%%%%%%%%%%%%%%%%%%%%%%%%%%%%%%%%%%%%%%%%%%

\subsection{Cross section combinations}

Both collaborations produced combinations of their $\ttbar$ cross section measurements performed in various channels. The combinations were performed by means of a combined likelihood function which was constructed from the ones of the individual measurements. Care was taken concerning the treatment of correlations between different sources of systematic uncertainty. In addition, cross checks using a simple averaging or the BLUE~\cite{Lyons:1988rp} method were performed to validate the result.

\begin{figure}[t]
\centering
%\begin{minipage}{0.49\linewidth}
\includegraphics[width=0.49\linewidth]{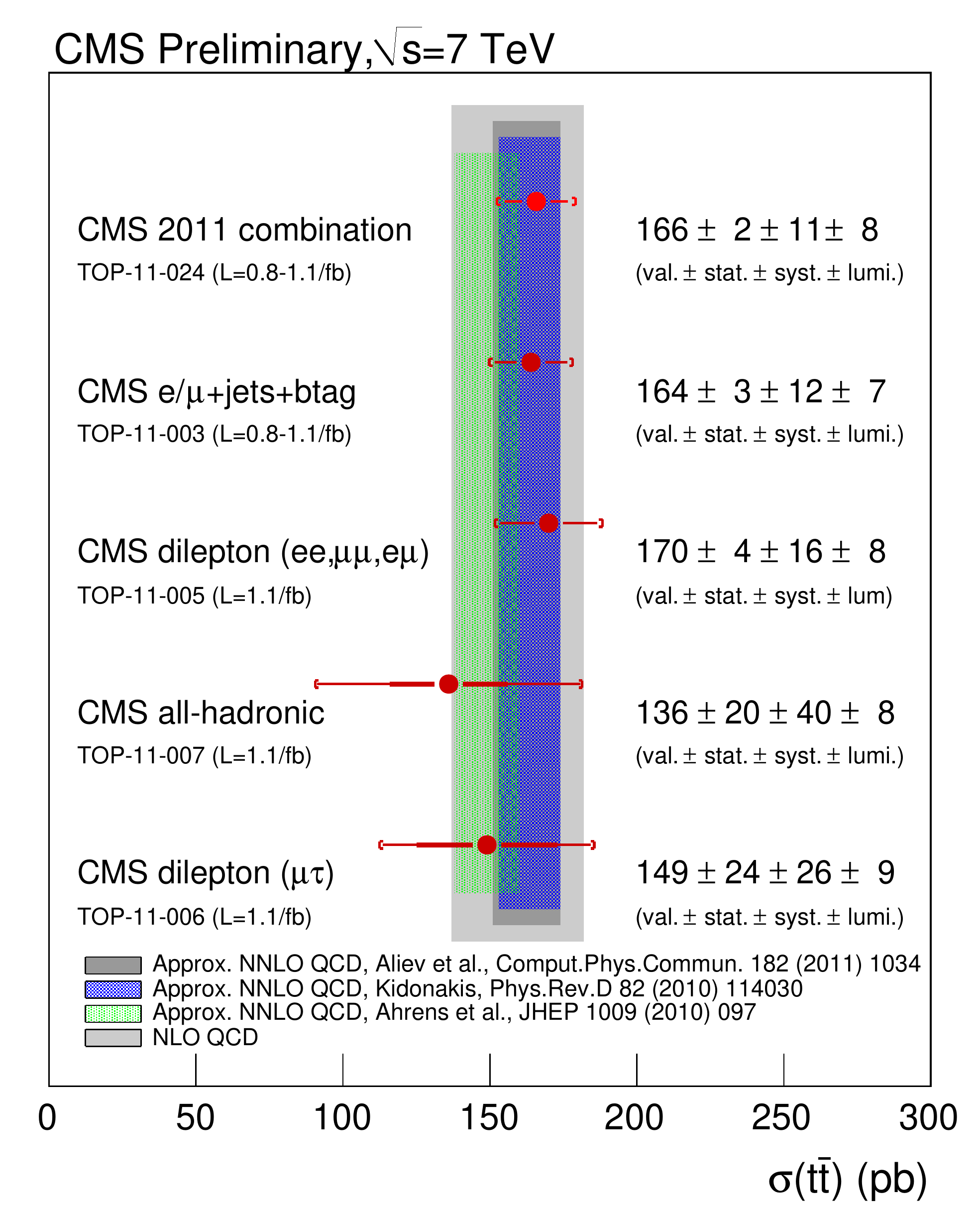}
%\end{minipage}
%\begin{minipage}{0.49\linewidth}
\includegraphics[width=0.49\linewidth]{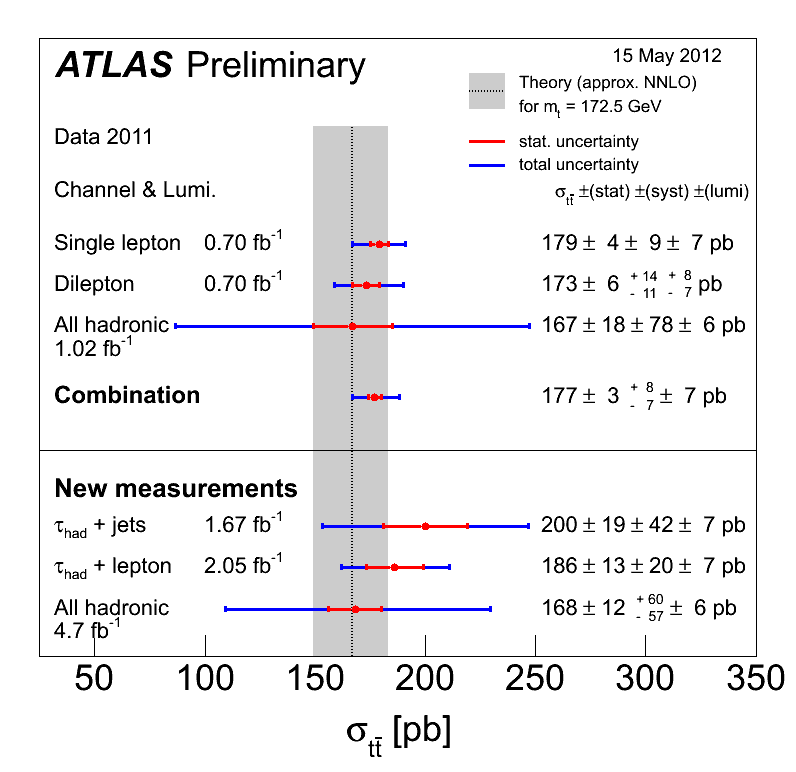}
%\end{minipage}
\caption{Combinations of CMS~\protect\cite{CMS-PAS-TOP-11-024} (left) and ATLAS~\protect\cite{ATLAS-CONF-2012-024,atlas-topsum-may12}
 (right) $\ttbar$ cross section measurements using 2011 data, compared with approximate NNLO theory calculations (Ref.~\protect\refcite{Aliev:2010zk} in the case of ATLAS). See section~\ref{sec:xstheo} for more details regarding the calculations.
}
\label{fig:xscombination}
\end{figure}

Following earlier combinations of measurements in the di-lepton and lepton+jets channels~\cite{CMS-PAS-TOP-11-001,Chatrchyan:2011yy}, the most recent CMS combination~\cite{CMS-PAS-TOP-11-024} was based on all individual preliminary cross section measurements using 2011 data corresponding to $\Lint = 0.8 - 1.1 \fbinv$, including the di-lepton channels with electrons and muons, the muon-tau di-lepton channel~\cite{CMS-PAS-TOP-11-006}, the $e,\mu$+jets channel and the hadronic channel.  The result of the combination (see   Fig.~\ref{fig:xscombination}, left) was
\begin{equation}
\sigma_{t\bar{t}} = 165.8 \pm 2.2 \stat \pm 10.6 \syst \pm 7.8 \lumi \rm\ pb \ .
\end{equation}

In the case of ATLAS, earlier and meanwhile superseded combinations of $\ttbar$ cross section measurements in the lepton+jets and di-lepton channels were reported in Refs.~\refcite{ATLAS-CONF-2011-040,ATLAS-CONF-2011-108}.
The most recent ATLAS combination~\cite{ATLAS-CONF-2012-024}
used the 2011 measurements in the di-lepton (combining only channels without $b$-tagging and not using track leptons), lepton+jets and hadronic channels (earlier preliminary version of measurement discussed in section~\ref{sec:xshadronic}), using up to $\Lint=1.02 \fbinv$ of data. The result (see Fig.~\ref{fig:xscombination}, right) was
\begin{equation}
\sigma_{t\bar{t}} = 177 \pm 3 \stat ^{+8}_{-7} \syst \pm 7 \lumi \rm\ pb \ .
\end{equation}

The relative uncertainties of these combinations correspond to 8\% (6\%) in the case of CMS (ATLAS), comparable with or better than the precision of the various approximate NNLO theory calculations.

\begin{figure}[t]
\centering
\includegraphics[width=0.8\textwidth]{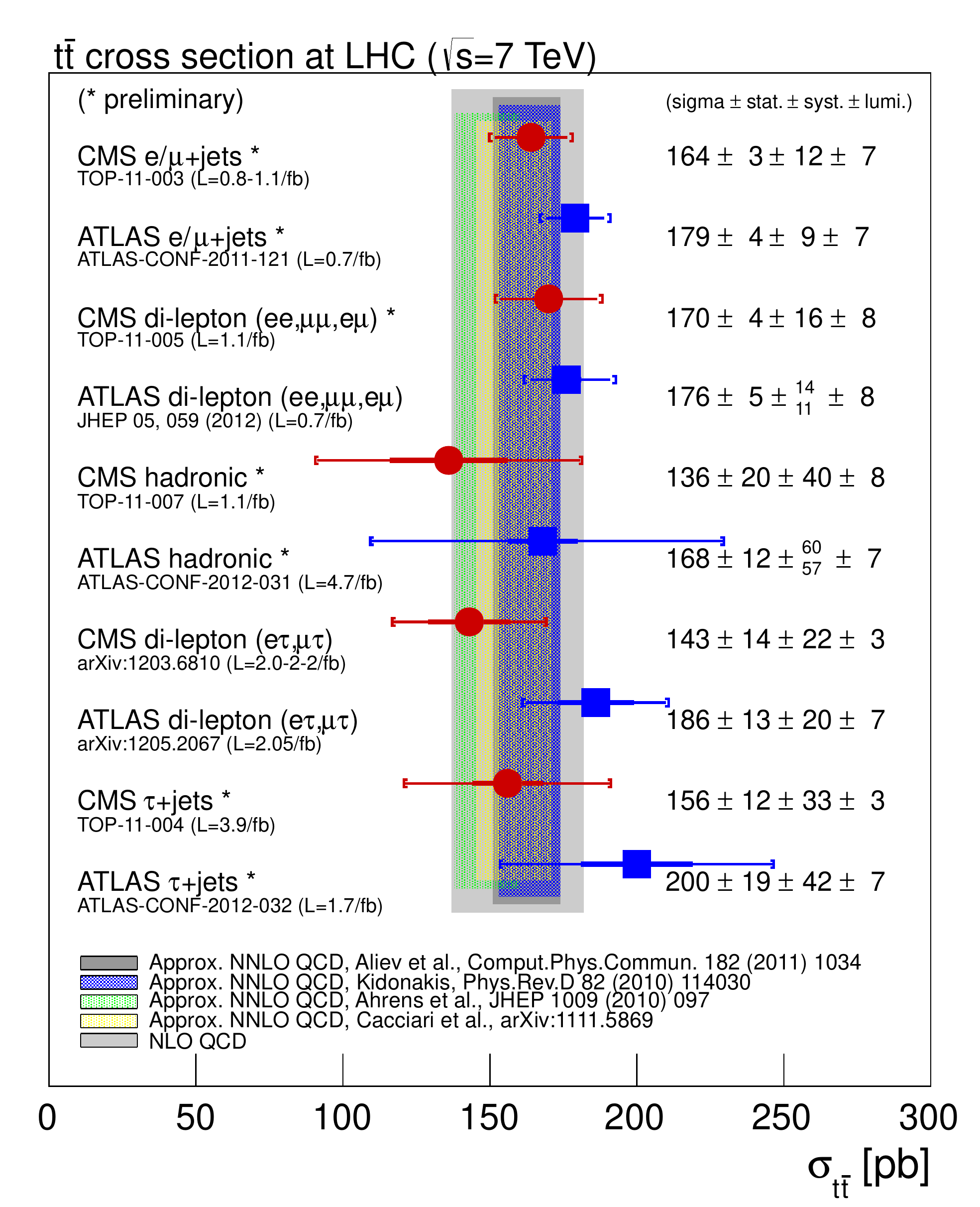}
\caption{Summary of the most precise measurements of $\sigma_{\ttbar}$ per decay mode and experiment, compared with several theory predictions at NLO and approximate NNLO QCD. See section~\ref{sec:xstheo} for more details regarding the calculations. }
\label{fig:summary_ttbar}
\end{figure}

A summary of the most precise $\sigma_{\ttbar}$ measurements per channel and experiment is shown in Fig.~\ref{fig:summary_ttbar}, compared with some of the most precise theory predictions. In general, very good agreement between measurements and theory is observed, which constitutes an important test of the SM in a new energy domain and in a different production mode, compared with the Tevatron. If a significant discrepancy between $\sigma_{\ttbar}$ measurements made in different decay channels would be found, it may be an indication of a new physics signature which impacts the various decay modes in different ways.

%%%%%%%%%%%%%%%%%%%%%%%%%%%%%%%%%%%%%%%%%%%%%%%%%%%%%%%%%%%%%%%%%%%%%%%%%%%%%%%

\subsection{Differential distributions and cross sections}
\label{sec:diffxs}

The large abundance of top quark pair production at LHC allows to measure not only the total cross section $\sigma_\ttbar$, but also differential cross sections $d\sigma_\ttbar / dX$ using relevant variables $X$, for instance those related to the kinematics of the top (anti-) quark or the $\ttbar$ system. They can be used to validate MC models as well as explicit higher order QCD calculations of top quark production. In addition, deviations could signal contributions from new physics. Cross sections may be quoted either after extrapolation to the full phase space (as done in the case of the total cross section), or only within the kinematic range where the decay products are measured within the detector (so-called visible phase space). In order to facilitate comparisons with theoretical models and other experiments, corrections need to be applied to the observed spectra, for which two basic choices exist: If the cross section is defined at the \textit{hadron level} (i.e., after particle decays and hadronization), only the detector response is corrected for. On the other hand, a \textit{parton level} cross section is defined at the level of partons before hadronization. The first definition is closer to what is measured experimentally and can be compared with MC simulations. However, the second definition may be needed in order to compare with fixed order QCD calculations.
The hadron level definition needs to include a prescription to reconstruct the top quark from its decay products.

\begin{figure}[t]
\centering
\begin{minipage}{0.52\linewidth}
\includegraphics[width=0.99\linewidth]{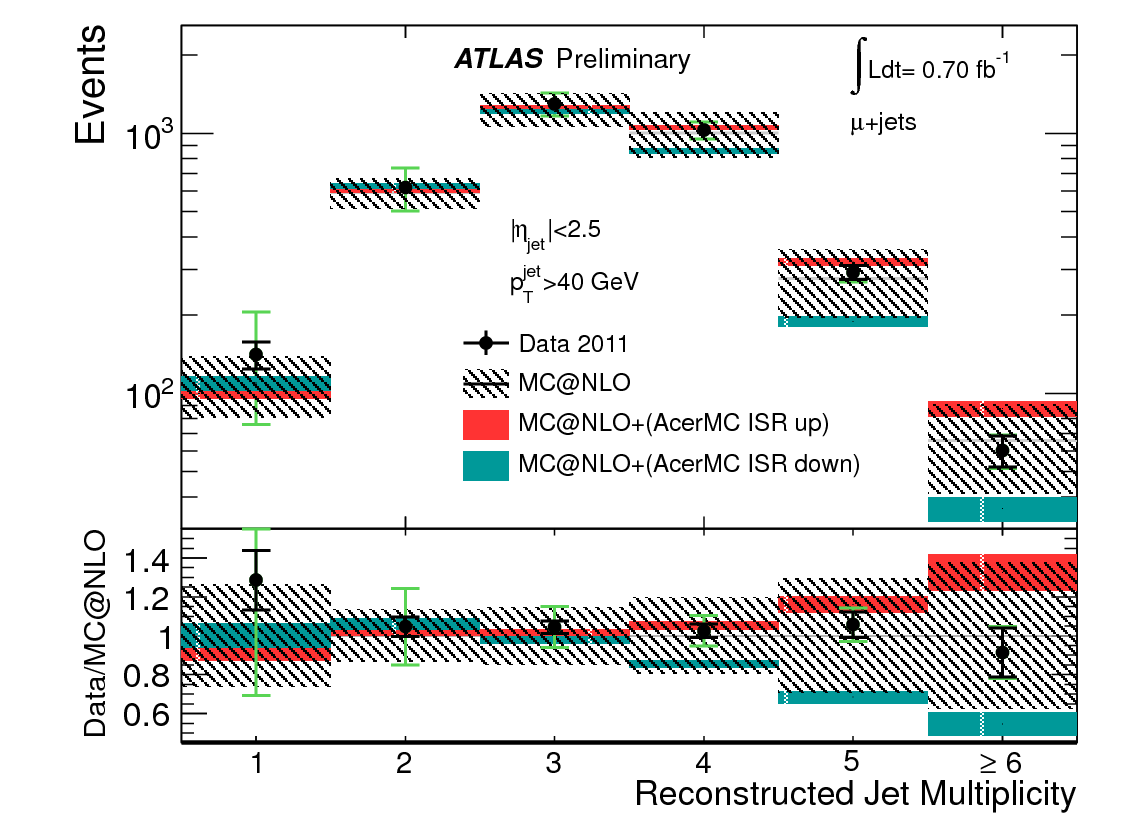}
\end{minipage}
\begin{minipage}{0.47\linewidth}
\includegraphics[width=0.99\linewidth]{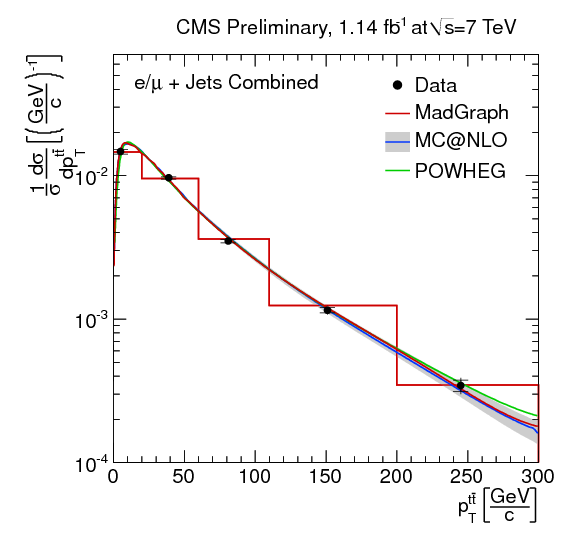}
\end{minipage}
\caption{
Left: Reconstructed jet multiplicity for jets with $p_T>40\GeV$ after background subtraction for $\ttbar$ events in the $\mu$+jets channel from ATLAS~\protect\cite{ATLAS-CONF-2011-142}.
Right: Unfolded and normalized differential top quark pair production cross section as a function of $p_T^\ttbar$ from CMS~\protect\cite{cmspas-top-11-013}.
}
\label{fig:jetmult}
\end{figure}

ATLAS  performed a measurement~\cite{ATLAS-CONF-2011-142} of the jet-multiplicity distribution in the lepton+jets channel using $\Lint = 0.70 \fbinv$ of data. Comparisons with models were however performed at the reconstructed level, no unfolding was attempted.
The event selection required exactly one isolated lepton and at least four  jets with $p_T>25 \GeV$, one of which had to be identified as $b$-jet. 
The background subtracted jet multiplicities were compared with the MC@NLO simulation for jet $p_T$ thresholds of 25, 40 and 60 GeV. In general, good agreement was observed between data and simulation (Fig.~\ref{fig:jetmult}, left). The impact of variations of the amount of ISR in the simulation was studied, but within the uncertainties, the data could not yet distinguish between different models.
See also section~\ref{sec:atlasisr} for an ATLAS measurement of $\ttbar$ production with a veto on additional jet activity.

CMS performed the first measurement of unfolded differential $\ttbar$ production cross sections at LHC using $\Lint=1.14\fbinv$ of data~\cite{cmspas-top-11-013}, using both the di-lepton and lepton+jets channels. The event selection was very similar to the total cross section measurements in these channels.
Cross sections were unfolded to the parton level within the visible phase space, defined in terms of ($p_T,\eta$)-requirements on the final state objects. In the case of the lepton+jets channel, they corresponded to $|\eta|<2.1$, $p_T>30\GeV$ for the lepton from the leptonic $W$-boson decay, and $|\eta|<2.4$, $p_T>30\GeV$ for the $b$-quarks from the top quark decay as well as the light quarks from the hadronic $W$-boson decay. In the di-lepton channel, the visible phase space was defined as $|\eta|<2.4$, $p_T>20\GeV$ for the leptons from the $W$-boson decays, while the requirements on the $b$-quarks were the same as for the lepton+jets channel.

Normalized cross sections of the form $\frac{1}{\sigma} \frac{d\sigma}{dX}$ were measured, which have the advantage of the cancellation of many systematic uncertainties. Since the cross sections were corrected to the parton level, the uncertainty due to the hadronization model had to be determined by comparing samples simulated with POWHEG and MC@NLO using either PYTHIA or HERWIG for hadronization, and it was found to be typically of the order of a few percent. The dominant systematic uncertainties on the normalized differential cross sections originated from the jet energy scale, the lepton selection, the $b$-tagging, and from signal modeling uncertainties. The cross sections were measured differentially as function of leptonic variables such as $p_{T,l}$, $y_l$, but also as function of kinematics of the top (anti-) quark $p_{T,t}$, $y_t$, or the $\ttbar$ system: $p_{T,\ttbar}$, $y_\ttbar$, $M_\ttbar$.
The measured cross sections were compared with MADGRAPH, POWHEG and MC@NLO. Within the uncertainties, good agreement was observed for all measured distributions. As an example, Fig.~\ref{fig:jetmult} (right) shows the differential cross section as function of $p_T^\ttbar$, the transverse momentum of the $\ttbar$ system, in the lepton+jets channel.

Future high statistics data samples will allow more precise measurements, also using smaller measurement intervals. In addition, double differential cross section measurements will become possible, which will allow more detailed comparisons with theory.

\section{Top Quark Mass}
\label{sec:mass}

As discussed in section~\ref{sec:theory}, the mass of the top quark $m_t$ is a fundamental parameter of the SM. At the Tevatron, $m_t$ has been measured precisely by the CDF and D0 experiments and the combined value is $m_t=173.2\pm 0.9 \GeV$~\cite{Lancaster:2011wr}.  For recent reviews on Tevatron $m_t$ measurements, see Refs.~\refcite{Fiedler:2010sy,Galtieri:2011yd}.

In the following, the initial direct measurements at the LHC of $m_t$, 
as well as indirect extractions  using the measured cross section, 
are discussed. In addition, a measurement of the mass difference between the top quark and its anti-quark is presented.
Previous studies on $m_t$ measurements at LHC based on simulation can be found in Refs.~\refcite{cmsnote-2006-077,cmsnote-2006-066,cmsnote-2006-058} for CMS and in Refs.~\refcite{atlastdr,ATL-PHYS-PUB-2010-004} for ATLAS.

\subsection{Direct measurements}

\subsubsection{Di-lepton channel}
\label{sec:cmsmassdil}

CMS measured the top quark mass in the di-lepton channel using $\Lint = 36 \pbinv$ of 2010 data~\cite{Chatrchyan:2011nb}. The event selection followed closely the one of the cross section measurement in this channel using the same dataset (see section~\ref{sec:cmsdil}), requesting at least two jets. The $b$-tagging information was used in order to improve the probability of choosing the correct jets in the reconstruction of the $\ttbar$ system.
For each $\ttbar$ event, its properties are described by 24 variables (the four-momenta of the six final state particles). Because of the presence of two neutrinos in the final state, the $m_t$ reconstruction in the di-lepton channel results in an under-constrained system.
By applying mass constraints, such as $m_W=80.4 \GeV$ or $m_t = m_{\bar{t}}$, in addition to the measured quantities, the number of unknown parameters can be reduced from 24 to one. The mass measurement was carried out using two different methods, originally developed at Tevatron, which differ in the way this remaining unknown is constrained. 
They are the Matrix Weighting Technique (MWT)~\cite{Abbott:1997fv} and the fully kinematic method (KIN)~\cite{Abulencia:2006js}. The improved methods AMWT (analytical MWT) and KINb (KIN using $b$-tagging) are discussed in the following.

In the KINb method, the kinematic equations describing the $\ttbar$ system are solved many times per event for each lepton-jet combination. Each time, the event is reconstructed by varying independently the jet momentum components and the $\MET$ direction within their resolutions. In parallel, the unmeasured value of $p_z^{\ttbar}$ is drawn randomly from a simulated distribution, to fully constrain the $\ttbar$ system. For each set of variations and each combination, there are up to four solutions of the kinematic equations, and the one with the smallest $M_\ttbar$ is accepted if $|m_t-m_{\bar{t}}|< 3 \GeV $. For each event, the jet-lepton combination with the largest number of solutions is chosen and the mass is found by fitting the peak of the $m_t$ distribution of all solutions from the event.

In the AMWT method, the mass of the top quark is used to fully constrain the $\ttbar$ system. The analytical method proposed in Ref.~\refcite{Sonnenschein:2006ud} is used to determine the neutrino momenta. For a given $m_t$ hypothesis, the constraints and the measured observables restrict the $p_T$ of the neutrinos to lie on ellipses in the $p_x$-$p_y$ plane, and their intersections provide the solutions that fulfill the constraints. With two possible lepton-jet combinations, there are up to eight solutions for the neutrino momenta for a given hypothesis of $m_t$. Each event is reconstructed many times scanning $m_t$ in 1 GeV intervals from 100 to 300 GeV. Solutions are often found for large mass intervals. In order to determine a preferred $m_t$ value, a weight~\cite{Dalitz:1991wa} is assigned to each solution which takes into account the PDF and the probability to observe a charged lepton of a particular energy given the assumed $m_t$. For each $m_t$ value, the weights are added for all solutions. Detector resolution effects are taken into account by varying the reconstructed momenta within their resolutions, and averaging the corresponding weights.  For each event, the $m_t$ with the highest averaged weight is taken. 

\begin{figure}[t]
\centering
\begin{minipage}{0.39\linewidth}
\includegraphics[width=0.99\linewidth]{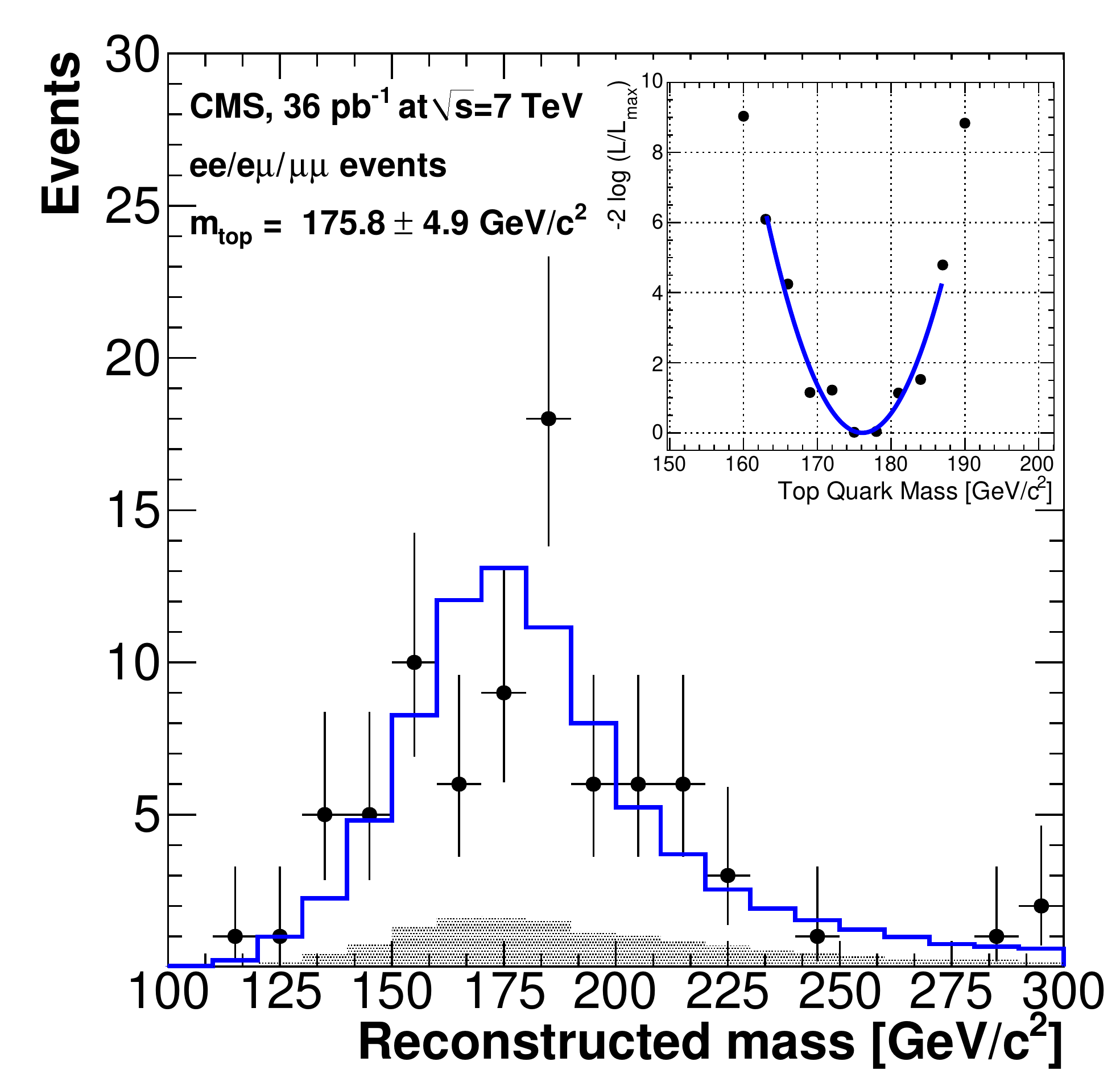}
\end{minipage}
\begin{minipage}{0.59\linewidth}
\vspace{2mm}
\includegraphics[width=0.99\linewidth]{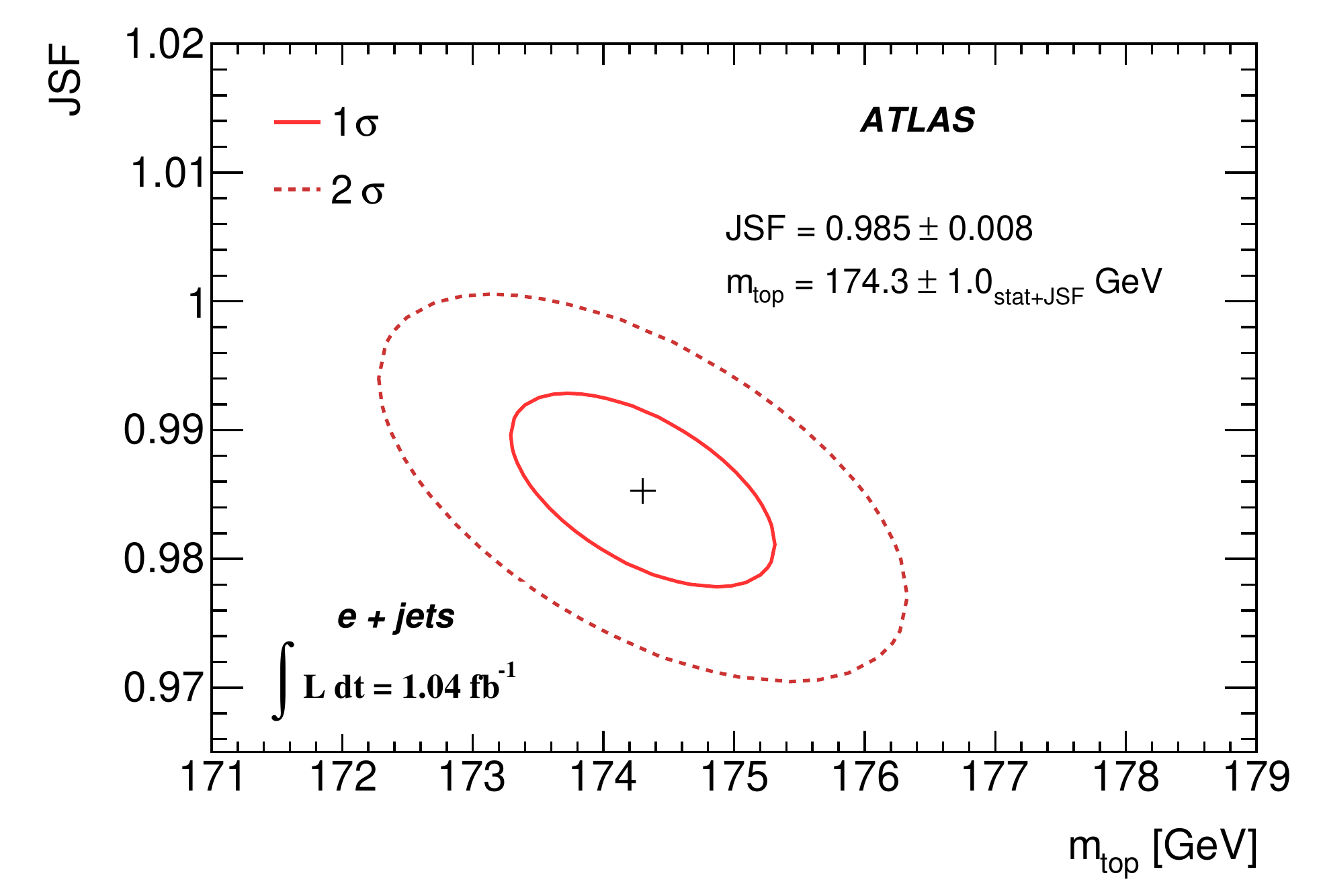}
\end{minipage}
\caption{Left: Reconstructed $m_t$ distribution using the AMWT  method in the di-lepton channel from CMS~\protect\cite{Chatrchyan:2011nb}, showing also the total background plus signal model, and the background-only shape. The inset shows the likelihood as function of $m_t$. Right: Simultaneous determination of $m_t$ and a jet energy scale factor JSF in the $\mu$+jets channel from ATLAS~\protect\cite{atlasmtopljets}.  }
\label{fig:massdilljets}
\end{figure}

In both methods, a maximum likelihood fit to the reconstructed $m_t$ distribution is performed to extract the final $m_t$ value, and a calibration is performed using pseudo-experiments.
Fig.~\ref{fig:massdilljets} (left) shows the $m_t$ distribution obtained with the AMWT method. The minimum value of the negative log-likelihood is taken as the measurement of $m_t$.
The results of the two measurements, which agreed with each other, were combined using the BLUE~\cite{Lyons:1988rp} method, taking into account the statistical correlation of 57\%. The result, in which the weight of the AMWT (KINb) measurements is 0.65 (0.35), was
\begin{equation}
m_t = 175.5 \pm 4.6 \stat \pm 4.6 \syst \GeV \ .
\end{equation}
The dominating systematic uncertainties were due to the $b$-tagging and the jet energy scale uncertainty, where the latter included an overall and a $b$-jet specific contribution~\cite{Chatrchyan:2011ds}, as well as due to signal modeling, simulation of pile-up and underlying event, and the PDF uncertainty.

An updated measurement using $\Lint =2.3 \fbinv$ of 2011 data was performed employing the KINb method~\cite{cmspas-top-11-016}, and using events with at least one $b$-tagged jet. The result was
\begin{equation}
m_t = 173.3 \pm 1.2 \stat ^{+2.5}_{-2.6} \syst \GeV \ ,
\end{equation}
which constitutes the most precise $m_t$ measurement in this channel to date.

%%%%%%%%%%%%%%%%%%%%%%%%%%%%%%%%%%%%%%%%%%%%%%%%%%%%%%%%%%%%%%%%%%%%%%%%%%%%%%%

\subsubsection{Lepton+jets channel}
\label{sec:massljets}

CMS performed a measurement~\cite{CMS-PAS-TOP-10-009} of $m_t$ in the lepton+jets channel using the 2010 dataset of $\Lint = 36 \pbinv$. The event selection corresponded to the one used in the cross section measurement without $b$-tagging using the same channel and dataset~\cite{Chatrchyan:2011ew}, discussed in section~\ref{sec:cmsljets}. Around 800 events with one electron or muon and at least four jets were selected.
A secondary vertex $b$-tagging algorithm was used in the mass measurement, but not in the event selection.
The top quark mass was measured using the \textit{Ideogram} method, which had previously been used for $m_W$ measurements at LEP~\cite{:2008xh} as well as for $m_t$ measurements at Tevatron \cite{Abazov:2007rk,Aaltonen:2006xc}. 

In this method, a constrained kinematic fit is performed for all 24 possible jet-parton assignments (including the ambiguity due to the $p_z$ of the neutrino, constrained by requiring $m_t=m_{\bar{t}}$). 
For each event, a likelihood to observe the event is calculated as function of the assumed value of $m_t$, consisting of two terms corresponding to the probability of the event to be either signal or background.
The signal probability density itself consists of two terms, corresponding to the correct (wrong) assignment of jets to partons. The signal and background probability densities are parameterized using analytic functions, derived from simulation. An overall sample likelihood is constructed by multiplying all event likelihoods. The measurement of $m_t$ and its uncertainty is then obtained from the position and shape of the maximum of the sample likelihood curve. The method is calibrated in simulation using pseudo-experiments.

The result of the measurement using 2010 data was
\begin{equation}
m_t = 173.1 \pm 2.1 \stat ^{+2.8}_{-2.5} \syst \GeV \ .
\end{equation}
The systematic uncertainty was largely dominated by the jet energy scale uncertainty. Other important contributions were due to the signal and background modeling.
The combination of this measurement with the one in the di-lepton channel discussed in section~\ref{sec:cmsmassdil} using the BLUE method yielded
\begin{equation}
m_t = 173.4 \pm 1.9 \stat \pm 2.7 \syst \GeV \ ,
\end{equation}
consistent with the world average.

A simultaneous measurement of $m_t$ and the jet energy scale was performed by CMS in the muon+jets channel, using the full 2011 dataset corresponding to $\Lint=4.7 \fbinv$~\cite{cmspas-top-11-015}. Around 2400 events were selected in the final state with one muon and at least four jets, of which at least two were required to be identified as $b$-jets.
The measurement was performed by employing a kinematic fit and extending the ideogram method discussed above. Two-dimensional (2d) event likelihoods were constructed as function of $m_t$ and a jet energy scale parameter. The latter was estimated in-situ using the reconstructed $W$-boson mass, before $m_W$ was constrained to the world average value in the kinematic fit. The most likely values for $m_t$ and the JES parameter were obtained by minimizing the 2d sample likelihood, obtained from multiplying the event likelihoods. The method was calibrated using pseudo-experiments, and the result for $m_t$ was
\begin{equation}
m_t = 172.6 \pm 0.6 {\rm\ (stat.+JES)} \pm 1.2 \syst \GeV \ ,
\end{equation}
while the JES parameter was found to be consistent with unity: $1.004 \pm 005 \stat \pm 0.012 \syst$. Important contributions to the systematic uncertainty came from the $b$-jet energy scale uncertainty and the signal modeling. The total uncertainty of this result is below $1\%$, an impressive achievement in this early phase of LHC operation. However, it should be noted that the uncertainty due to the modeling of color reconnection  and the underlying event have not been evaluated yet.
As a cross check, the analysis performed on 2010 data~\cite{CMS-PAS-TOP-10-009} discussed above was repeated on the 2011 dataset in the muon+jets channel. The result was $m_t=172.6 \pm 0.2 \stat \pm 1.8 \syst \GeV$, 
consistent with the main analysis.

%%%%%%%%%%%%%%%%%%%%%%%%%%%%%%%%%%%%%%%%%%%%%%%%%%%%%%%%%%%%%%%%%%%%%%%%%%%%%%%

ATLAS presented  measurements of $m_t$ in the lepton+jets channel with three different methods, using 2010 data ($\Lint = 35 \pbinv$)~\cite{ATLAS-CONF-2011-033} and also 2011 data ($\Lint = 1.04 \fbinv$)~\cite{atlasmtopljets}. In the analyses, events were used which contained exactly one isolated lepton and at least four jets, of which at least one was required to be identified as b-jet.

The first method (1d-$R_{32}$) is a one-dimensional (1d) template analysis which is based on the measurement of the mass ratio $R_{32}= m^{reco}_t / m^{reco}_W$. Using this ratio provides stability against jet energy scale variations and avoids the need for an in-situ calibration technique. For each event, the three jets which maximize their vectorially summed $p_T$ are assumed to originate from the hadronically decaying top quark. The two out of three jets originating from the hadronically decaying $W$-boson are found trivially in case one of the three jets is $b$-tagged, while otherwise the two jets with smallest $\Delta R$ are chosen. 
The reconstructed $R_{32}$ distribution can be described by the sum of signal and background templates.
An un-binned likelihood fit was performed to the data, from which the value of $m_t$ was determined. The calibration of the method was verified, and its systematic uncertainties evaluated, using pseudo-experiments with simulated samples. Using 2010 data, the result was
\begin{equation}
m_t = 169.3 \pm 4.0 \stat \pm 4.9 \syst \GeV \ ,
\end{equation}
corresponding to the most precise ATLAS measurement using 2010 data.
The systematic error is dominated by contributions due to the jet energy scale uncertainty, as well as the signal modeling and the background normalization.
In the 2011 version of this analysis, the jet triplet to calculate $R_{32}$ was chosen using a kinematic fit maximizing an event likelihood, which considerably improved the correct jet matching probability. 
The updated result was
$m_t = 174.4 \pm 0.9 \stat \pm 2.5 \syst \GeV$, less precise than the one based on the same dataset employing the 2d-analysis (see below).

The second method (2d-analysis) is a 2d template analysis in which $m_t$ and a global jet energy scale factor (JSF) are determined simultaneously, using the $m_t^{reco}$ and $m_W^{reco}$ distributions. The value of JSF is mainly determined from the difference between observed and predicted $m_W^{reco}$ distribution, which results in a reduced systematic uncertainty of $m_t$. Also, the known value of $m_W$ is used to improve on the resolution on $m_t$. For each event, a kinematic fit was performed imposing the hypothesis of a $W$-boson decay for each pair of non-$b$-tagged jets for which the value of $m_W^{reco}$ was within a window of approx three sigma around the peak of the distribution. The fit provides the best light jet combination per event and the corresponding parton scale factors for their jet energies. The hadronic top quark was then reconstructed by adding the $b$-jet to this jet pair, and $m_t^{reco}$ was calculated using the rescaled light jet energies.
Templates depending on $m_t$ and the JSF were constructed for $m_t^{reco}$, while those for $m_W^{reco}$ depend only on JSF. 
Fitting the templates to the distributions observed in data, the result using the 2010 dataset was
$m_t = 166.1 \pm 4.6 \stat \pm 4.4 \syst \GeV$,
where the measured JSF was $1.08^{+0.04}_{-0.06} \stat$ ($1.01^{+0.05}_{-0.05} \stat$) in the electron (muon) channel. The systematic uncertainty on $m_t$ due to the jet energy scale was significantly reduced, as expected, while the statistical error of the measurement was somewhat increased. 
In the 2011 version of this analysis, the kinematic fit was not used to select the best light jet combination, but was only used to determine the scale factors for the jet energies. Instead, every light jet pair was combined with every $b$-tagged jet and the triplet with the maximum $p_T$ defined the top quark candidate.  The result of the 2d maximum likelihood fit for the $\mu$+jets channel is shown in Fig.~\ref{fig:massdilljets} (right), and the value of $m_t$  was determined as
\begin{equation}
m_t = 174.5 \pm 0.6 \stat \pm 2.3 \syst \GeV \ .
\end{equation}
The statistical uncertainty is of the same size as the one of the Tevatron combination.
Important contributions to the systematic uncertainty originated from the relative jet energy scale uncertainty for $b$-jets with respect to light jets, as well as from various aspects of the signal modeling (choice of MC generator, amount of ISR/FSR). In particular, the uncertainties due to the modeling of the underlying event and of color reconnection effects were estimated to be of the order of 0.6 GeV each. The latter was obtained from using two different assumptions on the size of color reconnection.

A third method (1d-kinfit), was also applied to the 2010 dataset. It employed a 1d template analysis using a kinematic likelihood fit, which was used to relate the observed objects to parton level predictions, employing transfer functions obtained from simulation. It yielded a consistent, but less precise measurement compared with the other methods.

%%%%%%%%%%%%%%%%%%%%%%%%%%%%%%%%%%%%%%%%%%%%%%%%%%%%%%%%%%%%%%%%%%%%%%%%%%%%%%%

\subsubsection{Hadronic channel}

ATLAS measured $m_t$ in the hadronic channel using $\Lint = 2.04 \fbinv$ of data~\cite{ATLAS-CONF-2012-030}. The event selection matched the one used in the cross section measurement in the same channel (section~\ref{sec:xshadronic}). For each event and jet permutation, a $\chi^2$ function, which tests the compatibility of the reconstructed $m_{jj}$ and $m_{jjb}$ invariant masses with the expected values for $m_W$ and $m_t$, was minimized as a function of $m_W$ and $m_t$, and the combination with the smallest $\chi^2$ value was taken. For the determination of the two $m_{jjb}$ values per event which enter the $m_t$ extraction, the untagged jet energies were rescaled according to the ratio of measured and expected $m_W$ values. The value of $m_t$ was extracted using a binned maximum likelihood fit to the $m_{jjb}$ distribution, employing $m_t$-dependent signal templates from simulation, and background templates modeled using an event mixing technique. The result was
\begin{equation}
m_t = 174.9 \pm 2.1 \stat \pm 3.8 \syst \GeV \ ,
\end{equation}
and the systematic error was dominated by the jet energy scale uncertainty as well as the modeling of signal and background templates.

%%%%%%%%%%%%%%%%%%%%%%%%%%%%%%%%%%%%%%%%%%%%%%%%%%%%%%%%%%%%%%%%%%%%%%%%%%%%%%%

\subsubsection{Summary and outlook}

\begin{figure}[t]
\centering
\includegraphics[width=0.8\textwidth]{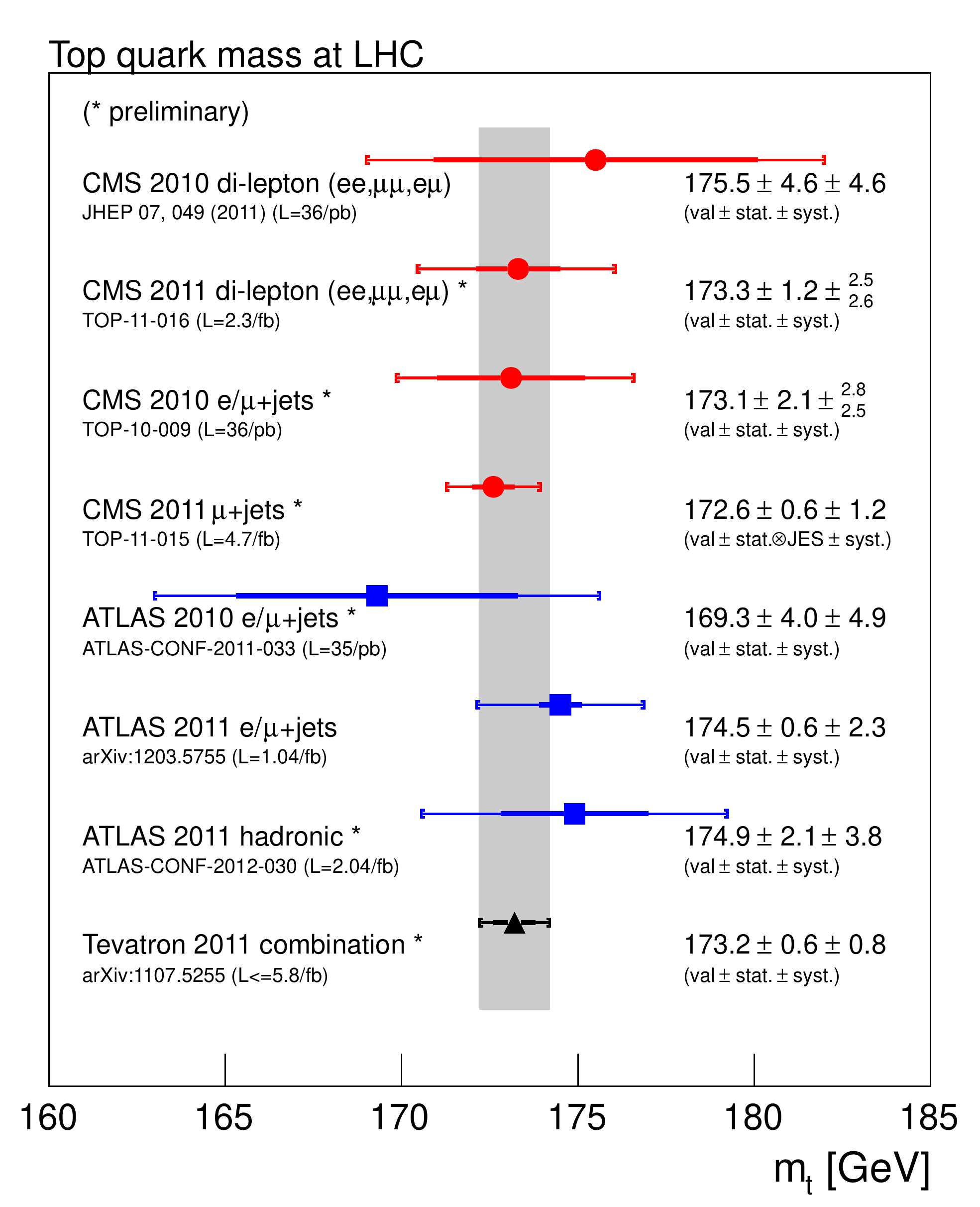}
\caption{Summary of measurements of $m_t$ performed at LHC, compared with the Tevatron average. }
\label{fig:summary_mtop}
\end{figure}

Fig.~\ref{fig:summary_mtop} shows a summary of the $m_t$ measurements at the LHC, compared with the current Tevatron average.
In addition, CMS performed a combination of their measurements~\cite{cms-pas-top-11-018}, which yielded
\begin{equation}
m_t = 172.6 \pm 0.4 \stat \pm 1.2 \syst \GeV \ .
\end{equation}

Future measurements of $m_t$ with the goal to reduce the total uncertainty well below one GeV need to improve considerably on the understanding of the uncertainties related to the modeling of top quark pair production in the simulation. This concerns in particular effects due to perturbative higher order corrections, but also the modeling of hadronization, the underlying event and color reconnection. For instance, the latter is expected to induce additional uncertainties of the size of the order of one GeV~\cite{Skands:2007zg}. Beyond merely studying the impact of such effects by comparing various models, the data themselves should be used to constrain these sources of uncertainty. This could be done for instance by performing measurements of $m_t$ in a differential way, e.g. as a function of $p_T$ of the top quark, or other variables. Such differential measurements become possible due to the large top quark event samples produced at LHC.

Regarding experimental uncertainties, one important area for potential improvement concerns  the knowledge of the jet energy scale. Given the already achieved accuracy, large improvements may not happen on a short time scale. On the other hand, methods to measure $m_t$ which simultaneously constrain the JES, ideally separately for light quark and $b$-jets, are very promising.
In addition, it will be important to apply different measurement techniques with complementary systematics. Examples are methods employing the correlation of $m_t$ with the lepton $p_T$ (which is expected to be less sensitive to modeling of underlying event and color reconnection) or the transverse decay length of a $b$-tagged jet, both having already been applied at the Tevatron~\cite{Abulencia:2006rz,Aaltonen:2009hd,Aaltonen:2011wt}.
Another method, described in Refs.~\refcite{Kharchilava:1999yj,cmsnote-2006-058}, uses leptonic $W$-boson decays and $J/\psi \rightarrow ll$ final states from the fragmentation of the $b$-quark, employing the correlation of $m_t$ with the invariant mass of the lepton from the $W$-boson decay and the $J/\psi$. All of these methods have the advantage that they are largely independent of the jet energy scale, and are feasible given the large samples of $\ttbar$ events available at LHC. 
Other ideas include making use of the invariant mass of the lepton and the b-jet~\cite{Biswas:2010sa}, for which NLO QCD predictions are available~\cite{Melnikov:2009dn}, or of the $\ttbar$ invariant mass distribution~\cite{Frederix:2007gi}.
As far as tools are concerned, an extension to NLO of the matrix element method~\cite{Fiedler:2010sg}, which was employed for $m_t$ measurements at Tevatron, has recently become available~\cite{Campbell:2012cz}.

Improved understanding is also needed on the theoretical meaning of the measured $m_t$, which is usually identified with $m_t^{(MC)} = \mtpole$, but which faces an uncertainty due to non-perturbative effects. In addition, future global electroweak fits using very precise measurements of $m_t$ and $m_W$ as (uncorrelated) inputs may need to consider the fact that often constraints on $m_W$ are applied in measurements of $m_t$, though this is not yet an issue given the present precision on $m_t$.

%%%%%%%%%%%%%%%%%%%%%%%%%%%%%%%%%%%%%%%%%%%%%%%%%%%%%%%%%%%%%%%%%%%%%%%%%%%%%%%

\subsection{Indirect mass measurement from the cross section}
\label{sec:massfromxsec}

Direct measurements of $m_t$, as the ones discussed in the previous sections, rely on experimental observables sensitive to $m_t$, as well as on a calibration of the method with the use of MC simulation which introduces modeling uncertainties. Moreover, the measurement is performed with respect to a definition of $m_t$ in the MC generator which usually does not exactly correspond to a specific renormalization scheme.
An alternative approach for the extraction of $m_t$ is provided by the dependence of the $\ttbar$ cross section on the top quark mass, $\sigma_{\ttbar}(m_t)$, as calculated in perturbative QCD. This dependence is several (about 4-5) times larger than the dependence of the experimental acceptance on the $m_t$ value used in the MC simulation in the $\sigma_\ttbar$ measurement. Therefore, experimental measurements of the cross section can be used for an indirect determination of $m_t$. Besides the complementarity, this method allows to extract $m_t$ in well-defined renormalization schemes used in the $\sigma_\ttbar$ calculation, in particular using the pole mass $\mtpole$ or $\overline{MS}$-mass $\mtmsbar$ definitions.

\begin{figure}[t]
\centering
\begin{minipage}{0.49\linewidth}
\centering
%\vspace{2mm}
\includegraphics[width=0.99\linewidth]{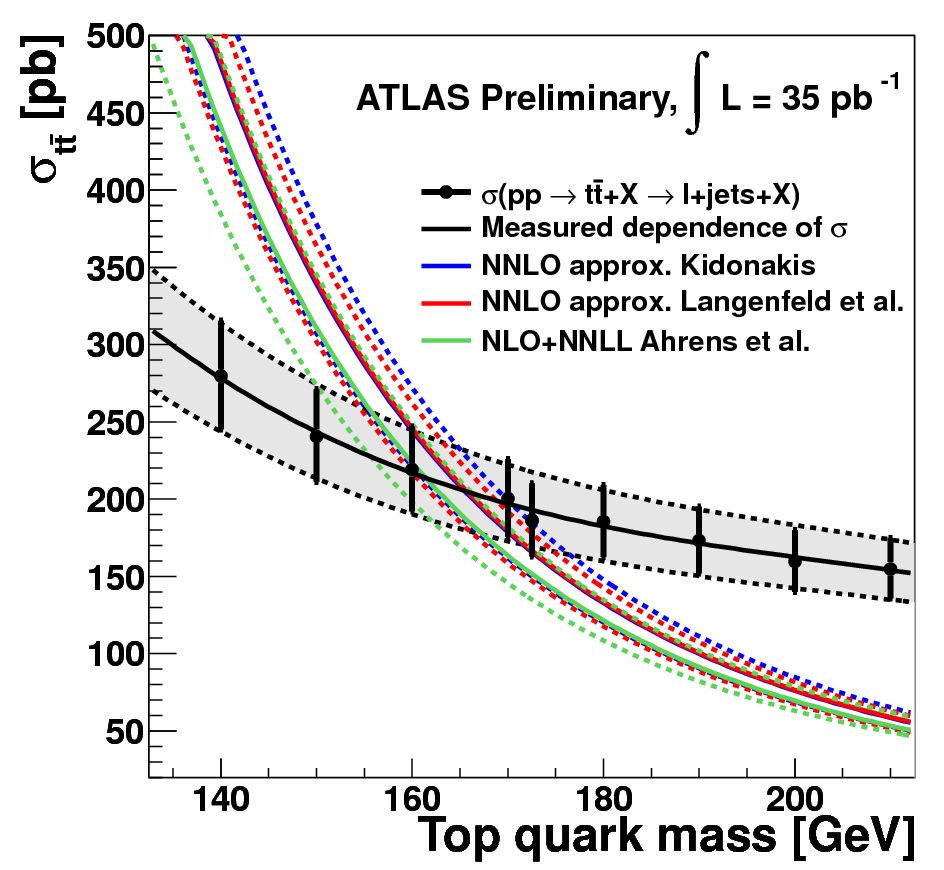}
\end{minipage}
\begin{minipage}{0.49\linewidth}
\centering
%\vspace{5mm}
\includegraphics[width=0.99\linewidth]{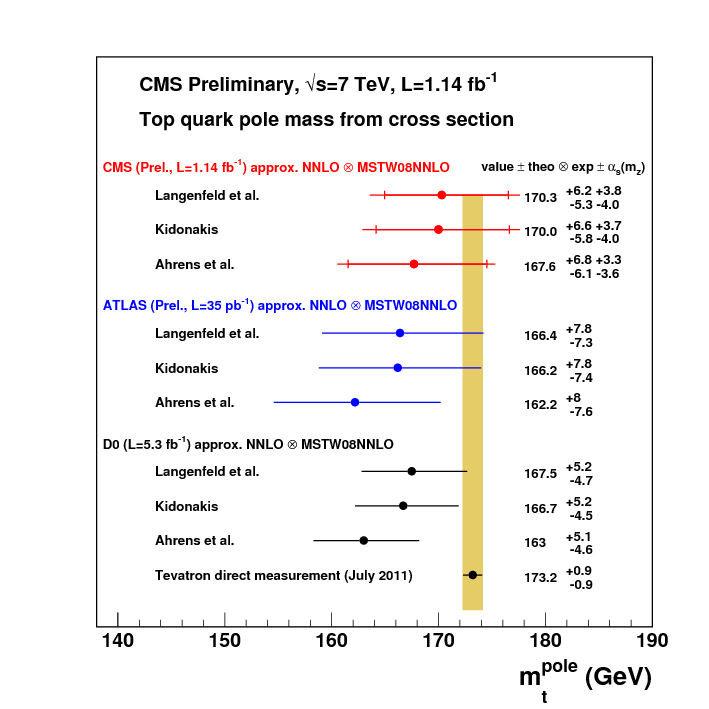}
\end{minipage}
\caption{Left: Dependence of the ATLAS $\ttbar$ cross section measurement, as well as of various theory calculations, on $m_t$~\protect\cite{ATLAS-CONF-2011-054}. Right: Summary of indirect measurements of $\mtpole$ from various experiments~\protect\cite{CMS-PAS-TOP-11-008}.}
\label{fig:mtopfromxs}
\end{figure}

Both ATLAS~\cite{ATLAS-CONF-2011-054} ($\Lint=35\pbinv$, using the cross section measurement in the lepton+jets channel with $b$-tagging) and CMS~\cite{CMS-PAS-TOP-11-008} ($\Lint=1.14\fbinv$, using the cross section measurement in the di-lepton channel) performed such extractions, following the technique used by D0 as described in Ref.~\refcite{Abazov:2011pta}. 
The mass value was extracted by maximizing the likelihood function
\begin{equation}
\mathcal{L}(m_t) \sim \int f_{exp}(\sigma_{\ttbar}|m_t) f_{th}(\sigma_{\ttbar}|m_t) d\sigma_{\ttbar} \ ,
\end{equation}
where $f_{exp}$ and $f_{th}$ are probability density functions parameterizing either the dependence of the experimentally measured cross section on $m_t$ via the acceptance, or the dependence of the theory cross section on $m_t$, respectively, including the corresponding uncertainties (Fig.\ref{fig:mtopfromxs}, left).

A compilation of $\mtpole$ determinations from ATLAS, CMS and D0 using the theory calculations of Langenfeld et al.~\cite{Langenfeld:2009wd}, Kidonakis~\cite{Kidonakis:2010dk} and Ahrens et al.~\cite{Ahrens:2010zv}, all interfaced with the MSTW08NNLO PDF, is shown in Fig.~\ref{fig:mtopfromxs} (right).
The uncertainties on the extracted $\mtpole$ values at LHC were of the order of $\mathcal{O}(7-8 \GeV)$. They included, in addition to the uncertainties in the theory cross section (due to factorization scale and PDF) and those in the cross section measurement, a contribution from the assumption of identifying the mass used in the MC generator with the pole mass $m_t^{(MC)} = \mtpole$. The latter was evaluated by varying $m_t^{(MC)}\pm 1 \GeV$. CMS also included an additional uncertainty from varying $\alpha_s(M_Z)$ in the PDF. The extracted $\mtpole$ values were found to be consistent with the direct measurements, as well as with the world average. 

CMS also performed an extraction of $m_t$ in the $\overline{MS}$ scheme and the results are $\mtmsbar= 163.1^{+6.8}_{-6.1} \GeV$ using the calculation from Langenfeld et al.~\cite{Langenfeld:2009wd}, and $\mtmsbar 159.8^{+7.3}_{-6.8} \GeV$ using the one from Ahrens et al.~\cite{Ahrens:2010zv}, respectively. In addition, $m_t$ extractions in both mass schemes were performed by CMS using the HERAPDF15NNLO PDF set, yielding $m_t$ values which are 1.2-1.5 GeV higher compared with the ones obtained when using MSTW08NNLO.

Future measurements of $\sigma_\ttbar$ with reduced uncertainty, as well as improved theory, in particular the full NNLO QCD calculation, will allow for more precise determinations of $m_t$ in a well defined renormalization scheme, and provide an important cross check of the direct measurements.

%%%%%%%%%%%%%%%%%%%%%%%%%%%%%%%%%%%%%%%%%%%%%%%%%%%%%%%%%%%%%%%%%%%%%%%%%%%%%%%

\subsection{Mass difference between $t$ and $\bar{t}$}

A difference in mass between particle and anti-particle would indicate a violation of CPT symmetry. Since the top quark has a very short lifetime and decays before it can hadronize, it is the only quark for which the quark - anti-quark mass difference can be directly measured. Previous measurements of the top quark - anti-quark mass difference were performed at the Tevatron~\cite{Aaltonen:2011wr,Abazov:2011ch}.

CMS presented a measurement of $\Delta m_t = m_t - m_{\bar{t}}$ using $\Lint = 4.96 \fbinv$ of 2011 data~\cite{top-11-019-paper}. The analysis used the lepton+jets channel and the ideogram method, as discussed in section~\ref{sec:massljets}, to measure $m_{t(\bar{t})}$ separately in the $\mu^+$+jets and $\mu^-$+jets sub-samples. The method was modified in this analysis to use a different kinematic fit and only the hadronic part of the $\ttbar$ decay.  Having obtained individual measurements of $m_t$ and $m_{\bar{t}}$, the two measurements were subtracted to calculate $\Delta m_t$.
Most of the systematic uncertainties important for a measurement of $m_t$
are expected to be significantly reduced or cancel out completely when measuring the $m_t$-$m_{\bar{t}}$ difference. 
On the other hand, the measurement could be influenced from systematic effects related to lepton charge identification or a possible asymmetry in jet response to $b$ and $\bar{b}$ quarks. The final result was
\begin{equation}
\Delta m_t = -0.44 \pm 0.46 \stat \pm 0.27 \syst \GeV \ ,
\end{equation}
where the precision of the measurement was limited by the statistical uncertainty, and the dominant contributions to the systematic uncertainty arose from the $b$- vs. $\bar{b}$-jet response, pile-up modeling, cross section difference between $W^+$ and $W^-$ background, and the statistical uncertainty on the difference in mass calibration between the two sub-samples.
The measured value is in agreement with the SM expectation. This represents the most precise measurement to date of this quantity.

\section{Other Properties of Top Quarks and their Interactions}
\label{sec:prop}

In this section, further measurements of the properties of top quarks and their interactions are discussed, such as measurements of the top quark charge, the ratio of branching fractions $R=\BR(t\rightarrow W b) / \BR(t\rightarrow W q) $,  $\ttbar$ spin correlation and $W$-boson polarization in top decays, the production of $\ttbar$ in association with a photon, the charge asymmetry in $\ttbar$ production, and finally a measurement of additional jet activity in $\ttbar$ events.

\subsection{Top quark charge}
\label{sec:charge}

In the SM, the top quark has a charge $Q=+\frac{2}{3}e$, which can be verified experimentally by measuring the charges of its decay products. In particular, the hypothesis that an exotic quark with charge $Q=-\frac{4}{3}e$ is produced can be tested. Previous measurements at the Tevatron~\cite{Abazov:2006vd,Aaltonen:2010js} have excluded this possibility at 95\% CL. A  study for the LHC based on simulation can be found in Ref.~\refcite{atlastdr}  

\begin{figure}[t]
\centering
\begin{minipage}{0.60\linewidth}
\centering
\vspace*{1mm}
\includegraphics[width=0.99\linewidth]{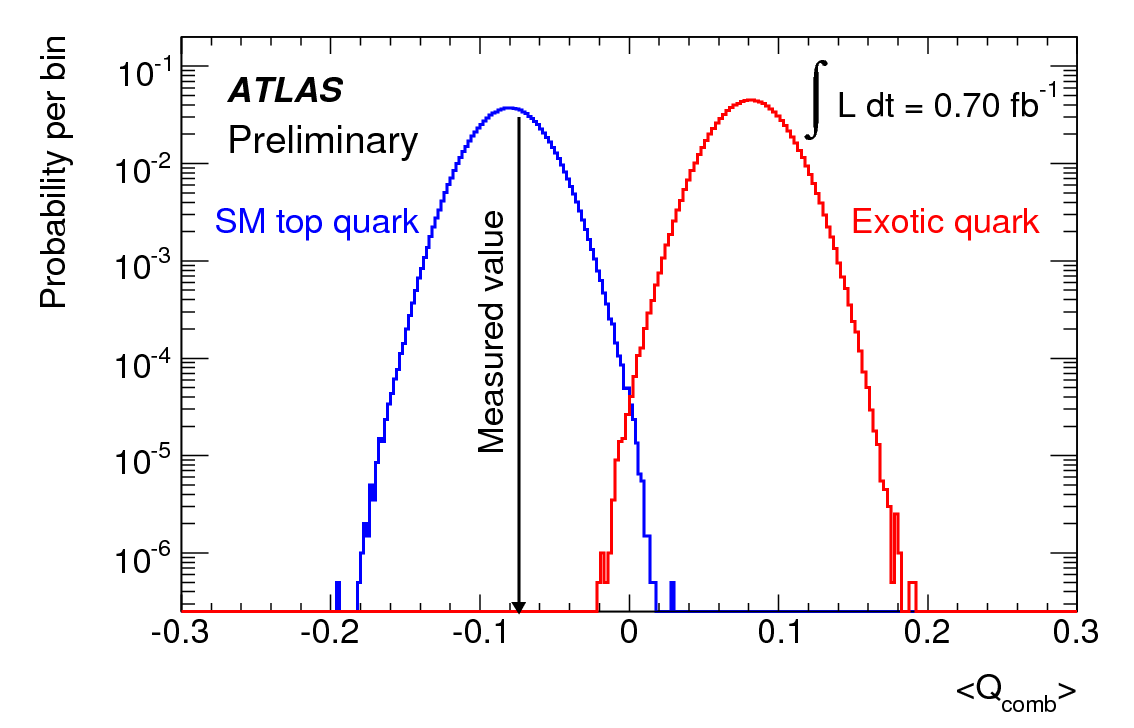}
\end{minipage}
\begin{minipage}{0.39\linewidth}
\centering
\includegraphics[width=0.99\linewidth]{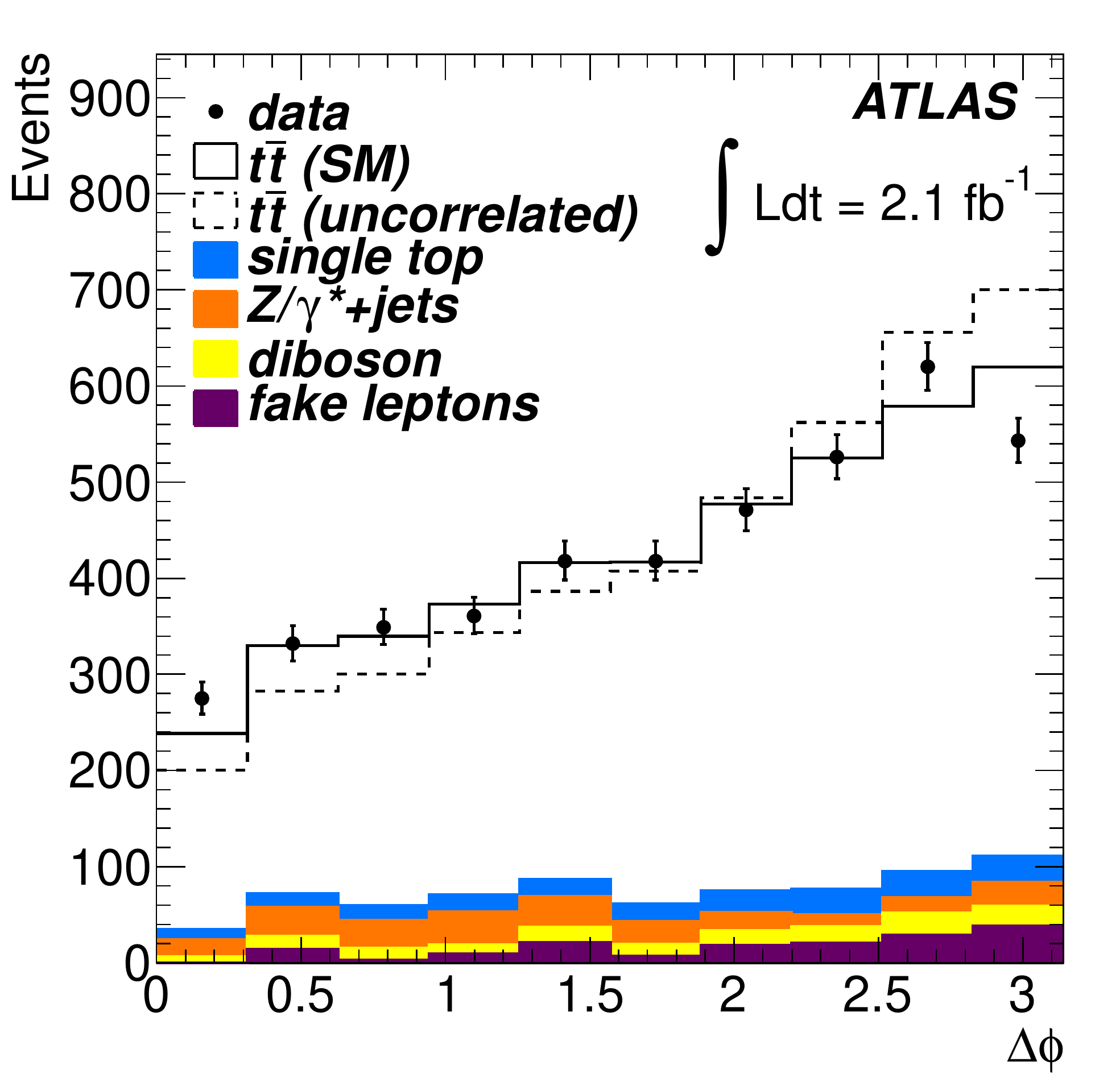}
\end{minipage}
\caption{Left: Distributions of the average charge product for the charge weighting method for SM and exotic top quark charge scenarios, compared with the measured value from ATLAS~\protect\cite{ATLAS-CONF-2011-141}. Right: Distribution of the difference in azimuthal angle of the two leptons in the ATLAS spin correlation measurement~\protect\cite{Collaboration:2012sm}.}
\label{fig:chargespin}
\end{figure}

ATLAS presented a measurement of the top quark charge using $\Lint=0.70\fbinv$ of 2011 data in the lepton+jets channel~\cite{ATLAS-CONF-2011-141}.  In the $t\rightarrow l \nu_l b$ decay mode, the measurement of the lepton charge is straightforward, but it must be combined with a charge measurement of the $b$-quark, which has fragmented to a jet, in order to obtain the top quark charge. Two methods were used for the $b$-quark charge determination. The first uses a weighted sum of the charges of tracks associated with the $b$-jet (track charge method), while the second employs the charge of the lepton produced in a semi-leptonic b-decay (soft lepton method). 
The soft lepton method uses the fact that the charge of the soft muon produced in the semi-leptonic decay $b\rightarrow \mu \nu X$ ($\BR\sim 11\%$) is equal to the $b$-quark charge. However, this correlation is diluted due to cascade decays involving a $c$-quark $b\rightarrow c \rightarrow \mu \nu X$ as well as due to $B^0$-meson oscillations.
Events were selected in a similar way as for ATLAS measurements of $m_t$ in the same channel, requiring one isolated lepton and at least four jets, one of which had to be $b$-tagged. In addition, for the track charge method, a second $b$-jet was required, while for the soft lepton method, a muon with $p_T>4 \GeV$ had to be identified in one of the jets. 
In both methods, the average charge product $<Q_{comb} = Q_l \cdot Q_{b-jet}>$ 
of lepton and $b$-jet was evaluated and the results were found to be consistent with standard model expectations. While statistical and systematic uncertainties were of similar size, the latter were dominated by the signal modeling in simulation. The (dis-) agreement of the measurement with the theoretically expected values for a top quark with charge $-\frac{4}{3}$ or $+\frac{2}{3}$ was statistically interpreted to derive an exclusion of the production of an exotic top quark with charge $-\frac{4}{3}$ (Fig.~\ref{fig:chargespin} left), and both charge reconstruction methods yielded an exclusion of the exotic scenario with more than five sigma CL.

CMS performed a similar measurement in the muon+jets channel, using a dataset corresponding to $\Lint = 4.6 \fbinv$~\cite{cmspas-top-11-031}. Events containing one isolated muon and at least four jets were selected, of which two were required to be $b$-tagged. Invariant mass constraints were applied in order to associate the $b$-jets with the leptonic and hadronic top decays. Similarly to the soft lepton method used by ATLAS, the $b$-quark charge was obtained requiring a soft muon in one of the $b$-jets ($p_T>4 \GeV$, $\Delta R <0.4$), originating from a semi-leptonic decay of the $B$-hadron. The $b$-quark leading to the other $b$-jet was defined to be of opposite charge. The overall precision of the measurement could be improved by applying the requirement $p_{T,rel}> 0.85 \GeV$ on the transverse momentum of the muon with respect to the jet axis. The $b$-charge assignment purity was found to be consistent between simulation and data, using a control sample enriched in QCD $b\bar{b}$ production. According to the charge, the events were assigned to two categories, $-\frac{4}{3}$ and $+\frac{2}{3}$, and an asymmetry was constructed, taking into account background and dilution factors. The systematic uncertainty was dominated by the $b$-jet charge assignment and the signal modeling. The measured asymmetry was found to be in agreement with the SM prediction, and the exotic top quark charge of $-\frac{4}{3}$ could be excluded with high significance. 

The cross section for the production of a $\ttbar$ pair in association with a photon is also sensitive to the top quark charge, see section~\ref{sec:ttgamma}.

%%%%%%%%%%%%%%%%%%%%%%%%%%%%%%%%%%%%%%%%%%%%%%%%%%%%%%%%%%%%%%%%%%%%%%%%%%%%%%%

\subsection{Measurement of R}
\label{sec:r}

As discussed in section~\ref{sec:topdecay}, a deviation from unity of the measured value of the ratio of branching fractions $R=\frac{\BR(t\rightarrow W b) }{ \BR(t\rightarrow W q) }$ could be an indication of a fourth generation of fermions, and a recent result from D0~\cite{Abazov:2011zk} yielded a value which is around $2.5\sigma$ smaller than unity. 

CMS performed a measurement of $R$ using a data sample corresponding to $\Lint = 2.2 \fbinv$ in the di-lepton channel~\cite{cmspas-top-11-029}. The event selection followed the one of the cross section measurement in this channel, and the dominant background from Drell-Yan production was estimated from data.
The value of $R$ was determined from a profile likelihood fit to the observed multiplicities of reconstructed $b$-jets, which took into account the $b$-tagging efficiency and mis-tagging rate, as well as contributions from non-top backgrounds and events with mis-assigned jets, obtained using a data-driven method. Events were considered not only with two, but also with three reconstructed jets, in order to constrain the contribution of jets from ISR. The result was
\begin{equation}
R = 1.00 \pm 0.04 \statsyst \ ,
\end{equation}
where the systematic uncertainty was dominated by the knowledge of the $b$-tagging efficiency, as well as the signal modeling. Using a Feldman-Cousins frequentist method, the 95\% CL interval was determined as $0.85<R<1.0$. The measurement is in very good agreement with the SM.

%%%%%%%%%%%%%%%%%%%%%%%%%%%%%%%%%%%%%%%%%%%%%%%%%%%%%%%%%%%%%%%%%%%%%%%%%%%%%%%

\subsection{$\ttbar$ spin correlation}
\label{sec:spincorr}

The lifetime of the top quark is more than one order of magnitude smaller than the timescale for hadronization. It is thus possible to measure the spin correlation between the $\ttbar$ pair from angular correlations of their decay products and test the corresponding SM predictions~\cite{Stelzer:1995gc,Bernreuther:2001rq,Bernreuther:2010ny,Uwer:2004vp,Mahlon:2010gw}. In addition, many scenarios for physics beyond the standard model  suggest different production and decay properties of top quarks, 
which would also show up in spin correlation effects that are
different than the SM predictions, see, e.g., Refs.~\cite{Kane:1991bg,Cheung:1996kc,Bernreuther:1997gs,Frederix:2007gi}.
Because of the different production mode and energy, spin correlation measurements at the Tevatron and the LHC are complementary. The most recent Tevatron measurement by D0 reports a $3.1\sigma$ evidence for spin correlation~\cite{Abazov:2011gi}.
Previous simulation studies on the potential of spin correlations measurements at LHC can be found in Refs.~\refcite{Hubaut,Hubauta,Hubaut2005,atlastdr} (ATLAS) and~\refcite{cmsnote-2006-111} (CMS).

In standard model top quark decays, the angular distributions of the decay products of a polarized sample of top quarks are given by $\frac{1}{N} \frac{dN}{d \cos(\theta_i)} = \frac{1}{2} [1+\alpha_i \cos(\theta_i)]$, where $\theta_i$ is  the angle between the direction of the decay particle in the top quark rest frame and the spin quantization direction, and $\alpha_i$ is a coefficient expressing the spin analyzing power of the decay particle. Charged leptons and down-type quarks from the W-boson decay have values of $\alpha_i$ close to one and are thus the most effective spin analyzers. However, jets from up-type and down-type quarks can not be distinguished well experimentally. Therefore, spin correlations can be well measured using di-lepton events. A correlation coefficient is defined as the fractional difference in the number of events where the spins of the top quark pair are aligned and those where they have opposite alignment:
$A = \frac{N(\uparrow\uparrow)+N(\downarrow\downarrow) - N(\uparrow\downarrow) - N(\downarrow\uparrow)}{N(\uparrow\uparrow)+N(\downarrow\downarrow) + N(\uparrow\downarrow) + N(\downarrow\uparrow)}$. It requires the definition of two spin analyzing vectors for the top and anti-top quarks. In the \textit{helicity basis}, they are given by the direction of flight of the top quark in the $\ttbar$ center-of-mass frame, and defined such that the spin analyzing vectors of top and anti-top have opposite sign. The \textit{LHC maximal basis}~\cite{Uwer:2004vp} on the other hand describes a basis for which $A$ is maximal for top pairs produced by gluon fusion.

ATLAS performed a test of the hypothesis that the size of the $\ttbar$ spin correlation is as expected from the SM, as opposed to the hypothesis that the spins are uncorrelated, using a sample of di-lepton events corresponding to $\Lint = 2.1 \fbinv$ of 2011 data~\cite{Collaboration:2012sm}. The event selection and background estimation was performed in a very similar way compared with the cross section measurement in the di-lepton channel without $b$-tagging and using the same dataset, see section~\ref{sec:atlasdil}. Simulated $\ttbar$ samples with and without SM spin correlation were generated using MC@NLO.

The measurement of the spin correlation used the absolute difference in azimuthal angle between the two charged leptons, $\Delta\phi = |\phi_{l^+} - \phi_{l^-}  |$, which has the advantage that no full reconstruction of the $\ttbar$ system is required. No requirement on $M_\ttbar$ was applied.
Although this observable is less discriminating between SM-correlated and uncorrelated events~\cite{Bernreuther:2010ny} than the corresponding $\Delta\phi$ for low-energy di-lepton events with an upper limit on $M_\ttbar$~\cite{Mahlon:2010gw}, it can be measured in a more straightforward way.
The correlation coefficient $A$ was derived by performing a maximum likelihood fit to the measured $\Delta\phi$ distribution (Fig.~\ref{fig:chargespin},  right), using a linear superposition of that expected from simulations with and without spin correlation. The resulting measurement of the relative fraction of events with spin correlation yielded a value of
$f_{SM} = 1.30 \pm 0.14 \stat ^{+0.27}_{-0.22} \syst$.
The systematic uncertainties were evaluated using pseudo-experiments and dominated by contributions from the jet energy scale, resolution and efficiency, from fake leptons as well as from signal modeling.

The result was also transformed into the correlation coefficient $A_{hel.}=0.40 \pm 0.04 \stat ^{+0.08}_{-0.07} \syst$ using the helicity basis, or alternatively into 
$A_{max.}=0.57 \pm 0.06 \stat ^{+0.12}_{-0.10} \syst$ using the LHC maximal basis. These values are in good agreement with the parton level SM predictions of $A_{hel.}=0.31$ (from Ref.~\refcite{Bernreuther:2010ny}) and $A_{max.} =0.44 $ (using MC@NLO).  The hypothesis of zero spin correlation could be excluded with an observed (expected) significance of $5.1 \sigma$ $(4.2 \sigma)$.

%%%%%%%%%%%%%%%%%%%%%%%%%%%%%%%%%%%%%%%%%%%%%%%%%%%%%%%%%%%%%%%%%%%%%%%%%%%%%%%

\subsection{W polarization in top quark decays}
\label{sec:whelicity}

Since the top quark decays almost exclusively as $t\rightarrow Wb$, top quark production allows to study the properties of the $Wtb$ vertex, for instance by measuring the polarization of the $W$-bosons produced in top-quark decays. In the electroweak interaction, the $Wtb$ vertex has a V-A structure, where V (A) is the vector (axial-vector) contribution to the vertex. The polarization of the $W$-bosons produced in top quark decays can be either longitudinal (helicity 0), left- or right-handed (helicities $\pm 1$), and the corresponding partial widths are labeled $\Gamma_{0,R,L}$. Neglecting the $b$-quark mass, $\Gamma_R$ should be zero at tree level. The relative event rates or helicity fractions $F_i = \Gamma_i / \Gamma$ ($F_0+F_L+F_R =1$) are predicted in NNLO QCD as $F_0 = 0.687 \pm 0.005$, $F_L = 0.311 \pm 0.005$ and $F_R =
0.0017 \pm 0.0001$~\cite{Czarnecki:2010gb}.

Defining $\cts$ as the angle between the
direction of the charged lepton from the $W$-boson decay and the reversed direction of the $b$-quark from the top quark decay, both boosted into the $W$-boson rest frame, the differential decay rate is
\begin{equation}
\frac{1}{\Gamma}\frac{d\Gamma}{d\cts} =
\frac{3}{8} (1+\cts)^2 F_R +
\frac{3}{8} (1-\cts)^2 F_L +
\frac{3}{4} (1-\cos^2\theta^*) F_0 \ .
\end{equation}
This can be used to measure the helicity fractions $F_i$ from studying the shape of the $\cts$ distribution. Another method to obtain information about the helicity fractions is to measure angular asymmetries 
$A_\pm = \frac{N(\cts>z) - N(\cts<z)}{N(\cts>z) + N(\cts<z)}$. If one defines $z=\mp(2^{2/3}-1)$, then the asymmetry $A_+$ ($A_-$) depends only on $F_R$ ($F_L$) and $F_0$, respectively, and the helicity fractions can be determined in a straightforward way. The SM NNLO QCD values for the asymmetries are $A_+ = 0.537 \pm 0.004$ and $A_- = -0.841 \pm 0.006$~\cite{Czarnecki:2010gb}.

Deviations of the values of the helicity fractions $F_i$ or angular asymmetries $A_\pm$ from their SM predictions could be due to new physics beyond the SM contributing to the $Wtb$ vertex, arising from new interactions involving the top quark at higher energies. 
In an effective field theory approach~\cite{Buchmuller:1985jz,AguilarSaavedra:2008zc,Zhang:2010dr},
new physics above the electroweak symmetry breaking scale can be parameterized in terms of an effective Lagrangian in a model independent way using dimension-six operators which are invariant under the SM gauge symmetry. After electroweak symmetry breaking, the resulting general form of the $Wtb$ Lagrangian is:
\begin{equation}
\mathcal{L}_{Wtb} = 
-\frac{g}{\sqrt{2}} \bar{b} \gamma^\mu (V_L P_L + V_R P_R) t W_\mu^-
-\frac{g}{\sqrt{2}} \bar{b} \frac{i\sigma^{\mu\nu}q_\nu}{M_W}    (g_L P_L + g_R P_R) t W_\mu^-
+ h.c. \ ,
\end{equation}
where
\begin{equation}
V_L = V_{tb} + C_{\phi q}^{(3,3+3)} \frac{v^2}{\Lambda^2}  \ ,
V_R = \frac{1}{2} C_{\phi\phi}^{33*} \frac{v^2}{\Lambda^2} \ ,
g_L = \sqrt{2} C_{dW}^{33*} \frac{v^2}{\Lambda^2}  \ ,
g_R = \sqrt{2} C_{uW}^{33} \frac{v^2}{\Lambda^2} \ .
\end{equation}
Here, $\Lambda$ is the new physics scale and the $C$ are the effective operators coefficients. The anomalous couplings $V_R$, $g_L$ and $g_R$ are absent in the standard model at tree level, while the SM coupling $V_{tb}$ receives a correction from the operator $C_{\phi q}^{(3,3+3)}$. In the presence of such anomalous $Wtb$ couplings, the helicity fractions and angular asymmetries would deviate from their SM expectations, such that constraints on these new couplings can be obtained from the helicity measurements.
Previous feasibility studies for $W$-boson helicity and anomalous coupling measurements at LHC can be found in Refs.~\refcite{Hubauta,Hubaut2005,atlastdr}.

\begin{figure}[t]
\centering
\begin{minipage}{0.44\linewidth}
\centering
\vspace*{5mm}
\includegraphics[width=0.99\linewidth]{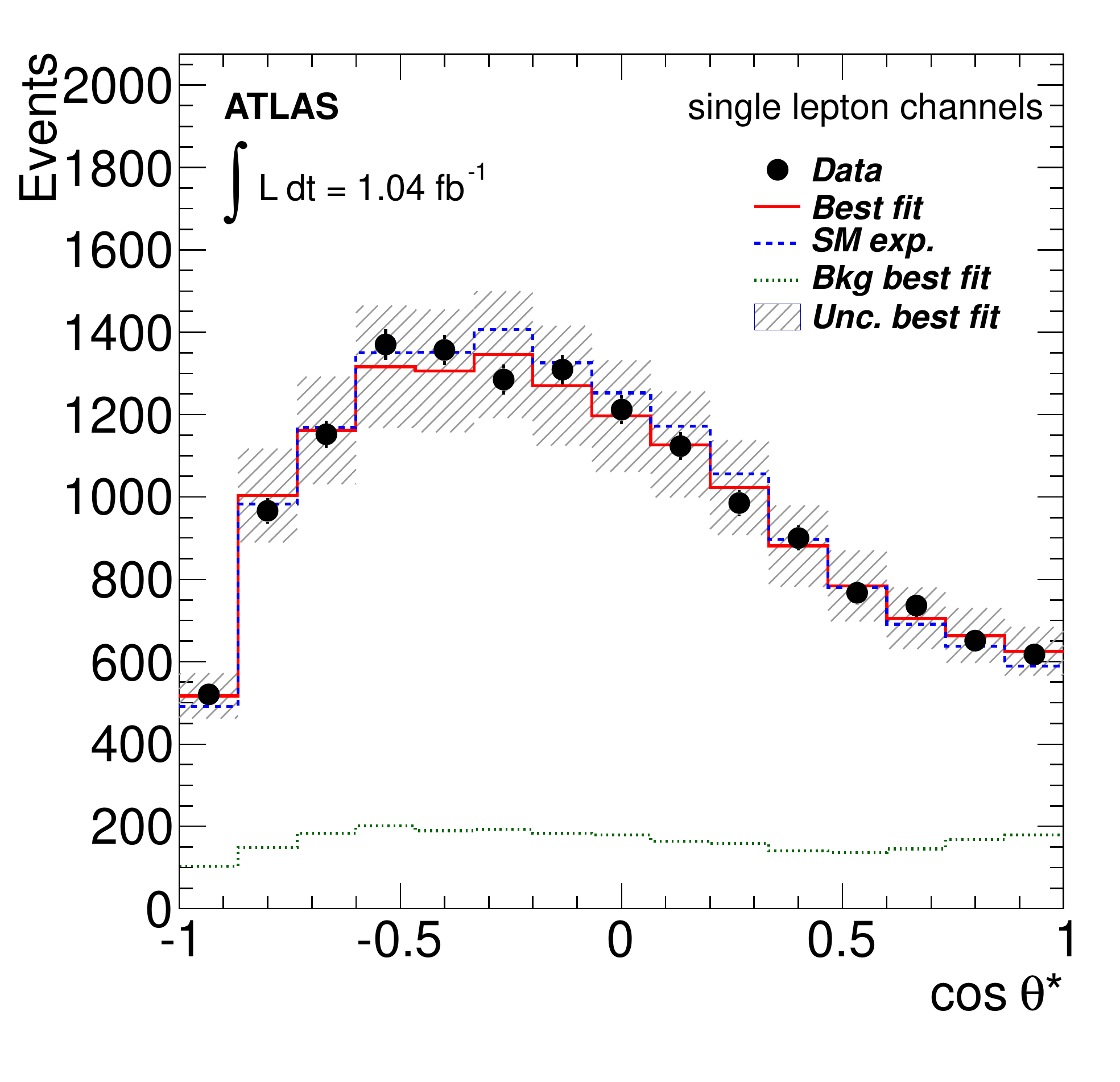}
\end{minipage}
\begin{minipage}{0.55\linewidth}
\centering
\includegraphics[width=0.99\linewidth]{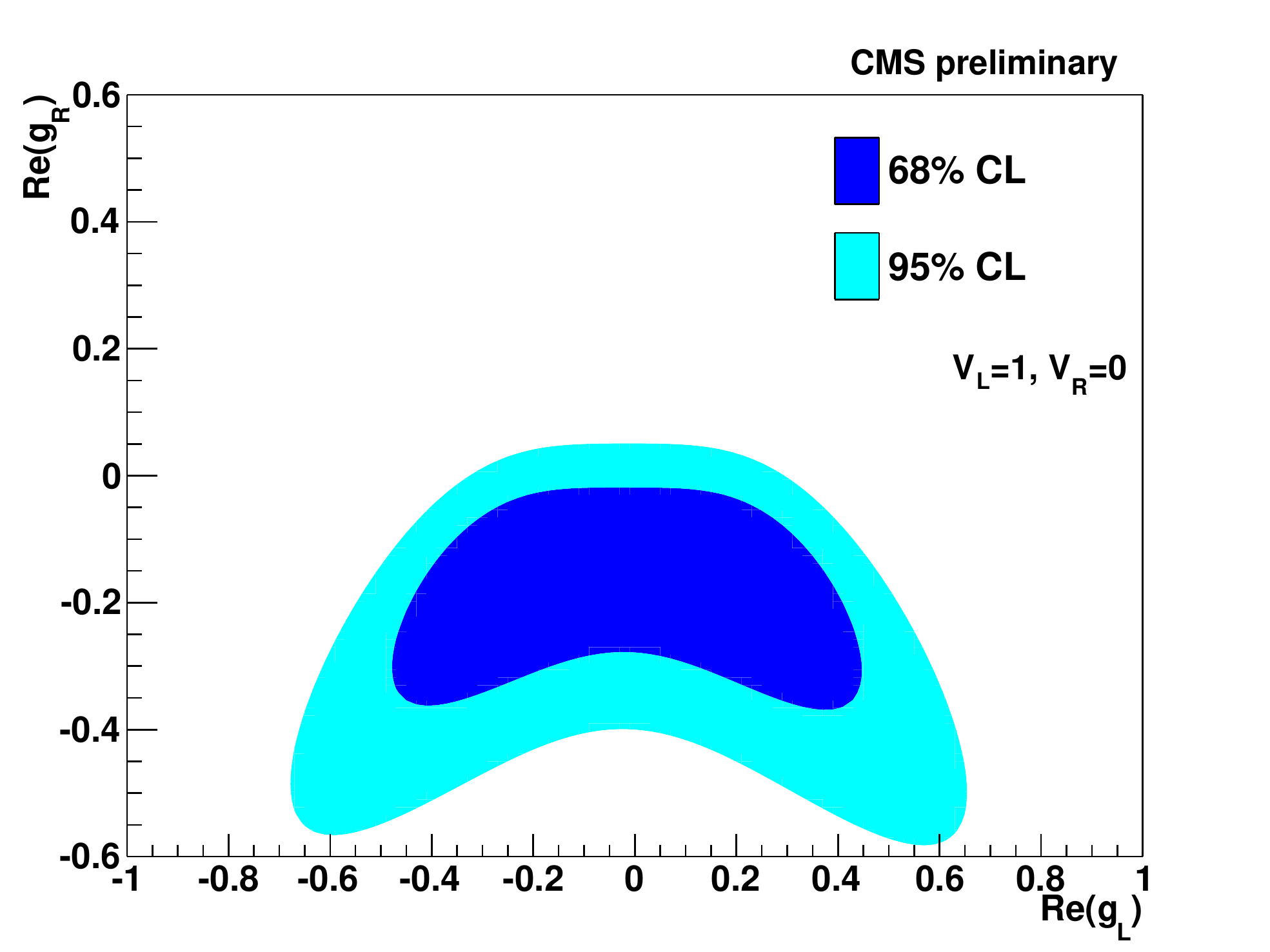}
\end{minipage}
\caption{Left: ATLAS fit to the observed $\cts$ distribution in the lepton+jets channel, from which the $W$-boson helicity fractions are determined~\protect\cite{Collaboration:2012ky}. Right: Allowed regions of the $Wtb$ anomalous couplings $(g_L,g_R)$, determined from the CMS measurements of the helicity fractions~\protect\cite{cmspas-top-11-020}. }
\label{fig:whelicity}
\end{figure}

ATLAS reported measurements~\cite{Collaboration:2012ky} of helicity fractions and angular asymmetries in top pair production at LHC using $\Lint = 1.04 \fbinv$ and employing both the lepton+jets and di-lepton channels. An earlier result using 2010 data can be found in Ref.~\refcite{ATLAS-CONF-2011-037}.
The event selections in the lepton+jets and di-lepton channels were very similar to the ones used in the $\ttbar$ cross section measurements using the same datasets, as discussed in previous sections. 
Top quark pairs were reconstructed in the single lepton channel by maximizing a kinematic likelihood function (similar as done in the top mass measurement in the same channel using 2011 data, see section~\ref{sec:massljets}). In the di-lepton channel, a set of kinematic equations is solved using the minimum $m_{l_1j_1}+m_{l_2j_2}$ to identify the jet-lepton pairing, and imposing $m_t$ and $p_z$ of the neutrinos as kinematic constraints. For the measurement of the angular asymmetries, a $\chi^2$ minimization technique was used in case of the lepton+jets channel, while in the di-lepton channel, the same method as described above was employed.
The $W$-boson helicity fractions were measured by performing a binned maximum likelihood fit to the observed $\cts$ distribution (Fig.~\ref{fig:whelicity} left), using signal templates corresponding to the three helicity states (considering efficiencies and acceptances) as well as various background templates. In the fit, the normalizations of the background templates and the corresponding uncertainties were set to the expected values. 
The angular asymmetries were measured by subtracting the expected background from the observed event numbers above and below the value of $z$ in the $\cts$ distributions and applying an iterative unfolding procedure.
% to correct for detector and reconstruction effects.

\begin{table}[t]
\tbl{ATLAS~\protect\cite{Collaboration:2012ky} and CMS~\protect\cite{cmspas-top-11-020} measurements of $W$-boson helicity fractions, as well as derived limits on anomalous $Wtb$ couplings. The first (second) uncertainty corresponds to the statistical (systematic) error.}
{
\begin{tabular}{lll}
\toprule
Measurement & ATLAS (l+jets + di-lepton) & CMS (l+jets, prelim.)\\
\colrule
$F_0$           & $0.67 \pm 0.03 \pm 0.06$ & $0.57 \pm 0.07 \pm 0.05$ \\ 
$F_L$           & $0.32 \pm 0.02 \pm 0.03$ & $0.39 \pm 0.05 \pm 0.03$ \\ 
$F_R$           & $0.01 \pm 0.01 \pm 0.04$ & $0.04 \pm 0.04 \pm 0.04$ \\ 
\colrule
$F_0$ ($F_R=0$) & $0.66 \pm 0.03 \pm 0.04$ & $0.64\pm0.03 \pm 0.05$ \\ 
\colrule
$\RE V_R$  & $\in [-0.20,0.23]$ & -- \\ 
$\RE g_L$  & $\in [-0.14,0.11]$ & -- \\ 
$\RE g_R$  & $\in [-0.08,0.04]$ & $-0.07 \pm 0.05 ^{+0.07}_{-0.08}$  \\ 
$\frac{\RE (C_{uW}^{33})}{\Lambda^2} \ [\TeV^{-2}] $  & $\in [-0.9,2.3]$ & $-0.81\pm 0.62 ^{+0.85}_{-0.95}$ \\

\botrule
\end{tabular}
\label{tab:whel} 
}
\end{table}

The measured helicity fractions are presented in Table~\ref{tab:whel}, combining the individual measurements in the lepton+jets and di-lepton channels. The final results were in turn obtained by combining the extractions of the helicity fractions using both the $\cts$ template fit as well as the angular asymmetry measurements, which were found to be consistent with each other. 
The dominating contribution to the systematic uncertainties originated from the signal and background modeling (choice of MC generator, amount of ISR/FSR, etc.), mis-identified leptons, jet energy scale uncertainty, as well as uncertainties associated with the $\ttbar$ reconstruction and template fit methods. 
Fixing $F_R$ at zero and using the template fit method only yielded the result also shown in Table~\ref{tab:whel}.
The measured values of the angular asymmetries were 
$A_+ = 0.52 \pm 0.02 \pm 0.03$  and $A_- = -0.84 \pm 0.01 \pm 0.02$ in the lepton+jets channel, and 
$A_+ = 0.56 \pm 0.02 \pm 0.04$ and $A_- = -0.84 \pm 0.02 \pm 0.04$ in the di-lepton channel, respectively.

The combined measurements of the helicity fractions were used to derive limits on anomalous couplings $g_L$ and $g_R$ (setting $V_L=1$, $V_R=0$) using the TOPFIT program~\cite{AguilarSaavedra:2006fy,AguilarSaavedra:2010nx}.  Allowing only one coupling at a time to be non-zero, the 95\% CL intervals on single anomalous couplings were obtained. The results are listed in Table~\ref{tab:whel}. These limits are consistent with the standard model, and improve the ones obtained by D0~\cite{Abazov:2009ky}.
A 95\% CL limit on the operator coefficient $C_{uW}^{33}$ is also obtained from the $F_L$ measurement where $F_R$ was fixed to zero in the context of Ref.~\refcite{Zhang:2010dr}, see Table~\ref{tab:whel}.

A similar measurement was performed in the muon+jets channel by CMS, based on $\Lint = 2.2 \fbinv$ of 2011 data~\cite{cmspas-top-11-020}.  Events which contain one isolated muon and at least four jets were selected, one of which was required to be $b$-tagged. The $\ttbar$ system was reconstructed by means of a kinematic fit, imposing $m_t$ and $m_W$ mass constraints and considering the $b$-tagging information. 
The helicity fractions were measured by means of a binned maximum likelihood fit to the reconstructed $\cts$ distribution.
The systematic uncertainties were dominated by contributions due to background normalization and signal modeling. The results for the simultaneous fit of $F_0$ and $F_L$, as well as for the fit of $F_0$ under the assumption that $F_R=0$, are presented in Table~\ref{tab:whel}.

The measured helicity fractions were used to set limits on anomalous $Wtb$ couplings. Allowing only $g_R$ to deviate from zero (no contribution to $F_R$), the fit results for 
$\RE g_R$ and correspondingly $\frac{\RE (C_{uW}^{33})}{\Lambda^2}$ are in good agreement with the SM (see Table~\ref{tab:whel}). 
The allowed region in the $(\RE g_L,\RE g_R)$ plane when allowing also $g_L$ to be non-zero is shown in Fig.~\ref{fig:whelicity} (right).

For both ATLAS and CMS, the results are consistent with, and have comparable or better uncertainties than the measurements at the Tevatron~\cite{Abazov:2010jn,Aaltonen:2010ha,Aaltonen:2012rz,Aaltonen:2012tk}.
Measurements of the $W$-boson helicity can also be combined with single top quark production measurements in order to constrain the $Wtb$ vertex~\cite{AguilarSaavedra:2011ct,arxiv:1204.2332}.

%%%%%%%%%%%%%%%%%%%%%%%%%%%%%%%%%%%%%%%%%%%%%%%%%%%%%%%%%%%%%%%%%%%%%%%%%%%%%%%

\subsection{$\ttbar$ production in association with a photon}
\label{sec:ttgamma}

The electroweak couplings of the top quark can be studied by investigating events where top quark pairs are produced in association with a gauge boson, in particular using $\ttbar\gamma$, $\ttbar W$ or $\ttbar Z$ events. While the latter two topologies require substantial amounts of integrated luminosity to become accessible at the LHC, the cross section for $\ttbar\gamma$ production in $pp$ collisions is large enough to be measurable already with the presently available LHC data sample: the LO cross section (for $p_{T,\gamma}>8 \GeV$) times branching ratio for top pairs decaying in the di-lepton and lepton+jets channels is $0.84 \rm\ pb$ at $\sqrt{s}=7 \TeV$. However, more detailed tests of the couplings at the $\ttbar\gamma$ vertex are only possible with larger integrated luminosities~\cite{Baur:2004uw}.
$\ttbar\gamma$ production can be classified into \textit{radiative top quark production}, where photons are radiated from off-shell top quarks or incoming partons, and \textit{radiative top quark decay}, where a photon is radiated from an on-shell top quark or its decay products including the W-boson. Both types of processes interfere with each other.
CDF reported first evidence for $\ttbar\gamma$ production at the Tevatron in Ref.~\refcite{Aaltonen:2011sp}.

ATLAS presented a first measurement of the $\ttbar\gamma$ production cross section at LHC  using $\Lint = 1.04\fbinv$ of data~\cite{ATLAS-CONF-2011-153}.
Photons, which can convert into $e^+e^-$ pairs by interacting with detector material in front of the calorimeter, were identified by applying requirements on shower shape and hadronic leakage, and their energy scale and resolution was measured using $Z\rightarrow ee$ events.
$\ttbar\gamma$ events were simulated using WHIZARD~\cite{Kilian:2007gr}, which takes all contributing LO diagrams into account, and overlap with photon radiation in the inclusive $\ttbar$ sample is removed.

Events in the lepton+jets channel were selected by requiring one isolated electron or muon, at least four jets, and at least one $b$-tagged jet, similarly to the inclusive $\ttbar$ cross section measurement. In addition, at least one well identified photon with $p_T>15\GeV$ not close to a jet was requested, with an additional veto applied in the $e$+jets channel to reject $Z\rightarrow ee$ events with an electron misidentified as photon. The selection yielded around 120 events, while around 50 signal events were expected.
The cross section was measured by performing a template fit to a track-based photon isolation variable, exploiting the fact that prompt photons tend to be isolated, while photons produced from hadron decays are usually found close to a jet. The templates for signal and various backgrounds were determined from data. The most important sources of background were hadrons faking photons in $\ttbar$ events, electrons faking photons in di-lepton $\ttbar$ events and remaining background contributions from $\ttbar$ events. Non-$\ttbar$ backgrounds included real or fake photons in $W$+jets or QCD multi-jet events.
The result, corresponding to the $\ttbar\gamma$ cross section, for photons with $p_T>8 \GeV$, times branching ratio into the di-lepton and lepton+jets channels, was
\begin{equation}
\sigma_{\ttbar\gamma} \cdot {\rm BR} = 2.0 \pm 0.5 \stat \pm 0.7 \syst \pm 0.08 \lumi \ pb .
\end{equation}
The systematic uncertainty was dominated by signal modeling, jet energy scale, $b$-tagging performance, photon identification efficiency and the modeling of the background templates. The observed (expected) significance of the measurement is $2.7\sigma$  $(3.0\pm0.9 \sigma)$,  and the result is consistent with the SM prediction including NLO corrections from Ref.~\refcite{Melnikov:2011ta}, evaluated for $\sqrt{s}= 7 \TeV$.

%%%%%%%%%%%%%%%%%%%%%%%%%%%%%%%%%%%%%%%%%%%%%%%%%%%%%%%%%%%%%%%%%%%%%%%%%%%%%%%

\subsection{Charge asymmetry in $\ttbar$ production}
\label{sec:ac}

In the context of $\ttbar$ production in (anti-) proton collisions, the term \textit{charge asymmetry} usually refers to a difference in the rapidity (or other differential) distributions of top quarks and anti-quarks. In the SM and at LO QCD, the charge asymmetry is exactly zero. At NLO QCD, a small asymmetry appears~\cite{Kuhn:1998jr,Kuhn:1998kw} in the case of $q\bar{q}$ annihilation. It is due to terms in the squared matrix element which are odd under the exchange of top quark and anti-quark and originate from interference between Born and box diagrams in the $q\bar{q}\rightarrow \ttbar$ process, between initial and final state radiation in the $q\bar{q}\rightarrow \ttbar g$ process, and between amplitudes which have a relative sign difference under the exchange of $t$ and $\bar{t}$ in the $qg\rightarrow \ttbar q$ process.
Measuring this asymmetry probes perturbative QCD predictions and allows to test many new physics models where top quark pairs are produced through the exchange of new heavy particles, for instance axigluons~\cite{Antunano:2007da,Frampton:2009rk}, $Z'$-bosons~\cite{Rosner:1996eb} or Kaluza-Klein excitations of gluons~\cite{Ferrario:2008wm}. Only in the case of $s$-channel production and decay into $\ttbar$ they would lead to distortions of the top pair invariant mass distribution (see section~\ref{sec:mttbar}), but the existence of such new physics contributions to $\ttbar$ production could result in a charge asymmetry which deviates from the SM expectation.

At the $p\bar{p}$ collider Tevatron, the charge asymmetry manifests as a forward-backward asymmetry in the rapidity difference $\Delta y = y_t - y_{\bar{t}}$ between top quarks and anti-quarks, due to the asymmetric initial state, such that 
\begin{equation}
A = \frac{N^+ - N^-}{N^+ + N^-} \ ,
\label{eq:ac}
\end{equation}
becomes non-zero, where $N^+$ ($N^-$) corresponds to the number of events with positive (negative) values of $\Delta y$.
Recent measurements of this quantity by CDF~\cite{Aaltonen:2011kc,cdf-conf-10436,cdf-conf-10584,cdf-conf-10807}
 and D0~\cite{Abazov:2011rq} yielded results which were significantly larger than the SM prediction of about $8\%$~\cite{Kuhn:1998jr,Kuhn:1998kw,Bernreuther:2010ny,Kuhn:2011ri,Ahrens:2011uf,Hollik:2011ps}. For large values of the $\ttbar$ invariant mass $M_\ttbar>450 \GeV$, CDF found an even larger deviation, which however has not been confirmed by D0.

There is no forward-backward asymmetry at the LHC, since the initial state is symmetric. However, due to the proton PDF, the incoming quarks have on average more momentum compared with the anti-quarks, which means that the charge asymmetry results in a rapidity distribution of top quarks which is slightly broader than that of top anti-quarks. At large rapidities, more top quarks than anti-quarks are produced, while in the central region, more top anti-quarks than quarks are produced.
This asymmetry can be quantified by considering the difference in absolute (pseudo-) rapidities $\Delta|\eta|=|\eta_t|-|\eta_{\bar{t}}|$~\cite{Diener:2009ee} or $\Delta|y|=|y_t|-|y_{\bar{t}}|$, respectively. Alternatively, the variable $\Delta y^2 = (y_t - y_{\bar{t}}) \cdot (y_t + y_{\bar{t}}) = y^2_t - y^2_{\bar{t}} $ has been proposed~\cite{Jung:2011zv}, which multiplies the Tevatron observable with a factor accounting for the boost of the $\ttbar$ system. The corresponding values of the charge asymmetries at $\sqrt{s}=7 \TeV$ as predicted by the SM (defined analogously as in Eq.~\ref{eq:ac}) are $A^\eta_C=0.0136\pm 0.0008$, $A^{y^2}_C = 0.0115 \pm 0.0006$ (as calculated in Ref.~\refcite{Kuhn:2011ri}) and $A^y_C\sim 0.006$ (MC@NLO). They are much smaller compared with the forward-backward asymmetry at the Tevatron, also because $q\bar{q}$ initial states are much less frequent in $\ttbar$ production the LHC.

Many models attempting to explain  the Tevatron forward-backward asymmetry in terms of new physics have been proposed. For many of them, predictions for the charge asymmetry at the LHC are also available. While it is impossible to give full account on all models proposed, we refer to Refs.~\refcite{Kamenik:2011wt,AguilarSaavedra:2012ma} and references therein for an overview. However, any proposed model must respect constraints obtained not only from $\ttbar$ total cross section and invariant mass distribution (see section~\ref{sec:mttbar}) measurements, but also those from precision electroweak measurements, as well as from searches for resonances in di-jet production or same-sign top quark production (see section~\ref{sec:samesign}).

CMS presented a first measurement of the charge asymmetry at LHC 
using $\Lint = 1.09 \fbinv$ of 2011 data in Ref.~\refcite{:2011hk}. 
Events were selected in the lepton+jets channel requiring exactly one isolated  electron or muon and at least four jets, at least one of which had to be $b$-tagged, in a manner very  similar to the cross section measurement in the same channel. The selection yielded around 12700 events. Backgrounds, which contaminated the selected sample at the level of around $20\%$, were estimated from data using a technique similar to that applied in an earlier cross section measurement discussed in section~\ref{sec:cmsljets}. The $\ttbar$ final state was reconstructed solving the jet assignment by means of a probability variable which accounts for the known masses of the two top quarks and the hadronically decaying $W$-boson, as well as the $b$-tagging information. The background subtracted raw asymmetries, obtained from the reconstructed $\Delta|\eta|$ and $\Delta y^2 $ distributions, were corrected for inefficiencies and migrations using a regularized unfolding procedure. The method and its linearity were validated using pseudo-experiments.
The results for the unfolded asymmetries were
\begin{eqnarray}
A^\eta_C   = -0.017 \pm 0.032 \stat ^{+0.025}_{-0.036} \syst \ , \\
A^{y^2}_C  = -0.013 \pm 0.028 \stat ^{+0.029}_{-0.031} \syst \ . \nonumber
\end{eqnarray}
The most important contribution to the systematic uncertainties were due to the signal modeling, estimated by changing the factorization scale and matching threshold, as well as the amount of initial and final state radiation.  The result is consistent with the NLO QCD theory prediction.

\begin{figure}[t]
\centering
\begin{minipage}{0.49\linewidth}
\centering
%\vspace{2mm}
\includegraphics[width=0.99\linewidth]{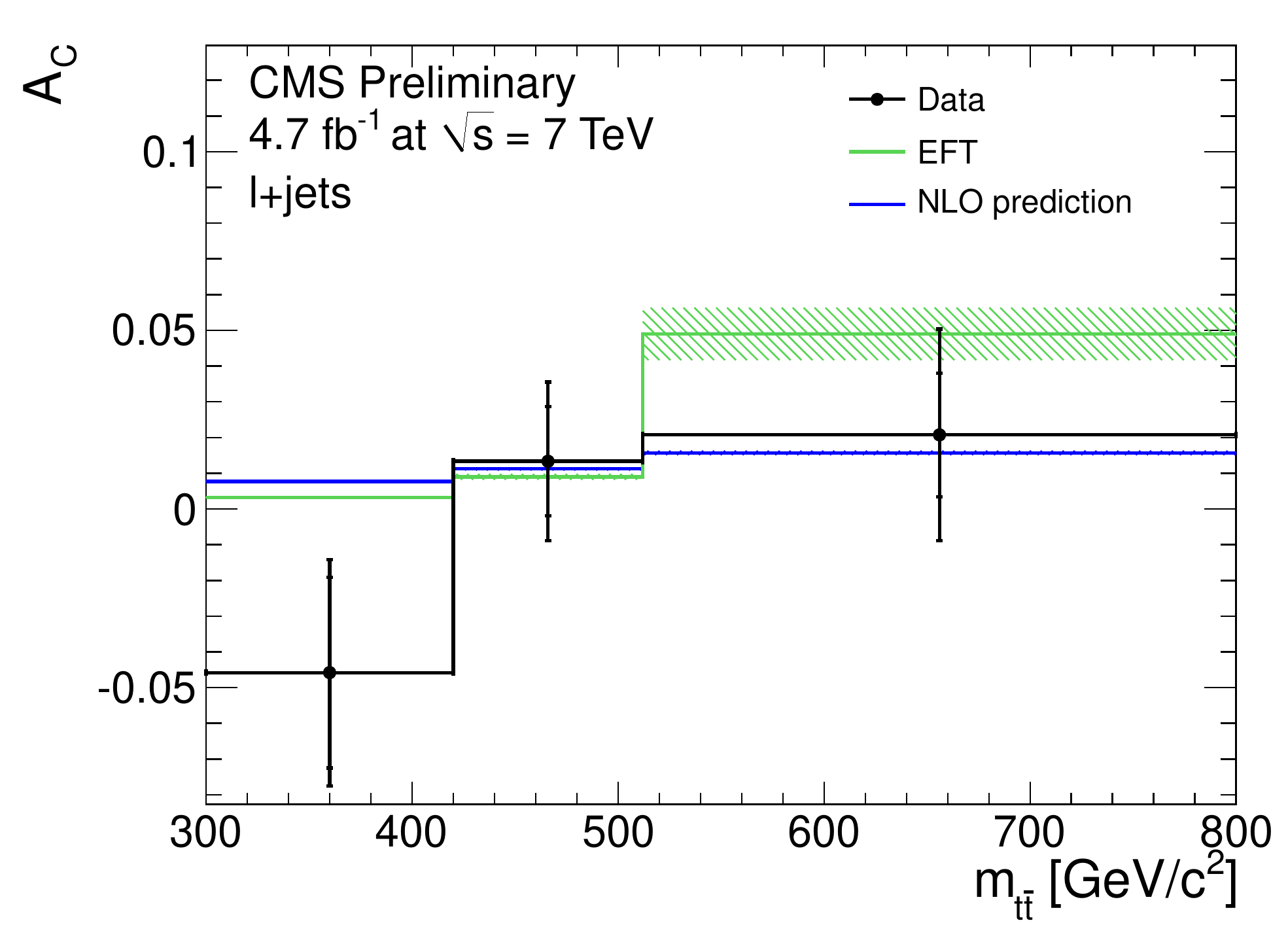}
\end{minipage}
\begin{minipage}{0.49\linewidth}
\centering
\includegraphics[width=0.99\linewidth]{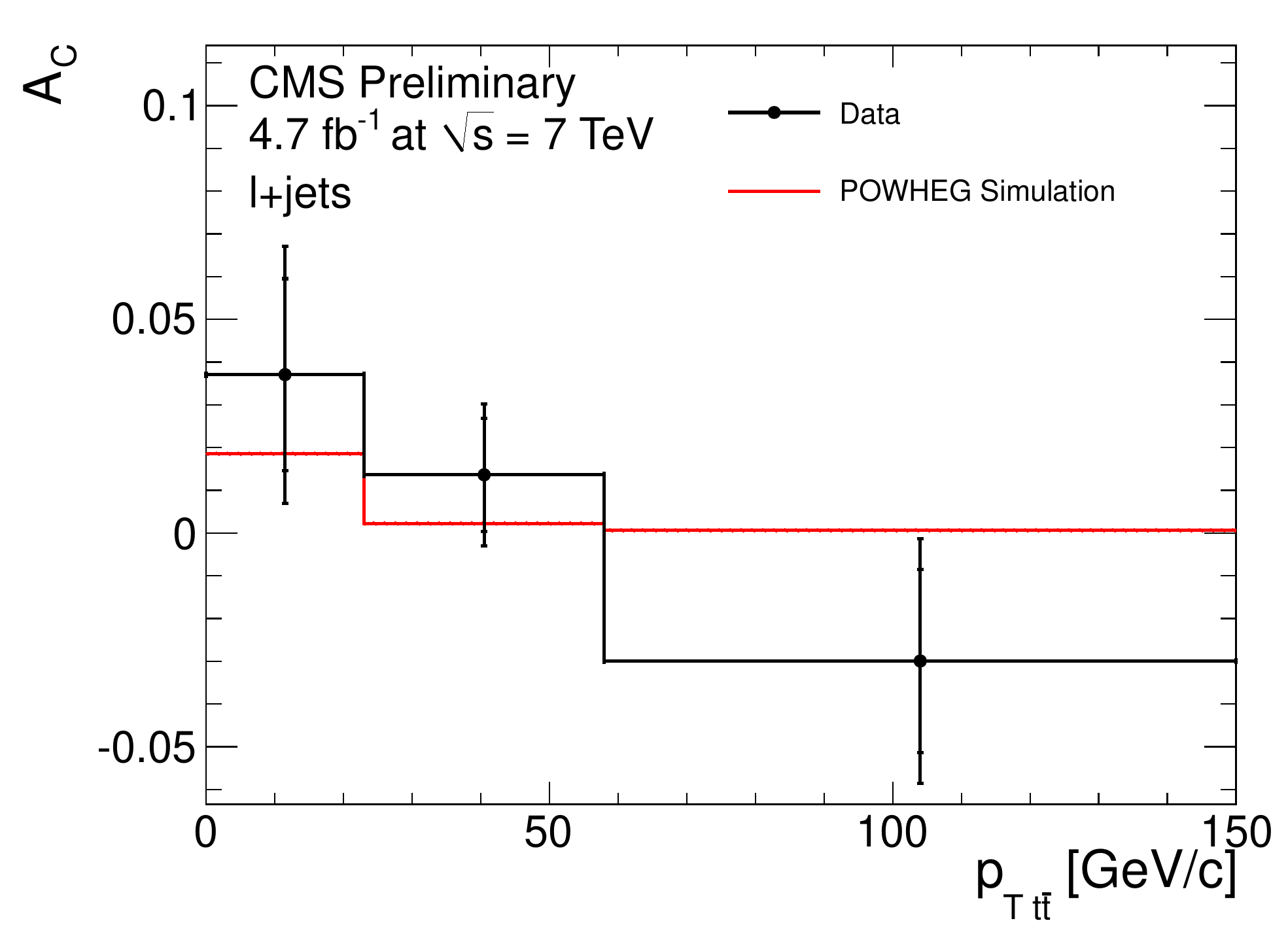}
\end{minipage}
\caption{Charge asymmetry as a function of $M_\ttbar$ and $p_{T,\ttbar}$ measured by CMS~\protect\cite{cmspas-top-11-030}, compared with predictions from the SM as well as from an effective field theory model~\protect\cite{Gabrielli:2011zw,gabrielli-lh11}.}
\label{fig:chargeasymcms}
\end{figure}

CMS also measured the charge asymmetry using $\Delta|y|$ as sensitive variable, both inclusively as well as differentially, with the full 2011 data sample of $\Lint=4.7\fbinv$~\cite{cmspas-top-11-030}, and employing techniques very similar to their first measurement. The differential measurements were performed as a function of absolute rapidity ($|y_\ttbar|$), transverse momentum ($p_{T,\ttbar}$) and invariant mass ($M_\ttbar$) of the $\ttbar$ pair, employing a two-dimensional regularized unfolding method.
The variables chosen enhance the asymmetry in certain regions of phase space.
In the SM, the asymmetry is expected to depend on  $M_\ttbar$  because the ratio of $q\bar{q}$ to $gg$ induced initial states is increasing with mass. Furthermore, a new heavy particle with different couplings to top quarks and anti-quarks may also result in a mass-dependent asymmetry.
For the same kinematic reason, an increased asymmetry is expected at large $|y_\ttbar|$ values. At larger values of $p_{T,\ttbar}$, the negative contribution due to ISR-FSR interference is enhanced. 
The linearity of the 2d-unfolding method was validated using pseudo experiments. Various changes with respect to the previous measurement were applied in the evaluation of systematic uncertainties. Predictions at NLO QCD using POWHEG were preferred to MADGRAPH to simulate the $\ttbar$ signal, eliminating the need for a systematic uncertainty due to the ME+PS matching scale.  The uncertainty due to the hadronization and shower modeling was evaluated using the difference between PYTHIA and HERWIG. Various re-weighting scenarios were used to assess the uncertainty due the assumed dependence on the differential measurement variable. The ISR/FSR systematic was not separately evaluated and assumed to be mostly covered by the factorization scale variation.
The result for the inclusive measurement was
\begin{equation}
A_C^y = 0.004 \pm 0.010 \stat \pm 0.012 \syst \ ,
\end{equation}
in good agreement with the SM theory value. Also the differential measurements are, within the present uncertainties, consistent with SM theory expectations. Fig.~\ref{fig:chargeasymcms} shows the unfolded asymmetry as a function of $M_\ttbar$ and $p_{T,\ttbar}$, compared with SM NLO QCD theory as well as with an effective field theory model~\cite{Gabrielli:2011zw,gabrielli-lh11} which is able to explain the large asymmetry measured by CDF.

\begin{figure}[t]
\centering
\begin{minipage}{0.49\linewidth}
\centering
\vspace{2mm}
\includegraphics[width=0.99\linewidth]{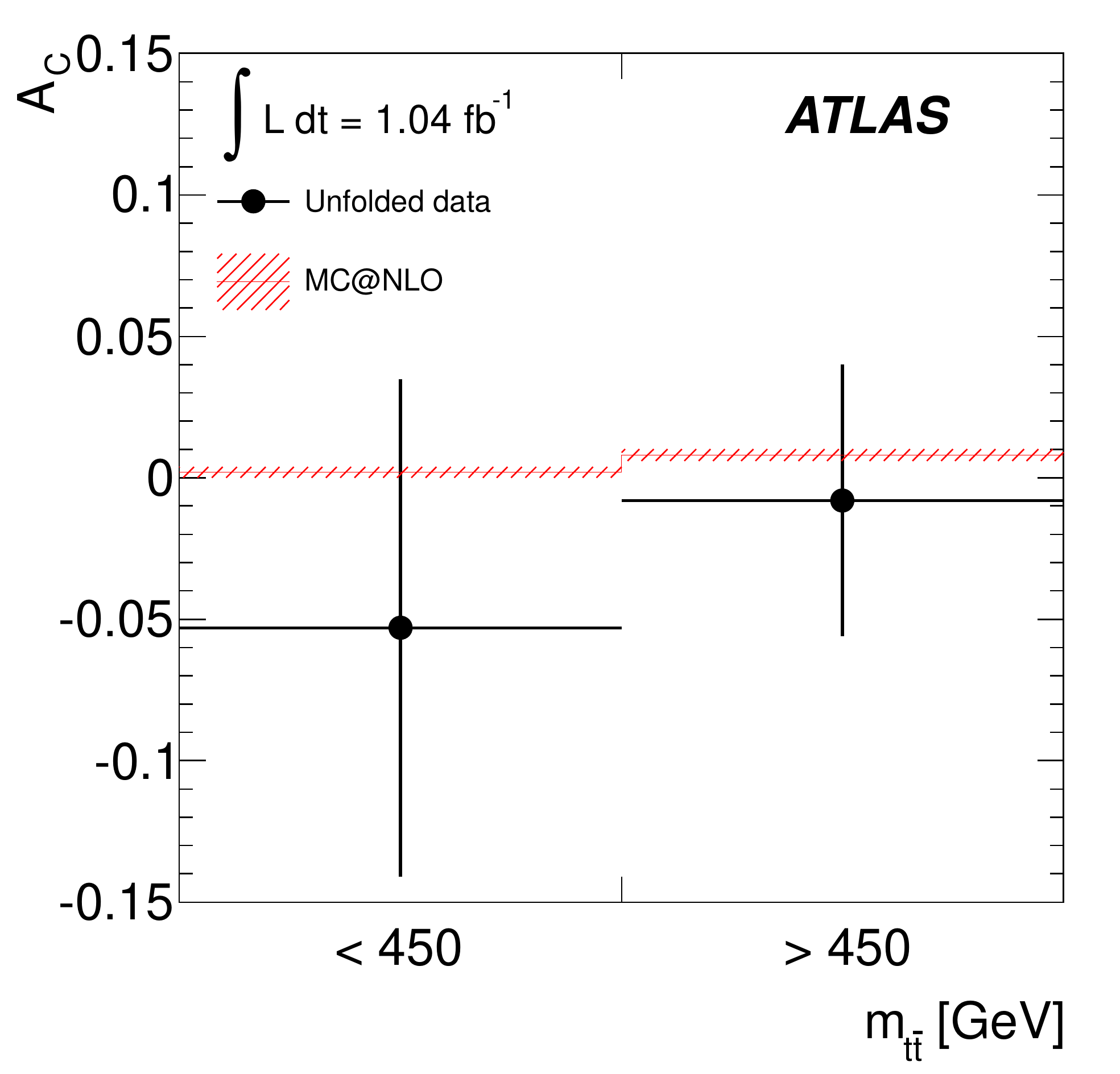}
\end{minipage}
\begin{minipage}{0.49\linewidth}
\centering
\includegraphics[width=0.99\linewidth]{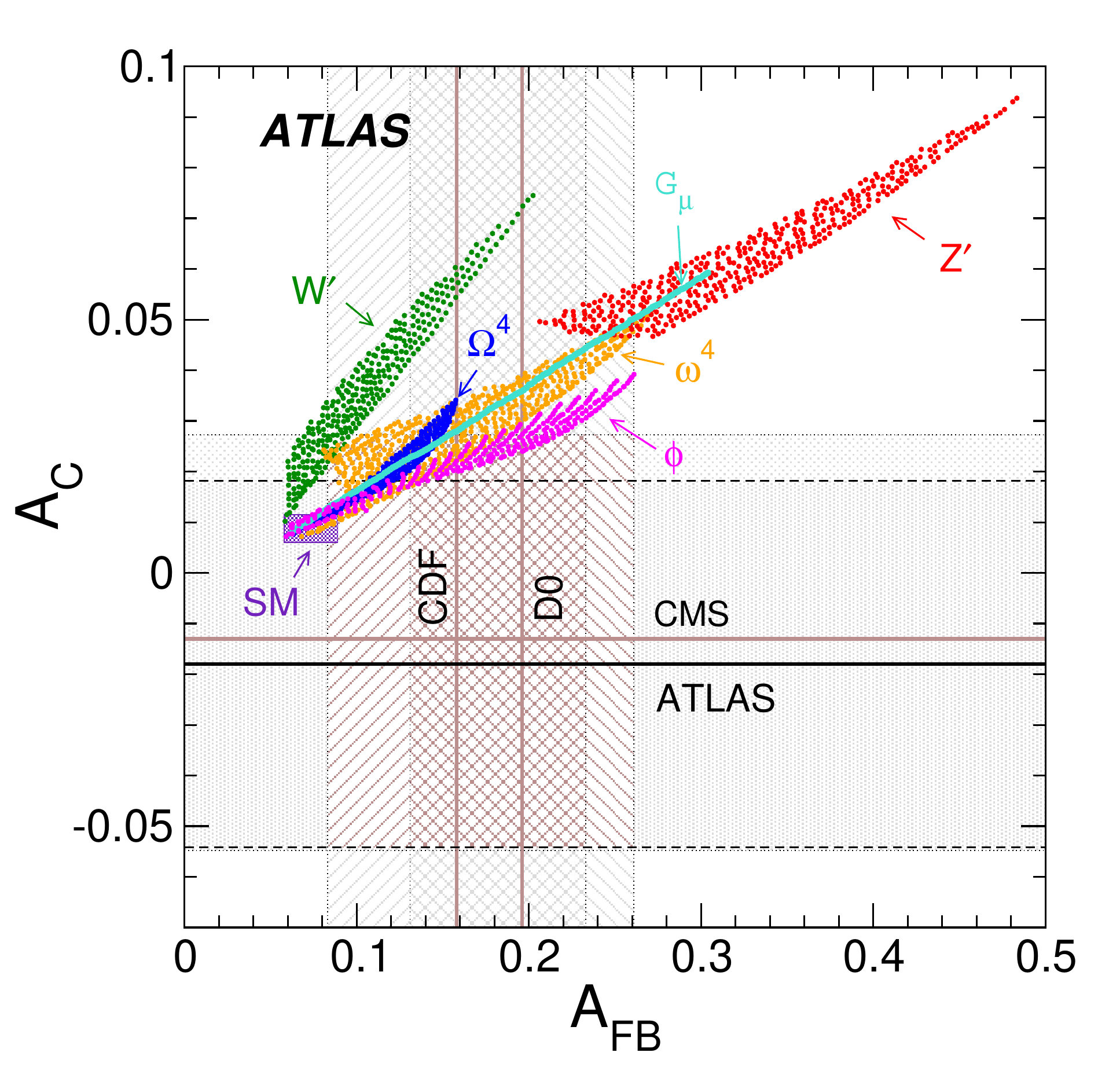}
\end{minipage}
\caption{Left: Charge asymmetry in two $M_\ttbar$ intervals as measured by ATLAS~\protect\cite{Collaboration:2012ug}. Right: Measured $A_{FB}$ and $A_C$ values from Tevatron and LHC, compared~\protect\cite{Collaboration:2012ug} to predictions from various new physics models~\protect\cite{AguilarSaavedra:2011hz,AguilarSaavedra:2011ug}. 
}
\label{fig:chargeasymatlas}
\end{figure}

ATLAS measured the charge asymmetry $A_C^y$ based on $\Lint = 1.04 \fbinv$ of data~\cite{Collaboration:2012ug}. The lepton+jets channel was used, selecting events with one isolated lepton and at least four jets, one of which was required to be $b$-tagged. Event selection and background determination were very similar to the ones used in the cross section measurement discussed in section~\ref{sec:atlasljets}.  The $\ttbar$ final state was reconstructed using a kinematic likelihood method, and the measured $\Delta|y|$ distribution was unfolded for acceptance and migration effects using an iterative Bayesian method. The measurement yielded
\begin{equation}
A^y_C   = -0.018 \pm 0.028 \stat \pm 0.023 \syst \ , 
\end{equation}
where the systematic uncertainty was mostly due to the signal modeling in simulation. The result is consistent with the SM prediction. The asymmetry was also measured differentially in two intervals of the invariant mass of the $\ttbar$ system ($M_\ttbar$), employing a 2d unfolding in ($A_C$,$M_\ttbar$). Within the uncertainties, no deviation from the SM was found (Fig.~\ref{fig:chargeasymatlas} left).

The relation between the Tevatron $A_{FB}$ and the LHC $A_C$ values is model dependent. Fig.~\ref{fig:chargeasymatlas} (right) shows correlated predictions  for various new physics scenarios, compared with the experimental measurements made at the two colliders. More details on the models, which include a new $Z'$ or $W'$ boson with right-handed couplings, a heavy axigluon, a new scalar doublet and a charge 4/3 scalar, can be found in Refs.~\refcite{AguilarSaavedra:2011hz,AguilarSaavedra:2011ug}. Scans over ranges of masses or coupling were performed, considering consistency with $\ttbar$  section measurements and new physics contributions at high $M_\ttbar$ values (see section~\ref{sec:mttbar}). Models invoking a new flavor-changing $Z'$ and $W'$ boson are disfavored.

In the SM, the size of the forward-backward asymmetry decreases with increasing $p_{T,\ttbar}$, and it turns negative at around $p_{T,\ttbar}\sim 25 \GeV$, due to the negative contribution from $\ttbar$+1 jet events. The $p_{T,\ttbar}$ distribution observed by D0~\cite{Abazov:2011rq} is somewhat softer than the NLO SM prediction, which possibly hints at an effect which may be correlated with the increased asymmetry. So far LHC data do not seem to indicate a problem related to the modeling of the $p_{T,\ttbar}$ distribution (see section~\ref{sec:diffxs}), but besides the different energy regime and production mode, the experimental precision is still limited. It remains to be seen if mis-modeling of $\ttbar$ production could potentially serve as  (partial) explanation for the discrepancy with the SM observed at the Tevatron. In this context, it is also interesting to note that even general purpose LO MC generators with coherent parton or dipole showering may exhibit an asymmetry~\cite{Abazov:2011rq,Skands:2012mm}.

In the future, more precise and differential measurements (focusing on kinematic regions in which the predicted asymmetry is enhanced), as well as additional observables, for instance a lepton pair asymmetry in di-leptonic $\ttbar$ decays for which SM predictions exist~\cite{Bernreuther:2010ny}, are needed in order to clarify whether the LHC data are consistent with the SM. 
There are certain differences between ATLAS and CMS in the evaluation of systematic uncertainties related to the modeling of signal and background, for instance due to the amount of QCD radiation (so-called ISR/FSR or scale uncertainties). Some work is needed, and already ongoing, to understand these differences, aiming at a more homogeneous treatment. This concerns also several other measurements in the top quark sector.
The theory prediction for the charge asymmetry in the SM must be updated to NLO QCD~\footnote{The LO asymmetry arises only from NLO QCD contributions to the total $\ttbar$ cross section.}. This may soon be achieved using techniques based on the recent NNLO QCD calculation of the $q\bar{q}\rightarrow \ttbar$ total cross section~\cite{Baernreuther:2012ws} (neglecting the small contributions from $qg$ initial states to the asymmetry).

%%%%%%%%%%%%%%%%%%%%%%%%%%%%%%%%%%%%%%%%%%%%%%%%%%%%%%%%%%%%%%%%%%%%%%%%%%%%%%%

\subsection{Additional jet activity in $\ttbar$ production}
\label{sec:atlasisr}

Many measurements presented in this review suffer from significant systematic uncertainties due to the modeling of additional quark and gluon radiation in $\ttbar$ production.
Experimental data are needed to validate the MC models and reduce these uncertainties.

ATLAS performed a measurement of $\ttbar$ production with a veto on additional jet activity using $\Lint = 2.05 \fbinv$ of 2011 data in the di-lepton channel~\cite{Aad:2012jr}. A clean sample of $\ttbar$ events was obtained by selecting events with two isolated high $p_T$ leptons, two $b$-tagged jets and large $\MET$, following closely previous cross section measurements in this final state. A \textit{gap fraction}, defined as the ratio of $\ttbar$ events with no additional jet above a given $p_T$ threshold, was used as sensitive variable. This gap fraction was measured differentially as a function of the $p_T$ threshold and in various rapidity intervals of the extra jet. Alternatively, the gap fraction was also expressed in terms of the scalar sum of all additional jets, not just the leading one. The measurements were corrected for detector effects and compared with various NLO (MC@NLO, POWHEG) and ME+PS LO (ALPGEN, SHERPA, ACERMC) generators. In the measurement of the gap fractions, many systematic uncertainties canceled, the remaining ones were dominated by jet energy scale and $b$-tagging, as well as the unfolding to particle level and the background modeling.

\begin{figure}[t]
\centering
\begin{minipage}{0.49\linewidth}
\centering
%\vspace{2mm}
\includegraphics[width=0.9\linewidth]{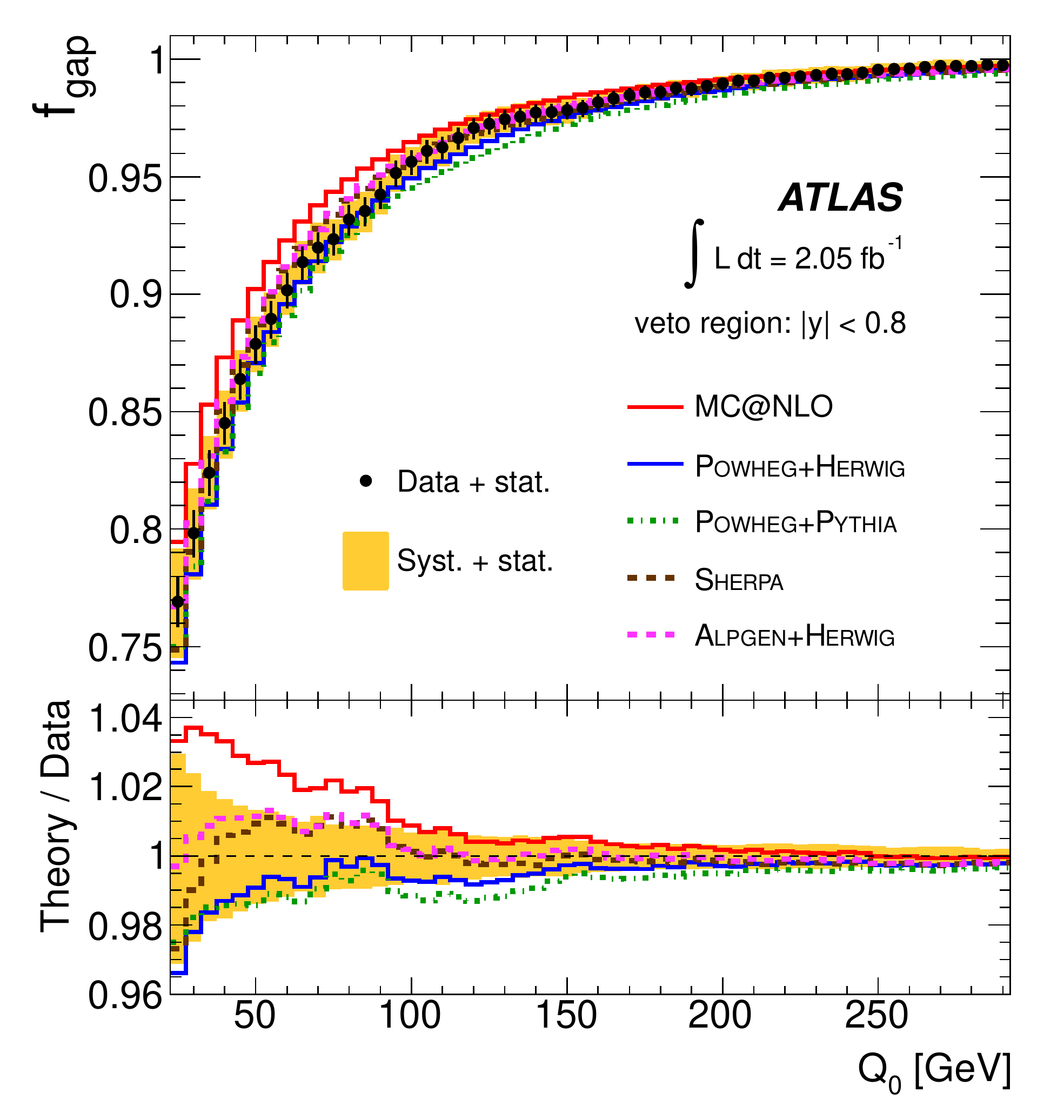}
\end{minipage}
\begin{minipage}{0.49\linewidth}
\centering
\includegraphics[width=0.9\linewidth]{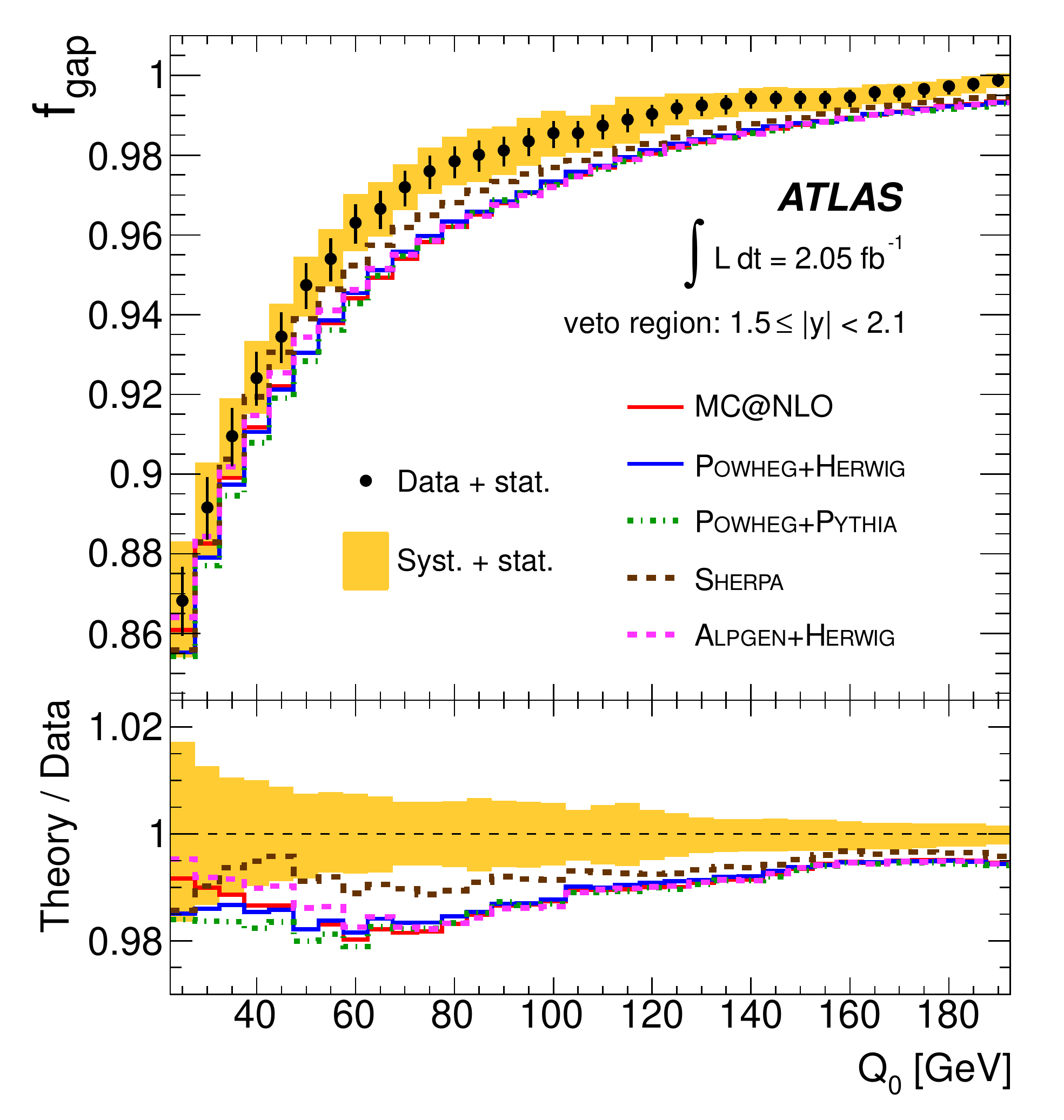}
\end{minipage}
\caption{ATLAS measurements of the fraction of $\ttbar$ events with no additional jet above $p_T>Q_0$, for the two rapidity intervals $|y|<0.8$ (left) and $1.5<|y|<2.1$ (right)~\protect\cite{Aad:2012jr}.}
\label{fig:jetveto}
\end{figure}

The experimental uncertainty on the gap fraction is comparable to or smaller than the spread between the various MC models, demonstrating the sensitivity of the measurement. It is also smaller than the difference between the usual variations in the amount of ISR/FSR, implying that these may be reduced in future systematic studies. In the central rapidity region $|y|<0.8$, the gap fraction predicted by MC@NLO is too large compared with the data, as well as with, e.g., ALPGEN (Fig.~\ref{fig:jetveto}, left), indicating that MC@NLO produces too few jets (see also Ref.~\refcite{Mangano:2006rw}). On the other hand, all MC models undershoot the gap fraction at large rapidities $1.5<|y|<2.1$ (Fig.~\ref{fig:jetveto}, right), which implies that they produce too much extra jet activity in the forward region. To improve the level of description, higher order QCD corrections may be needed, for instance those which account for the top quark decay products~\cite{Denner:2010jp,Bevilacqua:2010qb} or NLO QCD calculations of $\ttbar$ production with one or two additional jets~\cite{Dittmaier:2007wz,Dittmaier:2008uj,Melnikov:2010iu,Bevilacqua:2010ve,Bevilacqua:2011aa,Melnikov:2011qx}.

\section{Electroweak Production of Single Top Quarks}
\label{sec:singletop}

In this section, measurements of the cross section for the electroweak production of single top quarks in the $t-$ and $tW$-channels are presented, as well as the search for single top quark production in the $s$-channel. Single top quark topologies involve fewer jets compared with $\ttbar$ production, and the signal to background ratio is generally significantly smaller. This may explain that the discovery of single top quark production (measuring the combined $t$- and $s$-channel modes) at the Tevatron could only be announced in 2009~\cite{Abazov:2009ii,Aaltonen:2009jj}. For a review on Tevatron measurements of single top quark production, see Ref.~\refcite{Heinson:2010xh}.
At LHC, the signal-to-background ratio is generally more favorable. 

%%%%%%%%%%%%%%%%%%%%%%%%%%%%%%%%%%%%%%%%%%%%%%%%%%%%%%%%%%%%%%%%%%%%%%%%%%%%%%%

\subsection{t-channel}

At the LHC, the $t-$channel mode of single top quark production has the largest cross section and the cleanest signature, with a light quark jet recoiling against the top quark. Feasibility studies for the LHC based on simulation can be found in Refs.~\refcite{cmsnote-2006-086,cmspas-top-09-005,atlastdr,ATL-PHYS-PUB-2010-003}.

\subsubsection{ATLAS measurements}
\label{sec:atlastchannel35pb}

Several measurements of the $t$-channel single top quark production cross section were performed by ATLAS~\cite{ATLAS-CONF-2011-027,ATLAS-CONF-2011-088,Collaboration:2012ux}, using various techniques and data samples ranging from $\Lint =35 \pbinv$ to $1.04\fbinv$.
Events were selected requiring one isolated electron or muon with $p_T>20 \GeV$ (raised to $p_T>25 \GeV$ using 2011 data), exactly two jets with $p_T>25 \GeV$ (three-jet events were added in the most recent 2011 analysis) and extended pseudo-rapidity coverage of $|\eta|<4.9$ (tightened to $|\eta|<4.5$ using 2011 data), exactly one $b$-tagged jet and $\MET>25\GeV$, $M_T>60 \GeV - \MET$.

A cut-based analysis was performed using $\Lint =35 \pbinv$ of 2010 data~\cite{ATLAS-CONF-2011-027}. In order to enrich the single top signal, additional selection requirements were placed on the $m_t$ estimator calculated using the leading $b$-jet and imposing a $W$-boson mass constraint, $130 < m_{l \nu b} < 210 \GeV$, and on the pseudo-rapidity of the non-$b$-tagged jet $|\eta_{light}|>2.5$. 
32 events were selected.
The main backgrounds, originating from QCD multi-jet and $W$+jets events, were estimated using data-driven methods. 
The cross section was measured as a counting experiment using a likelihood function which incorporated systematic uncertainties via a set of nuisance parameters. The result was
\begin{equation}
\sigma_{t-ch.} = 53 ^{+27}_{-24} \stat ^{+38}_{-27} \syst  {\rm\ pb} = 53 ^{+46}_{-36} \statsyst {\rm\ pb} \ ,
\end{equation}
where the systematic uncertainty was dominated by contributions from jet energy scale and $b$-tagging calibration, signal modeling and background estimation.
A cross check using a likelihood ratio method based on five discriminating variables yielded a consistent, but less precise, result.

An updated result based on $\Lint = 156 \pbinv$ of 2011 data extracted the cross section by employing a neural network~\cite{ATLAS-CONF-2011-088}.
The NN used in total 22 discriminating input variables corresponding to kinematic properties of individual jets and leptons and their combinations.
The most important variables were $m_{l\nu b}$, $|\eta_{light}|$ and the invariant mass of the two jets $m_{12}$. The analysis combined a three-layer feed-forward NN with complex preprocessing, using Bayesian regularization techniques~\cite{neurobayes,Feindt:2006pm}. The resulting NN output variable peaked at large (small) values for signal (background), respectively. In order to enrich the signal, a requirement $>0.86$ was imposed on the NN output variable. In total 134 events passed this selection, with a S/B of 1.14. The cross section was measured using a profile likelihood method. The result was
\begin{equation}
\sigma_{t-ch.} = 76 ^{+41}_{-21} \statsyst \rm\ pb \ ,
\end{equation}
corresponding to an observed (expected) significance of $6.2\sigma$ ($5.7\sigma$). The dominating contributions to the systematic uncertainty were due to the jet energy scale, $b$-tagging and signal modeling. A cross check analysis using a cut-based approach yielded a consistent result, but with a reduced expected sensitivity of $4.4\sigma$. 

\begin{figure}
\centering
\begin{minipage}{0.49\linewidth}
\centering
%\vspace{2mm}
\includegraphics[width=0.99\linewidth]{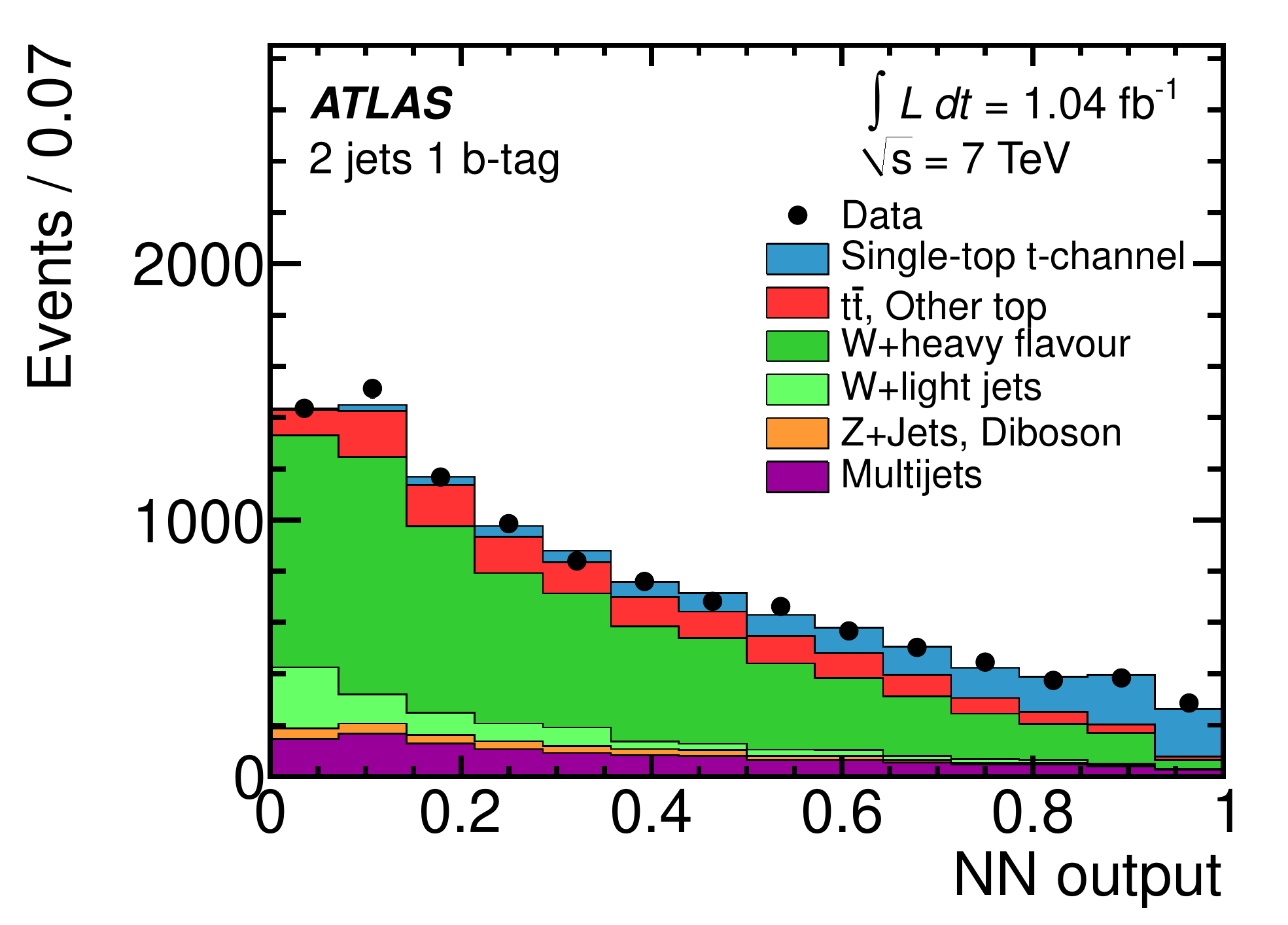}
\end{minipage}
\begin{minipage}{0.49\linewidth}
\centering
%\vspace{2mm}
\includegraphics[width=0.99\linewidth]{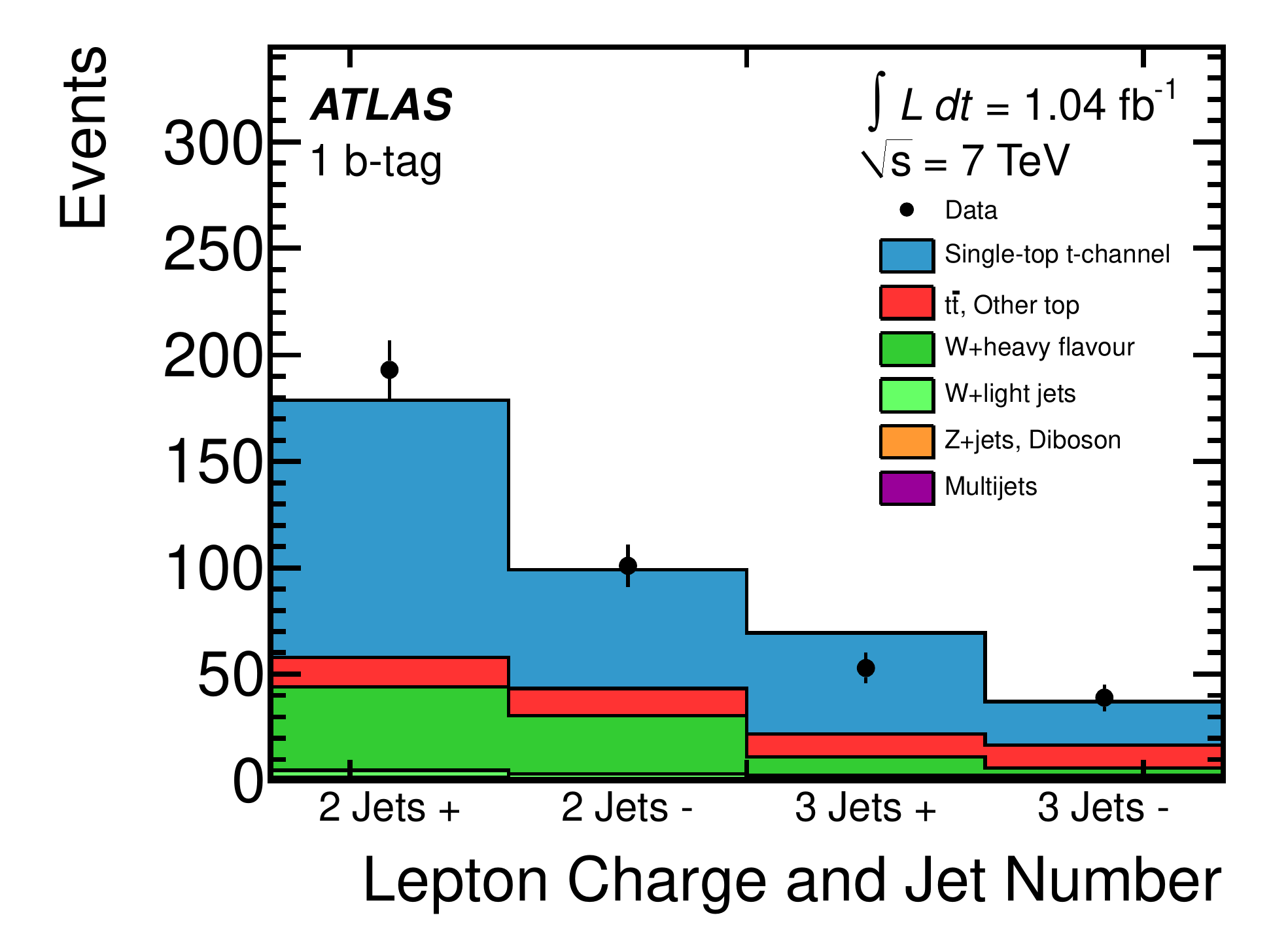}
\end{minipage}
\caption{Neural Network output distribution in the two-jet sample (left), and event counts per lepton charge and jet multiplicity (right), both from the ATLAS $t$-channel single top cross section measurement based on $\Lint = 1.04 \fbinv$ of 2011 data~\protect\cite{Collaboration:2012ux}.}
\label{fig:atlastchannel}
\end{figure}

The most recent ATLAS measurement of the $t$-channel single top cross section was based on $\Lint = 1.04 \fbinv$ of 2011 data~\cite{Collaboration:2012ux}.
The measurement was performed simultaneously in the samples with either two or three jets, both with one $b$-tagged jet, employing a neural network. In the two-jet sample, 12 input variables were used, the most discriminating being the top-mass estimator $m_{l\nu b}$, the pseudo-rapidity and the $E_T$ of the leading untagged jet. In the three-jet sample, 18 variables were used, the most discriminating ones being the invariant mass of the two leading jets $m_{12}$, $m_{l\nu b}$, and the absolute pseudo-rapidity difference of the leading and lowest $p_T$ jet. The QCD multi-jet background was estimated from a template fit to the $\MET$ distribution, while the shape of the $W$+jets background was taken from simulation, with the relative contributions of light and heavy flavors being determined in-situ from the data. The cross section was measured by means of a simultaneous maximum likelihood fit to the NN output distribution (Fig.~\ref{fig:atlastchannel}, left) in the two- and three-jet samples. The result was
\begin{equation}
\sigma_{t-ch.} = 83 \pm 4 \stat ^{+20}_{-19} \syst \rm\ pb \ ,
\end{equation}
corresponding to an observed (expected) significance of $7.2\sigma$ $(6.0\sigma)$. The most important systematic uncertainties were due to the jet energy scale and $b$-tagging efficiency knowledge, as well as signal and background modeling. The latter included an additional contribution due to an observed mis-modeling of jets a large $|\eta|$.
%5
An alternative cross section measurement was performed using a cut-based method which employed a tighter event selection in order to increase the signal significance. It yielded a consistent, but less precise result. Nevertheless, by splitting the samples by lepton charge (Fig.~\ref{fig:atlastchannel}, right), it was used to measure the single top quark and anti-quark cross sections separately, yielding $\sigma_{t-ch.}(t) = 59 ^{+18}_{-16} \rm\ pb$ and $\sigma_{t-ch.}(\bar{t}) = 33 ^{+13}_{-12} \rm\ pb$.
Finally, the value of $|V_{tb}|$ was extracted from the ratio of the measured and the theory cross sections and assuming $|V_{tb}|\gg|V_{ts}|,|V_{td}|$, independently of assumptions on the number of quark generations or on the unitarity of the CKM matrix. The result was
\begin{equation}
|V_{tb}| = \sqrt{\sigma_{exp} / \sigma_{th}} = 1.13 \, ^{+0.14}_{-0.13} \exper \pm 0.02 \ther \ .
\end{equation}
Restricting possible values of $|V_{tb}|$ to the interval $[0,1]$, the lower limit on  $|V_{tb}|$ was set at $|V_{tb}|>0.75$ at $95\%$ CL.

\subsubsection{CMS measurements}
\label{sec:singletop-tch-cms}

CMS has measured the $t$-channel single top quark cross section using the 2010 dataset corresponding to $\Lint = 36\pbinv$~\cite{Chatrchyan:2011vp}. Two complementary approaches were pursued, one performing a simultaneous fit of two discriminating variables (2D), the other one combining many input variables using a Boosted Decision Tree (BDT). 

Events were selected by requiring exactly one isolated muon (electron) with $p_T>20 \ (30) \GeV$ and exactly two jets with $p_T>30 \GeV$, one of which $b$-tagged using a track counting algorithm with a tight working point (see section~\ref{sec:btag}). In addition, events with a second $b$-jet (using a loose working point) were vetoed in the 2D analysis. To reduce contributions from QCD multi-jet production, a requirement $M_T> 40 \ (50) \GeV$ was applied in the muon (electron) channel. 184 (221) events were selected in the 2D (BDT) analysis, respectively, corresponding to a signal purity of $16-18\%$.
The background from QCD multi-jet production was estimated from a fit to the $M_T$ distribution. In the 2D analysis, the $W$+jets background was obtained from two control regions: one where the tight $b$-tagging requirement is not fulfilled (dominated by $W$+light jets events), and a subset of this sample where one jet passes a loose, but not tight, $b$-tagging requirement. Fits to the $M_T$ distributions were performed in both samples to determine $W$+light scale factors, while the heavy flavor scale factor was taken from the $\ttbar$ cross section measurement.

\begin{figure}
\centering
\begin{minipage}{0.51\linewidth}
\centering
\vspace{4mm}
\includegraphics[width=0.99\linewidth]{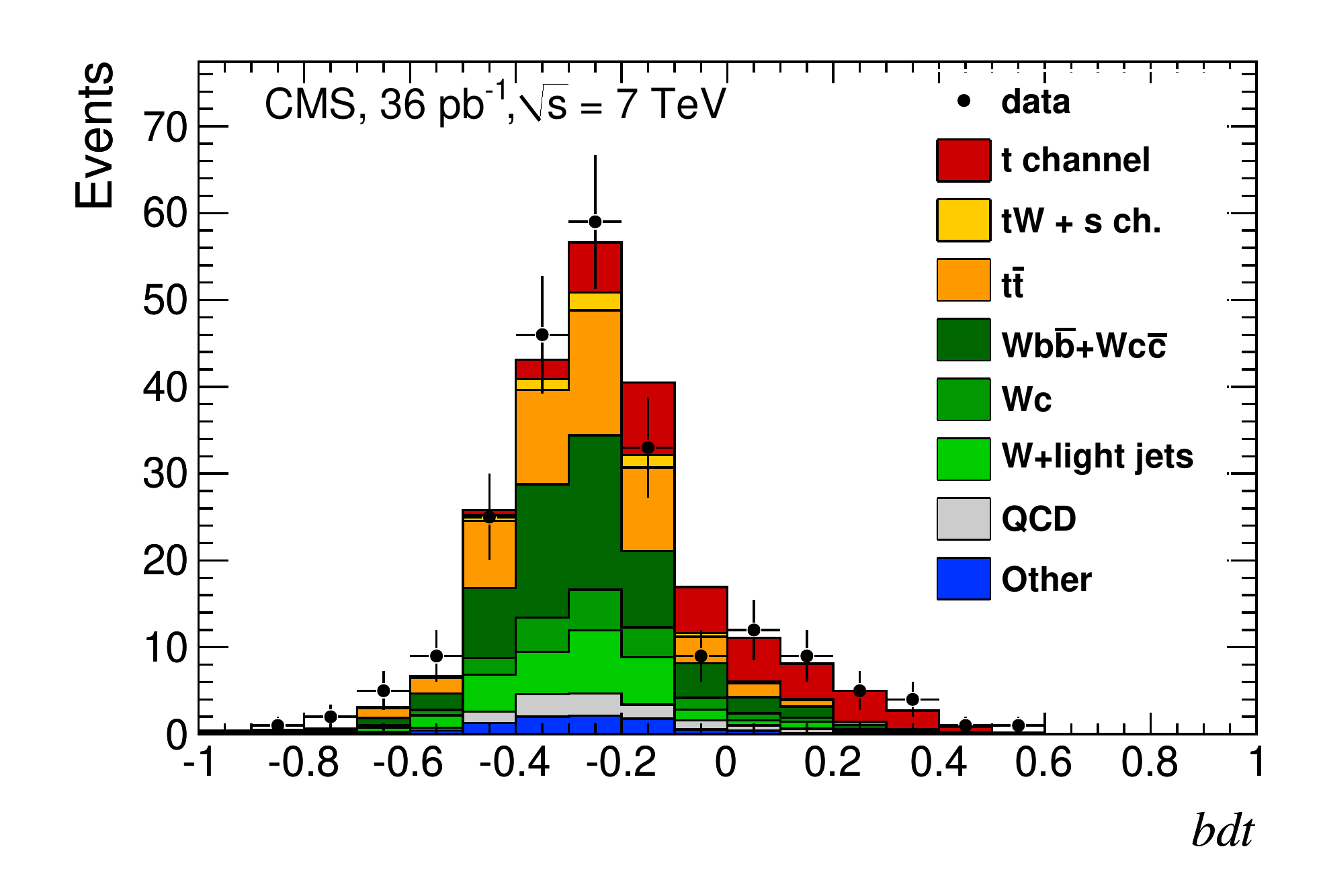}
\end{minipage}
\begin{minipage}{0.48\linewidth}
\centering
%\vspace{2mm}
\includegraphics[width=0.99\linewidth]{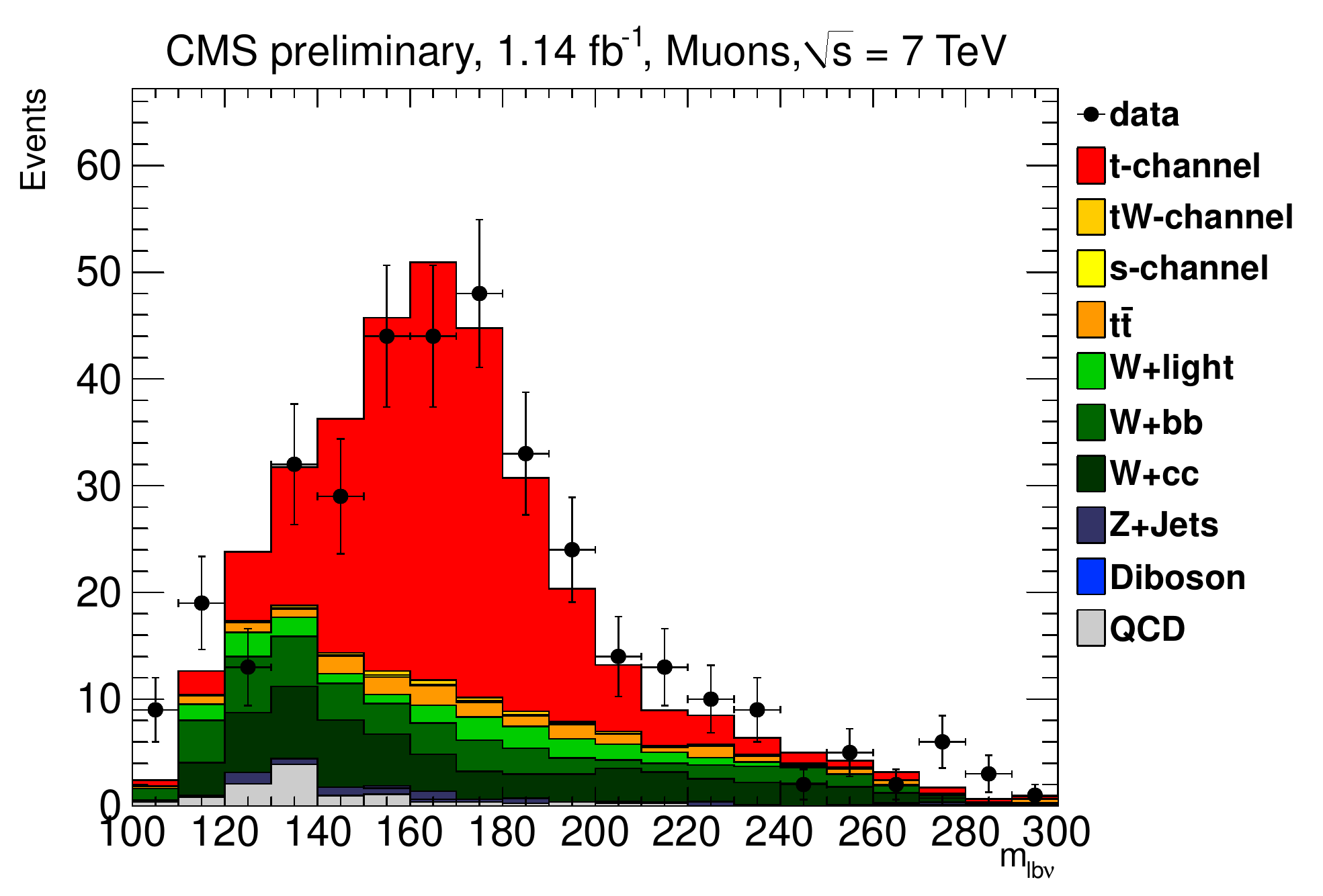}
\end{minipage}
\caption{Left: Distribution of the BDT classifier variable used in the CMS measurement of the single top quark $t$-channel cross section with 2010 data~\protect\cite{Chatrchyan:2011vp}. Right: Reconstructed mass $m_{l\nu b}$ in the CMS 2011 measurement~\protect\cite{cmspas-top-11-021}. In both cases, the simulations were scaled to the cross section fit results. }
\label{fig:cmstchannel}
\end{figure}

In the 2D analysis, a two-dimensional maximum likelihood was performed to the distributions of $|\eta_{light}|$ and $\cts$, defined here as the angle between the outgoing lepton direction and the spin axis, approximated by the direction of the light jet in the top-quark rest frame. Due to the V-A structure of the electroweak interaction, this variable shows a positive slope for single top signal, while it is expected to be flat for background. The $\eta_{light}$ distribution is interpreted as the pseudo-rapidity of the light jet recoiling against the single top quark, such that signal events are expected to accumulate on average at larger values compared with background.
In the fit, the total background contribution was allowed to float, while its composition was fixed according to the expected contributions. The shapes of the $W$+jets and QCD multi-jet backgrounds were taken from data, while the ones of the signal and the other backgrounds were from simulation. The cross section was measured as
\begin{equation}
\sigma_{t-ch.} = 124.2 \pm 33.8 \stat ^{+30.0}_{-33.9} \syst {\rm\ pb} \ .
\end{equation}

In the BDT analysis, 37 well modeled variables were combined into a single BDT classifier variable. The choice of variables was inspired by the Tevatron measurement from Ref.~\refcite{Abazov:2008kt}, optimized for LHC kinematics. The most discriminating variables were the lepton momentum, the mass of the system formed by the $W$-boson and the two jets, the $p_T$ of the system formed by the two jets, the $p_T$ of the $b$-tagged jet and the reconstructed top quark mass $m_{l \nu b}$. The BDT classifier variable was validated using simulated signal and $\ttbar$, as well as data-driven W+jets samples. The cross section was measured from the BDT classifier distribution (Fig.~\ref{fig:cmstchannel}, left), using a Bayesian approach in which background normalizations and systematic uncertainties were treated as nuisance parameters, as
\begin{equation}
\sigma_{t-ch.} = 78.7 \pm 25.4 \stat ^{+13.2}_{-14.6} \syst {\rm\ pb} \ .
\end{equation}

In both analyses, the systematic uncertainties were dominated by contributions from the imperfect knowledge of the jet energy scale and $b$-tagging calibrations, as well as the signal and background modeling. The results from both measurements were combined using the BLUE technique, considering a statistical correlation of $51\%$ and assuming that most of the systematic uncertainties were fully correlated. The combined measurement was
\begin{equation}
\sigma_{t-ch.} = 83.6 \pm 29.8 \statsyst \pm 3.3 \lumi {\rm\ pb} \ ,
\end{equation}
where the weight of the BDT analysis in the combination was $89\%$. The observed significance was $3.7\sigma$ ($3.5\sigma$) for the 2D (BDT) analysis, while $2.1\pm1.1\sigma$ ($2.9\pm1.0\sigma$) were expected.

Similarly to the ATLAS measurement and assuming $|V_{tb}|\gg|V_{ts}|,|V_{td}|$, an effective value of the CKM matrix element $|V_{tb}|$ was determined as 
\begin{equation}
|V_{tb}| = 1.14 \pm 0.22 \exper \pm 0.02 \ther \ .
\end{equation}
Using the assumption that $0 \leq |V_{tb}|^2 \leq 1$, a lower bound of $|V_{tb}|> 0.62 \ (0.68)$ was inferred at $95\%$ CL from the 2D (BDT) analysis, respectively.

CMS presented an updated measurement using 2011 data corresponding up to $\Lint = 1.51 \fbinv$ in the lepton+jets channel~\cite{cmspas-top-11-021}. Events were selected in a similar way as for the 2010 measurement, but employing a $b$-tagging requirement already at the trigger level. The signal region was defined by the requirements of two jets, one of them $b$-tagged, and a reconstructed top mass in the range $130 < m_{l\nu b} < 220 \GeV$. QCD multi-jet background was estimated from data, and the shape of the $W$+jets background was estimated from the  $m_{l\nu b}$ sideband region. The cross section was extracted from a maximum likelihood fit to the $|\eta_{light}|$ distribution, resulting in
\begin{equation}
\sigma_{t-ch.} = 70.2 \pm 5.2 \stat \pm 10.4 \syst \pm 3.4 \lumi {\rm\ pb} \ .
\end{equation}
Due to the much larger dataset the systematic uncertainties could be significantly better constrained compared with the previous measurement. They were dominated by contributions originating from the $W$+jets background estimation, the jet energy scale and the modeling of signal and top pair background. Fig.~\ref{fig:cmstchannel} (right) shows the $m_{l\nu b}$ distribution after the fit, which exhibits a clear peak due to single top quark production. Finally, the CKM matrix element $|V_{tb}|$ was extracted as
\begin{equation}
|V_{tb}| = 1.04 \pm 0.09 \exper \pm 0.02 \ther \ .
\end{equation}

Both the ATLAS and CMS measurements of $t$-channel single top quark production are in good agreement with the SM prediction (see section~\ref{sec:stoptheo}).

%%%%%%%%%%%%%%%%%%%%%%%%%%%%%%%%%%%%%%%%%%%%%%%%%%%%%%%%%%%%%%%%%%%%%%%%%%%%%%%

\subsection{tW-channel}

Single top quark production in the $tW$-channel has not been observed at the Tevatron. As discussed in section~\ref{sec:mcsingletop}, this mode interferes at NLO QCD with top quark pair production, and several methods have been implemented in current MC generators to allow an unambiguous signal definition.
According to the decays of the two $W$-bosons, single top production in the $tW$-channel can be studied either in the di-lepton channel, where both $W$-bosons decay into a charged lepton and a neutrino, or in the lepton+jets channel, where one of the $W$-bosons decays into lepton and neutrino, while the other one decays into two jets.
Older feasibility studies based on simulation can be found in Refs.~\refcite{cmsnote-2006-084,atlastdr,atl-phys-pub-2009-001}.

\begin{figure}[t]
\centering
\begin{minipage}{0.49\textwidth}
\centering
\vspace{2mm}
\includegraphics[width=0.99\textwidth]{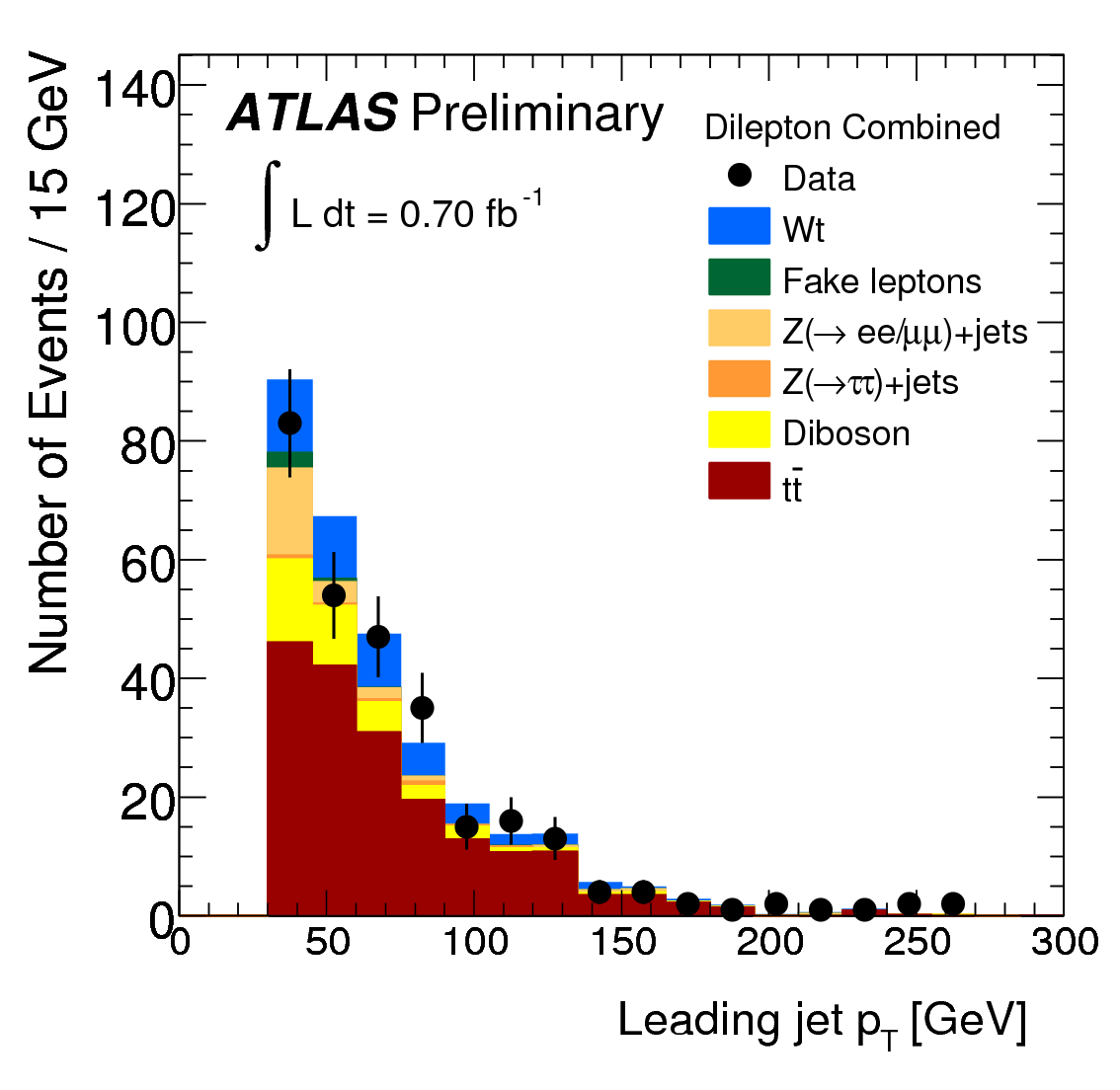}
\end{minipage}
\begin{minipage}{0.48\textwidth}
\centering
\includegraphics[width=0.99\textwidth]{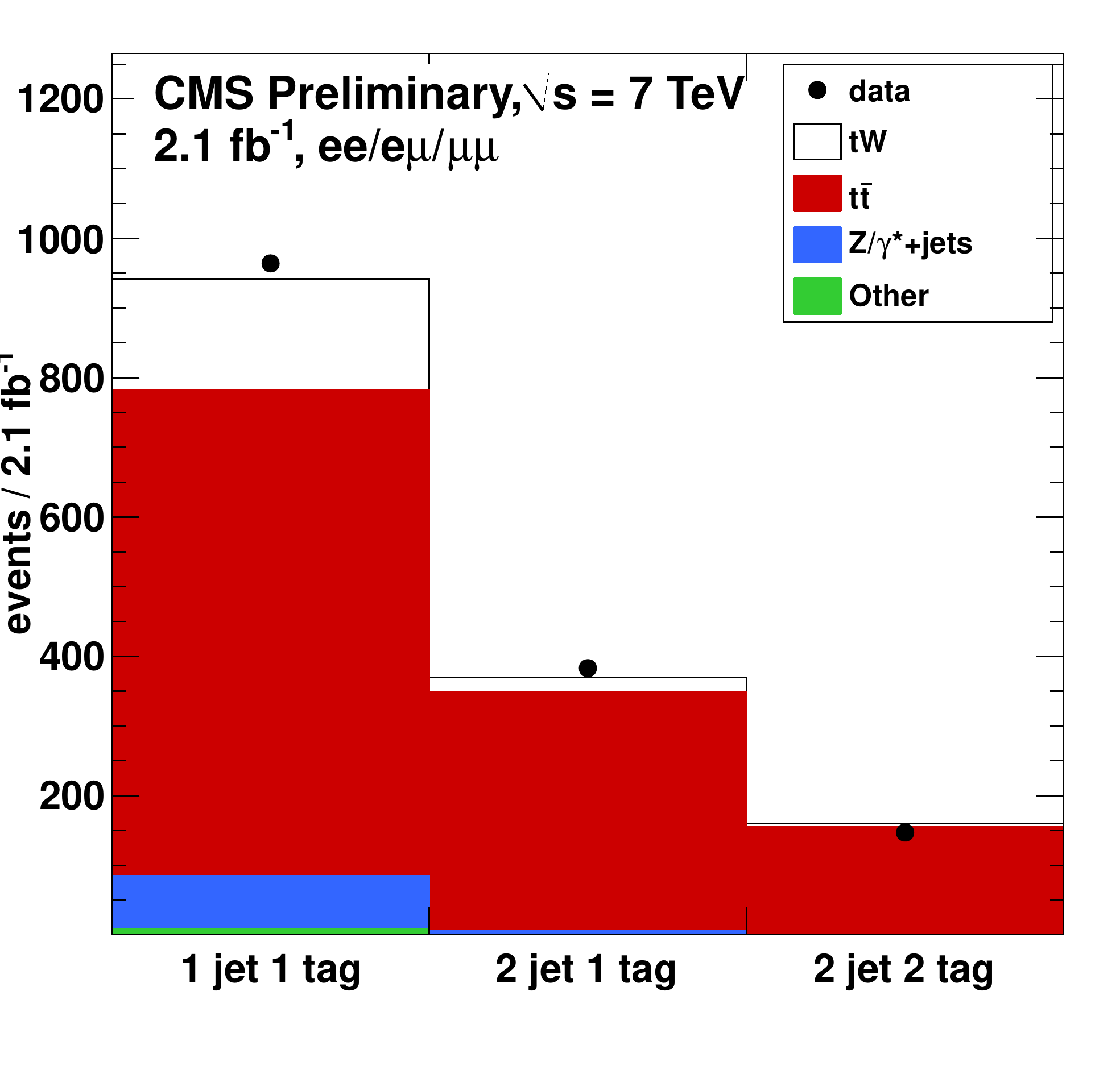}
\end{minipage}
\caption{Left: Leading jet $p_T$ for events passing the single top $tW$-channel event selection in ATLAS~\protect\cite{ATLAS-CONF-2011-104}. Right: Event yields in the signal (one jet, one $b$-tag) and control (two jets, one or two $b$-tags) regions in the CMS single top $tW$-channel measurement~\protect\cite{CMS-PAS-TOP-11-022}.   }
\label{fig:twchannel}
\end{figure}

Following a first study using 2010 data~\cite{ATLAS-CONF-2011-027}, a search for single top quark production in the $tW$-channel in the di-lepton channel was performed by ATLAS using $\Lint=0.70\fbinv$ of data~\cite{ATLAS-CONF-2011-104}. The event selection required exactly two isolated leptons of opposite charge with $p_T>25 \GeV$, one or more jets with $p_T>30\GeV$, $\MET>50\GeV$ and $|M(ll)-m_{Z}|>10 \GeV$ (only in the $ee$,$\mu\mu$ channels). There was no $b$-tagging requirement.
In order to reduce the $Z\rightarrow \tau \tau$ background, an additional requirement was placed on the angle between each lepton $l_i$ and the missing transverse energy: $\sum_i\Delta\phi(l_i,\MET) >2.5$.
Backgrounds with one (mostly from $W$+jets) or two (mostly from QCD multi-jet production) fake leptons were estimated using a matrix method. Drell-Yan backgrounds were estimated using control regions.
The event selection yielded 1073 events, while 78 (1033) signal (background) events were expected, the background mostly originating from $\ttbar$ production (see Fig.~\ref{fig:twchannel}, left). The $tW$-channel signal region was defined by requiring exactly one jet. A control region used to estimate the $\ttbar$  background was defined by requiring at least two jets, and the resulting scale factor was applied to the signal region, assuming that the $\ttbar$ simulation correctly models the jet multiplicity. The signal region contained 287 events, which matched the sum of expected signal ($47.1\pm 3.4$) and background ($246\pm23$) yields.
The cross section was estimated by maximizing a likelihood function which also included parameterized effects of systematic uncertainties as nuisance parameters. The result was
\begin{equation}
\sigma_{tW-ch.} = 14.4 \, ^{+5.3}_{-5.1} \stat ^{+9.7}_{-9.4} \syst \rm\ pb \ .
\end{equation}
The systematic uncertainty was dominated by the jet energy scale and resolution, top pair background and signal modeling.
The background-only hypothesis is rejected at the $1.2\sigma$ level, and the $95\%$ CL observed (expected) cross section upper limit is 39.1 (40.6) pb.

CMS performed a search for single top production in the $tW$-channel using $\Lint = 2.1 \fbinv$ collected in 2011~\cite{CMS-PAS-TOP-11-022}. Events were selected in the di-lepton channel, containing exactly two isolated leptons of opposite charge with $p_T>20\GeV$, $|M(ll)-m_{Z}|>10 \GeV$, $\MET>30 \GeV$ (the latter two requirements were applied only in the $ee$, $\mu\mu$ channels), and exactly one jet with $p_T>30\GeV$ which had to be $b$-tagged. To further reduce the contribution from $\ttbar$ background, events with additional $b$-jets with $p_T>20 \GeV$ were vetoed. The $p_{T,{\rm system}}$ variable, corresponding to the $p_T$ of the system formed by the two leptons, the jet and the $\MET$ was required to be $p_{T,{\rm system}}<60\GeV$, and $H_T$, the scalar sum of the transverse momenta of leptons, jet and $\MET$ was required to be $H_T > 160 \GeV$. In total 964 events were selected, with an expectation of 152 signal events.
Background from Drell-Yan production was estimated from data in the same way as done for the $\ttbar$ cross section measurement (see section~\ref{sec:cmsdil}). Two control regions, defined by the presence of at least two jets, either one or both $b$-tagged, were used to constrain the $\ttbar$ background in the cross section measurement (see Fig.~\ref{fig:twchannel}, right).
The cross section was measured performing a counting experiment, assuming Poisson statistics, simultaneously in the signal and the two sideband regions. A likelihood function was maximized, which incorporated effects of systematic uncertainties as nuisance parameters, in order to determine the normalizations of the single top $tW$-channel signal as well as the $\ttbar$ background.  The resulting cross section was
\begin{equation}
\sigma_{tW-ch.} = 22 \, ^{+7}_{-9} \statsyst \rm\ pb \ ,
\end{equation}
and the observed (expected) significance at 95\% CL was $2.7\sigma$ ($1.8\pm 0.9 \sigma$). Important sources of systematic uncertainty were due to the jet energy scale and $b$-tagging calibrations, as well as the signal and background modeling.

The results from both experiments are in good agreement with the SM prediction, even though the measurement precision is still limited.

%%%%%%%%%%%%%%%%%%%%%%%%%%%%%%%%%%%%%%%%%%%%%%%%%%%%%%%%%%%%%%%%%%%%%%%%%%%%%%%

\subsection{s-channel}

Even though the observation of single top quark production at the Tevatron was based on the combination of $t$- and $s$-channels, the $s$-channel mode has not been observed individually. Despite the small cross section of $4.6 \rm\ pb$~\cite{Kidonakis:2010tc} at the LHC for $\sqrt{s}=7 \TeV$, this mode is interesting since it is sensitive to various new physics models, for instance $W'$ bosons or a charged Higgs boson $H^\pm$~\cite{Tait:2000sh}. At leading order, the partonic final state consists of a top-quark and a $b$-quark.
A feasibility study for the LHC using simulation can be found in Ref.~\refcite{atlastdr}.

ATLAS performed an initial search for single top quark production in the $s$-channel using a data sample of $\Lint=0.70\fbinv$~\cite{ATLAS-CONF-2011-118}. The analysis considered the final state where the $W$-boson from the top quark decays leptonically. Events were selected which contained exactly one isolated electron or muon with $p_T>25 \GeV$, $\MET>25 \GeV$, $M_T(W)> 60 \GeV - \MET$ and exactly two jets with $p_T>25 \GeV$. In this event sample, either one or both jets could be required to be $b$-tagged using a secondary vertex algorithm. The most important backgrounds, originating from $W$+jets and QCD multi-jet production, were estimated from data. 
Other backgrounds, including $\ttbar$ production, were estimated from simulation.
A further refinement of the event selection was applied in order to enrich the signal and suppress backgrounds. This included the requirement that both jets  had to be $b$-tagged, as well as several requirements on kinematic variables, such as $M_T(W)$, the top mass reconstructed using either the first or second jet, the $p_T$ of the system formed by the two jets, and the opening angle between the two jets, or between the leading jet and the lepton. Optimizing the selection resulted in an improvement of the number of expected signal events divided by the square root of the number of expected background events, $S/\sqrt{B}$ from 0.26 to 0.98, corresponding to an expected number of 16 (269) signal (background) events. 296 events were observed in data, in good agreement with expectation.
A limit on $s$-channel single top quark production was obtained using a counting experiment and minimizing a likelihood function which included nuisance parameters for systematic uncertainties. The resulting observed (expected) $95\%$ CL upper limit is 26.5 (20.5) pb, about five times larger than the SM expectation.

\subsection{Outlook}

The near future will see differential measurements of the single top quark cross section in the $t$-channel, more precise measurement of the $tW$ associated production mode, and potentially the observation of the $s$-channel mode. Since the measurements will be more and more limited by the knowledge of systematic uncertainties, it will be important to constrain these even more using in-situ techniques, as well as to further improve the background rejection using multi-variate methods.

Measurements of single top quark production provide important information in the context of the SM.
The latter concerns in particular the determination of CKM matrix elements $|V_{tq}|$ by combining the information from measurements of single top quark cross sections with the measured value of $R$~\cite{Alwall:2006bx,Lacker:2012ek}, i.e. without the assumption that $R=1$.
In addition, constraints on the $b$-quark PDF may be possible from differential single top quark cross section measurements.

Beyond the SM, measurements of single top quark production in all channels are important because of their different sensitivity to new physics contributions~\cite{Tait:2000sh}. For example, a deficit in all channels would be an indication that $|V_{tb}|<1$. An excess in the $s$-channel may be the hint of the production of a charged resonance, while a $t$-channel excess may indicate a contribution from FCNC $ug\rightarrow t$ production (see section~\ref{sec:fcnc}).

\section{Search for New Physics using Top Quarks}
\label{sec:topbsm}

There are many scenarios of new physics beyond the SM which involve top quarks, and the current status of corresponding searches at the LHC are discussed in the following. Topics include new particles decaying into $\ttbar$ pairs, FCNC in top quark decays and production, anomalous $\MET$ in $\ttbar$ production, same-sign top quark production, a potential fourth generation of quarks and other searches for new physics using top quarks.

\subsection{$\ttbar$ invariant mass distribution}
\label{sec:mttbar}

There are many extensions of the SM predicting new interactions which have enhanced couplings to top quarks, resulting in new particles that would decay dominantly into $\ttbar$ pairs and may, depending on the width of the new particle, show up as resonances in the top-quark pair invariant mass distribution $M_\ttbar$. In many models, the couplings to the first two generations may be small, such that the new resonance would not necessarily show up in other topologies. Besides the simple picture of a resonance shape on top of the SM background, also more complicated effects on the shape of the $M_\ttbar$ may be possible, which could be caused e.g. by wide resonances, interference effects,  associated production of scalar particles or additional undetected decay products.

New particles coupling predominantly to top quarks could be realized in many different ways (see, e.g., Ref.~\refcite{Frederix:2007gi} and references therein for an overview).
They could be spin-0 scalars or pseudo-scalars, for instance in super-symmetric (SUSY) or Two-Higgs-Doublet (2HDM) models. They could also be spin-1 vector or axial-vector particles, for example a leptophobic topcolor $Z'$ boson, a Kaluza-Klein (KK) gluon in the Randall-Sundrum (RS) model or an axigluon. Finally, models involving spin-2 gravitons have also been proposed. 
Searches at Tevatron and LHC typically use the topcolor $Z'$ model~\cite{Hill:1993hs,Hill:1994hp,Harris:1999ya,Harris:2011ez} as a proxy for a narrow $\ttbar$ resonance of width $\Gamma_{Z'}/m_{Z'} \sim 1\%$ or $3\%$, while a KK gluon in the RS model of warped extra dimensions~\cite{Randall:1999ee,Agashe:2006hk,Lillie:2007yh} is used as proxy for a broader resonance, often assuming a width of $10\%$.
Other models were proposed in order to explain the larger than predicted $A_{FB}$ at the Tevatron (see section~\ref{sec:ac}), see, e.g., Refs.~\refcite{Antunano:2007da,Frampton:2009rk,Bai:2011ed,Gresham:2011pa,Djouadi:2011aj}, as well as model-independent studies~\cite{Delaunay:2011gv,AguilarSaavedra:2011vw}.

The most stringent limits on a narrow topcolor $Z'$ decaying into $\ttbar$ obtained at the Tevatron~\cite{Abazov:2011gv,Aaltonen:2011ts} provide an exclusion for masses below $m_\ttbar \sim 900 \GeV$.
Feasibility studies, based on simulation, on searches for resonances in the top-quark pair invariant mass distribution can be found in Refs.~\refcite{cmspas-top-09-009,cmspas-exo-09-008,cmspas-exo-09-002} (CMS) and in Refs.~\refcite{Cogneras2006,atlastdr,atl-phys-pub-2009-081,ATL-PHYS-PUB-2010-008} (ATLAS).

In the following, first the analyses focusing on the invariant mass range where $M_\ttbar$ is smaller than around $1 \TeV$ are discussed (so-called \textit{low-mass regime}). In this kinematic domain, the final state topologies resemble standard $\ttbar$ final states. Then, analyses focusing on the \textit{high-mass regime}, corresponding to $M_\ttbar$ significantly larger than $1 \TeV$, are presented. At high mass, decay products of the top quarks tend to be collimated, such that dedicated reconstruction algorithms become necessary.

\subsubsection{Low-mass analyses}

\begin{figure}[t]
\centering
\begin{minipage}{0.49\textwidth}
\centering
%\vspace{2mm}
\includegraphics[width=0.99\textwidth]{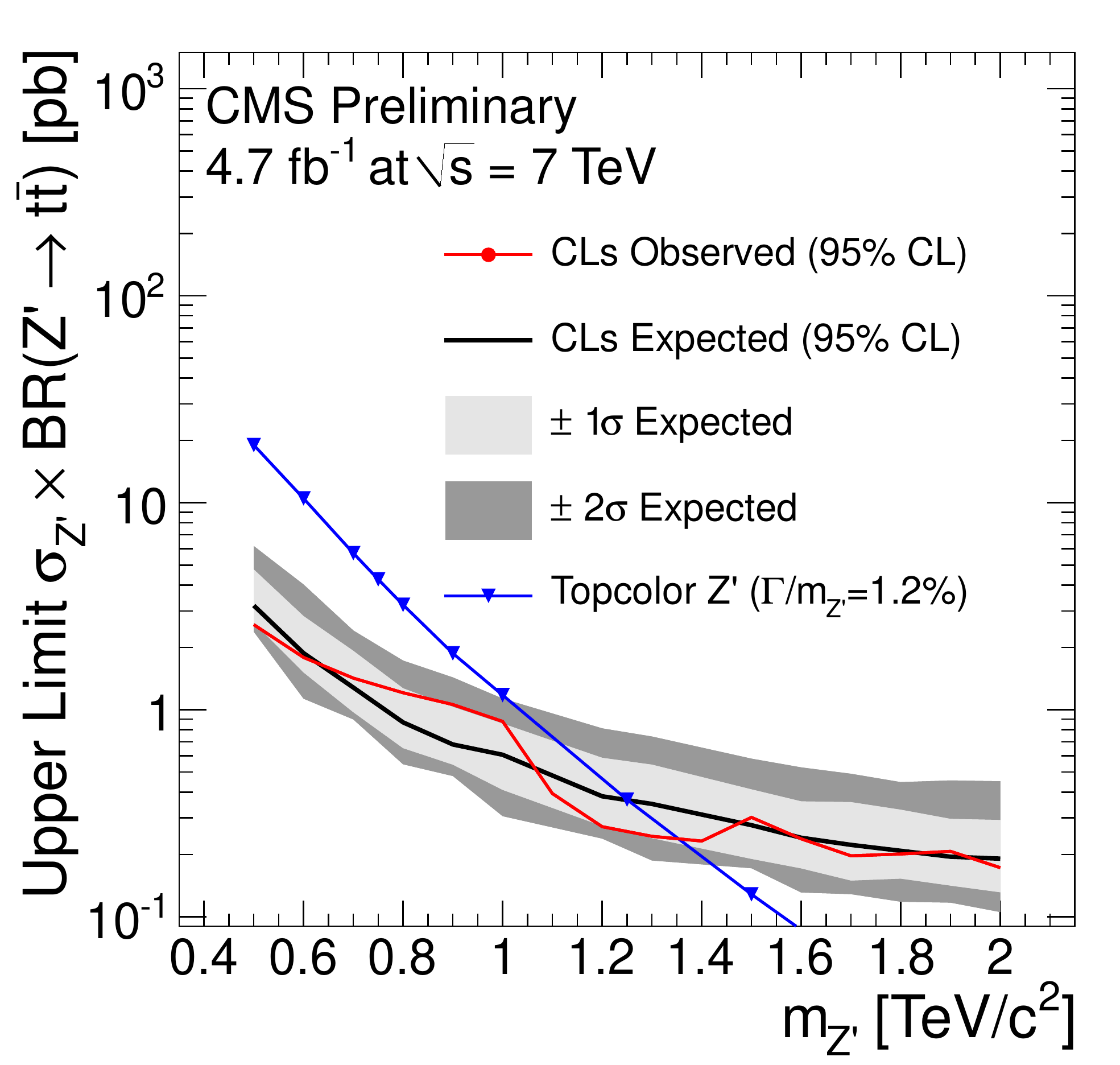}
\end{minipage}
\begin{minipage}{0.49\textwidth}
\centering
\vspace{2mm}
\includegraphics[width=0.99\textwidth]{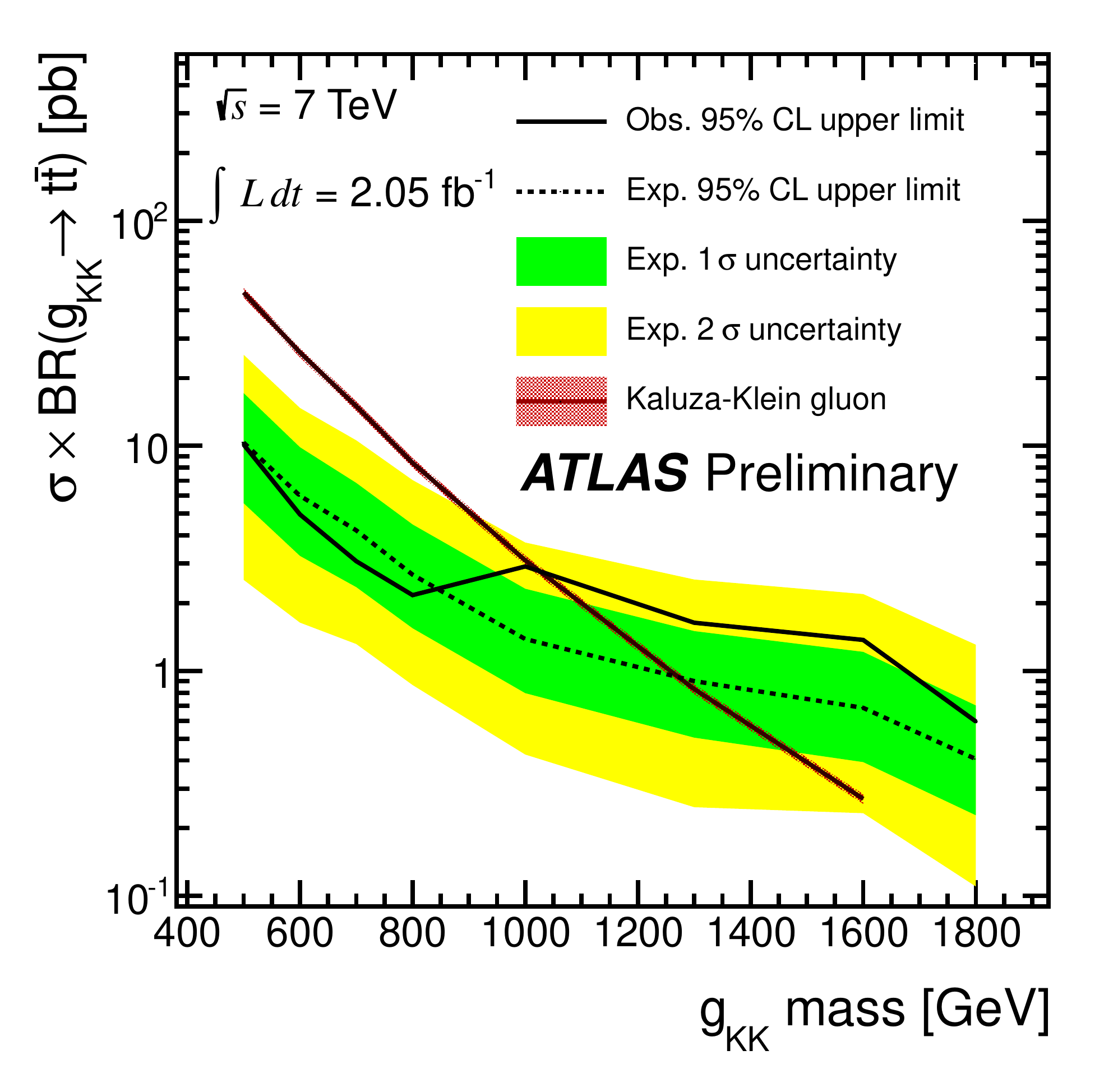}
\end{minipage}
\caption{Observed and expected upper limits on $\sigma_{Z'} \cdot \BR (Z'\rightarrow \ttbar)$ (CMS~\protect\cite{cmspas-top-11-009}, left) and  $\sigma_{g_{KK}} \cdot \BR (g_{KK}\rightarrow \ttbar)$ (ATLAS~\protect\cite{ATLAS-CONF-2012-029}, right), as function of mass.  }
\label{fig:mttbarlowmass}
\end{figure}

CMS presented a first search for new resonances decaying into a $\ttbar$ pair in the lepton+jets channel using $\Lint = 36\pbinv$ of 2010 data~\cite{CMS-PAS-TOP-10-007}, which was subsequently updated to the full 2011 dataset of $\Lint = 4.7 \fbinv$~\cite{cmspas-top-11-009}. Only the latter measurement will be discussed in the following.
Events were selected requiring the presence of exactly one isolated lepton ($e$ or $\mu$), $\MET>20\GeV$ and at least three jets with $p_T>50\GeV$. The leading jet had to satisfy $p_T>70 \GeV$. If four jets with $p_T>50\GeV$ were found, additional jets with  $p_T>30 \GeV$ were also considered. For each lepton flavor, events were assigned to one out of four categories according to their jet multiplicity (j) and the number of identified $b$-jets (t) using a secondary vertex algorithm, the categories being 3j1t, 4j0t, 4j1t and 4j2t. This improved the sensitivity of the measurement due to the different S/B in each category.
The $\ttbar$ invariant mass $M_{\ttbar}$ was reconstructed using a $\chi^2$ criterion which employed various mass and momentum constraints. The observed mass distributions showed no excess above SM expectations.
Shape and normalization of the background from QCD multi-jet production were determined from data, and the shape of the $W+$jets background was also estimated from a control region. The shape of the other backgrounds, in particular from SM top quark pair, single top quark and $Z$+jets production, were taken from simulation, while their normalization was determined during the statistical evaluation, which was performed simultaneously in the binned $M_{\ttbar}$ distributions in all eight categories. The impact of systematic uncertainties on rate and shape of the signal and background templates was included with nuisance parameters. 
The $\rm CL_S$ criterion~\cite{Read:2002hq,Junk:1999kv} was applied to set limits on the production of new particles as a function of mass.
The observed 95\% CL upper limits on the cross section times branching ratio of a new heavy particle range from about 6 pb at 400 GeV to less than 1 pb above 900 GeV. A  topcolor $Z'$ with width $1\% \ (10\%)$ is excluded for masses below $1.3 \ (1.7) \TeV$, while a KK gluon is excluded with a mass below $1.4 \TeV$. As an example, the limit on a $Z'$ with a width of $1.2\%$ is shown in Fig.~\ref{fig:mttbarlowmass} (left).

CMS also searched for a narrow $Z'$ resonance in the di-lepton channel using the full 2011 dataset ($\Lint = 5.0 \fbinv$)~\cite{cmspas-top-11-010}. A standard di-lepton event selection was used, requiring at least one out of two jets to be $b$-tagged. Backgrounds from Drell-Yan and QCD multi-jet production were estimated in a data-driven way. The $\ttbar$ invariant mass was calculated from all final state objects, assuming $p_z=0$ for the two neutrinos. A neural network was used to search for a resonance, and $95\%$ CL upper limits on $\sigma_{Z'} \cdot \BR(Z' \rightarrow \ttbar)$ for a narrow topcolor $Z'$ with a width of $1.2\%$ were set as function of $m_{Z'}$ using the $\rm CL_S$ method. Masses below $m_{Z'}<1.1 \TeV$ were excluded.

ATLAS performed an initial search for $\ttbar$ resonances in the lepton+jets channel using 2010 data~\cite{ATLAS-CONF-2011-070}, which was subsequently updated using first $\Lint = 200 \pbinv$~\cite{ATLAS-CONF-2011-087} and most recently $\Lint = 2.05\fbinv$~\cite{ATLAS-CONF-2012-029} of 2011 data. Since the analyses are very similar, only the most recent result will be discussed in the following. 
Lepton+jets events were selected following the criteria used in the SM $\ttbar$ cross section measurement, see section~\ref{sec:atlasljets}, requiring at least four jets, one of which had to be identified as $b$-jet. In case one of the jets had a jet mass above $60 \GeV$ it was assumed to be due to jet merging and the event was accepted even if it contained only three jets.
$W$+jets and QCD multi-jet backgrounds were estimated from data. The $\ttbar$ system was reconstructed using an algorithm which disfavors jets far away from the remainder of the activity in the event. No excess was found in the reconstructed $M_{\ttbar}$ distribution and limits were derived using a Bayesian approach. 
The observed (expected) 95\% CL limit on  $\sigma_{Z'} \cdot \BR (Z'\rightarrow \ttbar)$ range from 9.3 (8.5) pb at $m_{Z'} = 500 \GeV$ to 0.95 (0.62) pb at $m_{Z'} = 1300 \GeV$, which allowed to exclude a  topcolor $Z'$ in the range $500<m_{Z'}<860 \GeV$.
The observed (expected) limit on $\sigma_{g_{KK}} \cdot \BR (g_{KK}\rightarrow \ttbar)$ ranges from 11.6 (10.3) pb at $m_{g_{KK}} = 500 \GeV$ to 1.6 (0.9) pb at $m_{g_{KK}} = 1300 \GeV$, which excluded $g_{KK}$ resonances with mass between 500 and 1025 GeV at 95\% CL, see Fig.~\ref{fig:mttbarlowmass} (right).

\begin{figure}[t]
\centering
\begin{minipage}{0.49\textwidth}
\centering
%\vspace{2mm}
\includegraphics[width=0.99\textwidth]{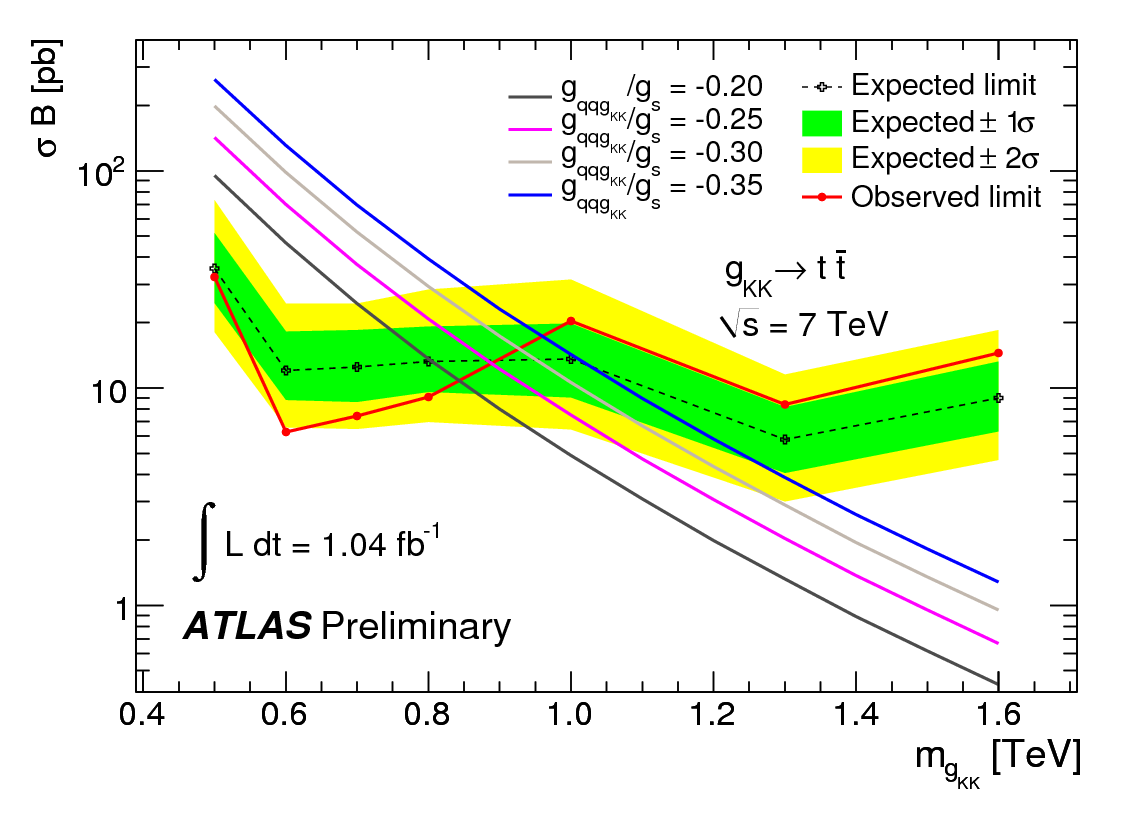}
\end{minipage}
\begin{minipage}{0.49\textwidth}
\centering
%\vspace{2mm}
\includegraphics[width=0.99\textwidth]{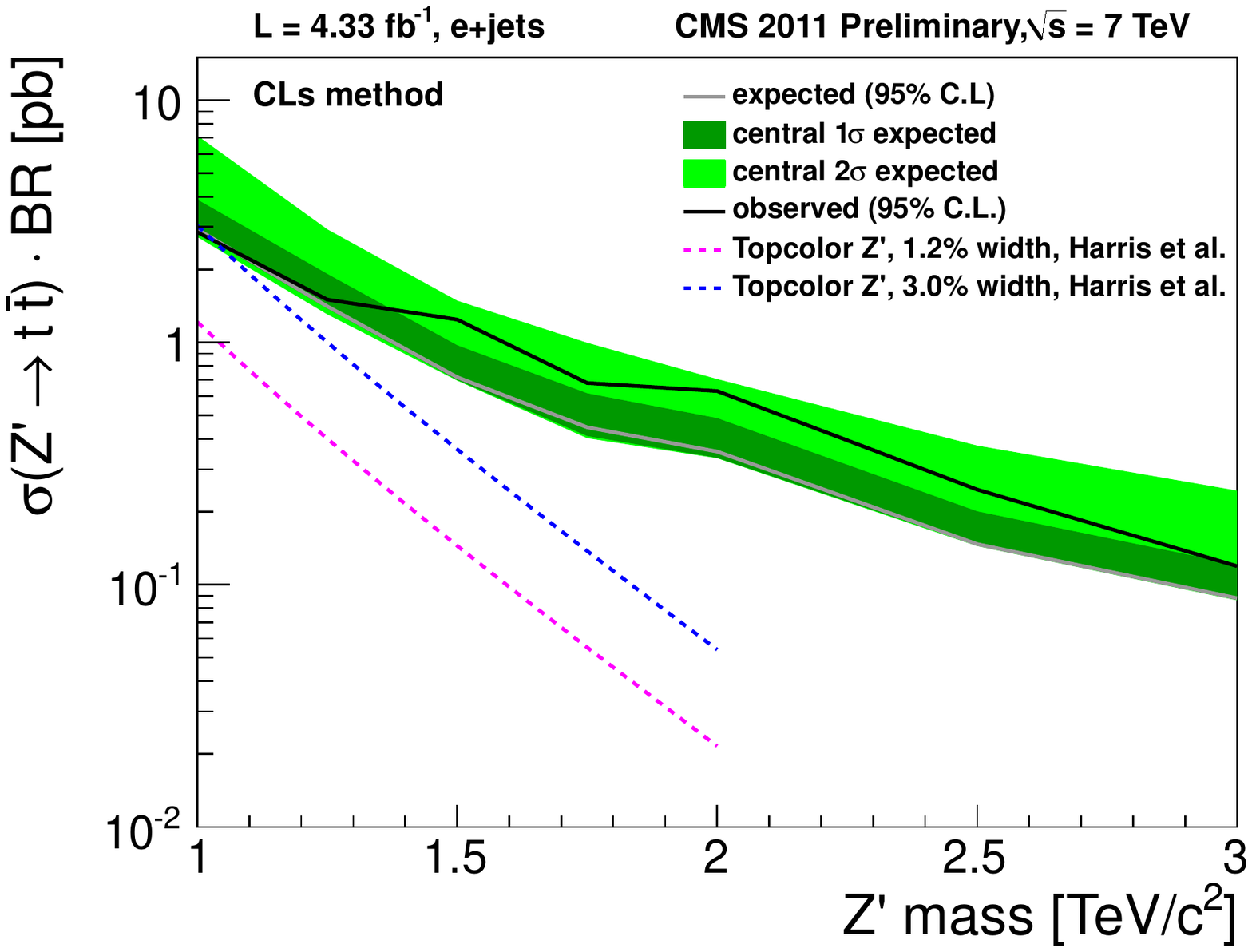}
\end{minipage}
\caption{Expected and observed limits on cross section times branching ratio at 95\% CL and expected cross section for a KK gluon $g_{KK}$ (left, ATLAS~\protect\cite{ATLAS-CONF-2011-123}), and for a narrow heavy resonance compared with the cross section for a topcolor $Z'$ (right, CMS~\protect\cite{cmspas-exo-11-092}), both shown as function of the invariant mass of the resonance.    }
\label{fig:mttbarlowmassdilhighmass}
\end{figure}

A search for a KK-gluon resonance decaying into $\ttbar$ was performed by ATLAS using $\Lint=1.04\pbinv$ of data, employing the di-lepton final state~\cite{ATLAS-CONF-2011-123}. Despite the lower branching ratio compared with the lepton+jets channel, this has the advantage of less non-top background. Events were selected following the criteria used in the $\ttbar$ cross section measurement in the same channel, requiring exactly two opposite charge leptons and two or more jets. Backgrounds from $Z$+jets, $W$+jets and QCD multi-jet events were estimated using data-driven techniques also used in the $\ttbar$ cross section measurement. The analysis used the $H_T+\MET$ distribution, where $H_T$ is the sum of the transverse momenta of the two leptons and all reconstructed jets. Upper limits on the cross section times branching ratio for KK-gluons were derived as a function of the mass $m_{g_{KK}}$ using a Bayesian approach and for varying couplings of the KK gluon. The obtained limits are shown in Fig.~\ref{fig:mttbarlowmassdilhighmass} (left). Depending on the choice of coupling, lower mass limits in the range 0.84 to 0.96 TeV were set.

%%%%%%%%%%%%%%%%%%%%%%%%%%%%%%%%%%%%%%%%%%%%%%%%%%%%%%%%%%%%%%%%%%%%%%%%%%%%%%%

\subsubsection{High-mass analyses}

A search for heavy narrow resonances decaying into $\ttbar$ was performed by CMS in the muon+jets channel using a data sample corresponding to $\Lint = 1.14 \fbinv$~\cite{CMS-PAS-EXO-11-055}. The standard selection of events in the lepton+jets topology was modified in order to account for the fact that decay products of top quarks with high momentum tend to be collimated. In particular, the usual requirement that the muon ($p_T>35 \GeV$) from the W-boson decay be isolated was replaced with a two-dimensional requirement on the radial distance and relative $p_T$ with respect to the closest jet. To accommodate for merging of several close-by jets, only two jets were required in the selection, with $p_T>250 \ (50) \GeV$ for the first (second) jet. In addition, the scalar sum of muon $p_T$ and $\MET$ was required to fulfill $H_{T,lep}>150 \GeV$. In the reconstruction of the invariant $\ttbar$ mass, the assignment of jets to top quarks was based on topological criteria which favor back-to-back highly boosted top quark pairs. The $M_{\ttbar}$ resolution varied from 13\% at $M_{\ttbar}=1.0 \TeV$ to 7\% at 3 TeV. Background from QCD multi-jet events was modeled from data using a sideband region, while the shapes of the other backgrounds were taken from simulation. Upper limits on $\sigma_{Z'} \cdot \BR (Z'\rightarrow \ttbar)$ were obtained using a Bayesian method which evaluated two distributions, $H_{T,lep}$  for $H_{T,lep}<150 \GeV$ and $M_{\ttbar}$  for $H_{T,lep}> 150 \GeV$. The likelihood included terms for signal and backgrounds, as well as the impact of systematic uncertainties via nuisance parameters. 
The analysis could not yet exclude a topcolor $Z'$ with a natural width of 1\%, but a topcolor $Z'$ with a width of 3\% could be excluded for $805 < m_{Z'} < 935 \GeV$ and $960 < m_{Z'} < 1060 \GeV$.

Complementing the above search in the muon+jets channel, CMS performed a very similar search in the electron+jets channel using $\Lint=4.3\fbinv$ of data~\cite{cmspas-exo-11-092}. The event selection differed with respect to the transverse momentum requirements on the electron and the leading jets. In addition, the $95\%$ CL limits were derived using the $\rm CL_S$ method based on a likelihood ratio using the $M_\ttbar$ distribution. Using the topcolor $Z'$ model with a width of $1\%$, the obtained upper limit on cross section times branching ratio is $2.51 \ (0.62) \rm\ pb$ at $M_\ttbar= 1 \ (2) \TeV$, again still not excluding such a hypothetical particle. The resulting observed and expected 95\% CL limit as function of $m_{Z'}$ are shown in Fig.~\ref{fig:mttbarlowmassdilhighmass} (right), compared with the cross section of a topcolor $Z'$.

\begin{figure}[t]
\centering
\begin{minipage}{0.49\textwidth}
\centering
%\vspace{2mm}
\includegraphics[width=1.1\textwidth]{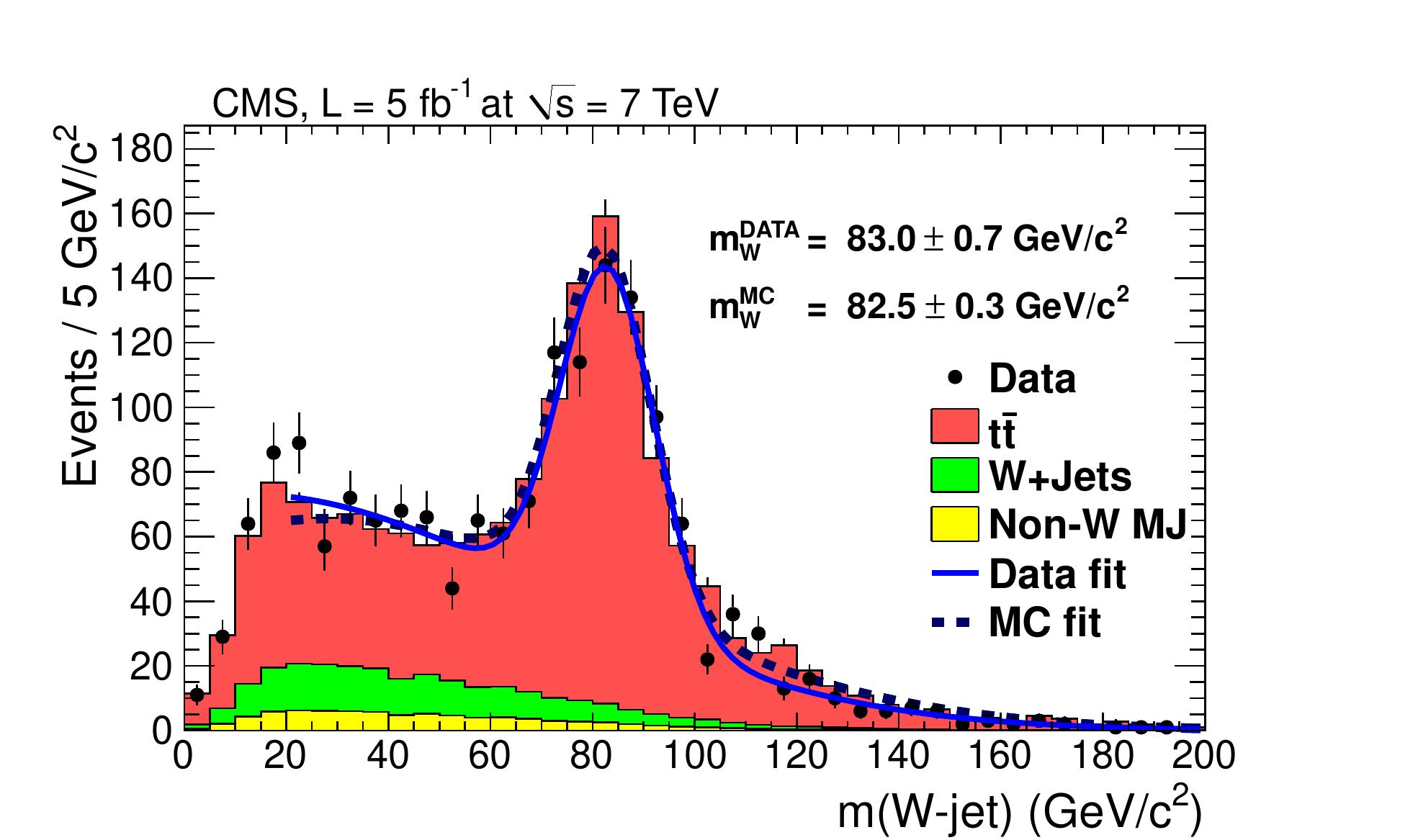}
\end{minipage}
\begin{minipage}{0.49\textwidth}
\centering
%\vspace{2mm}
\includegraphics[width=1.1\textwidth]{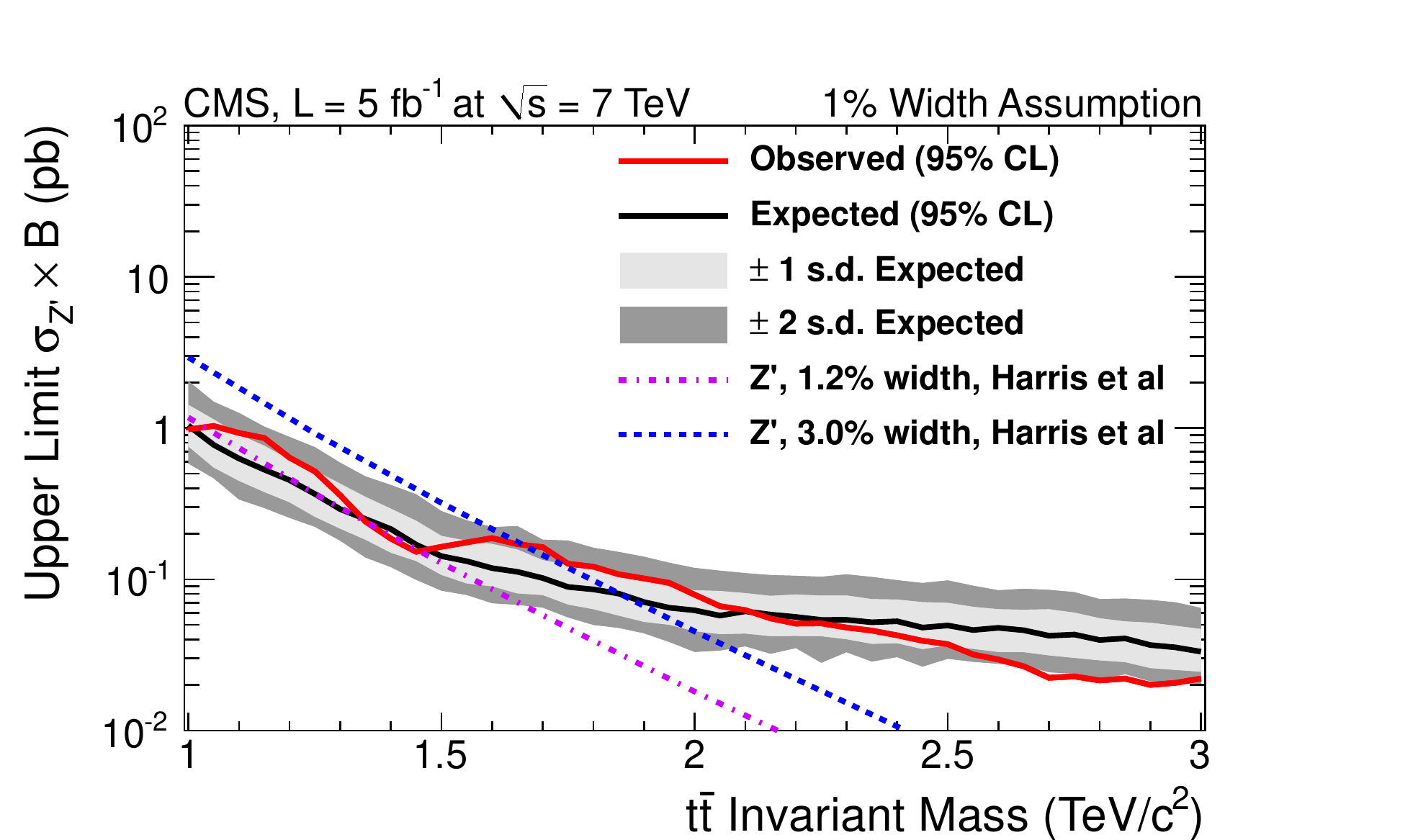}
\end{minipage}
\caption{Left: Mass of jets identified by the $W$ tagging algorithm in a lepton+jets $\ttbar$ sample. Right: Observed and expected limit on $\sigma_{Z'} \cdot \BR (Z'\rightarrow \ttbar)$ and cross sections for a topcolor $Z'$ and a KK gluon model, shown as a function of $\ttbar$ invariant mass~\protect\cite{exo-11-006-paper}.  }
\label{fig:mttbarhighmass}
\end{figure}

CMS also performed a search for narrow heavy resonances decaying into $\ttbar$ in the hadronic channel using $\Lint=5 \fbinv$ of data~\cite{exo-11-006-paper}. It was the first to apply dedicated algorithms to resolve the sub-structure of jets merged due to the boost of the top quark at large $M_{\ttbar}$. The hadronic channel has the advantage of a large branching fraction, but suffers from huge QCD multi-jet background. Jets were clustered using the Cambridge-Aachen (CA) jet clustering algorithm~\cite{Wobisch:1998wt,Dokshitzer:1997in} with an increased size parameter R=0.8 in order to accommodate the larger size of merged jets from boosted top decays. Two event categories were defined: ``type 1+1" events were required to contain at least two jets with $p_T>350\GeV$ using a dedicated \textit{top tagging} algorithm~\cite{jme-09-001,jme-10-013}, each supposed to consist of the decay products of the (anti-) top quark. Secondly, ``type 1+2" events were required to contain at least three jets, one with $p_T>350\GeV$ and reconstructed using the top tagging algorithm, and two jets with $p_T>200 \ (30) \GeV$ reconstructed using a \textit{$W$ tagging} algorithm. The top tagging algorithm uses the CA jets as input and applies two decomposition steps in order to identify three subjets originating from the decay of the top quark. It applies constraints on the jet mass to be consistent with the top quark mass, and the minimum pairwise subjet mass to be consistent with the $W$-boson mass. The $W$ tagging uses the jet pruning algorithm~\cite{Ellis:2009su,Ellis:2009me} which reclusters the CA jets and removes soft and wide-angle clusters. Two subjets with a roughly symmetric energy sharing are required, and the pruned jet mass must be consistent with the $W$-boson mass. The subjet energy scale was estimated in a sample of $\ttbar$ events in the lepton+jets channel with a boosted $W$-jet (see Fig.~\ref{fig:mttbarhighmass} left).  
The background from QCD multi-jet production, in which jets are misidentified by the top or $W$ tagging algorithms, was evaluated from data. The obtained limits, based on the combination of both samples are shown in Fig.~\ref{fig:mttbarhighmass} (right) for the assumption of a narrow topcolor $Z'$ model using widths of 1.2\% and 3.0\%. Sub-picobarn limits were obtained for a $Z'$ heavier than 1.1 TeV, excluding $m_{Z'}<1.6 \TeV$ ($m_{Z'}=1.3-1.5 \TeV$) in the case of a $Z'$ width of 3\% (1.2\%), while $m_{Z'}<2 \TeV$ was excluded for a width of 10\%. In addition, a KK-gluon model with $1.4< m_{g_{KK}} < 1.5 \TeV$ could also be excluded.

In the future, the $M_\ttbar$ distribution will be measured in more detail and the sensitivity to new physics contributions will be extended to even higher mass values, profiting from further refinements of dedicated reconstruction algorithms adapted to highly boosted final state topologies~\cite{Abdesselam:2010pt,Altheimer:2012mn}. Some work is needed to assess correlations between various searches performed using either different final state topologies (e.g. lepton+jets and di-lepton channels) and/or optimized for different mass ranges (low-mass and high-mass analyses), in order for each experiment to provide a combined limit with the best overall sensitivity as function of $M_\ttbar$. In the end, a limit combining inputs from both ATLAS and CMS will be the ultimate goal.

%%%%%%%%%%%%%%%%%%%%%%%%%%%%%%%%%%%%%%%%%%%%%%%%%%%%%%%%%%%%%%%%%%%%%%%%%%%%%%%

\subsection{FCNC in top quark decays and production}
\label{sec:fcnc}

Flavor changing neutral currents (FCNC) are forbidden in the SM at tree level, and are much smaller than the dominant decay mode at loop level. Several extensions of the SM predict increased branching fractions for FCNC decays of the top quark, such as the two-Higgs doublet model (2HDM), the minimal super-symmetric model (MSSM), topcolor assisted technicolor, super-symmetry with R-parity violation or models with warped extra dimensions (see Ref.~\refcite{AguilarSaavedra:2004wm} and references therein). Previous limits on the branching fractions BR($t\rightarrow q\gamma$), BR($t\rightarrow qZ$) and BR($t\rightarrow qg$) were obtained at HERA, LEP and Tevatron. The most precise upper limit on BR($t\rightarrow qZ$) has been set by D0 at 3.2\%~\cite{Abazov:2011qf}.
Older feasibility studies for the LHC can be found in Refs.~\refcite{Carvalho2005,Carvalho2005a,Carvalho2007,Cheng2006,cmsnote-2006-093,atlastdr}.

A search for the decay $t\rightarrow qZ$ was performed by ATLAS using 2010 data~\cite{ATLAS-CONF-2011-061} which was updated using $\Lint = 0.70\fbinv$ of 2011 data in Ref.~\refcite{ATLAS-CONF-2011-154}. Only the updated measurement will be discussed in the following.
The search considered only leptonic decays of the $Z$-boson from the FCNC top quark decay and the $W$-boson from the SM top decay, leading to a final state consisting of three isolated leptons, at least two jets and missing transverse energy. Signal events were simulated using the TOPREX generator~\cite{Slabospitsky:2002ag}. Events containing three isolated leptons ($e$ or $\mu$) were selected, at least two of which had to be of the same flavor and opposite charge and satisfy $|M(ll)-m_Z|<15 \GeV$, with $p_T > 25 \ (20) \GeV$ for the leading (non-leading) lepton(s). In addition, at least two jets were required with $p_T>30 \ (20) \GeV$ for the leading (sub-leading) jet and $\MET > 20 \GeV$. The $\ttbar$ system was reconstructed using a $\chi^2$ criterion to chose the assignment of leptons and jets. The reconstructed masses of the top quarks and the W-boson were required to be consistent with expectation. Backgrounds from di-boson production ($ZZ$ and $WZ$) were taken from simulation, while backgrounds with one or more fake leptons ($Z$+jets, $W$+jets, QCD multi-jet, SM $\ttbar$ and single top quark production) were estimated using various data-driven methods. Two events were selected, while the expected number of background events was $2.4^{+1.8}_{-0.3}$. An upper limit on the number of signal events was obtained using the modified frequentist likelihood method. The observed (expected) 95\% CL upper limit on BR($t\rightarrow qZ$) is 1.1\% ($1.3_{-0.5}^{+0.7}\%$). The dominating sources of systematic uncertainty were due to the $WZ$ background simulation, lepton trigger and identification, as well as jet energy scale calibration.

A similar search was performed by CMS using the full 2011 dataset corresponding to $\Lint = 4.6 \fbinv$~\cite{cmspas-top-11-028}, also using the three-lepton final state. Two leptons were required to be consistent with a $Z$-boson decay. In addition, the selection required two jets and $\MET >30 \GeV$. Finally, tight requirements on the reconstructed masses $M_{Zj}$ and $M_{Wb}$ were imposed, as well as a $b$-tagging requirement. As a cross check, a looser selection without $b$-tagging and mild requirements on $M_{Zj}$ and $M_{Wb}$ was employed as well. Backgrounds from Drell-Yan and $\ttbar$ production were estimated from data, while the remaining ones, including di-boson and single top quark production were obtained from simulation. Zero events were observed, while $0.6\pm 0.1 \pm 0.1$ events were expected. The corresponding $95\%$ CL upper limit of $\BR(t\rightarrow Z q)< 0.34\%$ was obtained using the $\rm CL_S$ method and is currently the world's best limit.

ATLAS performed a search for anomalous single top quark production through $qg\rightarrow t \rightarrow bW$ using $\Lint = 2.05 \fbinv$ of 2011 data~\cite{Collaboration:2012gd}. Since the $t\rightarrow qg$ decay mode is almost impossible to separate from QCD multi-jet production, better sensitivity can be achieved by searching for $qg\rightarrow t$ anomalous single top quark production. The search employed the final state consisting of a $b$-quark, a charged lepton and a neutrino. 
The PROTOS~\cite{AguilarSaavedra:2010rx} generator was used to simulate direct FCNC top quark production. About 26\,000 events were selected in data, while $24\,000\pm 7\,000$ were expected. A neural network based on 11 input variables was used to discriminate signal and background. An upper limit on $\sigma(qg\rightarrow t) \cdot \BR(t\rightarrow b l \nu)$ was set at 95\% CL as 3.9 pb, while the expected limit was 2.4 pb. Using the NLO QCD FCNC single top cross section, upper limits were also derived on generalized anomalous couplings, as well as on the branching fractions as $\BR(t\rightarrow ug) < 5.7  \cdot 10^{-5}$ (assuming $\BR(t\rightarrow cg ) = 0$) and 
$\BR(t\rightarrow cg) < 2.7  \cdot 10^{-4}$ (assuming $\BR(t\rightarrow ug ) = 0$). These limits are the most stringent to date on FCNC single top quark production.

%%%%%%%%%%%%%%%%%%%%%%%%%%%%%%%%%%%%%%%%%%%%%%%%%%%%%%%%%%%%%%%%%%%%%%%%%%%%%%%

\subsection{Anomalous $\MET$ in $\ttbar$ production}

Partners of the top quark with masses below around one TeV are often considered an elegant solution of the hierarchy problem. In particular, pair-produced exotic top partners $T\bar{T}$, each decaying into a top quark and a stable, neutral weakly interacting particle $A_0$, would correspond to the final state $T\bar{T}\rightarrow \ttbar A_0 A_0$, identical to top pair production but with larger $\MET$. Such a scenario may be realized for example in SUSY models with R-parity conservation (where $T$ is the stop quark and $A_0$ is the lightest super-symmetric particle), little Higgs models, models with  universal extra dimensions, or in models with third generation scalar leptoquarks.  Many of these models provide a mechanism for electroweak symmetry breaking and predict dark matter candidates, which can be identified indirectly through their large $\MET$ signature.

\begin{figure}[t]
\centering
\begin{minipage}{0.49\textwidth}
\centering
\vspace{3mm}
\includegraphics[width=1.0\textwidth]{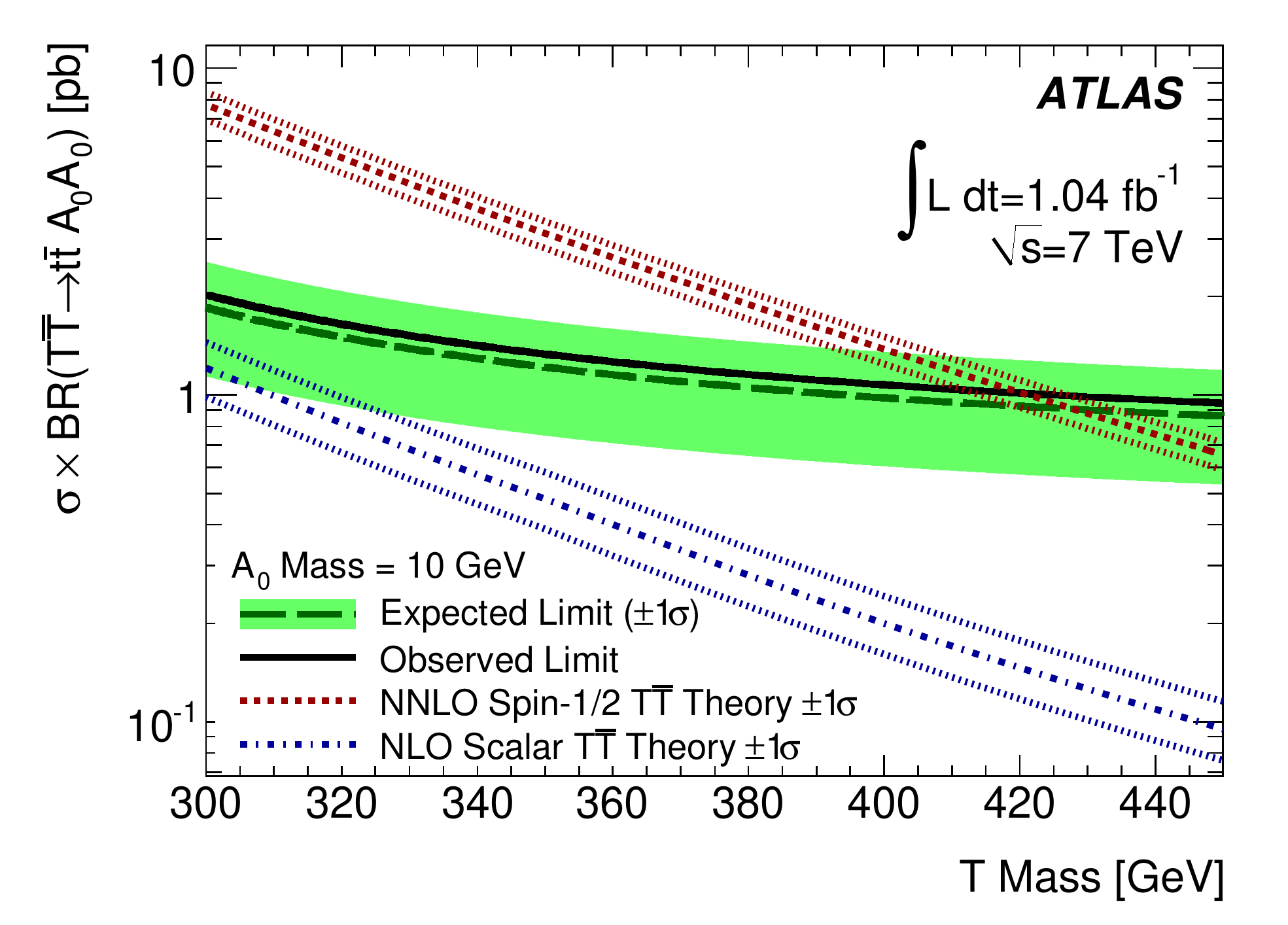}
\end{minipage}
\begin{minipage}{0.49\textwidth}
\centering
%\vspace{2mm}
\includegraphics[width=1.0\textwidth]{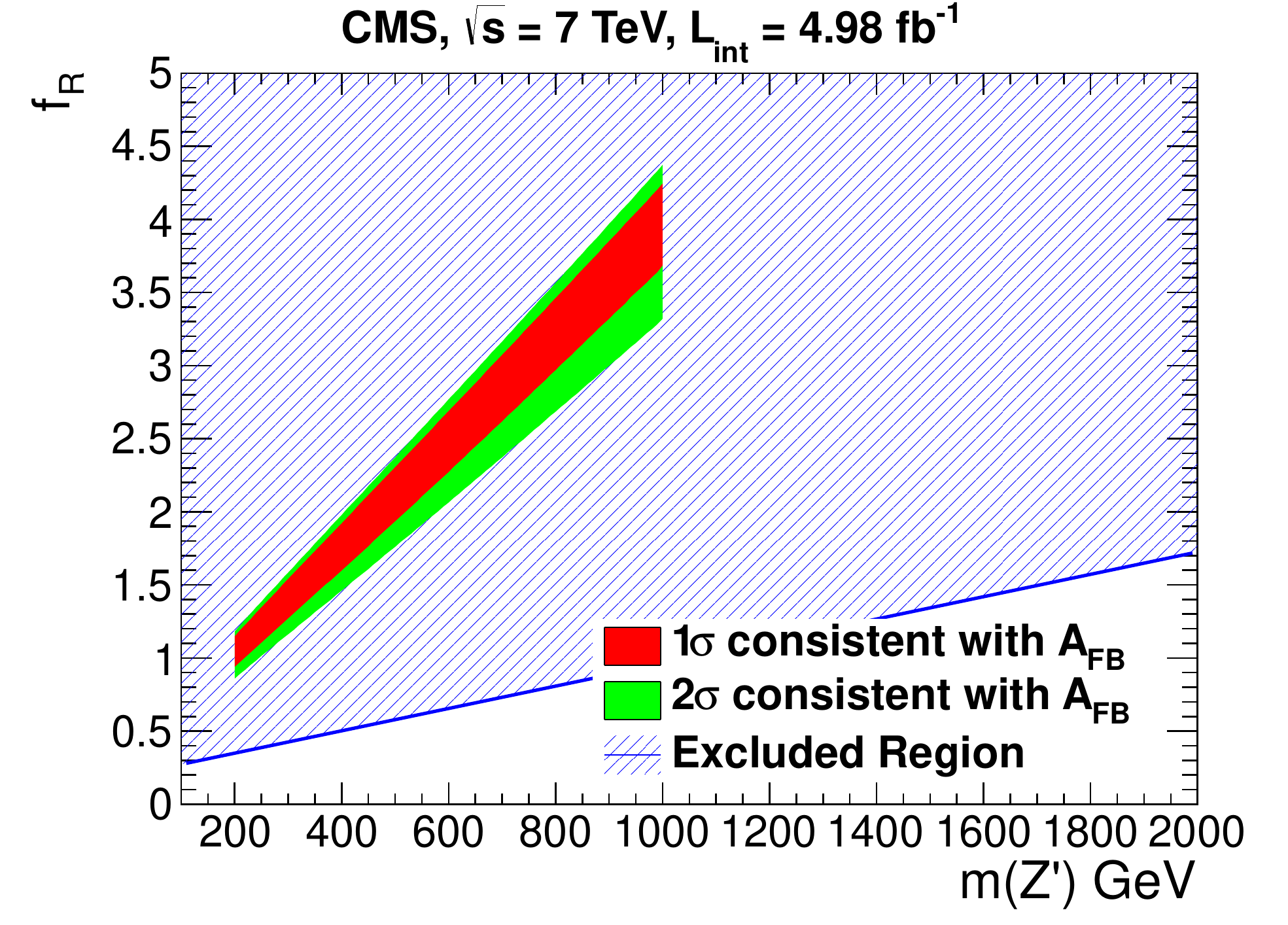}
\end{minipage}
\caption{Left: Excluded cross section times branching ratio of the process $T\bar{T}\rightarrow \ttbar A_0 A_0$ as function of $m_T$ for $m_{A^0}=10 \GeV$ in the ATLAS search~\protect\cite{Aad:2011wc}.
Right: Exclusion region at 95\% CL as function of $Z'$ mass and the right-handed coupling $f_R$ in the CMS search for same-sign top quark pairs~\protect\cite{Collaboration:2012sa}. Also shown is the region of parameter space consistent with the Tevatron measurements of $A_{FB}$ and $\sigma_{\ttbar}$ as inferred in Ref.~\protect\refcite{Berger:2011ua}.    }
\label{fig:topbsm}
\end{figure}

ATLAS performed a search for  $T\bar{T}\rightarrow \ttbar A_0 A_0$, using $\Lint=1.04\fbinv$ of data~\cite{Aad:2011wc}. The search was performed in the lepton+jets channel using similar criteria as for the $\ttbar$ cross section measurement, requiring one isolated lepton and at least four jets. To reduce $\ttbar$ and $W$+jets backgrounds, $\MET>100 \GeV$ and $M_T>150 \GeV$ were required in addition. The largest remaining background was found to be $\ttbar$ events in the di-lepton channel. The event yield in data was found to be in agreement with the expectation.  Thus, limits were obtained in the context of the model of fourth generation exotic up-type quarks from Ref.~\refcite{Alwall:2010jc}. Masses of  $m_T<420 \GeV$ and $m_{A^0}<140 \GeV$ were excluded at the 95\% CL (see Fig.~\ref{fig:topbsm} left), and a cross section times branching ratio of 1.1 pb was excluded for $m_T=420 \GeV$ and $m_{A^0}=10\GeV$. These limits are approximately valid also for other models, including pair production of scalar top quarks or third generation  leptoquarks.

%%%%%%%%%%%%%%%%%%%%%%%%%%%%%%%%%%%%%%%%%%%%%%%%%%%%%%%%%%%%%%%%%%%%%%%%%%%%%%%

\subsection{Same-sign top quark pair production}
\label{sec:samesign}

Many models which have been proposed to explain the larger than predicted forward-backward asymmetry in top quark pair production at the Tevatron (see section~\ref{sec:ac}) invoke FCNC in the top quark sector mediated by the $t$-channel exchange of a new massive $Z'$ boson (see, e.g., Ref.~\refcite{Jung:2009jz}). However, this type of interaction would also give rise to same-sign top quark pair production, for which in the case of $tt$ (as opposed to $\bar{t}\bar{t}$) production the cross section would be enhanced at LHC because of the large up-quark valence contribution to the proton PDF.

CMS has searched for like-sign top quark pair production using $\Lint=36\pbinv$ of data~\cite{Chatrchyan:2011dk}. The event selection and background estimation is similar to the one used for the $\ttbar$ cross section measurement in the di-lepton channel, except that both leptons were required to be of positive sign. Two events were selected, compared with $0.9\pm 0.6$ events expected for background. The 95\% CL upper limit on the  $pp\rightarrow tt$ cross section in the context of the FCNC $Z'$ model from Ref.~\refcite{Berger:2011ua} is 17 pb.
An update of this search using the full 2011 data sample of $\Lint = 4.98 \fbinv$ was performed~\cite{Collaboration:2012sa}, selecting events with two positively charged, like-sign leptons, at least two b-tagged jets and in addition $\MET >30 \GeV$, $H_T>80 \GeV$. Five events were observed, for $4.5\pm 1.7$ expected, which led to an improvement of the above limit in the context of the model from Ref.~\refcite{Berger:2011ua} to 0.61 pb.
Upper limits were also computed as function of the $Z'$ mass and the right-handed coupling $f_R$ at the $utZ'$ vertex. The region of parameter space consistent with the Tevatron $A_{FB}$ measurement~\cite{Berger:2011ua} is disfavored (see Fig.~\ref{fig:topbsm} right). 
In addition, limits on a model~\cite{BarShalom:2007pw} with maximal flavor violation involving a new scalar SU(2) doublet were also placed.

ATLAS has performed a search for prompt like-sign muon pairs using a dataset corresponding to $\Lint = 1.6\fbinv$~\cite{Collaboration:2012cg}. Events containing a pair of two muons of the same charge with $p_T>20 \GeV$ were selected. Standard model backgrounds with two same-sign prompt leptons (mostly di-boson production) were taken from simulation, while backgrounds from non-prompt muons were estimated from data. No excess over the SM background was observed, and limits were placed in a model independent way, as well as in the context of like-sign top quark pair production. The 95\% CL upper limit on the production of a pair of right-handed top quarks ranges from 2.2 to 3.7 pb, depending on the $Z'$ mass. These limits are approximately a factor six less stringent than the one reported above by CMS.

\begin{figure}[t]
\centering
\begin{minipage}{0.46\textwidth}
\centering
\vspace{1mm}
\includegraphics[width=0.99\textwidth]{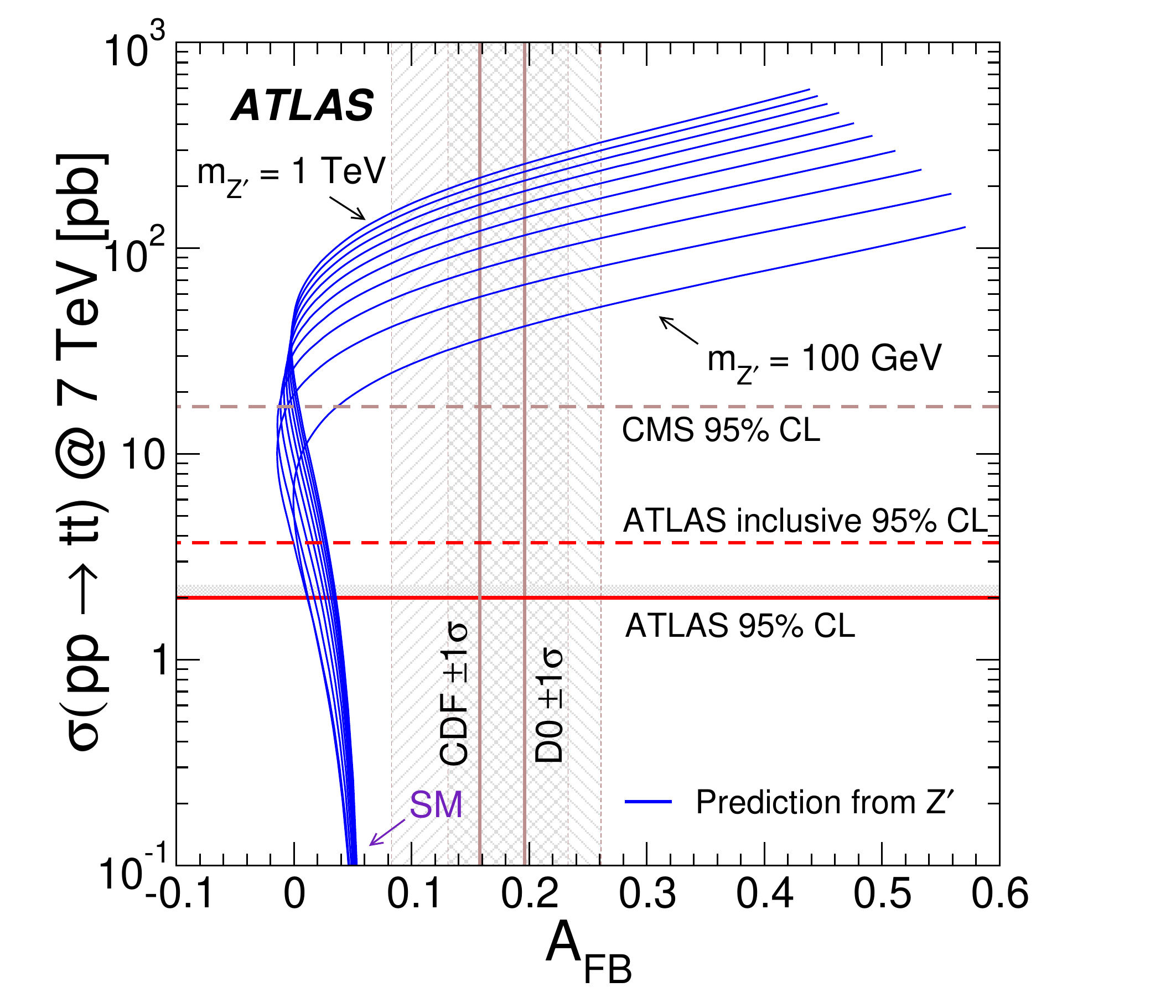}
\end{minipage}
\begin{minipage}{0.53\textwidth}
\centering
%\vspace{2mm}
\includegraphics[width=0.99\textwidth]{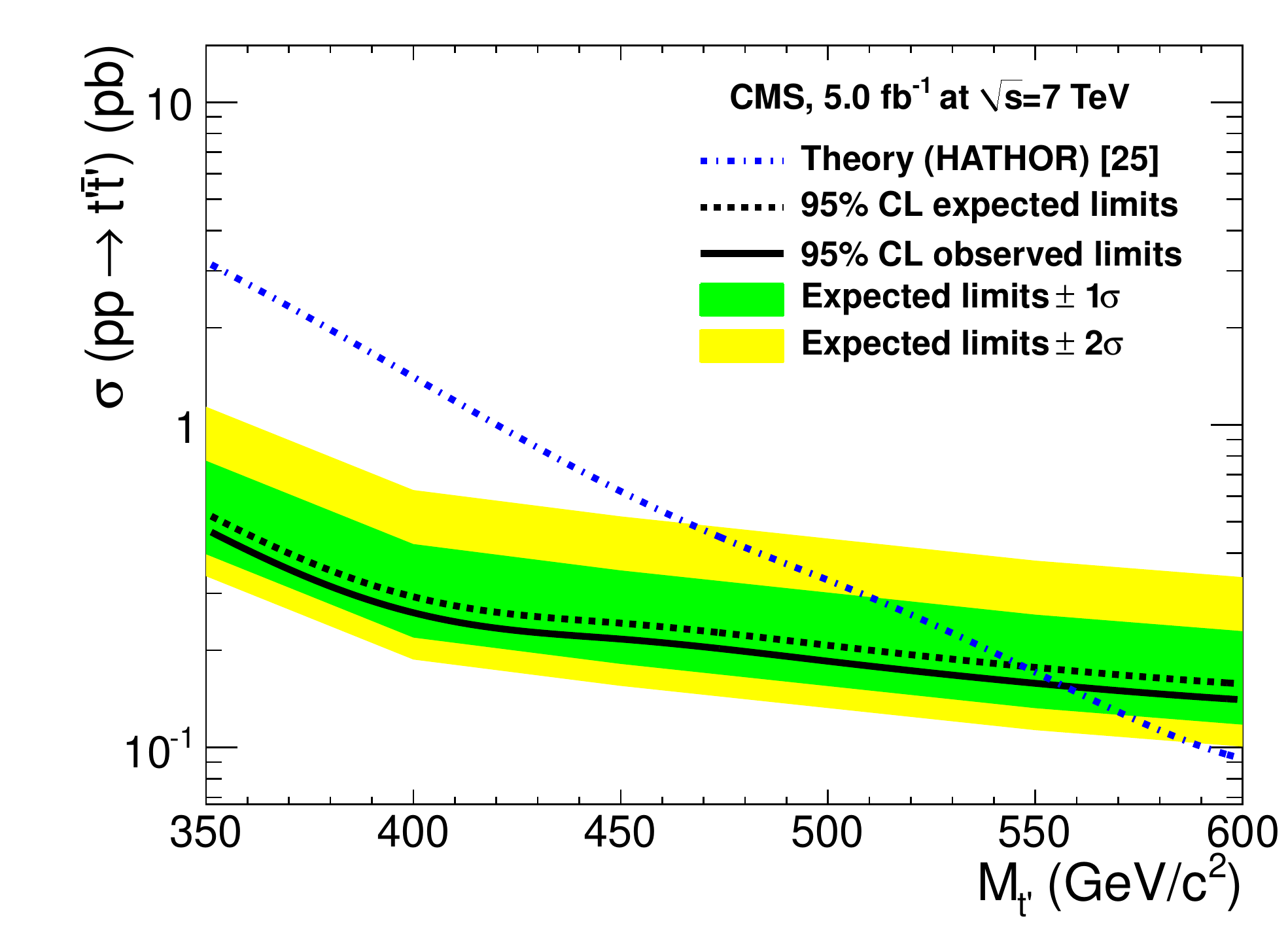}
\end{minipage}
\caption{Left: Contributions to the inclusive $A_{FB}$ at the Tevatron from the exchange of a $Z'$ with right-handed couplings (shown for various masses and couplings of the $Z'$), versus the same-sign top quark production cross section at LHC~\protect\cite{:2012bb}. Older limits from CMS~\protect\cite{Chatrchyan:2011dk} and ATLAS~\protect\cite{Collaboration:2012cg} are also shown. 
Right: 95\% CL upper limit on the $t'\bar{t}'$ production cross section as function of $m_{t'}$ determined by CMS~\protect\cite{exo-11-050}, compared with the theory prediction obtained using Ref.~\protect\refcite{Aliev:2010zk}. }
\label{fig:ssafb}
\end{figure}

Another search for same-sign top quark production was performed using the $ee$, $\mu\mu$ and $e\mu$ final states and $\Lint = 1.04\fbinv$\cite{:2012bb}. In the case of a heavy new particle mediating the $tt$ production, the upper limit at $95\%$ CL is 1.7 pb for each chirality. For light $Z'$ mediators, the limits range from 1.4 to 2.0 pb, depending on the $Z'$ mass. Limits were placed on generic classes of models with new particles mediating same-sign top quark production. In the case of a flavor-changing $Z'$-boson which was proposed to explain the large value of the $A_{FB}$ measurement at the Tevatron (see above), a significant part of parameter space is excluded (see Fig.~\ref{fig:ssafb}, left).

%%%%%%%%%%%%%%%%%%%%%%%%%%%%%%%%%%%%%%%%%%%%%%%%%%%%%%%%%%%%%%%%%%%%%%%%%%%%%%%

\subsection{Fourth generation of quarks}
\label{sec:fourthgen}

In principle, the number of generations of fermions in the standard model is not limited to three. Many searches for fourth generation fermions have been performed, for example at LEP and Tevatron, but so far all with negative outcome. Nevertheless, there is still considerable interest in a fourth generation, because its existence 
can provide sufficiently strong CP violation~\cite{Hou:2008xd} for
generating the baryon asymmetry of the universe at the electroweak scale,
may offer the heavy neutrino as a candidate for dark matter, would avoid the need for a light Higgs boson if the mass of the heavy quarks is large, and would relieve the tension in some flavor physics results.

CMS searched for a pair-produced new heavy top-like quark $t'$ which decays as $t'\rightarrow Wb$ in both the di-lepton~\cite{exo-11-050} (Fig.~\ref{fig:ssafb}, right) and lepton+jets\cite{cmspas-exo-11-099} channels using the full 2011 dataset, excluding masses below $m_{t'}=560 \GeV$. 
An inclusive search for fourth generation up- ($t'$) and down-type ($b'$) quarks of degenerate mass, produced singly or in pairs, was also performed by CMS~\cite{CMS-PAS-EXO-11-054}, yielding a lower limit of $m_{t'}=490 \GeV$.  
ATLAS searched for pair-produced heavy quarks decaying via $Q\rightarrow W q$ ($q$=$u,d,c,s,b$)~\cite{ATLAS-CONF-2011-022} in the di-lepton channel and set a lower limit of $m_Q=270 \GeV$, using only 2010 data. This search was updated using $\Lint = 1.04 \fbinv$ of 2011 data~\cite{Aad:2012bt}, yielding a lower limit of $m_Q=350 \GeV$.
Using the same data sample and assuming the exclusive $t'\rightarrow W b$ decay, ATLAS also placed a lower $95\%$ CL limit on the mass a $t'$ of $m_{t'}>404 \GeV$~\cite{Aad:2012xc}.
A search for a pair-produced heavy vector-like quarks which decay via FCNC as $T\rightarrow t Z$ was carried out by CMS~\cite{Chatrchyan:2011ay} and yielded a lower limit of $m_{T}=475 \GeV$.
ATLAS searched for a heavy vector-like quark which couples to light quarks~\cite{Aad:2011yn}, using both the charged current (CC) $Qq\rightarrow W q q'$ and the neutral current (NC) $Qq \rightarrow Z q q'$ modes, and lower limits on the mass of the new quark were set of $m=900 \ (760) \GeV$ from the CC (NC) process, respectively.
CMS also searched~\cite{Chatrchyan:2011em,Collaboration:2012ye} for pair-produced bottom-like quarks decaying as $b'\bar{b}' \rightarrow tW^- \bar{t}W^+$ using up to $\Lint = 4.9 \fbinv$ of data, and set a lower limit of $m_{b'}=611 \GeV$.
Similarly, ATLAS searched for $b'\bar{b}' \rightarrow tW^- \bar{t}W^+$ in the lepton+jets channel using $\Lint = 1.04 \fbinv$ of 2011 data and set a lower limit of $m_{b'}=480 \GeV$~\cite{Aad:2012us}. This was complemented by a search for events with same-sign di-lepton pairs, which yields a lower limit of $m_{b'}=450 \GeV$~\cite{:2012bb}.
Finally, a search for $b'\rightarrow Z b$ was carried out by ATLAS~\cite{arxiv:1204.1265} using $\Lint = 2.0 \fbinv$ of data, yielding a  lower limit of $m_{b'}>400 \GeV$.

These results generally improve upon previous limits obtained at the Tevatron (e.g. Refs.~\refcite{Aaltonen:2011tq,Abazov:2011vy,Aaltonen:2011vr}).
In the future, using larger datasets and profiting from the increased LHC center-of-mass energy, these limits can be further pushed to higher mass values.

%%%%%%%%%%%%%%%%%%%%%%%%%%%%%%%%%%%%%%%%%%%%%%%%%%%%%%%%%%%%%%%%%%%%%%%%%%%%%%

\subsection{Other new physics searches involving top quarks}
\label{sec:bsmother}

As discussed in sections~\ref{sec:intro} and~\ref{sec:taudilxs}, a light charged Higgs boson $H^\pm$ may be produced in top quark decays, provided that $m_H<m_t-m_b$, followed by the decay $H^\pm \rightarrow \tau \nu_\tau$. Both ATLAS~\cite{Aad:2012tj} ($\Lint = 4.6 \fbinv$) and CMS~\cite{arxiv:1205.5736} ($\Lint = 2.0-2.3 \fbinv$) have searched for charged Higgs production in top quark decays. Assuming $\BR(H^+ \rightarrow \tau^+ \nu_\tau)=1$, upper limits on the branching fraction $\BR (t \rightarrow H^+ b)$ in the range $5-1\%$   ($4-2\%$) were obtained for masses of the charged Higgs boson $m_{H^\pm}=90 \, (80) - 160 \GeV$ in the case of ATLAS (CMS), respectively. These results improve considerably upon earlier limits set at the Tevatron~\cite{Abulencia:2005jd,Abazov:2009aa}.

A new heavy $W'$-boson, predicted in several extensions of the SM such as models with extra dimensions or little Higgs models, could decay with a large BR as $W' \rightarrow tb$, provided that the leptonic decay modes are suppressed. This may be the case for instance for a right-handed $W'$, provided that the mass of the right-handed neutrino is heavy enough. Both CMS and ATLAS have set limits on the cross section times branching ratio $\sigma_{W'} \cdot \BR (W' \rightarrow tb)$ for a right-handed $W'$-boson with SM like couplings. While the ATLAS measurement~\cite{Aad:2012ej} based on $\Lint = 1.04 \fbinv$ excluded masses below 1.13 TeV, the preliminary CMS measurement~\cite{cmspas-exo-12-001} based on $\Lint = 5.0 \fbinv$ excluded masses below 1.85 TeV at 95\% CL. Both results improve upon previous Tevatron results~\cite{Aaltonen:2009qu,Abazov:2011xs}.

\section{Summary and Outlook}
\label{sec:summary}

The data sample corresponding to $\Lint \sim 5 \fbinv$ which was provided during the first two years of LHC operation at $\sqrt{s} = 7 \TeV$ resulted in a wealth of information about the properties of the top quark and its interactions in a new energy regime, complementing the Tevatron data. Precise (less than $7\%$) measurements of the $\ttbar$ cross section are challenging current theory calculations which approximate the full NNLO QCD result. Differential cross sections have been measured, thanks to the large abundance of top quark events. Measurements of the top quark mass with precision below $1\%$ are already competing with the Tevatron, thanks to impressive understanding of the detectors and systematic uncertainties, already after a relatively short running period, and the  measurement of the top quark - anti-quark mass difference is a world-best. Various other properties of top quarks and their interactions, in particular of the ratio of branching fractions $R$ (consistent with unity within $4\%$), the $W$-boson polarization in top decays, the $\ttbar$ spin correlation (established with more than $5\sigma$ significance) and the $\ttbar$ charge asymmetry $A_C$, are all in good agreement with the SM. Some tension with the SM observed at Tevatron in measurements of $R$ and the $\ttbar$ forward-backward asymmetry could not be confirmed by LHC results, though even more precise measurements are needed for a more definite statement. Single top quark production has been measured in the $t$- and $tW$-channels, in agreement with SM predictions. No signs of contributions from new physics have been found in the $M_\ttbar$ distribution and new heavy resonances decaying into $\ttbar$ pairs have been excluded for masses up to around 1.5 TeV, already exceeding corresponding Tevatron limits. Also searches for FCNC top quark decays and production, same-sign top quark production, fourth generation quarks and charged Higgs or $W'$-bosons so far have turned out negative, but have already provided more stringent exclusion limits than earlier ones obtained at the Tevatron.

From 2012, the LHC is running at the increased center-of-mass energy of $\sqrt{s}= 8 \TeV$, which results in an increase of the top quark pair and single top cross sections of $30$ to $50\%$, while the most important backgrounds are expected to increase less strongly with energy, leading to an improved S/B. In addition, the anticipated integrated luminosity delivered before the start of the first long LHC shutdown in 2013-2014 of $\Lint = 15-20 \fbinv$ will yield very large samples of top quark events.  

Precision measurements of $\ttbar$ and single top cross sections at $\sqrt{s}= 8 \TeV$ will be the first task, to confirm the energy dependence. In addition, ratios of cross sections such as $\sigma_\ttbar^{8 \rm\ TeV} / \sigma_\ttbar^{7 \rm\ TeV}$ can be measured in which many experimental and theoretical systematic uncertainties cancel to a large extent, including scale and PDF (except for the large-$x$ region, see below) uncertainties. To also eliminate the luminosity uncertainty, double ratios with respect to standard candles such as $Z$-boson production can be extracted: $(\sigma_\ttbar/\sigma_Z)^{8 \rm\ TeV} / (\sigma_\ttbar/\sigma_Z)^{7 \rm\ TeV}$. Such double ratios would also be fairly sensitive to contributions from new physics. Besides inclusive cross section measurements, the large data sample will also enable detailed measurements of differential cross sections. 

The competition between ever more precise theory calculations, hopefully soon to full NNLO QCD accuracy, and further reduction of experimental systematic uncertainties will continue. Also here the large dataset will help, and regions of phase space can be selected which are less sensitive with respect to the impact of dominating systematic uncertainties.
In addition, data can be used to constrain systematic uncertainties due to the modeling of top quark production in MC generators such as from the parton shower or the choice of factorization scales, in particular using differential distributions. A first example was presented in section~\ref{sec:atlasisr}. Studying ISR in a gluon induced process such as $\ttbar$ production is also beneficial for the modeling of Higgs production $gg\rightarrow H$. As discussed in section~\ref{sec:mass}, measurements of $m_t$ performed differentially in $p_T$ or other variables will help to constrain important uncertainties due to the modeling of soft physics such as the underlying event and color reconnection, and alternative mass measurements with uncorrelated systematic uncertainties will become possible.

Top quark cross sections can also be useful to constrain the PDF. In fact, a large contribution to the total uncertainty in the $\ttbar$ cross section is due to the imperfect knowledge of the PDF (see Table~\ref{tab:xs}, section~\ref{sec:xstheo}). In the case of $\ttbar$ production, constraints on the medium to large $x$ gluon distribution can be obtained by measurements of $\sigma_\ttbar$ as function of rapidity and $p_T$, and the ratio $\sigma_\ttbar / \sigma_Z$ profits from the fact that the PDF uncertainties for $\ttbar$ and $Z$-boson production are anti-correlated. Ratios
$\sigma_\ttbar^{8 \rm\ TeV} / \sigma_\ttbar^{7 \rm\ TeV}$ may provide sensitivity to the large-$x$ region. Precise differential measurements of $t$-channel single top quark production will eventually be used to place constraints on the $b$-quark PDF, using the cross section as function of the $b$-jet rapidity. Furthermore, the ratio of single top quark and top anti-quark cross sections, ideally as function of rapidity, is sensitive to the ratio of up and down quark PDF in the proton.

Measurements of $\ttbar$ production in association with a boson ($\gamma$, $W$, $Z$) will eventually allow to determine couplings of the top quark to bosons (a first result for $\ttbar+\gamma$ was discussed in section~\ref{sec:ttgamma}), even though the expected event numbers will be small for $\ttbar+W/Z$ ($\mathcal{O}(30-50)$ events for $20 \fbinv$).
In addition, the measurement of $\ttbar+\gamma$ production will yield limits on the production of excited top quarks which decay as $t^* \rightarrow t + \gamma$.

Measurements of $\ttbar$+jets are important tests of QCD (NLO QCD predictions exist for $\ttbar$+1 jet and $\ttbar$+2 jets, see section~\ref{sec:xstheo}), as well as being useful to constrain backgrounds in searches for new physics or the Higgs boson. The latter is particularly the case for the $\ttbar + b\bar{b}$ channel, which will pave the way to an observation of the Higgs boson in association with a $\ttbar$ pair ($\ttbar H$). 

Even though it is not a discovery channel, the $\ttbar H$ mode will be important as it can be used to measure the Yukawa coupling of the top quark. Previous sensitivity studies based on simulation~\cite{Ball:2007zza,atlastdr} indicated that very large amounts of integrated luminosity would be needed, and the analyses are plagued with small statistical significance of the signal and large systematic uncertainties due to the backgrounds. In the meantime however, significant improvements in the reconstruction of physics objects have been made, and advanced reconstruction algorithms for highly boosted final state topologies have become available. These developments, together with the more precise NLO QCD calculations of the dominating $\ttbar + b\bar{b}$ and $\ttbar$+2 jets backgrounds, have the potential of significant improvements to the sensitivity of the $\ttbar H$ search.

Finally, the cross section for the production of four top quarks $\ttbar\ttbar$ should be measured. While being very small in the SM ($\mathcal{O}(1 \rm\ fb)$), it may be considerably enhanced in certain new physics models, for instance SUSY in which pair-produced gluons dominantly decay into $\ttbar$ and a neutralino, topcolor or compositeness models.

In single top quark production, differential cross section measurements will become possible in the $t$- and $tW$-channels, and the $s$-channel mode could be observed for the first time at LHC. The single top quark and top anti-quark cross sections should be measured separately. Instead of exploiting the top quark polarization in electroweak single top quark production when measuring the cross section (e.g., by using $\cts$ as discriminating variable), the measurement could be decorrelated from $\cts$, which would allow the polarization to be measured. It is important to measure single top quark production in all channels, since contributions from new physics would affect the different production modes in a different way.

A global fit should be performed to constrain the properties of the $Wtb$ vertex, using measurements in single and pair production of top quarks. More detailed measurements of $\ttbar$ spin correlations are also needed.  Deviations from the SM expectation would indicate contributions from new physics.

Even more sensitive searches for FCNC top quark decays and same-sign top quark production must be performed, and the direct searches for quarks of a possible fourth generation, as well as indirect constraints from $|V_{tb}|$ determinations (single top cross sections, $R$ measurements) will continue.

Precise measurements of the $M_\ttbar$ distribution will allow to place even more stringent limits on new physics models, benefiting at high mass from sophisticated top tagging algorithms. If a resonance is observed, its spin-parity properties could be further investigated by measuring polar angle distributions in the corresponding mass region.

The top quark charge asymmetry $A_C$, which in the SM is around $10\%$ smaller at $\sqrt{s}=8 \TeV$ compared with its 7 TeV value, must be measured more precisely and differentially, in particularly by moving to kinematic regions in which the SM asymmetry, only present for quark-induced initial states, is enhanced. In addition, lepton asymmetries should be measured in di-lepton final states, since they are not affected from ambiguities due to the reconstruction of the $\ttbar$ system. SM NLO QCD theory predictions for the charge asymmetry are needed. Together with $A_{fb}$ measurements at the Tevatron, the parameter space for new physics models can be further constrained.

Precise measurements of $m_t$ are an important ingredient for global electroweak fits. Should  the Higgs boson be discovered, the values of $m_H$, $m_t$ and $m_W$ could be used to test the consistency of the SM.

The important effort to combine the most precise measurements of ATLAS and CMS on top quark physics has started already, beginning with the measurements of top quark pair and single top quark production cross sections. Others, such as a $m_t$ combination, should also include the Tevatron data.

Very exciting years are ahead for top quark physics, which will show if the data continue to be consistent with the SM expectations, or if hints for new physics will be found.

%%%%%%%%%%%%%%%%%%%%%%%%%%%%%%%%%%%%%%%%%%%%%%%%%%%%%%%%%%%%%%%%%%%%%%%%%%%%%%%

\section*{Acknowledgments}
\addcontentsline{toc}{section}{Acknowledgments}

The author is indebted to his colleagues of the ATLAS and CMS collaborations for producing an impressive menu of measurements in the field of top quark physics in a very short amount of time. This includes not only his friends from the top quark physics analysis groups of the two experiments, but also everybody who contributed to the construction, commissioning and operation of the detectors, as well as those who worked to provide well understood and calibrated physics objects, not to forget those who maintain the computing infrastructure. These results would not have been possible without the outstanding performance of the LHC machine, including its operational team. Finally, it is a pleasure to thank Werner Bernreuther, Roberto Chierici, Markus Cristinziani, Roberto Tenchini and Peter Uwer for providing useful feedback and suggestions on the manuscript.

%%%%%%%%%%%%%%%%%%%%%%%%%%%%%%%%%%%%%%%%%%%%%%%%%%%%%%%%%%%%%%%%%%%%%%%%%%%%%%%

\addcontentsline{toc}{section}{References}
\bibliography{bib/top,bib/top_lhc,bib/top_tev,bib/general}

\begin{thebibliography}{100}

\bibitem{Abe:1995hr}
CDF Collaboration, F.~Abe {\em et~al.}, {\em Phys. Rev. Lett.} {\bf 74}, 2626
  (1995), \href{http://arxiv.org/abs/hep-ex/9503002}{{\ttfamily
  arXiv:hep-ex/9503002 [hep-ex]}}.

\bibitem{Abachi:1995iq}
D0 Collaboration, S.~Abachi {\em et~al.}, {\em Phys. Rev. Lett.} {\bf 74}, 2632
   (1995), \href{http://arxiv.org/abs/hep-ex/9503003}{{\ttfamily
  arXiv:hep-ex/9503003 [hep-ex]}}.

\bibitem{Lancaster:2011wr}
CDF and D0 Collaborations, {TEVEWWG}, {\it {C}ombination of {CDF} and {D}0
  results on the mass of the top quark using up to 5.8~$fb^{-1}$ of data}
  (2011), \href{http://arxiv.org/abs/1107.5255}{{\ttfamily arXiv:1107.5255
  [hep-ex]}}.

\bibitem{TevatronElectroweakWorkingGroup:2012gb}
CDF and D0 Collaborations, {TEVEWWG}, {\it 2012 {U}pdate of the {C}ombination
  of {CDF} and {D}0 {R}esults for the {M}ass of the {W} {B}oson}  (2012),
  \href{http://arxiv.org/abs/1204.0042}{{\ttfamily arXiv:1204.0042 [hep-ex]}}.

\bibitem{Papucci:2011wy}
M.~Papucci, J.~T. Ruderman and A.~Weiler, {\it {N}atural {SUSY} {E}ndures}
  (2011), \href{http://arxiv.org/abs/1110.6926}{{\ttfamily arXiv:1110.6926
  [hep-ph]}}.

\bibitem{Wimpenny:1996dz}
S.~Wimpenny and B.~Winer, {\em Ann. Rev. Nucl. Part. Sci.} {\bf 46}, 149
  (1996).

\bibitem{Campagnari1997}
C.~Campagnari and M.~Franklin, {\em Rev. Mod. Phys.} {\bf 69}, 137  (1997),
  \href{http://arxiv.org/abs/hep-ex/9608003}{{\ttfamily arXiv:hep-ex/9608003}}.

\bibitem{Bhat1998}
P.~C. Bhat, H.~Prosper and S.~S. Snyder, {\em Int. J. Mod. Phys.} {\bf A13},
  5113  (1998), \href{http://arxiv.org/abs/hep-ex/9809011}{{\ttfamily
  hep-ex/9809011}}.

\bibitem{Tollefson:1999wt}
K.~Tollefson and E.~Varnes, {\em Ann. Rev. Nucl. Part. Sci.} {\bf 49}, 435
  (1999).

\bibitem{Chakraborty:2003iw}
D.~Chakraborty, J.~Konigsberg and D.~L. Rainwater, {\em Ann. Rev. Nucl. Part.
  Sci.} {\bf 53}, 301  (2003),
  \href{http://arxiv.org/abs/hep-ph/0303092}{{\ttfamily hep-ph/0303092}}.

\bibitem{Wagner2005}
W.~Wagner, {\em Rept. Prog. Phys.} {\bf 68}, 2409  (2005),
  \href{http://arxiv.org/abs/hep-ph/0507207}{{\ttfamily arXiv:hep-ph/0507207}}.

\bibitem{Quadt:2007jk}
A.~Quadt, {\em Eur. Phys. J.} {\bf C48}, 835  (2006).

\bibitem{Kehoe2008}
R.~Kehoe, M.~Narain and A.~Kumar, {\em Int. J. Mod. Phys.} {\bf A23}, 353
  (2008), \href{http://arxiv.org/abs/0712.2733}{{\ttfamily arXiv:0712.2733
  [hep-ex]}}.

\bibitem{Pleier:2008ig}
M.-A. Pleier, {\em Int. J. Mod. Phys.} {\bf A24}, 2899  (2009),
  \href{http://arxiv.org/abs/0810.5226}{{\ttfamily arXiv:0810.5226 [hep-ex]}}.

\bibitem{Incandela:2009pf}
J.~R. Incandela, A.~Quadt, W.~Wagner and D.~Wicke, {\em Prog. Part. Nucl.
  Phys.} {\bf 63}, 239  (2009),
  \href{http://arxiv.org/abs/0904.2499}{{\ttfamily arXiv:0904.2499 [hep-ex]}}.

\bibitem{Wicke:2010cg}
D.~Wicke, {\em Eur. Phys. J.} {\bf C71},   1627  (2011),
  \href{http://arxiv.org/abs/1005.2460}{{\ttfamily arXiv:1005.2460 [hep-ex]}}.

\bibitem{Deliot:2010ey}
F.~Deliot and D.~A. Glenzinski, {\it {Top Quark Physics at the Tevatron}}
  (2010), \href{http://arxiv.org/abs/1010.1202}{{\ttfamily arXiv:1010.1202
  [hep-ex]}}, acc. by Rev. Mod. Phys.

\bibitem{Lannon:2012fp}
K.~Lannon, F.~Margaroli and C.~Neu, {\it {M}easurements of the {P}roduction,
  {D}ecay and {P}roperties of the {T}op {Q}uark: {A} {R}eview}  (2012),
  \href{http://arxiv.org/abs/1201.5873}{{\ttfamily arXiv:1201.5873 [hep-ex]}}.

\bibitem{Bernreuther:2008ju}
W.~Bernreuther, {\em J. Phys.} {\bf G35},   083001  (2008),
  \href{http://arxiv.org/abs/0805.1333}{{\ttfamily arXiv:0805.1333 [hep-ph]}}.

\bibitem{Moch:2008qy}
S.~Moch and P.~Uwer, {\em Phys. Rev.} {\bf D78},   034003  (2008),
  \href{http://arxiv.org/abs/0804.1476}{{\ttfamily arXiv:0804.1476 [hep-ph]}}.

\bibitem{Kidonakis:2011ca}
N.~Kidonakis and B.~D. Pecjak, {\it {T}op-quark production and {QCD}}  (2011),
  \href{http://arxiv.org/abs/1108.6063}{{\ttfamily arXiv:1108.6063 [hep-ph]}},
  subm. to Eur. Phys. J. C.

\bibitem{Weinzierl:2012tc}
S.~Weinzierl, {\it {T}heoretical overview on top pair production and single top
  production}  (2012), \href{http://arxiv.org/abs/1201.4025}{{\ttfamily
  arXiv:1201.4025 [hep-ph]}}, to appear in the Proceedings of HCP2011, Paris
  (France).

\bibitem{herapdf15}
{\it {HERAPDF} 1.5 {NNLO}}
  \url{https://www.desy.de/h1zeus/combined_results/herapdftable}.

\bibitem{Watt:2012fj}
G.~Watt, {\em Nucl. Phys. Proc. Suppl.} {\bf 222-224}, 61  (2012),
  \href{http://arxiv.org/abs/1201.1295}{{\ttfamily arXiv:1201.1295 [hep-ph]}}.

\bibitem{Aliev:2010zk}
M.~Aliev, H.~Lacker, U.~Langenfeld, S.~Moch, P.~Uwer {\em et~al.}, {\em Comput.
  Phys. Commun.} {\bf 182}, 1034  (2011),
  \href{http://arxiv.org/abs/1007.1327}{{\ttfamily arXiv:1007.1327 [hep-ph]}}.

\bibitem{ATLAS-CONF-2011-121}
ATLAS Collaboration, {\it {M}easurement of the $t\bar{t}$ production
  cross-section in pp collisions at $\sqrt{s}=7 \rm\ {T}e{V}$ using kinematic
  information of lepton+jets events}, ATLAS-CONF-2011-121  (2011).

\bibitem{CMS-PAS-TOP-11-003}
CMS Collaboration, {\it {M}easurement of $t\bar{t}$ {P}air {P}roduction {C}ross
  {S}ection at $\sqrt{s}=7 \rm\ {T}e{V}$ using b-quark {J}et {I}dentification
  {T}echniques in {L}epton + {J}et {E}vents}, CMS-PAS-TOP-11-003  (2011).

\bibitem{Martin:2009iq}
A.~Martin, W.~Stirling, R.~Thorne and G.~Watt, {\em Eur. Phys. J.} {\bf C63},
  189  (2009), \href{http://arxiv.org/abs/0901.0002}{{\ttfamily arXiv:0901.0002
  [hep-ph]}}.

\bibitem{:2009wt}
H1 and ZEUS Collaborations, F.~Aaron {\em et~al.}, {\em JHEP} {\bf 1001},   109
   (2010), \href{http://arxiv.org/abs/0911.0884}{{\ttfamily arXiv:0911.0884
  [hep-ex]}}.

\bibitem{Lai:2010vv}
H.-L. Lai, M.~Guzzi, J.~Huston, Z.~Li, P.~M. Nadolsky {\em et~al.}, {\em Phys.
  Rev.} {\bf D82},   074024  (2010),
  \href{http://arxiv.org/abs/1007.2241}{{\ttfamily arXiv:1007.2241 [hep-ph]}}.

\bibitem{Ball:2011uy}
NNPDF Collaboration, R.~D. Ball {\em et~al.}, {\em Nucl. Phys.} {\bf B855}, 153
   (2012), \href{http://arxiv.org/abs/1107.2652}{{\ttfamily arXiv:1107.2652
  [hep-ph]}}.

\bibitem{Alekhin:2012ig}
S.~Alekhin, J.~Blumlein and S.~Moch, {\it {P}arton distribution functions and
  benchmark cross sections at {NNLO}}  (2012),
  \href{http://arxiv.org/abs/1202.2281}{{\ttfamily arXiv:1202.2281 [hep-ph]}}.

\bibitem{Nason:1987xz}
P.~Nason, S.~Dawson and R.~Ellis, {\em Nucl. Phys.} {\bf B303},   607  (1988).

\bibitem{Nason:1989zy}
P.~Nason, S.~Dawson and R.~K. Ellis, {\em Nucl. Phys.} {\bf B327}, 49  (1989),
  erratum-ibid.\ {\bf B335}, 260 (1990).

\bibitem{Beenakker:1988bq}
W.~Beenakker, H.~Kuijf, W.~van Neerven and J.~Smith, {\em Phys. Rev.} {\bf
  D40}, 54  (1989).

\bibitem{Beenakker:1993yr}
W.~Beenakker, A.~Denner, W.~Hollik, R.~Mertig, T.~Sack {\em et~al.}, {\em Nucl.
  Phys.} {\bf B411}, 343  (1994).

\bibitem{Bernreuther:2005is}
W.~Bernreuther, M.~Fuecker and Z.~Si, {\em Phys. Lett.} {\bf B633}, 54  (2006),
  \href{http://arxiv.org/abs/hep-ph/0508091}{{\ttfamily arXiv:hep-ph/0508091
  [hep-ph]}}.

\bibitem{Bernreuther:2006vg}
W.~Bernreuther, M.~Fuecker and Z.~Si, {\em Phys. Rev.} {\bf D74},   113005
  (2006), \href{http://arxiv.org/abs/hep-ph/0610334}{{\ttfamily
  arXiv:hep-ph/0610334 [hep-ph]}}.

\bibitem{Bernreuther:2008md}
W.~Bernreuther, M.~Fucker and Z.~Si, {\em Phys. Rev.} {\bf D78},   017503
  (2008), \href{http://arxiv.org/abs/0804.1237}{{\ttfamily arXiv:0804.1237
  [hep-ph]}}.

\bibitem{Kuhn:2005it}
J.~H. Kuhn, A.~Scharf and P.~Uwer, {\em Eur. Phys. J.} {\bf C45}, 139  (2006),
  \href{http://arxiv.org/abs/hep-ph/0508092}{{\ttfamily arXiv:hep-ph/0508092
  [hep-ph]}}.

\bibitem{Kuhn:2006vh}
J.~H. Kuhn, A.~Scharf and P.~Uwer, {\em Eur. Phys. J.} {\bf C51}, 37  (2007),
  \href{http://arxiv.org/abs/hep-ph/0610335}{{\ttfamily arXiv:hep-ph/0610335
  [hep-ph]}}.

\bibitem{Hollik:2007sw}
W.~Hollik and M.~Kollar, {\em Phys. Rev.} {\bf D77},   014008  (2008),
  \href{http://arxiv.org/abs/0708.1697}{{\ttfamily arXiv:0708.1697 [hep-ph]}}.

\bibitem{Bernreuther:2001rq}
W.~Bernreuther, A.~Brandenburg, Z.~Si and P.~Uwer, {\em Phys. Rev. Lett.} {\bf
  87},   242002  (2001), \href{http://arxiv.org/abs/hep-ph/0107086}{{\ttfamily
  arXiv:hep-ph/0107086 [hep-ph]}}.

\bibitem{Bernreuther:2004jv}
W.~Bernreuther, A.~Brandenburg, Z.~Si and P.~Uwer, {\em Nucl. Phys.} {\bf
  B690}, 81  (2004), \href{http://arxiv.org/abs/hep-ph/0403035}{{\ttfamily
  arXiv:hep-ph/0403035 [hep-ph]}}.

\bibitem{Melnikov:2009dn}
K.~Melnikov and M.~Schulze, {\em JHEP} {\bf 0908},   049  (2009),
  \href{http://arxiv.org/abs/0907.3090}{{\ttfamily arXiv:0907.3090 [hep-ph]}}.

\bibitem{Bernreuther:2010ny}
W.~Bernreuther and Z.~Si, {\em Nucl. Phys.} {\bf B837}, 90  (2010),
  \href{http://arxiv.org/abs/1003.3926}{{\ttfamily arXiv:1003.3926 [hep-ph]}}.

\bibitem{Campbell:2012uf}
J.~M. Campbell and R.~K. Ellis, {\it {T}op-quark processes at {NLO} in
  production and decay}  (2012),
  \href{http://arxiv.org/abs/1204.1513}{{\ttfamily arXiv:1204.1513 [hep-ph]}}.

\bibitem{Denner:2010jp}
A.~Denner, S.~Dittmaier, S.~Kallweit and S.~Pozzorini, {\em Phys. Rev. Lett.}
  {\bf 106},   052001  (2011), \href{http://arxiv.org/abs/1012.3975}{{\ttfamily
  arXiv:1012.3975 [hep-ph]}}.

\bibitem{Bevilacqua:2010qb}
G.~Bevilacqua, M.~Czakon, A.~van Hameren, C.~G. Papadopoulos and M.~Worek, {\em
  JHEP} {\bf 1102},   083  (2011),
  \href{http://arxiv.org/abs/1012.4230}{{\ttfamily arXiv:1012.4230 [hep-ph]}}.

\bibitem{Maestre:2012vp}
A.~Denner, S.~Dittmaier, S.~Kallweit, S.~Pozzorini and M.~Schulze, {\it
  {F}inite-width effects {I}n top-quark pair production and decay at the
  {LHC}}, in {\em {L}es {H}ouches 2011: {P}hysics at {T}e{V} {C}olliders {SM}
  and {NLO} {M}ultileg and {SM} {MC} {W}orking {G}roups {R}eport\/},  eds.
  J.~A. Maestre {\em et~al.} (2012).
\newblock p.~55.
\newblock \href{http://arxiv.org/abs/1203.6803}{{\ttfamily arXiv:1203.6803
  [hep-ph]}}.

\bibitem{Dittmaier:2007wz}
S.~Dittmaier, P.~Uwer and S.~Weinzierl, {\em Phys. Rev. Lett.} {\bf 98},
  262002  (2007), \href{http://arxiv.org/abs/hep-ph/0703120}{{\ttfamily
  arXiv:hep-ph/0703120 [hep-ph]}}.

\bibitem{Dittmaier:2008uj}
S.~Dittmaier, P.~Uwer and S.~Weinzierl, {\em Eur. Phys. J.} {\bf C59}, 625
  (2009), \href{http://arxiv.org/abs/0810.0452}{{\ttfamily arXiv:0810.0452
  [hep-ph]}}.

\bibitem{Melnikov:2010iu}
K.~Melnikov and M.~Schulze, {\em Nucl. Phys.} {\bf B840}, 129  (2010),
  \href{http://arxiv.org/abs/1004.3284}{{\ttfamily arXiv:1004.3284 [hep-ph]}}.

\bibitem{Melnikov:2011qx}
K.~Melnikov, A.~Scharf and M.~Schulze, {\em Phys. Rev.} {\bf D85},   054002
  (2012), \href{http://arxiv.org/abs/1111.4991}{{\ttfamily arXiv:1111.4991
  [hep-ph]}}.

\bibitem{Bevilacqua:2010ve}
G.~Bevilacqua, M.~Czakon, C.~Papadopoulos and M.~Worek, {\em Phys. Rev. Lett.}
  {\bf 104},   162002  (2010), \href{http://arxiv.org/abs/1002.4009}{{\ttfamily
  arXiv:1002.4009 [hep-ph]}}.

\bibitem{Bevilacqua:2011aa}
G.~Bevilacqua, M.~Czakon, C.~Papadopoulos and M.~Worek, {\em Phys. Rev.} {\bf
  D84},   114017  (2011), \href{http://arxiv.org/abs/1108.2851}{{\ttfamily
  arXiv:1108.2851 [hep-ph]}}.

\bibitem{Bredenstein:2010rs}
A.~Bredenstein, A.~Denner, S.~Dittmaier and S.~Pozzorini, {\em JHEP} {\bf
  1003},   021  (2010), \href{http://arxiv.org/abs/1001.4006}{{\ttfamily
  arXiv:1001.4006 [hep-ph]}}.

\bibitem{Baernreuther:2012ws}
P.~Baernreuther, M.~Czakon and A.~Mitov, {\it {P}ercent level precision physics
  at the {T}evatron: first genuine {NNLO} {QCD} corrections to q qbar
  -\&amp;gt; t tbar + {X}}  (2012),
  \href{http://arxiv.org/abs/1204.5201}{{\ttfamily arXiv:1204.5201 [hep-ph]}}.

\bibitem{Laenen:1991af}
E.~Laenen, J.~Smith and W.~van Neerven, {\em Nucl. Phys.} {\bf B369}, 543
  (1992).

\bibitem{Langenfeld:2009wd}
U.~Langenfeld, S.~Moch and P.~Uwer, {\em Phys. Rev.} {\bf D80},   054009
  (2009), \href{http://arxiv.org/abs/0906.5273}{{\ttfamily arXiv:0906.5273
  [hep-ph]}}.

\bibitem{Czakon:2009zw}
M.~Czakon, A.~Mitov and G.~F. Sterman, {\em Phys. Rev.} {\bf D80},   074017
  (2009), \href{http://arxiv.org/abs/0907.1790}{{\ttfamily arXiv:0907.1790
  [hep-ph]}}.

\bibitem{Cacciari:2011hy}
M.~Cacciari, M.~Czakon, M.~L. Mangano, A.~Mitov and P.~Nason, {\em Phys. Lett.}
  {\bf B710}, 612  (2012), \href{http://arxiv.org/abs/1111.5869}{{\ttfamily
  arXiv:1111.5869 [hep-ph]}}.

\bibitem{Kidonakis:2010dk}
N.~Kidonakis, {\em Phys. Rev.} {\bf D82},   114030  (2010),
  \href{http://arxiv.org/abs/1009.4935}{{\ttfamily arXiv:1009.4935 [hep-ph]}}.

\bibitem{Ahrens:2010zv}
V.~Ahrens, A.~Ferroglia, M.~Neubert, B.~D. Pecjak and L.~L. Yang, {\em JHEP}
  {\bf 1009},   097  (2010), \href{http://arxiv.org/abs/1003.5827}{{\ttfamily
  arXiv:1003.5827 [hep-ph]}}.

\bibitem{Beneke:2011mq}
M.~Beneke, P.~Falgari, S.~Klein and C.~Schwinn, {\em Nucl. Phys.} {\bf B855},
  695  (2012), \href{http://arxiv.org/abs/1109.1536}{{\ttfamily arXiv:1109.1536
  [hep-ph]}}.

\bibitem{Czakon:2011xx}
M.~Czakon and A.~Mitov, {\it {T}op++: a program for the calculation of the
  top-pair cross-section at hadron colliders}  (2011),
  \href{http://arxiv.org/abs/1112.5675}{{\ttfamily arXiv:1112.5675 [hep-ph]}}.

\bibitem{Ahrens:2011px}
V.~Ahrens, M.~Neubert, B.~D. Pecjak, A.~Ferroglia and L.~L. Yang, {\em Phys.
  Lett.} {\bf B703}, 135  (2011),
  \href{http://arxiv.org/abs/1105.5824}{{\ttfamily arXiv:1105.5824 [hep-ph]}}.

\bibitem{Moch:2012mk}
S.~Moch, P.~Uwer and A.~Vogt, {\it {O}n top-pair hadro-production at
  next-to-next-to-leading order}  (2012),
  \href{http://arxiv.org/abs/1203.6282}{{\ttfamily arXiv:1203.6282 [hep-ph]}}.

\bibitem{Brodsky:2012sz}
S.~J. Brodsky and X.-G. Wu, {\it {A}pplication of the {P}rinciple of {M}aximum
  {C}onformality to {T}op-{P}air {P}roduction}  (2012),
  \href{http://arxiv.org/abs/1204.1405}{{\ttfamily arXiv:1204.1405 [hep-ph]}}.

\bibitem{Alekhin:2009ni}
S.~Alekhin, J.~Blumlein, S.~Klein and S.~Moch, {\em Phys. Rev.} {\bf D81},
  014032  (2010), \href{http://arxiv.org/abs/0908.2766}{{\ttfamily
  arXiv:0908.2766 [hep-ph]}}.

\bibitem{Kidonakis:2011wy}
N.~Kidonakis, {\em Phys. Rev.} {\bf D83},   091503  (2011),
  \href{http://arxiv.org/abs/1103.2792}{{\ttfamily arXiv:1103.2792 [hep-ph]}}.

\bibitem{Kidonakis:2010tc}
N.~Kidonakis, {\em Phys. Rev.} {\bf D81},   054028  (2010),
  \href{http://arxiv.org/abs/1001.5034}{{\ttfamily arXiv:1001.5034 [hep-ph]}}.

\bibitem{Zhu:2010mr}
H.~X. Zhu, C.~S. Li, J.~Wang and J.~J. Zhang, {\em JHEP} {\bf 1102},   099
  (2011), \href{http://arxiv.org/abs/1006.0681}{{\ttfamily arXiv:1006.0681
  [hep-ph]}}.

\bibitem{Kidonakis:2010ux}
N.~Kidonakis, {\em Phys. Rev.} {\bf D82},   054018  (2010),
  \href{http://arxiv.org/abs/1005.4451}{{\ttfamily arXiv:1005.4451 [hep-ph]}}.

\bibitem{Harris:2002md}
B.~Harris, E.~Laenen, L.~Phaf, Z.~Sullivan and S.~Weinzierl, {\em Phys. Rev.}
  {\bf D66},   054024  (2002),
  \href{http://arxiv.org/abs/hep-ph/0207055}{{\ttfamily arXiv:hep-ph/0207055
  [hep-ph]}}.

\bibitem{Zhu:2002uj}
S.~Zhu, {\em Phys. Lett.} {\bf B524}, 283  (2002).

\bibitem{Stelzer:1998ni}
T.~Stelzer, Z.~Sullivan and S.~Willenbrock, {\em Phys. Rev.} {\bf D58},
  094021  (1998), \href{http://arxiv.org/abs/hep-ph/9807340}{{\ttfamily
  arXiv:hep-ph/9807340 [hep-ph]}}.

\bibitem{Jezabek:1988iv}
M.~Jezabek and J.~H. Kuhn, {\em Nucl. Phys.} {\bf B314},  ~1  (1989).

\bibitem{Abazov:2012vd}
D0 Collaboration, V.~M. Abazov {\em et~al.}, {\em Phys. Rev.} {\bf D85},
  091104  (2012), \href{http://arxiv.org/abs/1201.4156}{{\ttfamily
  arXiv:1201.4156 [hep-ex]}}.

\bibitem{Nakamura:2010zzi}
PDG Collaboration, K.~Nakamura {\em et~al.}, {\em J. Phys.} {\bf G37},   075021
   (2010).

\bibitem{Acosta:2005hr}
CDF Collaboration, D.~Acosta {\em et~al.}, {\em Phys. Rev. Lett.} {\bf 95},
  102002  (2005), \href{http://arxiv.org/abs/hep-ex/0505091}{{\ttfamily
  arXiv:hep-ex/0505091 [hep-ex]}}.

\bibitem{Abazov:2011zk}
D0 Collaboration, V.~Abazov {\em et~al.}, {\em Phys. Rev. Lett.} {\bf 107},
  121802  (2011), \href{http://arxiv.org/abs/1106.5436}{{\ttfamily
  arXiv:1106.5436 [hep-ex]}}.

\bibitem{Abazov:2011qf}
D0 Collaboration, V.~M. Abazov {\em et~al.}, {\em Phys. Lett.} {\bf B701}, 313
  (2011), \href{http://arxiv.org/abs/1103.4574}{{\ttfamily arXiv:1103.4574
  [hep-ex]}}.

\bibitem{:2010vi}
ALEPH and CDF and D0 and DELPHI and L3 and OPAL and SLD Collaborations,
  {LEPEWWG, TEVEWWG, SLD EW and HF groups}, {\it {P}recision {E}lectroweak
  {M}easurements and {C}onstraints on the {S}tandard {M}odel}  (2010),
  \href{http://arxiv.org/abs/1012.2367}{{\ttfamily arXiv:1012.2367 [hep-ex]}},
  updated for 2011 summer conferences, see
  \url{http://lepewwg.web.cern.ch/LEPEWWG}.

\bibitem{Baak:2011ze}
GFITTER Collaboration, M.~Baak, M.~Goebel, J.~Haller, A.~Hoecker, D.~Ludwig
  {\em et~al.}, {\it {Updated Status of the Global Electroweak Fit and
  Constraints on New Physics}}  (2011),
  \href{http://arxiv.org/abs/1107.0975}{{\ttfamily arXiv:1107.0975 [hep-ph]}},
  updated for 2011 summer conferences, see \url{http://gfitter.desy.de}.

\bibitem{CDFandD0:2011aa}
CDF and D0 Collaborations, {TEVNPH}, {\it {C}ombined {CDF} and {D}0 {U}pper
  {L}imits on {S}tandard {M}odel {H}iggs {B}oson {P}roduction with up to 8.6
  fb$^{-1}$ of {D}ata}  (2011),
  \href{http://arxiv.org/abs/1107.5518}{{\ttfamily arXiv:1107.5518 [hep-ex]}}.

\bibitem{Chatrchyan:2012tx}
CMS Collaboration, S.~Chatrchyan {\em et~al.}, {\em Phys. Lett.} {\bf B710}, 26
   (2012), \href{http://arxiv.org/abs/1202.1488}{{\ttfamily arXiv:1202.1488
  [hep-ex]}}.

\bibitem{:2012si}
ATLAS Collaboration, G.~Aad {\em et~al.}, {\em Phys. Lett.} {\bf B710}, 49
  (2012), \href{http://arxiv.org/abs/1202.1408}{{\ttfamily arXiv:1202.1408
  [hep-ex]}}.

\bibitem{Skands:2007zg}
P.~Z. Skands and D.~Wicke, {\em Eur. Phys. J.} {\bf C52}, 133  (2007),
  \href{http://arxiv.org/abs/hep-ph/0703081}{{\ttfamily arXiv:hep-ph/0703081
  [HEP-PH]}}.

\bibitem{Abazov:2006vd}
D0 Collaboration, V.~Abazov {\em et~al.}, {\em Phys. Rev. Lett.} {\bf 98},
  041801  (2007), \href{http://arxiv.org/abs/hep-ex/0608044}{{\ttfamily
  arXiv:hep-ex/0608044 [hep-ex]}}.

\bibitem{Aaltonen:2010js}
CDF Collaboration, T.~Aaltonen {\em et~al.}, {\em Phys. Rev. Lett.} {\bf 105},
   101801  (2010), \href{http://arxiv.org/abs/1006.4597}{{\ttfamily
  arXiv:1006.4597 [hep-ex]}}.

\bibitem{Baur:2001si}
U.~Baur, M.~Buice and L.~H. Orr, {\em Phys. Rev.} {\bf D64},   094019  (2001),
  \href{http://arxiv.org/abs/hep-ph/0106341}{{\ttfamily arXiv:hep-ph/0106341
  [hep-ph]}}.

\bibitem{atlasdet}
ATLAS Collaboration, G.~Aad {\em et~al.}, {\em JINST} {\bf 3},   S08003
  (2008).

\bibitem{cmsdet}
CMS Collaboration, R.~Adolphi {\em et~al.}, {\em JINST} {\bf 3},   S08004
  (2008).

\bibitem{vanderMeer:1968zz}
S.~van~der Meer, {\it {Calibration of the effective beam hight in the ISR}},
  CERN-ISR-PO-68-31  (1968).

\bibitem{Aad:2011dr}
ATLAS Collaboration, G.~Aad {\em et~al.}, {\em Eur. Phys. J.} {\bf C71},   1630
   (2011), \href{http://arxiv.org/abs/1101.2185}{{\ttfamily arXiv:1101.2185
  [hep-ex]}}.

\bibitem{ATLAS-CONF-2011-011}
ATLAS Collaboration, {\it {U}pdated {L}uminosity {D}etermination in pp
  {C}ollisions at $\sqrt{s}=7$ {T}e{V} using the {ATLAS} {D}etector},
  ATLAS-CONF-2011-011  (2011).

\bibitem{ATLAS-CONF-2011-116}
ATLAS Collaboration, {\it {L}uminosity {D}etermination in pp {C}ollisions at
  $\sqrt{s}=7$ {T}e{V} using the {ATLAS} {D}etector in 2011},
  ATLAS-CONF-2011-116  (2011).

\bibitem{CMS-DP-2011-002}
CMS Collaboration, {\it {A}bsolute luminosity normalization}, CMS-DP-2011-002
  (2011).

\bibitem{CMS-PAS-EWK-10-004}
CMS Collaboration, {\it {M}easurement of {CMS} {L}uminosity},
  CMS-PAS-EWK-10-004  (2010).

\bibitem{CMS-PAS-EWK-11-001}
CMS Collaboration, {\it {A}bsolute {C}alibration of {L}uminosity {M}easurement
  at {CMS}: {S}ummer 2011 {U}pdate}, CMS-PAS-EWK-11-001  (2011).

\bibitem{cmspas-smp-12-008}
CMS Collaboration, {\it {A}bsolute {C}alibration of the {L}uminosity
  {M}easurement at {CMS}: {W}inter 2012 {U}pdate}, CMS-PAS-SMP-12-008  (2012).

\bibitem{Bayatian:2006zz}
CMS Collaboration, G.~Bayatian {\em et~al.}, {\it {CMS} physics: {T}echnical
  design report; {V}olume {I}: {D}etector {P}erformance and {S}oftware},
  CERN-LHCC-2006-001  (2006).

\bibitem{Ball:2007zza}
CMS Collaboration, G.~L. Bayatian {\em et~al.}, {\em J. Phys.} {\bf G34}, 995
  (2007).

\bibitem{atlastdr}
ATLAS Collaboration, G.~Aad {\em et~al.}, {\it {E}xpected {P}erformance of the
  {ATLAS} {E}xperiment - {D}etector, {T}rigger and {P}hysics},
  CERN-OPEN-2008-020  (2009), \href{http://arxiv.org/abs/0901.0512}{{\ttfamily
  arXiv:0901.0512 [hep-ex]}}.

\bibitem{Collaboration:2010knc}
ATLAS Collaboration, G.~Aad {\em et~al.}, {\em JHEP} {\bf 1009},   056  (2010),
  \href{http://arxiv.org/abs/1005.5254}{{\ttfamily arXiv:1005.5254 [hep-ex]}}.

\bibitem{CMS-PAS-PFT-10-002}
CMS Collaboration, {\it {C}ommissioning of the {P}article-{F}low reconstruction
  in {M}inimum-{B}ias and {J}et {E}vents from pp {C}ollisions at 7 {T}e{V}},
  CMS-PAS-PFT-10-002  (2011).

\bibitem{Cacciari:2008gn}
M.~Cacciari, G.~P. Salam and G.~Soyez, {\em JHEP} {\bf 0804},   005  (2008),
  \href{http://arxiv.org/abs/0802.1188}{{\ttfamily arXiv:0802.1188 [hep-ph]}}.

\bibitem{Cacciari:2007fd}
M.~Cacciari and G.~P. Salam, {\em Phys. Lett.} {\bf B659}, 119  (2008),
  \href{http://arxiv.org/abs/0707.1378}{{\ttfamily arXiv:0707.1378 [hep-ph]}}.

\bibitem{Aad:2011mk}
ATLAS Collaboration, G.~Aad {\em et~al.}, {\em Eur. Phys. J.} {\bf C72},   1909
   (2012), \href{http://arxiv.org/abs/1110.3174}{{\ttfamily arXiv:1110.3174
  [hep-ex]}}.

\bibitem{CMS-PAS-EGM-10-004}
CMS Collaboration, {\it {E}lectron reconstruction and identification at
  $\sqrt{s} = 7$ {T}e{V}}, CMS-PAS-EGM-10-004  (2010).

\bibitem{:2011nx}
CMS Collaboration, S.~Chatrchyan {\em et~al.}, {\em JHEP} {\bf 1110},   132
  (2011), \href{http://arxiv.org/abs/1107.4789}{{\ttfamily arXiv:1107.4789
  [hep-ex]}}.

\bibitem{Aad:2011dm}
ATLAS Collaboration, G.~Aad {\em et~al.}, {\em Phys. Rev.} {\bf D85},   072004
  (2012), \href{http://arxiv.org/abs/1109.5141}{{\ttfamily arXiv:1109.5141
  [hep-ex]}}.

\bibitem{CMS-PAS-MUO-10-002}
CMS Collaboration, {\it {P}erformance of muon identification in pp collisions
  at $\sqrt{s}= 7$ {T}e{V}}, CMS-PAS-MUO-10-002  (2010).

\bibitem{ATLAS-CONF-2011-152}
ATLAS Collaboration, {\it {P}erformance of the {R}econstruction and
  {I}dentification of {H}adronic {T}au {D}ecays with {ATLAS}},
  ATLAS-CONF-2011-152  (2011).

\bibitem{Aad:2011fu}
ATLAS Collaboration, G.~Aad {\em et~al.}, {\em Phys. Lett.} {\bf B706}, 276
  (2012), \href{http://arxiv.org/abs/1108.4101}{{\ttfamily arXiv:1108.4101
  [hep-ex]}}.

\bibitem{tau-11-001}
CMS Collaboration, S.~Chatrchyan {\em et~al.}, {\em JINST} {\bf 7},   P01001
  (2012), \href{http://arxiv.org/abs/1109.6034}{{\ttfamily arXiv:1109.6034
  [physics.ins-det]}}.

\bibitem{Cacciari:2008gp}
M.~Cacciari, G.~P. Salam and G.~Soyez, {\em JHEP} {\bf 0804},   063  (2008),
  \href{http://arxiv.org/abs/0802.1189}{{\ttfamily arXiv:0802.1189 [hep-ph]}}.

\bibitem{Cacciari:2005hq}
M.~Cacciari and G.~P. Salam, {\em Phys. Lett.} {\bf B641}, 57  (2006),
  \href{http://arxiv.org/abs/hep-ph/0512210}{{\ttfamily arXiv:hep-ph/0512210
  [hep-ph]}}.

\bibitem{Cacciari:2011ma}
M.~Cacciari, G.~P. Salam and G.~Soyez, {\em Eur. Phys. J.} {\bf C72},   1896
  (2012), \href{http://arxiv.org/abs/1111.6097}{{\ttfamily arXiv:1111.6097
  [hep-ph]}}.

\bibitem{Chatrchyan:2011ds}
CMS Collaboration, S.~Chatrchyan {\em et~al.}, {\em JINST} {\bf 6},   P11002
  (2011), \href{http://arxiv.org/abs/1107.4277}{{\ttfamily arXiv:1107.4277
  [physics.ins-det]}}.

\bibitem{Aad:2011he}
ATLAS Collaboration, G.~Aad {\em et~al.}, {\it {J}et energy measurement with
  the {ATLAS} detector in proton-proton collisions at $\sqrt{s}=$ 7 {T}e{V}}
  (2011), \href{http://arxiv.org/abs/1112.6426}{{\ttfamily arXiv:1112.6426
  [hep-ex]}}.

\bibitem{cmsnote-2006-025}
J.~D'Hondt, S.~Lowette, J.~Heyninck and S.~Kasselmann, {\it {L}ight quark jet
  energy scale calibration using the {W} mass constraint in single-leptonic
  $t\bar{t}$ events}, CMS-NOTE-2006/025  (2006).

\bibitem{cmspas-top-07-004}
CMS Collaboration, {\it {M}easurement of jet energy scale corrections using top
  quark events}, CMS-PAS-TOP-07-004  (2007).

\bibitem{ATLAS-CONF-2010-054}
ATLAS Collaboration, {\it {J}et energy resolution and reconstruction
  efficiencies from in-situ techniques with the {ATLAS} {D}etector {U}sing
  {P}roton-{P}roton {C}ollisions at a {C}enter of {M}ass {E}nergy $\sqrt{s}$ =
  7 {T}e{V}}, ATLAS-CONF-2010-054  (2010).

\bibitem{Aad:2011re}
ATLAS Collaboration, G.~Aad {\em et~al.}, {\em Eur. Phys. J.} {\bf C72},   1844
   (2012), \href{http://arxiv.org/abs/1108.5602}{{\ttfamily arXiv:1108.5602
  [hep-ex]}}.

\bibitem{Chatrchyan:2011tn}
CMS Collaboration, S.~Chatrchyan {\em et~al.}, {\em JINST} {\bf 6},   P09001
  (2011), \href{http://arxiv.org/abs/1106.5048}{{\ttfamily arXiv:1106.5048
  [physics.ins-det]}}.

\bibitem{CMS-PAS-BTV-09-001}
CMS Collaboration, {\it {A}lgorithms for b {J}et identification in {CMS}},
  CMS-PAS-BTV-09-001  (2009).

\bibitem{ATLAS-CONF-2010-091}
ATLAS Collaboration, {\it {P}erformance of {I}mpact {P}arameter-{B}ased
  b-tagging {A}lgorithms with the {ATLAS} {D}etector using {P}roton-{P}roton
  {C}ollisions at $\sqrt{s}$= 7 {T}e{V}}, ATLAS-CONF-2010-091  (2010).

\bibitem{CMS-PAS-BTV-11-002}
CMS Collaboration, {\it {S}tatus of b-tagging and vertexing tools for 2011 data
  analysis}, CMS-PAS-BTV-11-002  (2011).

\bibitem{ATLAS-CONF-2010-042}
ATLAS Collaboration, {\it {P}erformance of the {ATLAS} {S}econdary {V}ertex
  $b$-tagging {A}lgorithm in 7~{T}e{V} {C}ollision {D}ata}, ATLAS-CONF-2010-042
   (2010).

\bibitem{CMS-PAS-BTV-07-003}
CMS Collaboration, {\it {I}mpact of {T}racker {M}isalignment on the {CMS}
  b-{T}agging {P}erformance}, CMS-PAS-BTV-07-003  (2009).

\bibitem{ATLAS-CONF-2011-102}
ATLAS Collaboration, {\it {C}ommissioning of the {ATLAS} high-performance
  b-tagging algorithms in the 7 {T}e{V} collision data}, ATLAS-CONF-2011-102
  (2011).

\bibitem{cmspas-btv-11-004}
CMS Collaboration, {\it b-{J}et {I}dentification in the {CMS} {E}xperiment},
  CMS-PAS-BTV-11-004  (2012).

\bibitem{CMS-PAS-BTV-11-001}
CMS Collaboration, {\it {P}erformance of the b-jet identification in {CMS}},
  CMS-PAS-BTV-11-001  (2011).

\bibitem{ATLAS-CONF-2011-089}
ATLAS Collaboration, {\it {C}alibrating the b-{T}ag {E}fficiency and {M}istag
  {R}ate in $35 \rm\ pb^{-1}$ of {D}ata with the {ATLAS} {D}etector},
  ATLAS-CONF-2011-089  (2011).

\bibitem{ATLAS-CONF-2011-143}
ATLAS Collaboration, {\it b-{J}et {T}agging {E}fficiency {C}alibration using
  the {S}ystem8 {M}ethod}, ATLAS-CONF-2011-143  (2011).

\bibitem{ATLAS-CONF-2012-043}
ATLAS Collaboration, {\it {M}easurement of the b-tag {E}fficiency in a {S}ample
  of {J}ets {C}ontaining {M}uons with $5 {\rm\ fb}^{-1}$ of {D}ata from the
  {ATLAS} {D}etector}, ATLAS-CONF-2012-043  (2012).

\bibitem{cmsnote-2006-013}
S.~Lowette, J.~D'Hondt, J.~Heyninck and P.~Vanlaer, {\it {O}ffline
  {C}alibration of b-{J}et {I}dentification {E}fficiencies}, CMS-NOTE-2006/013
  (2006).

\bibitem{CMS-PAS-BTV-11-003}
CMS Collaboration, {\it {M}easurement of the b-tagging efficiency using
  $t\bar{t}$ events}, CMS-PAS-BTV-11-003  (2012).

\bibitem{Agostinelli:2002hh}
GEANT4 Collaboration, S.~Agostinelli {\em et~al.}, {\em Nucl. Instrum. Meth.}
  {\bf A506}, 250  (2003).

\bibitem{Buckley:2011ms}
A.~Buckley, J.~Butterworth, S.~Gieseke, D.~Grellscheid, S.~Hoche {\em et~al.},
  {\em Phys. Rept.} {\bf 504}, 145  (2011),
  \href{http://arxiv.org/abs/1101.2599}{{\ttfamily arXiv:1101.2599 [hep-ph]}}.

\bibitem{Sjostrand:2006za}
T.~Sjostrand, S.~Mrenna and P.~Z. Skands, {\em JHEP} {\bf 0605},   026  (2006),
  \href{http://arxiv.org/abs/hep-ph/0603175}{{\ttfamily arXiv:hep-ph/0603175
  [hep-ph]}}.

\bibitem{Sjostrand:2007gs}
T.~Sjostrand, S.~Mrenna and P.~Z. Skands, {\em Comput. Phys. Commun.} {\bf
  178}, 852  (2008), \href{http://arxiv.org/abs/0710.3820}{{\ttfamily
  arXiv:0710.3820 [hep-ph]}}.

\bibitem{Corcella:2000bw}
G.~Corcella, I.~Knowles, G.~Marchesini, S.~Moretti, K.~Odagiri {\em et~al.},
  {\em JHEP} {\bf 0101},   010  (2001),
  \href{http://arxiv.org/abs/hep-ph/0011363}{{\ttfamily arXiv:hep-ph/0011363
  [hep-ph]}}.

\bibitem{Corcella:2002jc}
G.~Corcella, I.~Knowles, G.~Marchesini, S.~Moretti, K.~Odagiri {\em et~al.},
  {\it {HERWIG} 6.5 release note}  (2002),
  \href{http://arxiv.org/abs/hep-ph/0210213}{{\ttfamily arXiv:hep-ph/0210213
  [hep-ph]}}.

\bibitem{Gieseke:2011na}
S.~Gieseke, D.~Grellscheid, K.~Hamilton, A.~Papaefstathiou, S.~Platzer {\em
  et~al.}, {\it {H}erwig++ 2.5 {R}elease {N}ote}  (2011),
  \href{http://arxiv.org/abs/1102.1672}{{\ttfamily arXiv:1102.1672 [hep-ph]}}.

\bibitem{Alwall:2007st}
J.~Alwall, P.~Demin, S.~de~Visscher, R.~Frederix, M.~Herquet {\em et~al.}, {\em
  JHEP} {\bf 0709},   028  (2007),
  \href{http://arxiv.org/abs/0706.2334}{{\ttfamily arXiv:0706.2334 [hep-ph]}}.

\bibitem{Alwall:2011uj}
J.~Alwall, M.~Herquet, F.~Maltoni, O.~Mattelaer and T.~Stelzer, {\em JHEP} {\bf
  1106},   128  (2011), \href{http://arxiv.org/abs/1106.0522}{{\ttfamily
  arXiv:1106.0522 [hep-ph]}}.

\bibitem{Mangano:2002ea}
M.~L. Mangano, M.~Moretti, F.~Piccinini, R.~Pittau and A.~D. Polosa, {\em JHEP}
  {\bf 0307},   001  (2003),
  \href{http://arxiv.org/abs/hep-ph/0206293}{{\ttfamily arXiv:hep-ph/0206293
  [hep-ph]}}.

\bibitem{Gleisberg:2003xi}
T.~Gleisberg, S.~Hoeche, F.~Krauss, A.~Schalicke, S.~Schumann {\em et~al.},
  {\em JHEP} {\bf 0402},   056  (2004),
  \href{http://arxiv.org/abs/hep-ph/0311263}{{\ttfamily arXiv:hep-ph/0311263
  [hep-ph]}}.

\bibitem{Gleisberg:2008ta}
T.~Gleisberg, S.~Hoeche, F.~Krauss, M.~Schonherr, S.~Schumann {\em et~al.},
  {\em JHEP} {\bf 0902},   007  (2009),
  \href{http://arxiv.org/abs/0811.4622}{{\ttfamily arXiv:0811.4622 [hep-ph]}}.

\bibitem{Mangano:2006rw}
M.~L. Mangano, M.~Moretti, F.~Piccinini and M.~Treccani, {\em JHEP} {\bf 0701},
    013  (2007), \href{http://arxiv.org/abs/hep-ph/0611129}{{\ttfamily
  arXiv:hep-ph/0611129 [hep-ph]}}.

\bibitem{Schalicke:2005nv}
A.~Schalicke and F.~Krauss, {\em JHEP} {\bf 0507},   018  (2005),
  \href{http://arxiv.org/abs/hep-ph/0503281}{{\ttfamily arXiv:hep-ph/0503281
  [hep-ph]}}.

\bibitem{Frixione:2003ei}
S.~Frixione, P.~Nason and B.~R. Webber, {\em JHEP} {\bf 0308},   007  (2003),
  \href{http://arxiv.org/abs/hep-ph/0305252}{{\ttfamily arXiv:hep-ph/0305252
  [hep-ph]}}.

\bibitem{Nason:2004rx}
P.~Nason, {\em JHEP} {\bf 0411},   040  (2004),
  \href{http://arxiv.org/abs/hep-ph/0409146}{{\ttfamily arXiv:hep-ph/0409146
  [hep-ph]}}.

\bibitem{Alioli:2010xd}
S.~Alioli, P.~Nason, C.~Oleari and E.~Re, {\em JHEP} {\bf 1006},   043  (2010),
  \href{http://arxiv.org/abs/1002.2581}{{\ttfamily arXiv:1002.2581 [hep-ph]}}.

\bibitem{Alioli:2011as}
S.~Alioli, S.-O. Moch and P.~Uwer, {\em JHEP} {\bf 1201},   137  (2012),
  \href{http://arxiv.org/abs/1110.5251}{{\ttfamily arXiv:1110.5251 [hep-ph]}}.

\bibitem{Kardos:2011qa}
A.~Kardos, C.~Papadopoulos and Z.~Trocsanyi, {\em Phys. Lett.} {\bf B705}, 76
  (2011), \href{http://arxiv.org/abs/1101.2672}{{\ttfamily arXiv:1101.2672
  [hep-ph]}}.

\bibitem{Frederix:2011zi}
R.~Frederix, S.~Frixione, V.~Hirschi, F.~Maltoni, R.~Pittau {\em et~al.}, {\em
  Phys. Lett.} {\bf B701}, 427  (2011),
  \href{http://arxiv.org/abs/1104.5613}{{\ttfamily arXiv:1104.5613 [hep-ph]}}.

\bibitem{Hamilton:2010wh}
K.~Hamilton and P.~Nason, {\em JHEP} {\bf 1006},   039  (2010),
  \href{http://arxiv.org/abs/1004.1764}{{\ttfamily arXiv:1004.1764 [hep-ph]}}.

\bibitem{Hoche:2010kg}
S.~Hoche, F.~Krauss, M.~Schonherr and F.~Siegert, {\em JHEP} {\bf 1108},   123
  (2011), \href{http://arxiv.org/abs/1009.1127}{{\ttfamily arXiv:1009.1127
  [hep-ph]}}.

\bibitem{Pumplin:2002vw}
J.~Pumplin, D.~Stump, J.~Huston, H.~Lai, P.~M. Nadolsky {\em et~al.}, {\em
  JHEP} {\bf 0207},   012  (2002),
  \href{http://arxiv.org/abs/hep-ph/0201195}{{\ttfamily arXiv:hep-ph/0201195
  [hep-ph]}}.

\bibitem{Davidson:2010rw}
N.~Davidson, G.~Nanava, T.~Przedzinski, E.~Richter-Was and Z.~Was, {\em Comput.
  Phys. Commun.} {\bf 183}, 821  (2012),
  \href{http://arxiv.org/abs/1002.0543}{{\ttfamily arXiv:1002.0543 [hep-ph]}}.

\bibitem{Boos:2006af}
E.~Boos, V.~Bunichev, L.~Dudko, V.~Savrin and A.~Sherstnev, {\em Phys. Atom.
  Nucl.} {\bf 69}, 1317  (2006).

\bibitem{Kersevan:2004yg}
B.~P. Kersevan and E.~Richter-Was, {\it {T}he {M}onte {C}arlo event generator
  {A}cer{MC} version 2.0 with interfaces to {PYTHIA} 6.2 and {HERWIG} 6.5}
  (2004), \href{http://arxiv.org/abs/hep-ph/0405247}{{\ttfamily
  arXiv:hep-ph/0405247 [hep-ph]}}.

\bibitem{Belyaev:1998dn}
A.~Belyaev, E.~Boos and L.~Dudko, {\em Phys. Rev.} {\bf D59},   075001  (1999),
  \href{http://arxiv.org/abs/hep-ph/9806332}{{\ttfamily arXiv:hep-ph/9806332
  [hep-ph]}}.

\bibitem{White:2009yt}
C.~D. White, S.~Frixione, E.~Laenen and F.~Maltoni, {\em JHEP} {\bf 0911},
  074  (2009), \href{http://arxiv.org/abs/0908.0631}{{\ttfamily arXiv:0908.0631
  [hep-ph]}}.

\bibitem{Frixione:2008yi}
S.~Frixione, E.~Laenen, P.~Motylinski, B.~R. Webber and C.~D. White, {\em JHEP}
  {\bf 0807},   029  (2008), \href{http://arxiv.org/abs/0805.3067}{{\ttfamily
  arXiv:0805.3067 [hep-ph]}}.

\bibitem{Tait:1999cf}
T.~M. Tait, {\em Phys. Rev.} {\bf D61},   034001  (2000),
  \href{http://arxiv.org/abs/hep-ph/9909352}{{\ttfamily arXiv:hep-ph/9909352
  [hep-ph]}}.

\bibitem{Whalley:2005nh}
M.~Whalley, D.~Bourilkov and R.~Group, {\it {T}he {L}es {H}ouches accord {PDF}s
  ({LHAPDF}) and {LHAGLUE}}  (2005),
  \href{http://arxiv.org/abs/hep-ph/0508110}{{\ttfamily arXiv:hep-ph/0508110
  [hep-ph]}}.

\bibitem{Botje:2011sn}
M.~Botje, J.~Butterworth, A.~Cooper-Sarkar, A.~de~Roeck, J.~Feltesse {\em
  et~al.}, {\it {T}he {PDF}4{LHC} {W}orking {G}roup {I}nterim
  {R}ecommendations}  (2011), \href{http://arxiv.org/abs/1101.0538}{{\ttfamily
  arXiv:1101.0538 [hep-ph]}}.

\bibitem{:2011cq}
D0 Collaboration, V.~M. Abazov {\em et~al.}, {\em Phys. Lett.} {\bf B704}, 403
  (2011), \href{http://arxiv.org/abs/1105.5384}{{\ttfamily arXiv:1105.5384
  [hep-ex]}}.

\bibitem{cdf-conf-9913}
CDF Collaboration, {\it {C}ombination of {CDF} top quark pair production
  measurements with up to $4.6 \rm\ fb^{-1}$}, CDF-NOTE-9913  (2009).

\bibitem{ATLAS-CONF-2010-063}
ATLAS Collaboration, {\it {S}earch for top pair candidate events in {ATLAS} at
  $\sqrt{s}=7 \rm\ {T}e{V}$}, ATLAS-CONF-2010-063  (2010).

\bibitem{CMS-PAS-TOP-10-004}
CMS Collaboration, {\it {S}election of {T}op-{L}ike {E}vents in the {D}ilepton
  and {L}epton-plus-{J}ets {C}hannels in {E}arly 7 {T}e{V} {D}ata},
  CMS-PAS-TOP-10-004  (2010).

\bibitem{ATLAS-CONF-2010-087}
ATLAS Collaboration, {\it {B}ackground studies for top-pair production in
  lepton plus jets final states in $\sqrt{s}=7 \rm\ {T}e{V}$ {ATLAS} data},
  ATLAS-CONF-2010-087  (2010).

\bibitem{cmsnote-2006-077}
M.~Davids {\em et~al.}, {\it {M}easurement of top-pair cross section and
  top-quark mass in the di-lepton and full-hadronic channels with {CMS}},
  CMS-NOTE-2006/077  (2006).

\bibitem{cmspas-top-08-001}
CMS Collaboration, {\it {E}xpectations for observation of top quark pair
  production in the dilepton final state with the first $10 \rm\ pb^{-1}$ of
  {CMS} data}, CMS-PAS-TOP-08-001  (2008).

\bibitem{cmspas-top-08-002}
CMS Collaboration, {\it {M}easurements of the $t\bar{t}$ cross section in the
  dilepton channels with ${L}=100 \rm\ pb^{-1}$ using the {CMS} detector},
  CMS-PAS-TOP-08-002  (2008).

\bibitem{cmspas-top-09-002}
CMS Collaboration, {\it {E}xpectations for observation of top quark pair
  production in the dilepton final state with early data at 10 {T}e{V}},
  CMS-PAS-TOP-09-002  (2009).

\bibitem{atl-phys-pub-2009-086}
ATLAS Collaboration, {\it {P}rospects for measuring top pair production in the
  dilepton channel with early {ATLAS} data at $\sqrt{s}=10 \rm\ {T}e{V}$},
  ATL-PHYS-PUB-2009-086  (2009).

\bibitem{Khachatryan:2010ez}
CMS Collaboration, V.~Khachatryan {\em et~al.}, {\em Phys. Lett.} {\bf B695},
  424  (2011), \href{http://arxiv.org/abs/1010.5994}{{\ttfamily arXiv:1010.5994
  [hep-ex]}}.

\bibitem{Chatrchyan:2011nb}
CMS Collaboration, S.~Chatrchyan {\em et~al.}, {\em JHEP} {\bf 07},   049
  (2011), \href{http://arxiv.org/abs/1105.5661}{{\ttfamily arXiv:1105.5661
  [hep-ex]}}.

\bibitem{CMS-PAS-TOP-11-005}
CMS Collaboration, {\it {M}easurement of the $t\bar{t}$ production cross
  section in the dilepton channel in pp collisions at $\sqrt{s} = 7 \rm\
  {T}e{V}$ with a luminosity of 1.14/fb}, CMS-PAS-TOP-11-005  (2011).

\bibitem{:2012kg}
ATLAS Collaboration, G.~Aad {\em et~al.}, {\em JHEP} {\bf 1205},   059  (2012),
  \href{http://arxiv.org/abs/1202.4892}{{\ttfamily arXiv:1202.4892 [hep-ex]}}.

\bibitem{Aad:2010ey}
ATLAS Collaboration, G.~Aad {\em et~al.}, {\em Eur. Phys. J.} {\bf C71},   1577
   (2011), \href{http://arxiv.org/abs/1012.1792}{{\ttfamily arXiv:1012.1792
  [hep-ex]}}.

\bibitem{Aad:2011yb}
ATLAS Collaboration, G.~Aad {\em et~al.}, {\em Phys. Lett.} {\bf B707}, 459
  (2012), \href{http://arxiv.org/abs/1108.3699}{{\ttfamily arXiv:1108.3699
  [hep-ex]}}.

\bibitem{cmsnote-2006-024}
J.~D'Hondt {\em et~al.}, {\it {E}lectron and muon reconstruction in single
  leptonic $t\bar{t}$ events}, CMS-NOTE-2006/024  (2006).

\bibitem{cmsnote-2006-064}
J.~D'Hondt, J.~Heyninck and S.~Lowette, {\it {M}easurement of the cross section
  of single leptonic $t\bar{t}$ events}, CMS-NOTE-2006/064  (2006).

\bibitem{cmspas-top-08-005}
CMS Collaboration, {\it {O}bservability of {T}op {Q}uark {P}air {P}roduction in
  the {S}emileptonic {M}uon {C}hannel with the first $10 \rm\ pb^{-1}$of {CMS}
  {D}ata}, CMS-PAS-TOP-08-005  (2008).

\bibitem{cmspas-top-09-003}
CMS Collaboration, {\it {P}rospects for the first {M}easurement of the
  $t\bar{t}$ {C}ross {S}ection in the {M}uon plus {J}ets {C}hannel at
  $\sqrt{s}=10$ {T}e{V} with the {CMS} {D}etector}, CMS-PAS-TOP-09-003  (2009).

\bibitem{cmspas-top-09-004}
CMS Collaboration, {\it {P}lans for an early measurement of the $t\bar{t}$
  cross section in the electron+jets channel at 10 {T}e{V}}, CMS-PAS-TOP-09-004
   (2009).

\bibitem{cmspas-top-09-010}
CMS Collaboration, {\it {E}xpectation for a measurement of the $t\bar{t}$
  production cross section in the muon+jets final state using a multivariate
  technique}, CMS-PAS-TOP-09-010  (2009).

\bibitem{atl-phys-pub-2009-087}
ATLAS Collaboration, {\it {P}rospects for the top pair production cross-section
  at $\sqrt{s}=10$ {T}e{V} in the single lepton channel in {ATLAS}},
  ATL-PHYS-PUB-2009-087  (2009).

\bibitem{Berends:1990ax}
F.~A. Berends, H.~Kuijf, B.~Tausk and W.~Giele, {\em Nucl. Phys.} {\bf B357},
  32  (1991).

\bibitem{Aad:2012qf}
ATLAS Collaboration, G.~Aad {\em et~al.}, {\em Phys. Lett.} {\bf B711}, 244
  (2012), \href{http://arxiv.org/abs/1201.1889}{{\ttfamily arXiv:1201.1889
  [hep-ex]}}.

\bibitem{Chatrchyan:2011ew}
CMS Collaboration, S.~Chatrchyan {\em et~al.}, {\em Eur. Phys. J.} {\bf C71},
  1721  (2011), \href{http://arxiv.org/abs/1106.0902}{{\ttfamily
  arXiv:1106.0902 [hep-ex]}}.

\bibitem{Chatrchyan:2011yy}
CMS Collaboration, S.~Chatrchyan {\em et~al.}, {\em Phys. Rev.} {\bf D84},
  092004  (2011), \href{http://arxiv.org/abs/1108.3773}{{\ttfamily
  arXiv:1108.3773 [hep-ex]}}.

\bibitem{cmspas-top-08-004}
CMS Collaboration, {\it {T}owards the measurement of the $t\bar{t}$ cross
  section in the $e\tau$ and $\mu\tau$ dilepton channel in pp collisions at
  $\sqrt{s} = 14$ {T}e{V}}, CMS-PAS-TOP-08-004  (2008).

\bibitem{Collaboration:2012vs}
CMS Collaboration, S.~Chatrchyan {\em et~al.}, {\it {M}easurement of the top
  quark pair production cross section in pp collisions at $\sqrt{s}=$ 7 {T}e{V}
  in dilepton final states containing a tau}  (2012),
  \href{http://arxiv.org/abs/1203.6810}{{\ttfamily arXiv:1203.6810 [hep-ex]}}.

\bibitem{ATLAS-CONF-2012-031}
ATLAS Collaboration, {\it {M}easurement of the $t\bar{t}$ production cross
  section in the all-hadronic channel in $4.7 \rm\ fb^{-1}$ of pp collisions at
  $\sqrt{s} = 7 \rm\ {T}e{V}$ with the {ATLAS} detector}, ATLAS-CONF-2012-031
  (2012).

\bibitem{arxiv:1205.2067}
ATLAS Collaboration, G.~Aad {\em et~al.}, {\it {Measurement of the top quark
  pair cross section with ATLAS in pp collisions at $\sqrt{s}=$ 7 TeV using
  final states with an electron or a muon and a hadronically decaying tau
  lepton}}  (2012), \href{http://arxiv.org/abs/1205.2067}{{\ttfamily
  arXiv:1205.2067 [hep-ex]}}, subm. to Phys. Lett. B.

\bibitem{ATLAS-CONF-2012-032}
ATLAS Collaboration, {\it {M}easurement of the $t\bar{t}$ production cross
  section in the final state with a hadronically decaying tau lepton and jets
  using the {ATLAS} detector}, ATLAS-CONF-2012-032  (2012).

\bibitem{cmspas-top-11-004}
CMS Collaboration, {\it {M}easurement of the $t\bar{t}$ production cross
  section in the tau+jets channel in $pp$ collisions at $\sqrt{s}=7$ {T}e{V}},
  CMS-PAS-TOP-11-004  (2012).

\bibitem{CMS-PAS-TOP-11-007}
CMS Collaboration, {\it {M}easurement of the $t\bar{t}$ production cross
  section in the fully hadronic decay channel in pp collisions at 7 {T}e{V}},
  CMS-PAS-TOP-11-007  (2011).

\bibitem{ATLAS-CONF-2011-066}
ATLAS Collaboration, {\it {S}earch for $t\bar{t}$ production in the
  all-hadronic channel in {ATLAS} with $\sqrt{s} = 7 \rm\ {T}e{V}$ data},
  ATLAS-CONF-2011-066  (2011).

\bibitem{ATLAS-CONF-2011-140}
ATLAS Collaboration, {\it {M}easurement of $t\bar{t}$ production in the
  all-hadronic channel in 1.02/fb of pp collisions at $\sqrt{s}=7 \rm\ {T}e{V}$
  with the {ATLAS} detector}, ATLAS-CONF-2011-140  (2011).

\bibitem{Lyons:1988rp}
L.~Lyons, D.~Gibaut and P.~Clifford, {\em Nucl. Instrum. Meth.} {\bf A270},
  110  (1988).

\bibitem{CMS-PAS-TOP-11-001}
CMS Collaboration, {\it {C}ombination of top pair production cross sections in
  pp collisions at 7 {T}e{V} and comparisons with theory}, CMS-PAS-TOP-11-001
  (2011).

\bibitem{CMS-PAS-TOP-11-024}
CMS Collaboration, {\it {C}ombination of top pair production cross section
  measurements}, CMS-PAS-TOP-11-024  (2011).

\bibitem{CMS-PAS-TOP-11-006}
CMS Collaboration, {\it {F}irst measurement of the top quark pair production
  cross section in the dilepton channel with tau leptons in the final state in
  pp collisions at sqrt(s)=7~{T}e{V}}, CMS-PAS-TOP-11-006  (2011).

\bibitem{ATLAS-CONF-2011-040}
ATLAS Collaboration, {\it {A} combined measurement of the top quark pair
  production cross-section using dilepton and single-lepton final states},
  ATLAS-CONF-2011-040  (2011).

\bibitem{ATLAS-CONF-2011-108}
ATLAS Collaboration, {\it {M}easurement of the top quark pair production
  cross-section based on a statistical combination of measurements of dilepton
  and single-lepton final states at $\sqrt{s} = 7 \rm\ {T}e{V}$ with the
  {ATLAS} detector}, ATLAS-CONF-2011-108  (2011).

\bibitem{ATLAS-CONF-2012-024}
ATLAS Collaboration, {\it {S}tatistical combination of top quark pair
  production cross-section measurements using dilepton, single-lepton, and
  all-hadronic final states at $\sqrt{s} = 7 \rm\ {T}e{V}$ with the {ATLAS}
  detector}, ATLAS-CONF-2012-024  (2012).

\bibitem{atlas-topsum-may12}
ATLAS Collaboration, {\it {T}op {P}air {C}ross {S}ection {S}ummary {P}lot,
  {M}ay 2012}
  \url{https://twiki.cern.ch/twiki/bin/view/AtlasPublic/CombinedSummaryPlots}.

\bibitem{ATLAS-CONF-2011-142}
ATLAS Collaboration, {\it {R}econstructed jet multiplicities from the top-quark
  pair decays and associated jets in pp collisions at $\sqrt{s}=7 \rm\ {T}e{V}$
  measured with the {ATLAS} detector at the {LHC}}, ATLAS-CONF-2011-142
  (2011).

\bibitem{cmspas-top-11-013}
CMS Collaboration, {\it {M}easurement of {T}op {Q}uark {P}air {D}ifferential
  {C}ross {S}ections at $\sqrt{s}=$ 7 {T}e{V}}, CMS-PAS-TOP-11-013  (2012).

\bibitem{Fiedler:2010sy}
F.~Fiedler, {\it {P}recision {M}easurements of the {T}op {Q}uark {M}ass}
  (2010), \href{http://arxiv.org/abs/1003.0521}{{\ttfamily arXiv:1003.0521
  [hep-ex]}}.

\bibitem{Galtieri:2011yd}
A.~B. Galtieri, F.~Margaroli and I.~Volobouev, {\em Rept. Prog. Phys.} {\bf
  75},   056201  (2012), \href{http://arxiv.org/abs/1109.2163}{{\ttfamily
  arXiv:1109.2163 [hep-ex]}}.

\bibitem{cmsnote-2006-066}
J.~D'Hondt, J.~Heyninck and S.~Lowette, {\it {T}op {Q}uark mass measurement in
  single-leptonic $t\bar{t}$ events}, CMS-NOTE-2006/066  (2006).

\bibitem{cmsnote-2006-058}
R.~Chierici and A.~Dierlamm, {\it {D}etermination of the top mass in exclusive
  ${J}/\psi$ decays}, CMS-NOTE-2006/058  (2006).

\bibitem{ATL-PHYS-PUB-2010-004}
ATLAS Collaboration, {\it {P}rospects for the {M}easurement of the
  {T}op-{Q}uark {M}ass using the {T}emplate {M}ethod with early {ATLAS}
  {D}ata}, ATL-PHYS-PUB-2010-004  (2010).

\bibitem{Abbott:1997fv}
D0 Collaboration, B.~Abbott {\em et~al.}, {\em Phys. Rev. Lett.} {\bf 80}, 2063
   (1998), \href{http://arxiv.org/abs/hep-ex/9706014}{{\ttfamily
  arXiv:hep-ex/9706014 [hep-ex]}}.

\bibitem{Abulencia:2006js}
CDF Collaboration, A.~Abulencia {\em et~al.}, {\em Phys. Rev.} {\bf D73},
  112006  (2006), \href{http://arxiv.org/abs/hep-ex/0602008}{{\ttfamily
  arXiv:hep-ex/0602008 [hep-ex]}}.

\bibitem{Sonnenschein:2006ud}
L.~Sonnenschein, {\em Phys. Rev.} {\bf D73},   054015  (2006),
  \href{http://arxiv.org/abs/hep-ph/0603011}{{\ttfamily arXiv:hep-ph/0603011
  [hep-ph]}}, erratum-ibid. {\bf D78}, 079902 (2008).

\bibitem{Dalitz:1991wa}
R.~Dalitz and G.~R. Goldstein, {\em Phys. Rev.} {\bf D45}, 1531  (1992).

\bibitem{cmspas-top-11-016}
CMS Collaboration, {\it {M}easurement of the top quark mass in the dilepton
  channel in pp collisions at $\sqrt{s}=7 \rm\ {T}e{V}$}, CMS-PAS-TOP-11-016
  (2012).

\bibitem{atlasmtopljets}
ATLAS Collaboration, G.~Aad {\em et~al.}, {\it {M}easurement of the top quark
  mass with the template method in the top-antitop lepton + jets channel using
  {ATLAS} data}  (2012), \href{http://arxiv.org/abs/1203.5755}{{\ttfamily
  arXiv:1203.5755 [hep-ex]}}.

\bibitem{CMS-PAS-TOP-10-009}
CMS Collaboration, {\it {M}easurement of the top quark mass in the l+jets
  channel}, CMS-PAS-TOP-10-009  (2011).

\bibitem{:2008xh}
DELPHI Collaboration, J.~Abdallah {\em et~al.}, {\em Eur. Phys. J.} {\bf C55},
  1  (2008), \href{http://arxiv.org/abs/0803.2534}{{\ttfamily arXiv:0803.2534
  [hep-ex]}}.

\bibitem{Abazov:2007rk}
D0 Collaboration, V.~Abazov {\em et~al.}, {\em Phys. Rev.} {\bf D75},   092001
  (2007), \href{http://arxiv.org/abs/hep-ex/0702018}{{\ttfamily
  arXiv:hep-ex/0702018 [hep-ex]}}.

\bibitem{Aaltonen:2006xc}
CDF Collaboration, T.~Aaltonen {\em et~al.}, {\em Phys. Rev. Lett.} {\bf 98},
  142001  (2007), \href{http://arxiv.org/abs/hep-ex/0612026}{{\ttfamily
  arXiv:hep-ex/0612026 [hep-ex]}}.

\bibitem{cmspas-top-11-015}
CMS Collaboration, {\it {M}easurement of the top quark mass in the muon+jets
  channel}, CMS-PAS-TOP-11-015  (2012).

\bibitem{ATLAS-CONF-2011-033}
ATLAS Collaboration, {\it {M}easurement of the {T}op-{Q}uark {M}ass using the
  {T}emplate {M}ethod in pp {C}ollisions at $\sqrt{s}=7 \rm\ {T}e{V}$ with the
  {ATLAS} detector}, ATLAS-CONF-2011-033  (2011).

\bibitem{ATLAS-CONF-2012-030}
ATLAS Collaboration, {\it {D}etermination of the {T}op {Q}uark {M}ass with a
  {T}emplate {M}ethod in the {A}ll-{H}adronic {D}ecay {C}hannel using $2.04
  \rm\ fb^{-1}$ of {ATLAS} {D}ata}, ATLAS-CONF-2012-030  (2012).

\bibitem{cms-pas-top-11-018}
CMS Collaboration, {\it {T}op mass combination {M}oriond 2012}
  \url{https://twiki.cern.ch/twiki/bin/view/CMSPublic/PhysicsResultsTOPSummaryPlots}.

\bibitem{Abulencia:2006rz}
CDF Collaboration, A.~Abulencia {\em et~al.}, {\em Phys. Rev.} {\bf D75},
  071102  (2007), \href{http://arxiv.org/abs/hep-ex/0612061}{{\ttfamily
  arXiv:hep-ex/0612061 [hep-ex]}}.

\bibitem{Aaltonen:2009hd}
CDF Collaboration, T.~Aaltonen {\em et~al.}, {\em Phys. Rev.} {\bf D81},
  032002  (2010), \href{http://arxiv.org/abs/0910.0969}{{\ttfamily
  arXiv:0910.0969 [hep-ex]}}.

\bibitem{Aaltonen:2011wt}
CDF Collaboration, T.~Aaltonen {\em et~al.}, {\em Phys. Lett.} {\bf B698}, 371
  (2011), \href{http://arxiv.org/abs/1101.4926}{{\ttfamily arXiv:1101.4926
  [hep-ex]}}.

\bibitem{Kharchilava:1999yj}
A.~Kharchilava, {\em Phys. Lett.} {\bf B476}, 73  (2000),
  \href{http://arxiv.org/abs/hep-ph/9912320}{{\ttfamily arXiv:hep-ph/9912320
  [hep-ph]}}.

\bibitem{Biswas:2010sa}
S.~Biswas, K.~Melnikov and M.~Schulze, {\em JHEP} {\bf 1008},   048  (2010),
  \href{http://arxiv.org/abs/1006.0910}{{\ttfamily arXiv:1006.0910 [hep-ph]}}.

\bibitem{Frederix:2007gi}
R.~Frederix and F.~Maltoni, {\em JHEP} {\bf 0901},   047  (2009),
  \href{http://arxiv.org/abs/0712.2355}{{\ttfamily arXiv:0712.2355 [hep-ph]}}.

\bibitem{Fiedler:2010sg}
F.~Fiedler, A.~Grohsjean, P.~Haefner and P.~Schieferdecker, {\em Nucl. Instrum.
  Meth.} {\bf A624}, 203  (2010),
  \href{http://arxiv.org/abs/1003.1316}{{\ttfamily arXiv:1003.1316 [hep-ex]}}.

\bibitem{Campbell:2012cz}
J.~M. Campbell, W.~T. Giele and C.~Williams, {\it {T}he {M}atrix {E}lement
  {M}ethod at {N}ext-to-{L}eading {O}rder}  (2012),
  \href{http://arxiv.org/abs/1204.4424}{{\ttfamily arXiv:1204.4424 [hep-ph]}}.

\bibitem{ATLAS-CONF-2011-054}
ATLAS Collaboration, {\it {D}etermination of the {T}op-{Q}uark {M}ass from the
  $t\bar{t}$ {C}ross {S}ection {M}easurement in pp {C}ollisions at $\sqrt{s}=7
  \rm\ {T}e{V}$ with the {ATLAS} detector}, ATLAS-CONF-2011-054  (2011).

\bibitem{CMS-PAS-TOP-11-008}
CMS Collaboration, {\it {D}etermination of the {T}op {Q}uark {M}ass from the
  $t\bar{t}$ {C}ross {S}ection at $\sqrt{s} = 7$ {T}e{V}}, CMS-PAS-TOP-11-008
  (2011).

\bibitem{Abazov:2011pta}
D0 Collaboration, V.~M. Abazov {\em et~al.}, {\em Phys. Lett.} {\bf B703}, 422
  (2011), \href{http://arxiv.org/abs/1104.2887}{{\ttfamily arXiv:1104.2887
  [hep-ex]}}.

\bibitem{Aaltonen:2011wr}
CDF Collaboration, T.~Aaltonen {\em et~al.}, {\em Phys. Rev. Lett.} {\bf 106},
   152001  (2011), \href{http://arxiv.org/abs/1103.2782}{{\ttfamily
  arXiv:1103.2782 [hep-ex]}}.

\bibitem{Abazov:2011ch}
D0 Collaboration, V.~M. Abazov {\em et~al.}, {\em Phys. Rev.} {\bf D84},
  052005  (2011), \href{http://arxiv.org/abs/1106.2063}{{\ttfamily
  arXiv:1106.2063 [hep-ex]}}.

\bibitem{top-11-019-paper}
CMS Collaboration, S.~Chatrchyan {\em et~al.}, {\it {M}easurement of the mass
  difference between top and antitop quarks}  (2012),
  \href{http://arxiv.org/abs/1204.2807}{{\ttfamily arXiv:1204.2807 [hep-ex]}}.

\bibitem{ATLAS-CONF-2011-141}
ATLAS Collaboration, {\it {M}easurement of the top quark charge in pp
  collisions at $\sqrt{s}=7 \rm\ {T}e{V}$ in the {ATLAS} experiment},
  ATLAS-CONF-2011-141  (2011).

\bibitem{Collaboration:2012sm}
ATLAS Collaboration, G.~Aad {\em et~al.}, {\em Phys. Rev. Lett.} {\bf 108},
  212001  (2012), \href{http://arxiv.org/abs/1203.4081}{{\ttfamily
  arXiv:1203.4081 [hep-ex]}}.

\bibitem{cmspas-top-11-031}
CMS Collaboration, {\it {C}onstraints on the {T}op-{Q}uark {C}harge from
  {T}op-{P}air {E}vents}, CMS-PAS-TOP-11-031  (2012).

\bibitem{cmspas-top-11-029}
CMS Collaboration, {\it {F}irst measurement of ${B}(t \rightarrow {W}b) / {B}(t
  \rightarrow {W}q)$ in the dilepton channel in pp collisions at $\sqrt{s}=$ 7
  {T}e{V}}, CMS-PAS-TOP-11-029  (2012).

\bibitem{Stelzer:1995gc}
T.~Stelzer and S.~Willenbrock, {\em Phys. Lett.} {\bf B374}, 169  (1996),
  \href{http://arxiv.org/abs/hep-ph/9512292}{{\ttfamily arXiv:hep-ph/9512292
  [hep-ph]}}.

\bibitem{Uwer:2004vp}
P.~Uwer, {\em Phys. Lett.} {\bf B609}, 271  (2005),
  \href{http://arxiv.org/abs/hep-ph/0412097}{{\ttfamily arXiv:hep-ph/0412097
  [hep-ph]}}.

\bibitem{Mahlon:2010gw}
G.~Mahlon and S.~J. Parke, {\em Phys. Rev.} {\bf D81},   074024  (2010),
  \href{http://arxiv.org/abs/1001.3422}{{\ttfamily arXiv:1001.3422 [hep-ph]}}.

\bibitem{Kane:1991bg}
G.~L. Kane, G.~Ladinsky and C.~Yuan, {\em Phys. Rev.} {\bf D45}, 124  (1992).

\bibitem{Cheung:1996kc}
K.-M. Cheung, {\em Phys. Rev.} {\bf D55}, 4430  (1997),
  \href{http://arxiv.org/abs/hep-ph/9610368}{{\ttfamily arXiv:hep-ph/9610368
  [hep-ph]}}.

\bibitem{Bernreuther:1997gs}
W.~Bernreuther, M.~Flesch and P.~Haberl, {\em Phys. Rev.} {\bf D58},   114031
  (1998), \href{http://arxiv.org/abs/hep-ph/9709284}{{\ttfamily
  arXiv:hep-ph/9709284 [hep-ph]}}.

\bibitem{Abazov:2011gi}
D0 Collaboration, V.~M. Abazov {\em et~al.}, {\em Phys. Rev. Lett.} {\bf 108},
   032004  (2012), \href{http://arxiv.org/abs/1110.4194}{{\ttfamily
  arXiv:1110.4194 [hep-ex]}}.

\bibitem{Hubaut}
ATLAS Collaboration, F.~Hubaut, E.~Monnier and P.~Pralavorio, {\it {ATLAS}
  sensitivity to $t\bar{t}$ spin correlation in the semileptonic channel},
  ATL-PHYS-PUB-2005-001  (2005).

\bibitem{Hubauta}
ATLAS Collaboration, F.~Hubaut, E.~Monnier, P.~Pralavorio, B.~Resende and
  C.~Zhu, {\it {P}olarization studies in $t\bar{t}$ semileptonic events with
  {ATLAS} full simulation}, ATL-PHYS-PUB-2006-022  (2006).

\bibitem{Hubaut2005}
F.~Hubaut, E.~Monnier, P.~Pralavorio, K.~Smolek and V.~Simak, {\em Eur. Phys.
  J.} {\bf C44S2}, 13  (2005),
  \href{http://arxiv.org/abs/hep-ex/0508061}{{\ttfamily arXiv:hep-ex/0508061}}.

\bibitem{cmsnote-2006-111}
M.~Baarmand, H.~Mermerkaya and I.~Vodopiyanov, {\it {M}easurement of {S}pin
  {C}orrelation in {T}op {Q}uark {P}air {P}roduction in {S}emi-{L}eptonic
  {F}inal {S}tate}, CMS-NOTE-2006/111  (2006).

\bibitem{Czarnecki:2010gb}
A.~Czarnecki, J.~G. Korner and J.~H. Piclum, {\em Phys. Rev.} {\bf D81},
  111503  (2010), \href{http://arxiv.org/abs/1005.2625}{{\ttfamily
  arXiv:1005.2625 [hep-ph]}}.

\bibitem{Buchmuller:1985jz}
W.~Buchmuller and D.~Wyler, {\em Nucl. Phys.} {\bf B268},   621  (1986).

\bibitem{AguilarSaavedra:2008zc}
J.~Aguilar-Saavedra, {\em Nucl. Phys.} {\bf B812}, 181  (2009),
  \href{http://arxiv.org/abs/0811.3842}{{\ttfamily arXiv:0811.3842 [hep-ph]}}.

\bibitem{Zhang:2010dr}
C.~Zhang and S.~Willenbrock, {\em Phys. Rev.} {\bf D83},   034006  (2011),
  \href{http://arxiv.org/abs/1008.3869}{{\ttfamily arXiv:1008.3869 [hep-ph]}}.

\bibitem{Collaboration:2012ky}
ATLAS Collaboration, G.~Aad {\em et~al.}, {\it {Measurement of the W boson
  polarization in top quark decays with the ATLAS detector}}  (2012),
  \href{http://arxiv.org/abs/1205.2484}{{\ttfamily arXiv:1205.2484 [hep-ex]}},
  subm. to JHEP.

\bibitem{ATLAS-CONF-2011-037}
ATLAS Collaboration, {\it {M}easurement of the {W}-boson polarisation in top
  quark decays in pp collision data at $\sqrt{s}=$ 7 {T}e{V} using the {ATLAS}
  detector}, ATLAS-CONF-2011-037  (2011).

\bibitem{cmspas-top-11-020}
CMS Collaboration, {\it {M}easurement of the {W} boson polarization in
  semileptonic top pair decays with the {CMS} detector at the {LHC}},
  CMS-PAS-TOP-11-020  (2012).

\bibitem{AguilarSaavedra:2006fy}
J.~Aguilar-Saavedra, J.~Carvalho, N.~F. Castro, F.~Veloso and A.~Onofre, {\em
  Eur. Phys. J.} {\bf C50}, 519  (2007),
  \href{http://arxiv.org/abs/hep-ph/0605190}{{\ttfamily arXiv:hep-ph/0605190
  [hep-ph]}}.

\bibitem{AguilarSaavedra:2010nx}
J.~Aguilar-Saavedra and J.~Bernabeu, {\em Nucl. Phys.} {\bf B840}, 349  (2010),
  \href{http://arxiv.org/abs/1005.5382}{{\ttfamily arXiv:1005.5382 [hep-ph]}}.

\bibitem{Abazov:2009ky}
D0 Collaboration, V.~Abazov {\em et~al.}, {\em Phys. Rev. Lett.} {\bf 102},
  092002  (2009), \href{http://arxiv.org/abs/0901.0151}{{\ttfamily
  arXiv:0901.0151 [hep-ex]}}.

\bibitem{Abazov:2010jn}
D0 Collaboration, V.~M. Abazov {\em et~al.}, {\em Phys. Rev.} {\bf D83},
  032009  (2011), \href{http://arxiv.org/abs/1011.6549}{{\ttfamily
  arXiv:1011.6549 [hep-ex]}}.

\bibitem{Aaltonen:2010ha}
CDF Collaboration, T.~Aaltonen {\em et~al.}, {\em Phys. Rev. Lett.} {\bf 105},
   042002  (2010), \href{http://arxiv.org/abs/1003.0224}{{\ttfamily
  arXiv:1003.0224 [hep-ex]}}.

\bibitem{Aaltonen:2012rz}
CDF and D0 Collaborations, T.~Aaltonen {\em et~al.}, {\it {Combination of CDF
  and D0 measurements of the $W$ boson helicity in top quark decays}}  (2012),
  \href{http://arxiv.org/abs/1202.5272}{{\ttfamily arXiv:1202.5272 [hep-ex]}}.

\bibitem{Aaltonen:2012tk}
CDF Collaboration, T.~Aaltonen {\em et~al.}, {\it {$W$ boson polarization
  measurement in the $t\bar{t}$ dilepton channel using the CDF II Detector}}
  (2012), \href{http://arxiv.org/abs/1205.0354}{{\ttfamily arXiv:1205.0354
  [hep-ex]}}.

\bibitem{AguilarSaavedra:2011ct}
J.~Aguilar-Saavedra, N.~Castro and A.~Onofre, {\em Phys. Rev.} {\bf D83},
  117301  (2011), \href{http://arxiv.org/abs/1105.0117}{{\ttfamily
  arXiv:1105.0117 [hep-ph]}}.

\bibitem{arxiv:1204.2332}
D0 Collaboration, V.~M. Abazov {\em et~al.}, {\it {C}ombination of searches for
  anomalous top quark couplings with $5.4 \rm\ fb^{-1}$ of $p\bar{p}$
  collisions}  (2012), \href{http://arxiv.org/abs/1204.2332}{{\ttfamily
  arXiv:1204.2332 [hep-ex]}}, subm. to Phys. Lett. B.

\bibitem{Baur:2004uw}
U.~Baur, A.~Juste, L.~Orr and D.~Rainwater, {\em Phys. Rev.} {\bf D71},
  054013  (2005), \href{http://arxiv.org/abs/hep-ph/0412021}{{\ttfamily
  arXiv:hep-ph/0412021 [hep-ph]}}.

\bibitem{Aaltonen:2011sp}
CDF Collaboration, T.~Aaltonen {\em et~al.}, {\em Phys. Rev.} {\bf D84},
  031104  (2011), \href{http://arxiv.org/abs/1106.3970}{{\ttfamily
  arXiv:1106.3970 [hep-ex]}}.

\bibitem{ATLAS-CONF-2011-153}
ATLAS Collaboration, {\it {M}easurement of the inclusive $t\bar{t}\gamma$ cross
  section with the {ATLAS} detector}, ATLAS-CONF-2011-153  (2011).

\bibitem{Kilian:2007gr}
W.~Kilian, T.~Ohl and J.~Reuter, {\em Eur. Phys. J.} {\bf C71},   1742  (2011),
  \href{http://arxiv.org/abs/0708.4233}{{\ttfamily arXiv:0708.4233 [hep-ph]}}.

\bibitem{Melnikov:2011ta}
K.~Melnikov, M.~Schulze and A.~Scharf, {\em Phys. Rev.} {\bf D83},   074013
  (2011), \href{http://arxiv.org/abs/1102.1967}{{\ttfamily arXiv:1102.1967
  [hep-ph]}}.

\bibitem{Kuhn:1998jr}
J.~H. Kuhn and G.~Rodrigo, {\em Phys. Rev. Lett.} {\bf 81}, 49  (1998),
  \href{http://arxiv.org/abs/hep-ph/9802268}{{\ttfamily arXiv:hep-ph/9802268
  [hep-ph]}}.

\bibitem{Kuhn:1998kw}
J.~H. Kuhn and G.~Rodrigo, {\em Phys. Rev.} {\bf D59},   054017  (1999),
  \href{http://arxiv.org/abs/hep-ph/9807420}{{\ttfamily arXiv:hep-ph/9807420
  [hep-ph]}}.

\bibitem{Antunano:2007da}
O.~Antunano, J.~H. Kuhn and G.~Rodrigo, {\em Phys. Rev.} {\bf D77},   014003
  (2008), \href{http://arxiv.org/abs/0709.1652}{{\ttfamily arXiv:0709.1652
  [hep-ph]}}.

\bibitem{Frampton:2009rk}
P.~H. Frampton, J.~Shu and K.~Wang, {\em Phys. Lett.} {\bf B683}, 294  (2010),
  \href{http://arxiv.org/abs/0911.2955}{{\ttfamily arXiv:0911.2955 [hep-ph]}}.

\bibitem{Rosner:1996eb}
J.~L. Rosner, {\em Phys. Lett.} {\bf B387}, 113  (1996),
  \href{http://arxiv.org/abs/hep-ph/9607207}{{\ttfamily arXiv:hep-ph/9607207
  [hep-ph]}}.

\bibitem{Ferrario:2008wm}
P.~Ferrario and G.~Rodrigo, {\em Phys. Rev.} {\bf D78},   094018  (2008),
  \href{http://arxiv.org/abs/0809.3354}{{\ttfamily arXiv:0809.3354 [hep-ph]}}.

\bibitem{Aaltonen:2011kc}
CDF Collaboration, T.~Aaltonen {\em et~al.}, {\em Phys. Rev.} {\bf D83},
  112003  (2011), \href{http://arxiv.org/abs/1101.0034}{{\ttfamily
  arXiv:1101.0034 [hep-ex]}}.

\bibitem{cdf-conf-10436}
CDF Collaboration, {\it {M}easurement of the {F}orward {B}ackward {A}symmetry
  in {T}op {P}air {P}roduction in the {D}ilepton {D}ecay {C}hannel using $5.1
  \rm\ fb^{-1}$}, CDF-NOTE-10436  (2011).

\bibitem{cdf-conf-10584}
CDF Collaboration, {\it {C}ombination of the {F}orward-{B}ackward {A}symmetry
  in the {T}op {P}air {P}roduction from {L}+{J} and {DIL} {C}hannels using $5
  \rm\ fb^{-1}$}, CDF-NOTE-10584  (2011).

\bibitem{cdf-conf-10807}
CDF Collaboration, {\it {S}tudy of the {T}op {Q}uark {P}roduction {A}symmetry
  and {I}ts {M}ass and {R}apidity {D}ependence in the {F}ull {R}un {II}
  {T}evatron {D}ataset}, CDF-NOTE-10807  (2011).

\bibitem{Abazov:2011rq}
D0 Collaboration, V.~M. Abazov {\em et~al.}, {\em Phys. Rev.} {\bf D84},
  112005  (2011), \href{http://arxiv.org/abs/1107.4995}{{\ttfamily
  arXiv:1107.4995 [hep-ex]}}.

\bibitem{Kuhn:2011ri}
J.~H. Kuhn and G.~Rodrigo, {\em JHEP} {\bf 1201},   063  (2012),
  \href{http://arxiv.org/abs/1109.6830}{{\ttfamily arXiv:1109.6830 [hep-ph]}}.

\bibitem{Ahrens:2011uf}
V.~Ahrens, A.~Ferroglia, M.~Neubert, B.~D. Pecjak and L.~L. Yang, {\em Phys.
  Rev.} {\bf D84},   074004  (2011),
  \href{http://arxiv.org/abs/1106.6051}{{\ttfamily arXiv:1106.6051 [hep-ph]}}.

\bibitem{Hollik:2011ps}
W.~Hollik and D.~Pagani, {\em Phys. Rev.} {\bf D84},   093003  (2011),
  \href{http://arxiv.org/abs/1107.2606}{{\ttfamily arXiv:1107.2606 [hep-ph]}}.

\bibitem{Diener:2009ee}
R.~Diener, S.~Godfrey and T.~A. Martin, {\em Phys. Rev.} {\bf D80},   075014
  (2009), \href{http://arxiv.org/abs/0909.2022}{{\ttfamily arXiv:0909.2022
  [hep-ph]}}.

\bibitem{Jung:2011zv}
S.~Jung, A.~Pierce and J.~D. Wells, {\em Phys. Rev.} {\bf D83},   114039
  (2011), \href{http://arxiv.org/abs/1103.4835}{{\ttfamily arXiv:1103.4835
  [hep-ph]}}.

\bibitem{Kamenik:2011wt}
J.~F. Kamenik, J.~Shu and J.~Zupan, {\it {Review of new physics effects in
  $t\bar{t}$ production}}  (2011),
  \href{http://arxiv.org/abs/1107.5257}{{\ttfamily arXiv:1107.5257 [hep-ph]}}.

\bibitem{AguilarSaavedra:2012ma}
J.~Aguilar-Saavedra, {\it {O}verview of models for the $t\bar{t}$ asymmetry}
  (2012), \href{http://arxiv.org/abs/1202.2382}{{\ttfamily arXiv:1202.2382
  [hep-ph]}}.

\bibitem{:2011hk}
CMS Collaboration, S.~Chatrchyan {\em et~al.}, {\em Phys. Lett.} {\bf B709}, 28
   (2012), \href{http://arxiv.org/abs/1112.5100}{{\ttfamily arXiv:1112.5100
  [hep-ex]}}.

\bibitem{cmspas-top-11-030}
CMS Collaboration, {\it {D}ifferential measurements of the charge asymmetry in
  top quark pair prodcution}, CMS-PAS-TOP-11-030  (2012).

\bibitem{Gabrielli:2011zw}
E.~Gabrielli, M.~Raidal and A.~Racioppi, {\it {Implications of the effective
  axial-vector coupling of gluon on top-quark charge asymmetry at the LHC}}
  (2011), \href{http://arxiv.org/abs/1112.5885}{{\ttfamily arXiv:1112.5885
  [hep-ph]}}.

\bibitem{gabrielli-lh11}
E.~Gabrielli, A.~Giammanco, M.~Raidal and A.~Racioppi, {\it {E}ffective
  axial-vector coupling of the gluon and top-quark charge asymmetry at the
  {LHC}}, in {\em {L}es {H}ouches 2011: {P}hysics at {T}e{V} {C}olliders {N}ew
  {P}hysics {W}orking {G}roup {R}eport\/},  eds. G.~Brooijmans {\em et~al.}
  (2012).
\newblock p. 163.
\newblock \href{http://arxiv.org/abs/1203.1488}{{\ttfamily arXiv:1203.1488
  [hep-ph]}}.

\bibitem{Collaboration:2012ug}
ATLAS Collaboration, G.~Aad {\em et~al.}, {\it {M}easurement of the charge
  asymmetry in top quark pair production in pp collisions at $\sqrt{s}=$ 7
  {T}e{V} using the {ATLAS} detector}  (2012),
  \href{http://arxiv.org/abs/1203.4211}{{\ttfamily arXiv:1203.4211 [hep-ex]}}.

\bibitem{AguilarSaavedra:2011hz}
J.~Aguilar-Saavedra and M.~Perez-Victoria, {\em Phys. Rev.} {\bf D84},   115013
   (2011), \href{http://arxiv.org/abs/1105.4606}{{\ttfamily arXiv:1105.4606
  [hep-ph]}}.

\bibitem{AguilarSaavedra:2011ug}
J.~Aguilar-Saavedra and M.~Perez-Victoria, {\em JHEP} {\bf 1109},   097
  (2011), \href{http://arxiv.org/abs/1107.0841}{{\ttfamily arXiv:1107.0841
  [hep-ph]}}.

\bibitem{Skands:2012mm}
P.~Z. Skands, B.~R. Webber and J.~Winter, {\it {QCD Coherence and the Top Quark
  Asymmetry}}  (2012), \href{http://arxiv.org/abs/1205.1466}{{\ttfamily
  arXiv:1205.1466 [hep-ph]}}.

\bibitem{Aad:2012jr}
ATLAS Collaboration, G.~Aad {\em et~al.}, {\it {Measurement of $t \bar{t}$
  production with a veto on additional central jet activity in pp collisions at
  $\sqrt{s}=$ 7 TeV using the ATLAS detector}}  (2012),
  \href{http://arxiv.org/abs/1203.5015}{{\ttfamily arXiv:1203.5015 [hep-ex]}}.

\bibitem{Abazov:2009ii}
D0 Collaboration, V.~Abazov {\em et~al.}, {\em Phys. Rev. Lett.} {\bf 103},
  092001  (2009), \href{http://arxiv.org/abs/0903.0850}{{\ttfamily
  arXiv:0903.0850 [hep-ex]}}.

\bibitem{Aaltonen:2009jj}
CDF Collaboration, T.~Aaltonen {\em et~al.}, {\em Phys. Rev. Lett.} {\bf 103},
   092002  (2009), \href{http://arxiv.org/abs/0903.0885}{{\ttfamily
  arXiv:0903.0885 [hep-ex]}}.

\bibitem{Heinson:2010xh}
A.~Heinson, {\em Mod. Phys. Lett.} {\bf A25}, 309  (2010),
  \href{http://arxiv.org/abs/1002.4167}{{\ttfamily arXiv:1002.4167 [hep-ex]}}.

\bibitem{cmsnote-2006-086}
P.~Yeh {\em et~al.}, {\it {S}earch for {W}-associated {P}roduction of {S}ingle
  {T}op {Q}uarks in {CMS}}, CMS-NOTE-2006/086  (2006).

\bibitem{cmspas-top-09-005}
CMS Collaboration, {\it {P}rospects for the measurement of the single-top
  t-channel cross section in the muon channel with $200 \rm\ pb^{-1}$ at 10
  {T}e{V}}, CMS-PAS-TOP-09-005  (2009).

\bibitem{ATL-PHYS-PUB-2010-003}
ATLAS Collaboration, {\it {S}trategy to {S}earch for {S}ingle-{T}op {E}vents
  using early {D}ata of the {ATLAS} {D}etector at the {LHC}},
  ATL-PHYS-PUB-2010-003  (2010).

\bibitem{ATLAS-CONF-2011-027}
ATLAS Collaboration, {\it {S}earches for {S}ingle {T}op-{Q}uark {P}roduction
  with the {ATLAS} {D}etector in pp {C}ollisions at $\sqrt{s}=$ 7 {T}e{V}},
  ATLAS-CONF-2011-027  (2011).

\bibitem{ATLAS-CONF-2011-088}
ATLAS Collaboration, {\it {O}bservation of t {C}hannel {S}ingle {T}op-{Q}uark
  {P}roduction in pp {C}ollisions at $\sqrt{s} = 7$~{T}e{V} with the {ATLAS}
  detector}, ATLAS-CONF-2011-088  (2011).

\bibitem{Collaboration:2012ux}
ATLAS Collaboration, G.~Aad {\em et~al.}  (2012),
  \href{http://arxiv.org/abs/1205.3130}{{\ttfamily arXiv:1205.3130 [hep-ex]}},
  subm. to Phys. Lett. B.

\bibitem{neurobayes}
M.~Feindt, {\it {A} {N}eural {B}ayesian {E}stimator for {C}onditional
  {P}robability {D}ensities}  (2004),
  \href{http://arxiv.org/abs/physics/0402093}{{\ttfamily arXiv:physics/0402093
  [physics]}}.

\bibitem{Feindt:2006pm}
M.~Feindt and U.~Kerzel, {\em Nucl. Instrum. Meth.} {\bf A559}, 190  (2006).

\bibitem{Chatrchyan:2011vp}
CMS Collaboration, S.~Chatrchyan {\em et~al.}, {\em Phys. Rev. Lett.} {\bf
  107},   091802  (2011), \href{http://arxiv.org/abs/1106.3052}{{\ttfamily
  arXiv:1106.3052 [hep-ex]}}.

\bibitem{cmspas-top-11-021}
CMS Collaboration, {\it {M}easurement of the single top t-channel cross section
  in pp collisions at $\sqrt{s}=$ 7 {T}e{V} using 2011 data},
  CMS-PAS-TOP-11-021  (2012).

\bibitem{Abazov:2008kt}
D0 Collaboration, V.~M. Abazov {\em et~al.}, {\em Phys. Rev.} {\bf D78},
  012005  (2008), \href{http://arxiv.org/abs/0803.0739}{{\ttfamily
  arXiv:0803.0739 [hep-ex]}}.

\bibitem{cmsnote-2006-084}
V.~Abramov {\em et~al.}, {\it {S}election of single top events with the {CMS}
  detector at {LHC}}, CMS-NOTE-2006/084  (2006).

\bibitem{atl-phys-pub-2009-001}
ATLAS Collaboration, {\it {P}rospects for associated single top quark
  production cross-section measurements in the dilepton decay mode with
  {ATLAS}}, ATL-PHYS-PUB-2009-001  (2009).

\bibitem{ATLAS-CONF-2011-104}
ATLAS Collaboration, {\it {S}earch for {W}+t single-top events in the
  dileptonic channel at {ATLAS}}, ATLAS-CONF-2011-104  (2011).

\bibitem{CMS-PAS-TOP-11-022}
CMS Collaboration, {\it {S}earch for single top t{W} associated production in
  the dilepton decay channel in pp collisions at $\sqrt{s}=7 \rm\ {T}e{V}$},
  CMS-PAS-TOP-11-022  (2011).

\bibitem{Tait:2000sh}
T.~M. Tait and C.-P. Yuan, {\em Phys. Rev.} {\bf D63},   014018  (2000),
  \href{http://arxiv.org/abs/hep-ph/0007298}{{\ttfamily arXiv:hep-ph/0007298
  [hep-ph]}}.

\bibitem{ATLAS-CONF-2011-118}
ATLAS Collaboration, {\it {S}earch for s-channel single top-quark production in
  pp collisions at $\sqrt{s}=7 \rm\ {T}e{V}$}, ATLAS-CONF-2011-118  (2011).

\bibitem{Alwall:2006bx}
J.~Alwall, R.~Frederix, J.-M. Gerard, A.~Giammanco, M.~Herquet {\em et~al.},
  {\em Eur. Phys. J.} {\bf C49}, 791  (2007),
  \href{http://arxiv.org/abs/hep-ph/0607115}{{\ttfamily arXiv:hep-ph/0607115
  [hep-ph]}}.

\bibitem{Lacker:2012ek}
H.~Lacker, A.~Menzel, F.~Spettel, D.~Hirschbuhl, J.~Luck {\em et~al.}, {\it
  {Model-independent extraction of $|V_{tq}|$ matrix elements from top-quark
  measurements at hadron colliders}}  (2012),
  \href{http://arxiv.org/abs/1202.4694}{{\ttfamily arXiv:1202.4694 [hep-ph]}}.

\bibitem{Hill:1993hs}
C.~T. Hill and S.~J. Parke, {\em Phys. Rev.} {\bf D49}, 4454  (1994),
  \href{http://arxiv.org/abs/hep-ph/9312324}{{\ttfamily arXiv:hep-ph/9312324
  [hep-ph]}}.

\bibitem{Hill:1994hp}
C.~T. Hill, {\em Phys. Lett.} {\bf B345}, 483  (1995),
  \href{http://arxiv.org/abs/hep-ph/9411426}{{\ttfamily arXiv:hep-ph/9411426
  [hep-ph]}}.

\bibitem{Harris:1999ya}
R.~M. Harris, C.~T. Hill and S.~J. Parke, {\it {C}ross-section for topcolor
  ${Z}'$ decaying to top-antitop}  (1999),
  \href{http://arxiv.org/abs/hep-ph/9911288}{{\ttfamily arXiv:hep-ph/9911288
  [hep-ph]}}.

\bibitem{Harris:2011ez}
R.~M. Harris and S.~Jain, {\it {Cross Sections for Leptophobic Topcolor Z'
  decaying to top-antitop}}  (2011),
  \href{http://arxiv.org/abs/1112.4928}{{\ttfamily arXiv:1112.4928 [hep-ph]}}.

\bibitem{Randall:1999ee}
L.~Randall and R.~Sundrum, {\em Phys. Rev. Lett.} {\bf 83}, 3370  (1999),
  \href{http://arxiv.org/abs/hep-ph/9905221}{{\ttfamily arXiv:hep-ph/9905221
  [hep-ph]}}.

\bibitem{Agashe:2006hk}
K.~Agashe, A.~Belyaev, T.~Krupovnickas, G.~Perez and J.~Virzi, {\em Phys. Rev.}
  {\bf D77},   015003  (2008),
  \href{http://arxiv.org/abs/hep-ph/0612015}{{\ttfamily arXiv:hep-ph/0612015
  [hep-ph]}}.

\bibitem{Lillie:2007yh}
B.~Lillie, L.~Randall and L.-T. Wang, {\em JHEP} {\bf 0709},   074  (2007),
  \href{http://arxiv.org/abs/hep-ph/0701166}{{\ttfamily arXiv:hep-ph/0701166
  [hep-ph]}}.

\bibitem{Bai:2011ed}
Y.~Bai, J.~L. Hewett, J.~Kaplan and T.~G. Rizzo, {\em JHEP} {\bf 1103},   003
  (2011), \href{http://arxiv.org/abs/1101.5203}{{\ttfamily arXiv:1101.5203
  [hep-ph]}}.

\bibitem{Gresham:2011pa}
M.~I. Gresham, I.-W. Kim and K.~M. Zurek, {\em Phys. Rev.} {\bf D83},   114027
  (2011), \href{http://arxiv.org/abs/1103.3501}{{\ttfamily arXiv:1103.3501
  [hep-ph]}}.

\bibitem{Djouadi:2011aj}
A.~Djouadi, G.~Moreau and F.~Richard, {\em Phys. Lett.} {\bf B701}, 458
  (2011), \href{http://arxiv.org/abs/1105.3158}{{\ttfamily arXiv:1105.3158
  [hep-ph]}}.

\bibitem{Delaunay:2011gv}
C.~Delaunay, O.~Gedalia, Y.~Hochberg, G.~Perez and Y.~Soreq, {\em JHEP} {\bf
  1108},   031  (2011), \href{http://arxiv.org/abs/1103.2297}{{\ttfamily
  arXiv:1103.2297 [hep-ph]}}.

\bibitem{AguilarSaavedra:2011vw}
J.~Aguilar-Saavedra and M.~Perez-Victoria, {\em JHEP} {\bf 1105},   034
  (2011), \href{http://arxiv.org/abs/1103.2765}{{\ttfamily arXiv:1103.2765
  [hep-ph]}}.

\bibitem{Abazov:2011gv}
D0 Collaboration, V.~M. Abazov {\em et~al.}, {\em Phys. Rev.} {\bf D85},
  051101  (2012), \href{http://arxiv.org/abs/1111.1271}{{\ttfamily
  arXiv:1111.1271 [hep-ex]}}.

\bibitem{Aaltonen:2011ts}
CDF Collaboration, T.~Aaltonen {\em et~al.}, {\em Phys. Rev.} {\bf D84},
  072004  (2011), \href{http://arxiv.org/abs/1107.5063}{{\ttfamily
  arXiv:1107.5063 [hep-ex]}}.

\bibitem{cmspas-top-09-009}
CMS Collaboration, {\it {S}tudy of the top-pair invariant mass distribution in
  the semileptonic muon channel at 10 {T}e{V}}, CMS-PAS-TOP-09-009  (2009).

\bibitem{cmspas-exo-09-008}
CMS Collaboration, {\it {S}earch for heavy narrow $t\bar{t}$ resonances in
  muon-plus-jets final states with the {CMS} detector}, CMS-PAS-EXO-09-008
  (2009).

\bibitem{cmspas-exo-09-002}
CMS Collaboration, {\it {S}earch for {H}igh-{M}ass {R}esonances {D}ecaying into
  {T}op-{A}ntitop {P}airs in the {A}ll-{H}adronic {M}ode}, CMS-PAS-EXO-09-002
  (2009).

\bibitem{Cogneras2006}
ATLAS Collaboration, E.~Cogneras and D.~Pallin, {\it {G}eneric $t\bar{t}$
  resonance search with the {ATLAS} detector}, ATL-PHYS-PUB-2006-033  (2006).

\bibitem{atl-phys-pub-2009-081}
ATLAS Collaboration, {\it {R}econstruction of {H}igh {M}ass $t\bar{t}$
  {R}esonances in the {L}epton+{J}ets {C}hannel}, ATL-PHYS-PUB-2009-081
  (2009).

\bibitem{ATL-PHYS-PUB-2010-008}
ATLAS Collaboration, {\it {P}rospects for top anti-top resonance searches using
  early {ATLAS} data.}, ATL-PHYS-PUB-2010-008  (2010).

\bibitem{CMS-PAS-TOP-10-007}
CMS Collaboration, {\it {S}earch for {R}esonances in {S}emi-leptonic {T}op-pair
  {D}ecays {C}lose to {P}roduction {T}hreshold}, CMS-PAS-TOP-10-007  (2011).

\bibitem{cmspas-top-11-009}
CMS Collaboration, {\it {A} {S}earch for {R}esonances in {S}emileptonic {T}op
  {P}air {P}roduction at $\sqrt{s}=$ 7 {T}e{V}}, CMS-PAS-TOP-11-009  (2012).

\bibitem{Read:2002hq}
A.~L. Read, {\em J. Phys.} {\bf G28}, 2693  (2002).

\bibitem{Junk:1999kv}
T.~Junk, {\em Nucl. Instrum. Meth.} {\bf A434}, 435  (1999),
  \href{http://arxiv.org/abs/hep-ex/9902006}{{\ttfamily arXiv:hep-ex/9902006
  [hep-ex]}}.

\bibitem{cmspas-top-11-010}
CMS Collaboration, {\it {S}earch for ${Z}'$ resonances decaying to top-antitop
  pairs in fully leptonic final state in $pp$ collisions at $\sqrt{s} = 7$
  {T}e{V}}, CMS-PAS-TOP-11-010  (2012).

\bibitem{ATLAS-CONF-2012-029}
ATLAS Collaboration, {\it {A} {S}earch for $t\bar{t}$ {R}esonances in the
  {L}epton {P}lus {J}ets {C}hannel using 2.05 $fb^{-1}$}, ATLAS-CONF-2012-029
  (2012).

\bibitem{ATLAS-CONF-2011-070}
ATLAS Collaboration, {\it {A} {S}earch for {N}ew {H}igh-{M}ass {P}henomena
  {P}roducing {T}op {Q}uarks with the {ATLAS} {E}xperiment},
  ATLAS-CONF-2011-070  (2011).

\bibitem{ATLAS-CONF-2011-087}
ATLAS Collaboration, {\it {A} {S}earch for $t\bar{t}$ {R}esonances in the
  {L}epton {P}lus {J}ets {C}hannel in 200 pb$^{-1}$ of $pp$ {C}ollisions at
  $\sqrt{s} = 7$ {T}e{V}}, ATLAS-CONF-2011-087  (2011).

\bibitem{ATLAS-CONF-2011-123}
ATLAS Collaboration, {\it {A} {S}earch for $t\bar{t}$ {R}esonances in the
  {D}ilepton {C}hannel in 1.04/fb of pp collisions at $\sqrt{s}=7 \rm\ {T}e{V}$
  with the {ATLAS} experiment}, ATLAS-CONF-2011-123  (2011).

\bibitem{cmspas-exo-11-092}
CMS Collaboration, {\it {S}earch for high-mass resonances decaying to
  $t\bar{t}$ in the electron+jets channel}, CMS-PAS-EXO-11-092  (2012).

\bibitem{CMS-PAS-EXO-11-055}
CMS Collaboration, {\it {S}earch for heavy narrow resonances decaying to
  $t\bar{t}$ in the muon+jets channel}, CMS-PAS-EXO-11-055  (2011).

\bibitem{exo-11-006-paper}
CMS Collaboration, S.~Chatrchyan {\em et~al.}, {\it {S}earch for anomalous
  $t\bar{t}$ production in the highly-boosted all-hadronic final state}
  (2012), \href{http://arxiv.org/abs/1204.2488}{{\ttfamily arXiv:1204.2488
  [hep-ex]}}.

\bibitem{Wobisch:1998wt}
M.~Wobisch and T.~Wengler, {\it {H}adronization corrections to jet
  cross-sections in deep inelastic scattering}  (1998),
  \href{http://arxiv.org/abs/hep-ph/9907280}{{\ttfamily arXiv:hep-ph/9907280
  [hep-ph]}}.

\bibitem{Dokshitzer:1997in}
Y.~L. Dokshitzer, G.~Leder, S.~Moretti and B.~Webber, {\em JHEP} {\bf 9708},
  001  (1997), \href{http://arxiv.org/abs/hep-ph/9707323}{{\ttfamily
  arXiv:hep-ph/9707323 [hep-ph]}}.

\bibitem{jme-09-001}
CMS Collaboration, {\it {A} {C}ambridge-{A}achen ({C}-{A}) based {J}et
  {A}lgorithm for boosted top-jet tagging}, CMS-PAS-JME-09-001  (2009).

\bibitem{jme-10-013}
CMS Collaboration, {\it {S}tudy of {J}et {S}ubstructure in $pp$ {C}ollisions at
  7 {T}e{V} in {CMS}}, CMS-PAS-JME-10-013  (2011).

\bibitem{Ellis:2009su}
S.~D. Ellis, C.~K. Vermilion and J.~R. Walsh, {\em Phys. Rev.} {\bf D80},
  051501  (2009), \href{http://arxiv.org/abs/0903.5081}{{\ttfamily
  arXiv:0903.5081 [hep-ph]}}.

\bibitem{Ellis:2009me}
S.~D. Ellis, C.~K. Vermilion and J.~R. Walsh, {\em Phys. Rev.} {\bf D81},
  094023  (2010), \href{http://arxiv.org/abs/0912.0033}{{\ttfamily
  arXiv:0912.0033 [hep-ph]}}.

\bibitem{Abdesselam:2010pt}
A.~Abdesselam, E.~B. Kuutmann, U.~Bitenc, G.~Brooijmans, J.~Butterworth {\em
  et~al.}, {\em Eur. Phys. J.} {\bf C71},   1661  (2011),
  \href{http://arxiv.org/abs/1012.5412}{{\ttfamily arXiv:1012.5412 [hep-ph]}}.

\bibitem{Altheimer:2012mn}
J.~L. Evans, M.~Ibe, S.~Shirai and T.~T. Yanagida, {\it {A} 125 {G}e{V} {H}iggs
  {B}oson and {M}uon g-2 in {M}ore {G}eneric {G}auge {M}ediation}  (2012),
  \href{http://arxiv.org/abs/1201.2611}{{\ttfamily arXiv:1201.2611 [hep-ph]}}.

\bibitem{AguilarSaavedra:2004wm}
J.~Aguilar-Saavedra, {\em Acta Phys. Polon.} {\bf B35}, 2695  (2004),
  \href{http://arxiv.org/abs/hep-ph/0409342}{{\ttfamily arXiv:hep-ph/0409342
  [hep-ph]}}.

\bibitem{Carvalho2005}
ATLAS Collaboration, J.~Carvalho, N.~Castro, A.~Onofre and F.~Veloso, {\it
  {S}tudy of the {ATLAS} sensitivity to {FCNC} decays in single top events},
  ATL-PHYS-PUB-2005-026  (2005).

\bibitem{Carvalho2005a}
ATLAS Collaboration, J.~Carvalho, N.~Castro, A.~Onofre and F.~Veloso, {\it
  {S}tudy of {ATLAS} sensitivity to {FCNC} top decays}, ATL-PHYS-PUB-2005-009
  (2005).

\bibitem{Carvalho2007}
ATLAS Collaboration, J.~Carvalho {\em et~al.}, {\em Eur. Phys. J.} {\bf C52},
  999  (2007), \href{http://arxiv.org/abs/0712.1127}{{\ttfamily arXiv:0712.1127
  [hep-ex]}}.

\bibitem{Cheng2006}
ATLAS Collaboration, T.~L. Cheng and P.~Teixeira-Dias, {\it {S}ensitivity of
  {ATLAS} to {FCNC} single top quark production}, ATL-PHYS-PUB-2006-029
  (2006).

\bibitem{cmsnote-2006-093}
K.~Karafasoulis {\em et~al.}, {\it {S}tudy of {F}lavour {C}hanging {N}eutral
  {C}urrents in {T}op {Q}uark {D}ecays with the {CMS} {D}etector},
  CMS-NOTE-2006/093  (2006).

\bibitem{ATLAS-CONF-2011-061}
ATLAS Collaboration, {\it {S}earch for {FCNC} top quark processes at 7 {T}e{V}
  with the {ATLAS} detector}, ATLAS-CONF-2011-061  (2011).

\bibitem{ATLAS-CONF-2011-154}
ATLAS Collaboration, {\it {A} search for {F}lavour {C}hanging {N}eutral
  {C}urrents in {T}op {Q}uark {D}ecays $t\rightarrow q {Z}$ at $\sqrt{s}=7$
  {T}e{V} in $0.70 \rm\ fb^{-1}$ of pp collision data collected with the
  {ATLAS} {D}etector}, ATLAS-CONF-2011-154  (2011).

\bibitem{Slabospitsky:2002ag}
S.~Slabospitsky and L.~Sonnenschein, {\em Comput. Phys. Commun.} {\bf 148}, 87
  (2002), \href{http://arxiv.org/abs/hep-ph/0201292}{{\ttfamily
  arXiv:hep-ph/0201292 [hep-ph]}}.

\bibitem{cmspas-top-11-028}
CMS Collaboration, {\it {S}earch for {F}lavor {C}hanging {N}eutral {C}urrents
  in {T}op {Q}uark {D}ecays in pp {C}ollisions at $\sqrt{s}=$ 7 {T}e{V}},
  CMS-PAS-TOP-11-028  (2012).

\bibitem{Collaboration:2012gd}
ATLAS Collaboration, G.~Aad {\em et~al.}, {\em Phys. Lett.} {\bf B712}, 351
  (2012), \href{http://arxiv.org/abs/1203.0529}{{\ttfamily arXiv:1203.0529
  [hep-ex]}}.

\bibitem{AguilarSaavedra:2010rx}
J.~Aguilar-Saavedra, {\em Nucl. Phys.} {\bf B837}, 122  (2010),
  \href{http://arxiv.org/abs/1003.3173}{{\ttfamily arXiv:1003.3173 [hep-ph]}}.

\bibitem{Aad:2011wc}
ATLAS Collaboration, G.~Aad {\em et~al.}, {\em Phys. Rev. Lett.} {\bf 108},
  041805  (2012), \href{http://arxiv.org/abs/1109.4725}{{\ttfamily
  arXiv:1109.4725 [hep-ex]}}.

\bibitem{Alwall:2010jc}
J.~Alwall, J.~L. Feng, J.~Kumar and S.~Su, {\em Phys. Rev.} {\bf D81},   114027
   (2010), \href{http://arxiv.org/abs/1002.3366}{{\ttfamily arXiv:1002.3366
  [hep-ph]}}.

\bibitem{Collaboration:2012sa}
CMS Collaboration, S.~Chatrchyan {\em et~al.}  (2012),
  \href{http://arxiv.org/abs/1205.3933}{{\ttfamily arXiv:1205.3933 [hep-ex]}},
  subm. to JHEP.

\bibitem{Berger:2011ua}
E.~L. Berger, Q.-H. Cao, C.-R. Chen, C.~S. Li and H.~Zhang, {\em Phys. Rev.
  Lett.} {\bf 106},   201801  (2011),
  \href{http://arxiv.org/abs/1101.5625}{{\ttfamily arXiv:1101.5625 [hep-ph]}}.

\bibitem{Jung:2009jz}
S.~Jung, H.~Murayama, A.~Pierce and J.~D. Wells, {\em Phys. Rev.} {\bf D81},
  015004  (2010), \href{http://arxiv.org/abs/0907.4112}{{\ttfamily
  arXiv:0907.4112 [hep-ph]}}.

\bibitem{Chatrchyan:2011dk}
CMS Collaboration, S.~Chatrchyan {\em et~al.}, {\em JHEP} {\bf 08},   005
  (2011), \href{http://arxiv.org/abs/1106.2142}{{\ttfamily arXiv:1106.2142
  [hep-ex]}}.

\bibitem{BarShalom:2007pw}
S.~Bar-Shalom and A.~Rajaraman, {\em Phys. Rev.} {\bf D77},   095011  (2008),
  \href{http://arxiv.org/abs/0711.3193}{{\ttfamily arXiv:0711.3193 [hep-ph]}}.

\bibitem{Collaboration:2012cg}
ATLAS Collaboration, G.~Aad {\em et~al.}, {\em Phys. Rev.} {\bf D88},   032004
  (2012), \href{http://arxiv.org/abs/1201.1091}{{\ttfamily arXiv:1201.1091
  [hep-ex]}}.

\bibitem{:2012bb}
ATLAS Collaboration, G.~Aad {\em et~al.}, {\em JHEP} {\bf 1204},   069  (2012),
  \href{http://arxiv.org/abs/1202.5520}{{\ttfamily arXiv:1202.5520 [hep-ex]}}.

\bibitem{exo-11-050}
CMS Collaboration, S.~Chatrchyan {\em et~al.}, {\it {Search for heavy, top-like
  quark pair production in the dilepton final state in $pp$ collisions at
  $\sqrt{s} = 7$ TeV}}  (2012),
  \href{http://arxiv.org/abs/1203.5410}{{\ttfamily arXiv:1203.5410 [hep-ex]}}.

\bibitem{Hou:2008xd}
W.-S. Hou, {\em Chin. J. Phys.} {\bf 47},   134  (2009),
  \href{http://arxiv.org/abs/0803.1234}{{\ttfamily arXiv:0803.1234 [hep-ph]}}.

\bibitem{cmspas-exo-11-099}
CMS Collaboration, {\it {S}earch for the pair production of a fourth-generation
  up-type $t'$ quark in events with a lepton and at least four jets},
  CMS-PAS-EXO-11-099  (2012).

\bibitem{CMS-PAS-EXO-11-054}
CMS Collaboration, {\it {I}nclusive search for a fourth generation of quarks
  with the {CMS} experiment}, CMS-PAS-EXO-11-054  (2011).

\bibitem{ATLAS-CONF-2011-022}
ATLAS Collaboration, {\it {S}earch for {F}ourth {G}eneration {Q}uarks
  {D}ecaying to ${W}^+q {W}^-q \rightarrow l^+ \nu q l^- \nu \bar{q}$ in pp
  collisions at $\sqrt{s}=$ 7 {T}e{V} with the {ATLAS} {D}etector},
  ATLAS-CONF-2011-022  (2011).

\bibitem{Aad:2012bt}
ATLAS Collaboration, G.~Aad {\em et~al.}, {\it {S}earch for pair-produced heavy
  quarks decaying to {W}q in the two-lepton channel at $\sqrt{s}=$ 7 {T}e{V}
  with the {ATLAS} detector}  (2012),
  \href{http://arxiv.org/abs/1202.3389}{{\ttfamily arXiv:1202.3389 [hep-ex]}}.

\bibitem{Aad:2012xc}
ATLAS Collaboration, G.~Aad {\em et~al.}, {\it {S}earch for pair production of
  a heavy quark decaying to a {W} boson and a b quark in the lepton+jets
  channel with the {ATLAS} detector}  (2012),
  \href{http://arxiv.org/abs/1202.3076}{{\ttfamily arXiv:1202.3076 [hep-ex]}}.

\bibitem{Chatrchyan:2011ay}
CMS Collaboration, S.~Chatrchyan {\em et~al.}, {\em Phys. Rev. Lett.} {\bf
  107},   271802  (2011), \href{http://arxiv.org/abs/1109.4985}{{\ttfamily
  arXiv:1109.4985 [hep-ex]}}.

\bibitem{Aad:2011yn}
ATLAS Collaboration, G.~Aad {\em et~al.}, {\em Phys. Lett.} {\bf B712}, 22
  (2012), \href{http://arxiv.org/abs/1112.5755}{{\ttfamily arXiv:1112.5755
  [hep-ex]}}.

\bibitem{Chatrchyan:2011em}
CMS Collaboration, S.~Chatrchyan {\em et~al.}, {\em Phys. Lett.} {\bf B701},
  204  (2011), \href{http://arxiv.org/abs/1102.4746}{{\ttfamily arXiv:1102.4746
  [hep-ex]}}.

\bibitem{Collaboration:2012ye}
CMS Collaboration, S.~Chatrchyan {\em et~al.}, {\it {Search for heavy
  bottom-like quarks in 4.9 inverse femtobarns of pp collisions at $\sqrt{s}=$
  7 TeV}}  (2012), \href{http://arxiv.org/abs/1204.1088}{{\ttfamily
  arXiv:1204.1088 [hep-ex]}}.

\bibitem{Aad:2012us}
ATLAS Collaboration, G.~Aad {\em et~al.}, {\it {Search for down-type fourth
  generation quarks with the ATLAS detector in events with one lepton and high
  transverse momentum hadronically decaying W bosons in $\sqrt{s}=$ 7 TeV pp
  collisions}}  (2012), \href{http://arxiv.org/abs/1202.6540}{{\ttfamily
  arXiv:1202.6540 [hep-ex]}}.

\bibitem{arxiv:1204.1265}
ATLAS Collaboration, G.~Aad {\em et~al.}, {\it {Search for pair production of a
  new quark that decays to a Z boson and a bottom quark with the ATLAS
  detector}}  (2012), \href{http://arxiv.org/abs/1204.1265}{{\ttfamily
  arXiv:1204.1265 [hep-ex]}}.

\bibitem{Aaltonen:2011tq}
CDF Collaboration, T.~Aaltonen {\em et~al.}, {\em Phys. Rev. Lett.} {\bf 107},
   261801  (2011), \href{http://arxiv.org/abs/1107.3875}{{\ttfamily
  arXiv:1107.3875 [hep-ex]}}.

\bibitem{Abazov:2011vy}
D0 Collaboration, V.~M. Abazov {\em et~al.}, {\it {Search for a fourth
  generation $t'$ quark in $p\bar{p}$ collisions at $\sqrt{s}=1.96$ TeV}}
  (2011), \href{http://arxiv.org/abs/1104.4522}{{\ttfamily arXiv:1104.4522
  [hep-ex]}}.

\bibitem{Aaltonen:2011vr}
CDF Collaboration Collaboration, T.~Aaltonen {\em et~al.}, {\em Phys. Rev.
  Lett.} {\bf 106},   141803  (2011),
  \href{http://arxiv.org/abs/1101.5728}{{\ttfamily arXiv:1101.5728 [hep-ex]}}.

\bibitem{Aad:2012tj}
ATLAS Collaboration, G.~Aad {\em et~al.}, {\it {Search for charged Higgs bosons
  decaying via H+ -\&amp;gt; tau nu in top quark pair events using pp collision
  data at sqrt(s) = 7 TeV with the ATLAS detector}}  (2012),
  \href{http://arxiv.org/abs/1204.2760}{{\ttfamily arXiv:1204.2760 [hep-ex]}}.

\bibitem{arxiv:1205.5736}
CMS Collaboration, S.~Chatrchyan {\em et~al.}, {\it {Search for a light charged
  Higgs boson in top quark decays in pp collisions at sqrt(s) = 7 TeV}}
  (2012), \href{http://arxiv.org/abs/1205.5736}{{\ttfamily arXiv:1205.5736
  [hep-ex]}}.

\bibitem{Abulencia:2005jd}
CDF Collaboration, A.~Abulencia {\em et~al.}, {\em Phys. Rev. Lett.} {\bf 96},
   042003  (2006), \href{http://arxiv.org/abs/hep-ex/0510065}{{\ttfamily
  arXiv:hep-ex/0510065 [hep-ex]}}.

\bibitem{Abazov:2009aa}
D0 Collaboration, V.~Abazov {\em et~al.}, {\em Phys. Lett.} {\bf B682}, 278
  (2009), \href{http://arxiv.org/abs/0908.1811}{{\ttfamily arXiv:0908.1811
  [hep-ex]}}.

\bibitem{Aad:2012ej}
ATLAS Collaboration, G.~Aad {\em et~al.}, {\it {Search for tb resonances in
  proton-proton collisions at sqrt(s) = 7 TeV with the ATLAS detector}}
  (2012), \href{http://arxiv.org/abs/1205.1016}{{\ttfamily arXiv:1205.1016
  [hep-ex]}}.

\bibitem{cmspas-exo-12-001}
CMS Collaboration, {\it {S}earch for {H}eavy ${W}'$ {B}oson resonances decaying
  to a {B}ottom {Q}uark and a {T}op {Q}uark at $\sqrt{s}=$ 7 {T}e{V}},
  CMS-PAS-EXO-12-001  (2012).

\bibitem{Aaltonen:2009qu}
CDF Collaboration, T.~Aaltonen {\em et~al.}, {\em Phys. Rev. Lett.} {\bf 103},
   041801  (2009), \href{http://arxiv.org/abs/0902.3276}{{\ttfamily
  arXiv:0902.3276 [hep-ex]}}.

\bibitem{Abazov:2011xs}
D0 Collaboration, V.~M. Abazov {\em et~al.}, {\em Phys. Lett.} {\bf B699}, 145
  (2011), \href{http://arxiv.org/abs/1101.0806}{{\ttfamily arXiv:1101.0806
  [hep-ex]}}.

\end{thebibliography}

%%%%%%%%%%%%%%%%%%%%%%%%%%%%%%%%%%%%%%%%%%%%%%%%%%%%%%%%%%%%%%%%%%%%%%%%%%%%%%%

\end{document}